\documentclass[12pt]{article}

\usepackage{amsmath}         % \
\usepackage{amssymb}         %  | AMS Packages for math
\usepackage{amsfonts}        % /
\usepackage{pifont}          % from Brando
\usepackage{graphicx}        % Graphics
\usepackage{bbm}             % for \mathbbm{1} (unit matrix)

\usepackage{color}         % for \color{...}, colored text
\usepackage{slashed}       % for \slashed{k}
\usepackage{framed}        % for remarks
\usepackage{subcaption}    % for subfigs
\usepackage{mathrsfs}      % for Weinberg-esque letters
\usepackage{ragged2e}      % for ragged right (for institutions)
\usepackage{lipsum}        % for block of text (formatting test)
\usepackage{paralist}      % for compactitem
\usepackage{multirow}      % for multiple row elements in a table
\usepackage{cite}          % cancels collref, but groups citations
\usepackage{flip-thesis-acronyms}
\usepackage{booktabs}      %  for tables
\usepackage{nicefrac}      %  for fractions in tables,
\usepackage{youngtab}		% Young Tableaux
\usepackage{graphicx}

\usepackage{arydshln} 		% For dashed lines in arrays and tables
\usepackage{appendix}       % for subappendices

\usepackage{bibspacing}
\usepackage[margin=2cm]{geometry} % for reasonable margins

\usepackage{scalefnt}			    % for \smaller
\graphicspath{{figures/}}	    % separate folder for figures

\renewcommand{\tilde}{\widetilde} 
\newcommand{\dbar}{d\mkern-6mu\mathchar'26}

\numberwithin{equation}{section}

\newenvironment{institutions}[1][2em]
  {\begin{list}{}{\setlength\leftmargin{#1}\setlength\rightmargin{#1}}\item[]\RaggedRight}
  {\end{list}}

\usepackage{fancyhdr}		% to put preprint number

\fancypagestyle{firststyle}
{
	\rhead{\footnotesize \texttt{UCI-TR-2016-01}}
   \fancyfoot{}
}

\usepackage{tikzfeynman}

\usepackage[
	colorlinks=true,
	citecolor=black,
	linkcolor=black,
	urlcolor=blue,
	hypertexnames=false]{hyperref}

\begin{document}

\thispagestyle{firststyle}

\begin{center}

	{	% TITLE GOES HERE 	
		\huge \bf 
%		ESHEP 2013 Lectures on\\
%		\vspace{.1em}
		Beyond the Standard Model
		\\
		\vspace{.1em}
		\large
		Lectures at the 2013 European School of High Energy Physics 
	}

	\vskip .7cm
	
	\renewcommand*{\thefootnote}{\fnsymbol{footnote}}
	% Sets footnotes to a sequence of symbols, for e-mails

	{	% AUTHORS GO HERE
		% I kind of hacked this together for efficient output.
		% First e-mail address is a footnote which includes
		% the e-mail addresses of all authors.
		\bf
		Csaba Cs\'aki$^{a}$\footnote{\tt
		 \href{mailto: csaki@cornell.edu}
		 {csaki@cornell.edu},
		 \;
		 $^\dag$\href{mailto:flip.tanedo@uci.edu}{flip.tanedo@uci.edu}
		 }
		and
		Philip Tanedo$^{b\dag}$
%		\footnote{\tt 	
%		 },
		%
	}  

	\renewcommand{\thefootnote}{\arabic{footnote}}
	\setcounter{footnote}{0}
	% Returns footnote to numbers and restarts counter

\begin{institutions}[2.25cm]
\footnotesize
    
	$^{a}$ {\it Department of Physics, \textsc{lepp}, Cornell University, Ithaca, \textsc{ny} 14853}
	\\
   
	\vspace*{0.05cm}

	$^{b}$ {\it Department of Physics \& Astronomy, University of California, Irvine, \textsc{ca} 92697} 

\end{institutions}

\end{center}

%%%%%%%%%%%%%%%%%%%%%%%%%%%%%%%%%%%%%%%%%%%%%%%%%%%%%%%%%%%%

\begin{abstract}
\noindent 
We introduce aspects of physics beyond the Standard Model focusing on supersymmetry, extra dimensions, and a composite Higgs as solutions to the Hierarchy problem.
Lectures at the European School of High Energy Physics, Par\'adf\"urd\H{o}, Hungary, 5 -- 18 June 2013.
\end{abstract}

\noindent{%
Appearing in the
\emph{Proceedings of the 2013 European School of High-Energy Physics, Par\'{a}df\"{u}rd\H{o}, Hungary, 5--18 June 2013}, edited by M.~Mulders and G.~Perez.
}

%%%%%%%%%%%%%%%%%%%%%%%%%%%%%%%%%%%%%%%%%%%%%%%%%%%%%%%%%%%%

\small
\setcounter{tocdepth}{2}
\tableofcontents
\normalsize

\vspace{3em}

This document is based on lectures by \textsc{c.c.} on physics beyond the Standard Model at the 2013 European School of High-Energy Physics. We present a pedagogical introduction to supersymmetry, extra dimensions, and composite Higgs. 
We provide references to useful review literature and refer to those for more complete citations to original papers on these topics. We apologize for any omissions in our citations or choice of topics. 

\section{The Hierarchy Problem}

At loop level, the Higgs mass receives corrections from self interactions, gauge loops, and fermion loops (especially the top quark). Diagrammatically,
%
%\begin{center}
\begin{align}
\begin{tikzpicture}[line width=1.5 pt, scale=1.5]
	\draw[scalarnoarrow] (-.8,0) -- (-.3,0);
	\draw[scalarnoarrow] (.3,0) -- (.8,0);
	\begin{scope}
    	\clip (0,0) circle (.3cm);
    	\foreach \x in {-.9,-.8,...,.3}
			\draw[line width=.5 pt] (\x,-.3) -- (\x+.6,.3);
  	\end{scope}
  	\draw[fermionnoarrow] (0,0) circle (.3);
  	\begin{scope}[shift={(2,0)}]
  	\node at (-.8,0) {\Large{=}};
    \draw[scalarnoarrow] (-.2,-.5) -- (1.3,-.5);
    \draw[scalarnoarrow] (.55,-.1) circle (.4);
  	\end{scope}
  	\begin{scope}[shift={(4.5,0)}]
  	\node at (-.8,0) {\Large{+}};
    \draw[scalarnoarrow] (-.2,-.5) -- (1.3,-.5);
    \draw[vector] (.55,-.1) circle (.4);
  	\end{scope}
  	\begin{scope}[shift={(7.8,0)}]
  	\node at (-1.4,0) {\Large{+}};
%    \draw[scalarnoarrow] (-.2,-.5) -- (1.3,-.5);
%    \draw[vector] (.55,-.1) circle (.4);
    \draw[fermion] (180:.4) arc (180:0:.4);
	\draw[fermion] (0:.4) arc (0:-180:.4);
	\draw[scalarnoarrow] (-.4,0) -- (-.9,0);
	\draw[scalarnoarrow] (.4,0) -- (.9,0);
  	\end{scope}
 \end{tikzpicture}
% \end{center}
\nonumber
\end{align}
These loops are quadratically divergent and go like $\int  d^4k\; (k^2-m^2)^{-1} \sim \Lambda^2$ for some cutoff scale $\Lambda$.  Explicitly, 
\begin{align}
\delta m_H^2 &= \frac{\Lambda^2}{32\pi^2}
\left[
    6\lambda 
    + \frac 14 \left(9g^2 + 3g'^2\right)
    - y_t^2
\right]
\end{align}
% Underbrace: give the scales.
%
If $\Lambda \gg 10 \text{ \TeV}$ (for example, $\Lambda \sim M_\text{Pl}$), then the quantum correction to the Higgs mass is much larger than the mass itself, $\delta m_H^2 \gg m_H^2$. This is the \textbf{Hierarchy problem}: the Higgs mass is quadratically sensitive to \textit{any} mass scale of new physics. This problem is specific to elementary scalars. 

Unlike scalars, the quantum corrections to fermion and gauge boson masses are proportional to the particle masses themselves. In this way, small fermion and gauge boson masses are technically natural: the loop corrections are suppressed by the smallness of the tree-level parameter. 
For fermions this is because of the appearance of a new chiral symmetry in the massless limit. For gauge bosons this is because gauge symmetry is restored in the massless limit. By dimensional analysis, the corrections to these mass parameters cannot be quadratically sensitive to the cutoff, $\Lambda$,
\begin{align}
\Delta m_e &\sim m_e \ln\left(\frac{\Lambda}{m_e}\right)\\
\Delta M_W^2 &\sim M_W^2 \ln\left(\frac{\Lambda}{M_W}\right).
\end{align}

The Hierarchy problem is independent of the renormalization scheme. It is sometimes argued that in dimensional regularization there are no quadratic divergences since the $1/\epsilon$ poles correspond to logarithmic divergences. 
This is fallacious. The Hierarchy problem isn't about the cancellation of divergences, it is about the separation of the electroweak and \UV scales.
\textit{Any} new physics coupled to the Higgs will reintroduce the quadratic dependence on the scale at which the new physics appears. 
For example, suppose new physics enters at the scale $m_S$ by a four-point interaction between the Higgs and an additional complex scalar, $\Delta \mathcal L \supset \lambda_S |H|^2 |S|^2$. The contribution to the Higgs mass from a loop of the $S$ particle is
\begin{align}
\delta m_H^2 &= \frac{\lambda_S}{16\pi^2}
\left[
\Lambda_\text{UV}^2 - 2m_S^2 \ln\left( \frac{\Lambda_{\text{UV}}}{m_S} \right) +  \left(\text{finite}\right)
\right].
\label{eq:SM:sensitivity:to:scale}
\end{align}
Suppose one chose to ignore the term quadratic in the loop regulator, $\Lambda_\text{UV}^2$---note that there's no justification to do this---the logarithmically divergent piece (corresponding to the $1/\epsilon$) and the finite pieces are proportional to the squared mass scale of the new physics, $m_S^2$. 
The regulator $\Lambda_\text{UV}$ is not a physical scale, but $m_S^2$ is the scale of new physics. The Higgs mass is quadratically sensitive to this scale, no matter how one chooses to regulate the loop.

\begin{figure}%[h]
%\begin{align}
\begin{center}
\begin{tikzpicture}[line width=1.5 pt, scale=.5]
	\draw[scalarnoarrow] (-3,0) -- (3,0);
	\draw[provector] (-1.5,0) -- (-1,2);
	\draw[antivector] (1.5,0) -- (1,2);
	\draw[fermionnoarrow] (0,2) circle (1);
	\node at (0,2) {$\Psi$};
\end{tikzpicture}
\qquad\qquad\qquad
\begin{tikzpicture}[line width=1.5 pt, scale=.5]
	\draw[scalarnoarrow] (-3,0) -- (3,0);
	\draw[provector] (0,0) to [bend left=45] (-1,3);
	\draw[antivector] (0,0) to [bend right=45] (1,3);
	\draw[fermionnoarrow] (0,3) circle (1);
	\node at (0,3) {$\Psi$};
\end{tikzpicture}
\end{center}
%\end{align}
\caption{Heuristic two-loop contributions to the Higgs mass from heavy fermions, $\Psi$. Even though the $\Psi$ do not directly couple to the Higgs, they reintroduce a quadratic sensitivity  to the new scale.}
\label{fig:SUSY:twoloop}
\end{figure}
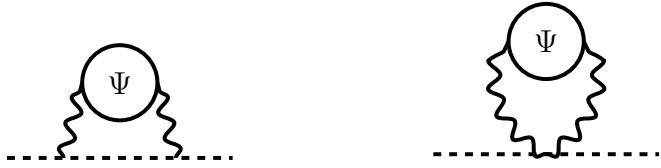

This quadratic sensitivity is true even if these new states are \textit{not} directly coupled to the Higgs but only interact with other Standard Model fields. For example, suppose there were a pair of heavy fermions $\Psi$ which are charged under the Standard Model gauge group but don't directly interact with the Higgs. One still expects two loop contributions to the Higgs mass from diagrams such as %:
those in Fig.~\ref{fig:SUSY:twoloop}.
These contributions are of the form
\begin{align}
\delta m_H^2 \sim \left(
\frac{g^2}{16\pi^2}
\right)^2
\left[
a \Lambda_\text{UV}^2 + 48 m_F^2 \ln \frac{\Lambda_\text{UV}}{m_F} + \left(\text{finite}\right)
\right].
\end{align}
This is indeed of the same form as (\ref{eq:SM:sensitivity:to:scale}). Note that in this case, the sensitivity to the new scale is softened by a loop factor.

% Wilsonian
The Higgs mass operator $|H|^2$ is a relevant and thus grows in the infrared.
From the Wilsonian perspective, the Hierarchy problem is the statement that is is difficult (finely tuned) to choose a renormalization group trajectory that flows to the correct Higgs mass.
In summary, the Hierarchy problem is the issue that the Higgs mass $m_H$ is sensitive to \textit{any} high scale in the theory, even if it only indirectly couples to the Standard Model. Thus na\"ively one would expect that $m_H$ should be on the order of the scale of new physics. In the Wilsonian picture, the Higgs mass is a relevant operator and so its importance grows towards the \textsc{ir}. Indeed, $m_H$ is the only relevant operator in the Standard Model. 

The implication of the Hierarchy problem is that there should to be new physics at the \TeV scale that eliminates the large loop contributions from above the \TeV scale\footnote{See \cite{deGouvea:2014xba} for a recent discussion of naturalness and fine-tuning in the post-Higgs era.}. In these lectures we explore some of options for the physics beyond the \SM that enforce naturalness. Before going into further detail, here is a brief overview of some of the possibilities for this to happen:
\begin{itemize}
\item \textbf{Supersymmetry}: relate the elementary scalar Higgs to fermions in such a way that the chiral symmetry protecting the fermion mass is extended to also protect the scalar mass.
\item \textbf{Gauge-Higgs unification}: relate the the elementary scalar Higgs to an elementary gauge field so that gauge symmetry also protects the Higgs mass.
\item \textbf{Technicolor}, \textbf{Higgsless}: there is no Higgs boson, just a dynamically generated condensate.
\item \textbf{Composite Higgs}, \textbf{warped extra dimensions}: There is a Higgs, but it is not elementary. At the \TeV scale the Higgs ``dissolves'': it becomes sensitive to large form factors that suppresses corrections.
\item \textbf{Pseudo-Goldstone Higgs}: The Higgs is a pseudo-Goldstone boson of a spontaneously broken symmetry. This gives some protection against quadratic divergences, usually removing the one-loop contribution. In practice one must still combine with additional mechanisms, such as \textbf{collective symmetry breaking}.
\item \textbf{Large extra dimensions}: The fundamental Planck scale is actually  $\sim$ \TeV and only appears much larger because gravity is diluted through its propagation in more directions.
\end{itemize}

\section{Supersymmetry}

Recall that under an infinitesimal transformation by an `ordinary' internal symmetry, a quantum field $\phi$ transforms as
\begin{align}
\varphi_i \to (\mathbbm{1}_{ij} + i\epsilon^a T^a_{ij}) \varphi_j,\label{eq:SUSY:symmetry}
\end{align}
where $\epsilon^a$ is an infinitesimal parameter, $T^a$ is the [bosonic] generator of the symmetry, and $i,j$ label the representation of $\phi$ with respect to this symmetry. These internal symmetries do not change the spin of $\phi$: bosons remain bosons and fermions remain fermions.
\textbf{Supersymmetry} (\SUSY) is a generalization of this `ordinary' symmetry where generator is now fermionic. Thus a \SUSY transformation changes fermions into bosons and vice versa. 

\begin{framed}
\noindent \footnotesize
\textbf{Further reading:} Wess and Bagger \cite{Wess:1992cp} is the canonical reference for the tools of supersymmetry. The text by Terning has a broad overview of \SUSY and its modern applications in particle physics. Additional reviews include \cite{Krippendorf:2010ui, Strassler:2003qg, Argyres:globalSUSY}. Key historical papers are collected in \cite{Shifman:2000rs} and a more personal account is presented in \cite{Kane:2000ew}. More formal topics in \SUSY that are beyond the scope of these lectures, but are key tools for model builders, can be found in \cite{Peskin:1997qi, Seiberg:1994pq, Intriligator:2007cp}.
\end{framed}

\subsection{The SUSY algebra}

The '60s were very successful for classifying hadrons based on Gell-Mann's SU(3) internal symmetry. Physicists then tried to enlarge this group to SU(6) so that it would include %SU(3)$_\text{Gell-Mann}\times$SU(3)$_\text{spin}$, 
\begin{align}
\text{SU(3)}_\text{Gell-Mann}\times\text{SU(2)}_\text{spin},
\end{align}
but they were unable to construct a viable relativistic model. Later this was understood to be a result of the Coleman-Mandula  `no go' theorem which states that one cannot construct a consistent quantum field theory based on a nontrivial combination of internal symmetries with space-time symmetry \cite{Coleman:1967ad}. The one exception came from Haag, Lopuszanski, and Sohnius: the only non-trivial combination of an internal and spacetime symmetry is to use a \textbf{graded Lie algebra} whose generators are fermionic \cite{Haag:1974qh}. Recall that fermionic objects obey anti-commutation relations rather than commutation relations. The main anti-commutation relation for \SUSY is:
\begin{align}
\left\{ Q^A_\alpha, \overline Q_{\dot{\alpha}_B} \right\}
=
2 P_\mu \sigma^\mu_{\alpha \dot{\beta}} \delta^A_B,
\label{eq:SUSY:SUSYalg}
\end{align}
where the $Q$ and $\overline Q$ are \SUSY generators (supercharges) and $P_\mu$ is the momentum operator. Here the $\alpha$ and $\dot{\alpha}$ are Lorentz indices while $A,B$ index the number of supercharges. For completeness, the rest of the algebra is
\begin{align}
	[M^{\mu\nu},M^{\rho\sigma}] &= i(M^{\mu\nu}\eta^{\nu\rho}+M^{\nu\rho}\eta^{\mu\sigma}-M^{\mu\rho}\eta^{\nu\sigma}-M^{\nu\sigma}\eta^{\mu\rho})\label{eq:SUSYalg:MM}
	\\
	[P^\mu,P^\nu] &= 0\label{eq:SUSYalg:PP}
	\\
	[M^{\mu\nu},P^\sigma]&=i(P^\mu\eta^{\nu\sigma}-P^\nu\eta^{\mu\sigma}\label{eq:SUSYalg:MP})
	\\
	[Q_\alpha^A,M^{\mu\nu}] &= (\sigma^{\mu\nu})_\alpha^{\phantom\alpha\beta}Q_\beta^A\label{eq:SUSYalg:QM}
	\\
	[Q_\alpha^A,P^\mu]&=0\label{eq:SUSYalg:QP}
	\\
	\{Q_\alpha^A,Q_\beta^B\} &= \epsilon_{\alpha\beta}Z^{AB}.\label{eq:SUSYalg:QQ}
%	\\
%	\{Q_\alpha,\overline Q_{\dot\beta}\} &= 2(\sigma^\mu)_{\alpha\dot\beta}P_\mu\label{eq:SUSYalg:QQbar}
\end{align}
The $Z^{AB}$ may appear for $\mathcal N>1$ and are known as \textbf{central charges}.
By the Coleman-Mandula theorem, we know that internal symmetry generators commute with the Poincar\'e generators. For example, the Standard Model gauge group commutes with the momentum, rotation, and boost operators. This carries over to the \SUSY algebra. For an internal symmetry generator $T_a$,
\begin{align}
	[T_a,Q_\alpha] &= 0.
\end{align}
This is true with one exception. The \SUSY generators come equipped with their own internal symmetry, called \textbf{$R$-symmetry}. For $\mathcal N=1$ there exists an automorphism of the supersymmetry algebra,
\begin{align}
	Q_\alpha &\rightarrow e^{it}Q_\alpha 
	%\label{eq:SUSYalg:RQ}
	&
	\overline Q_{\dot\alpha} &\rightarrow e^{-it}\overline Q_{\dot\alpha},\label{eq:SUSYalg:RQbar}
\end{align}
for some transformation parameter $t$.
This is a U$(1)$ internal symmetry. Applying this symmetry preserves the SUSY algebra. If $R$ is the generator of this U$(1)$, then its action on the \SUSY operators is given by
\begin{align}
	Q_\alpha &\rightarrow e^{-iRt}Q_\alpha e^{iRt}.\label{eq:SUSYalg:RQop}
\end{align}
By comparing the transformation of $Q$ under (\ref{eq:SUSYalg:RQop}), we find the corresponding algebra,
\begin{align}
	[Q_\alpha,R] &= \phantom{+}Q_\alpha
	%\\
	&
	[\overline Q_{\dot\alpha},R] &= -\overline Q_{\dot\alpha}.
\end{align}
Note that this means that different components of a \SUSY multiplet have different $R$ charge.
For $\mathcal N>1$ the $R$-symmetry group enlarges to U$(\mathcal N)$.

\subsection{Properties of supersymmetric theories}

Supersymmetric theories obey some key properties:

\begin{enumerate}
\item The number of fermionic degrees of freedom equals the number of bosonic degrees of freedom. To see this, first introduce an operator $(-)^{N_F}$ such that,
\begin{align}
(-)^{N_F} \left|q\right\rangle =&
\left\{
\begin{array}{ll}
+ \left|q\right\rangle & \quad\text{boson}\\
- \left|q\right\rangle & \quad\text{fermion}
\end{array}
\right.
\end{align}
where ${N_F}$ is the fermion number operator. Note that 
\begin{align}
(-)^{N_F} Q_\alpha^A \left| q \right\rangle = 
- Q_\alpha^A (-)^{N_F} \left|q\right\rangle
\end{align}
so that $(-)^{N_F}$ and the supercharges anticommute, $\left\{(-)^{N_F},Q_\alpha^A\right\}=0$. Next consider the operator in (\ref{eq:SUSY:SUSYalg}) weighted by $(-)^{N_F}$. When one sums over the states in a representation---which we write as a trace over the operator---one finds:
\begin{align}
    \text{Tr} \left[
        (-)^{N_F} \left\{ Q^A_\alpha, \overline Q^B_{\dot\beta}\right\}
    \right]
    = 
    \text{Tr}
     \left[
        -Q_\alpha^A(-)^{N_F}\overline Q^B_{\dot\beta}
        +
        (-)^{N_F} \overline Q^B_{\dot\beta} Q^A_\alpha
    \right]
    =
    0,
\end{align}
where in the last step we've used the cyclicity of the trace to convert the first term into the second term up to a minus sign. By (\ref{eq:SUSY:SUSYalg}) the left-hand side of this equation is simply $\text{Tr} \left[(-)^{N_F} 2 \sigma^\mu_{\alpha\dot\beta} P_\mu\right]$. Note that since Poincar\'e symmetry is assumed to be unbroken, $P_\mu$ is identical for each state in a representation. Thus we are left with the conclusion that
\begin{align}
\text{Tr}(-)^{N_F}=0,
\end{align}
which implies that there is an equal number of fermions and bosons. 

\item All states in a  supersymmetry multiplet (`supermultiplet' or \textbf{superfield}) have the same mass. This follows from the equivalence of $P_\mu$ acting on these states.

\item Energy for any state $\Psi$ is positive semi-definite $\langle \Psi | H | \Psi \rangle\geq 0$ and the energy for any vacuum with unbroken \SUSY vanishes exactly, $\langle 0 | H | 0 \rangle = 0$.

\end{enumerate}

\subsection{Classification of supersymmetry representations}

For the basic case of $\mathcal N=1$ \SUSY there is a single supercharge $Q$ and its conjugate $\overline Q$. The massless representations of this class of theories are separated into two cases:
\begin{itemize}
\item \textbf{(anti-)chiral superfield}: contains a complex scalar and a 2-component (Weyl) spinor.
\item \textbf{vector superfield}: contains a 2-component (Weyl) spinor and a gauge field.
\end{itemize}
These are the \textit{only} $\mathcal N=1$ representations that do not involve fields with spin greater than 1. 

\begin{framed}
\noindent \footnotesize \textbf{Multiplets when there is more supersymmetry}. If there are more \SUSY charges, e.g.\ $\mathcal{N}=2$, then the smallest representation is the \textbf{hypermultiplet} which contains a 4-component (Dirac) fermion and two complex scalars. For supersymmetric extensions of the \SM it is sufficient to focus only on the $\mathcal N=1$ case since this is the only case which admits the observed chiral fermions of the Standard Model.
\end{framed}

One can compare the number of bosonic and fermionic degrees of freedom in these representations. In the chiral superfield, the complex scalar carries 2 degrees of freedom while the complex Weyl spinor carries 4 degrees of freedom. Recall, however, that fermions only have two helicity states so that in fact only 2 of these fermionic degrees of freedom propagate on-shell. Since one of the key points of using fields to describe physical particles is that we can describe off-shell  propagation, we would like to also have supersymmetry hold off-shell. This requires adding two `dummy' scalar degrees of freedom, which we package in a non-propagating `auxiliary' complex field $F$:

\begin{center}
\begin{tabular}{lcc}
Field & off-shell degrees of freedom & on-shell degrees of freedom\\
\hline
scalar, $\phi$ & 2 & 2\\
fermion, $\psi$ & 4 & 2\\
auxiliary, $F$ & 2 & 0
\end{tabular}
\end{center}

For the vector superfield the Weyl spinor has 4 (2) off-(on-)shell degrees of freedom while the massless gauge boson has 3 (2)  off(on-)shell degrees of freedom after identifying gauge equivalent states. As in the chiral superfield, the number of on-shell degrees of freedom match automatically while the number of off-shell degrees of freedom require an additional non-propagating auxiliary field. In this case we introduce a real scalar, $D$:

\begin{center}
\begin{tabular}{lcc}
Field & off-shell degrees of freedom & on-shell degrees of freedom\\
\hline
fermion, $\psi$ & 4 & 2\\
gauge boson, $A_\mu$ & 3 & 2\\
auxiliary, $D$ & 1 & 0
\end{tabular}
\end{center}

\subsection{Superspace}

The most convenient way to describe $\mathcal N=1$ supersymmetric field theories is to use the \textbf{superspace} formalism. Here we understand the supersymmetry transformation generated by $Q$ and $\overline Q$ to be a spacetime transformation in an additional fermionic dimension. To do this, we introduce Weyl spinor superspace coordinates $\theta_\alpha$ and $\bar \theta^{\dot\alpha}$. Superfields are functions of $x$, $\theta$, and $\bar\theta$ and encode all of the off-shell degrees of freedom of a supermultiplet.

\begin{framed}
\noindent \footnotesize
\textbf{Weyl spinors and van der Waerden notation}. We assume familiarity with two-component Weyl spinors. These are the natural language for fermions in four-dimensions. We use the van der Waerden notation with dotted and undotted indices to distinguish the indices of left- and right-chiral spinors. Readers unfamiliar with this notation may consult \cite{Aitchison:SUSY, Wess:1992cp}. The encyclopedic `two component bible' is a useful reference for full details and as a template for doing calculations \cite{Dreiner:2008tw}.
\end{framed}

The \SUSY algebra tells us that the effect of a \SUSY transformation with infinitesimal parameters $\epsilon$ and $\bar\epsilon$ on a superspace coordinate $(x,\theta,\bar\theta)$ is
\begin{align}
(x^\mu,\theta,\bar\theta) \to 
(x^\mu + i \theta\sigma^\mu \bar\epsilon - i\epsilon \sigma^\mu \bar\theta,
\theta + \epsilon,
\bar\theta + \bar\epsilon
).
\label{eq:SUSY:superspace:coordinate:transform}
\end{align}
It is useful to define the superspace covariant derivatives,
\begin{align}
D_\alpha &= + \frac{\partial}{\partial \theta^\alpha}
+ i \sigma^{\mu}_{\alpha\dot\alpha}\bar\theta^{\dot\alpha}\partial_\mu
%\\
&
\overline D_{\dot\alpha} &= - \frac{\partial}{\partial\bar\theta_{\dot\alpha}} - i \theta^\alpha \sigma^\mu_{\alpha\dot\alpha} \partial_\mu.
\end{align}
These are `covariant derivatives' in that they anticommute with the \SUSY generators\footnote{One may be used to thinking of covariant derivatives as coming from local symmetries with some gauge field. Here, however, we consider only \emph{global} \SUSY. Geometrically, the covariant derivative comes from the fact that even rigid superspace carries torsion \cite{Gieres:1988qp}.}. They satisfy 
\begin{align}\{ D_\alpha, \overline D_{\dot \beta}\} &= -2i(\sigma^\mu)_{\alpha\dot\beta} \partial_\mu
&
\text{and}&
&
\{D_\alpha,D_\beta\}&=\{\overline D_{\dot \alpha},\overline D_{\dot \beta}\}=0
\label{eq:susy:covariant:rules}
\end{align}

By expanding in the fermionic coordinates, a generic superfield $F(x,\theta,\bar\theta)$ can be written in terms of component fields of different spin that propagate on ordinary spacetime,
%\begin{align}
%F(x,\theta,\bar\theta) =&\phantom{+} f(x) + \theta\psi(x) + \bar\theta\bar\chi(x) + \theta^2 M(x) + \bar\theta^2 N(x)
%\\
%& + \theta\sigma^\mu\bar\theta v_\mu(x) + \theta^2\bar\theta \bar\lambda(x) + \bar\theta^2\theta \xi + \theta^2\bar\theta^2 D(x).
%\end{align}
\begin{align*}
F(x,\theta,\bar\theta) =&\phantom{+} f(x) + \theta\psi(x) + \bar\theta\bar\chi(x) + \theta^2 M(x) + \bar\theta^2 N(x)
 + \theta\sigma^\mu\bar\theta v_\mu(x) + \theta^2\bar\theta \bar\lambda(x) + \bar\theta^2\theta \xi + \theta^2\bar\theta^2 D(x).
\end{align*}
This expansion is exact because higher powers of $\theta$ or $\bar\theta$ vanish identically because an anticommuting number $\theta_1$ satisfies $\left(\theta_1\right)^2=0$. As a sanity check, we are allowed quadratic terms in $\theta$ since it is a Weyl spinor and $\theta^2 = \theta^\alpha \theta_\alpha = \epsilon^{\alpha\beta}\theta_\beta\theta_\alpha =  2\theta_1\theta_2$. 

With modest effort, one can work out the transformation of each component of this general superfield by applying the transformation (\ref{eq:SUSY:superspace:coordinate:transform}), expanding all fields in $\theta$ and $\bar\theta$, and matching the coefficients of each term. Some of the terms require massaging by Fierz identities to get to the correct form.
Fortunately, the general superfield above is a reducible representation: some of these fields do not transform into one another. We can restrict to irreducible representations by imposing one of the following conditions:
\begin{align}
\text{chiral superfield} && \overline D_\alpha \Phi &= 0
\label{eq:SUSY:CSF:def}\\
\text{anti-chiral superfield} && D_{\dot\alpha}\overline\Phi &= 0\\
\text{vector (real) superfield} && V &= V^\dag
\\
\text{linear superfield} && 
    \overline{D}^2 L = D^2 L &=0
%    \\&& \text{and } L &= L^\dag
\end{align}
The chiral and anti-chiral superfields carry Weyl fermions of left- and right-handed helicity respectively. 
It is convenient to write all anti-chiral superfields into chiral superfields, for example by swapping the right-handed electron chiral superfield with a left-handed positron superfield. The field content is identical, one is just swapping which is the `particle' and which is the `anti-particle.' 

\begin{framed}
\noindent \footnotesize
\textbf{The linear superfield.}  The defining condition for this superfield includes a constraint that the vector component is divergence free, $\partial_\mu V^\mu=0$. It is thus a natural supersymmetrization of a conserved current. %In $d=3$ dimensions with $\mathcal N=2$ \SUSY, the linear superfield provides an alternate way to package the gauge field strength, $\int d^4 \theta \, L^2 \supset F_{\mu\nu}F^{\mu\nu}$. This has to do with the fact that in $d=3$, $*F$ is a conserved topological current \cite{Aharony:1997bx, deBoer:1997kr}.  In 4D the linear superfield encodes the field strength of a two-form potential and typically is only encountered when looking at the dimensional reduction of higher dimensional supergravity. 
% \flip{Comment on the supercurrent. See Weinberg III.}
We will not consider linear superfields further in these lectures.
\end{framed}

\subsection{Supersymmetric Lagrangians for chiral superfields}

One can check that because $\overline D_{\dot\alpha} (x^\mu + i \theta \sigma^\mu \bar\theta) = 0$, any function of $y^\mu = x^\mu + i \theta \sigma^\mu \bar\theta$ is automatically a chiral superfield (\CSF). Indeed, the most compact way of writing the components of a \CSF is
\begin{align}
\Phi(y,\theta) = \varphi(y) + \sqrt{2} \theta \psi(y) + \theta^2 F(y).
\end{align}
Again, we point out that this expansion is exact since higher powers of the Weyl spinor $\theta$ vanish by the antisymmetry of its components.
Under a \SUSY transformation with parameter $\epsilon$, the components of the \CSF each transform as
\begin{align}
\delta_\epsilon \varphi(x) &= \sqrt{2} \epsilon\psi(x)\\
\delta \psi(x) &= i\sqrt{2} \sigma^\mu \bar\epsilon \partial_\mu \varphi(x) + \sqrt{2} \epsilon F(x)\\
\delta_\epsilon F(x) &= i\sqrt{2} \bar\epsilon\bar\sigma^\mu\partial_\mu\psi(x).
\end{align}
Observe that the auxiliary field transforms into a total spacetime derivative. This is especially nice since a total derivative vanishes in the action and so the highest component of a \CSF is a candidate for a \SUSY-invariant term in the Lagrangian. Thus we arrive at our first way of constructing supersymmetric Lagrangian terms: write the $F$-term of a chiral superfield. 

To generate interesting interactions we don't want to write the $F$-terms of our fundamental fields---indeed, these are generally not even gauge invariant. Fortunately, one can check that a product of chiral superfields is itself a chiral superfield. Indeed, a general way of writing a supersymmetry Lagrangian term built out of chiral superfields is
\begin{align}
\mathcal L = \int d^2\theta \; W(\Phi) + \text{h.c.},
\end{align}
where $W$ is a holomorphic function of chiral superfields called the \textbf{superpotential}. Note that the integral over $d^2\theta$ is an ordinary fermionic integral that just picks out the highest component of $W$. Performing the fermionic integral gives Lagrangian terms
\begin{align}
\mathcal L = 
- \frac{\partial^2 W(\varphi)}{\partial \Phi_i\partial \Phi_j} \psi_i\psi_j - \sum_i \left| \frac{\partial W(\varphi)}{\partial \Phi_i}\right|^2.
\end{align}
Observe that the superpotential is evaluated on the scalar components of the superfields, $\Phi = \varphi$. One can check that restricting to renormalizable terms in the Lagrangian limits the mass dimension of the superpotential to $[W]\leq 3$.

\begin{framed}
\noindent \footnotesize
\textbf{Cancellation of quadratic divergences.}  
One can check from explicit calculations that the \SUSY formalism ensures the existence of superpartner particles with just the right couplings to cancel quadratic divergences. A more elegant way to see this, however, is to note that the symmetries of superspace itself prevent this. While it is beyond the scope of these lectures, the superpotential is not renormalized perturbatively---see, e.g.\ \cite{Intriligator:1995au, Peskin:1997qi} for details. The holomorphy of $W$ plays a key role in these arguments. The symmetries of the theory enforce the technical naturalness of parameters in $W$, including scalar masses. 
\end{framed}

Superpotential terms, however, do not include the usual kinetic terms for propagating fields. In fact, one can show that these terms appear in the $\theta^2\bar\theta^2$ term of the combination
\begin{align}
\left.\Phi^\dag \Phi\right|_{\theta^2\bar\theta^2}
&= 
FF^* + \frac 14 \varphi^* \partial^2 \varphi + \frac 14 \partial^2 \varphi^* \varphi - \frac 12 \partial_\mu \varphi^* \partial^\mu \varphi + \frac i2 \partial_\mu \bar\psi \bar\sigma^\mu \psi - \frac i2 \bar\psi \bar\sigma \partial_\mu \psi.
\label{eq:SUSY:CSF:kinetic:term}
\end{align}
Two immediate observations are in order:
\begin{enumerate}
\item The complex scalar $\varphi$ and Weyl fermion $\psi$ each have their canonical kinetic term. The non-propagating field, $F$, does not have any derivative terms: its equation of motion is algebraic and can be solved explicitly. This is precisely what is meant that $F$ is auxiliary.
\item $\Phi^\dag\Phi$ is not a chiral superfield. In fact, it's a real superfield and the $\theta^2\bar\theta^2$ component is the auxiliary $D$ field. Indeed, in the same way that the highest component of a \CSF transforms into a total derivative, the highest component of a real superfield also transforms into a total derivative and is a candidate term for the Lagrangian.
\end{enumerate}
We thus arrive at the second way to write supersymmetric Lagrangian terms: take the $D$-term of a real superfield. 
We may write this term as an integral over superspace, $\int d^4\theta \; \Phi^\dag\Phi$, where $d^4\theta = d^2\theta\, d^2\bar\theta$. 

More generally, we may write a generic real function $K(\Phi, \Phi^\dag)$ of chiral superfields, $\Phi$ and $\Phi^\dag$, whose $D$ term is supersymmetric contribution to the Lagrangian. This is called the \textbf{K\"ahler potential}.
The simplest K\"ahler potential built out of chiral superfields is precisely (\ref{eq:SUSY:CSF:kinetic:term}) and includes the necessary kinetic terms for the chiral superfield. One can check that restricting to renormalizable terms in the Lagrangian limits the mass dimension of the K\"ahler potential to $[K]\leq 2$. Combined with the condition that $K$ is real and the observation that chiral superfields are typically not gauge invariant, this usually restricts the K\"ahler potential to take the canonical form, $K=\Phi_i^\dag \Phi_i$.

The most general $\mathcal N=1$ supersymmetric Lagrangian for chiral superfields is thus
\begin{align}
\mathcal L = \int  d^4\theta\; K(\Phi,\Phi^\dag) + \left(\int d^2\theta\; W(\Phi) + \text{h.c.}\right).
\label{eq:SUSY:Lagrangian}
\end{align}
This expression is general, but renormalizability restricts the mass dimensions to be $[K]\leq 2$ and $[W]\leq 3$. For theories with more supersymmetry, e.g.\ $\mathcal N=2$, one must impose additional relations between $K$ and $W$.
Assuming a renormalizable supersymmetric theory of chiral superfields $\Phi_i$, we may plug in $K=\Phi_i^\dag \Phi_i$ and integrate out the auxiliary fields from (\ref{eq:SUSY:Lagrangian}). The result is
\begin{align}
\mathcal L = \partial_\mu \varphi_i^* \partial^\mu \varphi_i
+ i\bar\psi_i \bar \sigma^\mu \partial_\mu \psi_i
- \frac{\partial^2 W}{\partial \varphi_i\partial \varphi_j} \psi_i\psi_j - \sum_i \left| \frac{\partial W}{\partial \varphi_i}\right|^2.
\end{align}
Here the superpotential is assumed to be evaluated at its lowest component so that $W[\Phi_i(y,\theta)] \to W[\varphi_i(x)]$. Observe that dimension-2 terms in the superpotential link the mass terms of the Weyl fermion and the complex scalar. Further, dimension-3 terms in the superpotential connect Yukawa interactions to  quartic scalar couplings.

\subsection{Supersymmetric Lagrangians for vector superfields}

Until now, however, we have only described supersymmetric theories of complex scalars and fermions packaged as chiral superfields. In order to include the interactions of gauge fields we must write down \SUSY Lagrangians that include vector superfields. 

Suppose a set of chiral superfields $\Phi$ carry a U(1) charge such that $\Phi(x) \to \exp(-i\Lambda) \Phi(x)$. For an ordinary global symmetry this is an overall phase on each component of the chiral superfield. For a gauge symmetry, the transformation parameter is spacetime dependent, $\Lambda=\Lambda(x)$. Note, however, that this is now problematic because our definition of a chiral superfield, %(\ref{eq:SUSY:CSF:def}), 
$\overline D_\alpha \Phi = 0$, contains a spacetime derivative. It would appear that the na\"ive gauge transformation is not consistent with the irreducible \SUSY representations we've written because it does not preserve the chiral superfield condition.

This inconsistency is a relic of keeping $\Lambda(x)$ a function of spacetime rather than a function of the full superspace. We noted above that a function of $y^\mu = x^\mu + i \theta \sigma^\mu \bar\theta$ is a chiral superfield and, further, that a product of chiral superfields is also a chiral superfield. Thus a consistent way to include gauge transformations is to promote $\Lambda(x)$ to a chiral superfield $\Lambda(y)$ so that $\exp(-i\Lambda(y)) \Phi(y)$ is indeed chiral. In this way we see that supersymmetry has `complexified' the gauge group.

Under this complexified gauge transformation, the canonical K\"ahler potential term that contains the kinetic terms transforms to
\begin{align}
\Phi^\dag\Phi \to \Phi^\dag e^{-i(\Lambda - \Lambda^\dag)}\Phi.
\end{align}
For gauge theories one must modify the K\"ahler potential to accommodate this factor. This is unsurprising since gauging an ordinary quantum field theory requires one to modify the kinetic terms by promoting derivatives to covariant derivatives which include the gauge field. 
To correctly gauge a symmetry, we introduce a vector (real) superfield (\VSF) $V$ which transforms according to 
\begin{align}
V\to V+i(\Lambda-\Lambda^\dag)
\end{align}
and promote the K\"ahler potential to
\begin{align}
%\Phi^\dag\Phi \to \Phi^\dag e^V \Phi,
K(\Phi,\Phi^\dag)  =\Phi^\dag e^V \Phi.
\end{align}

A generic \VSF has many components, but many can be eliminated by partially gauge fixing to the \textbf{Wess-Zumino} gauge where
\begin{align}
V=& - \theta \sigma^\mu \bar\theta V_\mu(x)
+ i\theta^2\bar\theta\bar\lambda(x)
-i\bar\theta^2\theta\lambda(x) + \frac 12 \theta^2\bar\theta^2D(x).
\end{align}
here $V_\mu(x)$ is the gauge field of the local symmetry, $\lambda(x)$ and $\bar\lambda(x) = \lambda^\dag(x)$ are \textbf{gauginos}, and $D(x)$ is the auxiliary field needed to match off-shell fermionic and bosonic degrees of freedom. The two gauginos are the pair of two-component spinors that make up a Majorana four-component spinor. This gauge choice fixes the complex part of the `complexified' gauge symmetry, leaving the ordinary spacetime (rather than superspace) gauge redundancy that we are familiar with in quantum field theory.

We have not yet written a kinetic term for the vector superfield. A useful first step is to construct the chiral superfield,
\begin{align}
\mathcal{W}_\alpha =& -\frac 14 \overline D^{\dot\alpha} \overline D_{\dot \alpha} D_\alpha V\\
=& - i\lambda_\alpha(y)
+ \theta_\beta\left[
\delta_\alpha^\beta D(y)
- \frac i2 \left(\sigma^\mu\bar\sigma^\nu\right)^\beta_\alpha F_{\mu\nu}(y)
\right]
+ \theta^2\sigma^{\mu}_{\alpha\dot\alpha} \partial_\mu\bar\lambda^{\dot\alpha}(y).
\end{align}
One can see that $\mathcal W_\alpha$ is a chiral superfield because $\overline D_{\dot\beta} \mathcal W_\alpha = 0$ from the antisymmetry of the components of $\bar D$, (\ref{eq:susy:covariant:rules}).
Observe that unlike $\Phi$, the lowest component is a spin-1/2 field. Further, $\mathcal W$ contains the usual gauge field strength. Indeed, one can write the supersymmetric Yang Mills kinetic terms for the vector superfield as
\begin{align}
\mathcal L_{\text{SYM}} = \frac 14 \left.\mathcal W_\alpha \mathcal W^\alpha\right|_\theta^2 + \text{h.c.} = \frac 14 \int d^2\theta \mathcal W^2 + \text{h.c.}.
\end{align}
One can check that this gives the usual kinetic terms for the gauge field and gauginos as well as an auxiliary term. 
For completeness, the general form of the field strength superfield for a non-Abelian supersymmetric gauge theory is
\begin{align}
T^a \mathcal W^a_\alpha &= - \frac 14 \overline D^{\dot a} \overline D_{\dot a} e^{- T^a V^a} D_\alpha e^{T^a V^a}.
\end{align}
%One may check that in this case, the gauge kinetic term $\mathcal W^2\, d^2\theta + \text{h.c.}$ also includes a topological term, $F\tilde F$. 
Under a non-Abelian gauge transformation the chiral and vector superfields transform as
\begin{align}
\Phi &\to e^{-g T^a \Lambda^a}\Phi\\
e^{T^aV^a} &\to e^{T^a\Lambda^{a\dag}}e^{T^aV^a}e^{T^a\Lambda^a}.
\end{align}

The final form of the renormalizable, gauge-invariant, $\mathcal N=1$ supersymmetric Lagrangian is
\begin{align}
\mathcal L = \int d^4\theta \Phi_i^\dag e^{gV} \Phi_i
+ \int d^2\theta \left(\frac 14 \mathcal W_\alpha^a \mathcal W^{\alpha a} + \text{h.c.}\right)
+ \int d^2\theta \left(W(\Phi) + \text{h.c.}\right).
\end{align}

\begin{framed}
\noindent \footnotesize \textbf{Non-renormalization and the gauge kinetic term}.
Although $\mathcal W^2$ looks like it could be a superpotential term, it is important to treat it separately since it is the kinetic term for the gauge fields. 
%
%Further, there are well known arguments outside the scope of these lectures that the superpotential is not renormalized in perturbation theory---see, e.g.\ \cite{Intriligator:1995au, Peskin:1997qi} for details--- that do \emph{not} hold for the $\mathcal W^2$ term. 
%
Further the arguments that the superpotential is not renormalized in perturbation theory do \emph{not} hold for the $\mathcal W^2$ term. 
Indeed, the prefactor of $\mathcal W^2$ can be identified with the [holomorphic] gauge coupling, which is only corrected perturbatively at one loop order. One way to see this is to note that for non-Abelian theories, the gauge kinetic term $\mathcal W^2\, d^2\theta + \text{h.c.}$ also includes a topological term, $F\tilde F$, which we know is related to anomalies. Another way to see this is the note that the simplest demonstration of non-renormalization of the superpotential makes use of holomorphy and the global symmetries of $W$: the vector (real) superfield from which $\mathcal W_\alpha$ is built, however, is not holomorphic and its fields cannot carry have the U(1) global symmetries used in the proof.% A final way to see this is through the observation that $\int d^4x \overline D^2 f(x,\theta,\bar\theta) = \int d^4x\, d^2\theta f(x,\theta,\bar\theta$ so that indeed $\mathcal L_\text{SYM}$ can be converted into a $d^4\theta$ term which is not protected against renormalization.
% 
%% A third way: this gauge kinetic term could have been written in the Kahler potential all along.
\end{framed}

\subsection{Example: SUSY QED}

As a simple example, consider the supersymmetric version of quantum electrodynamics, \textsc{sqed}. 
In ordinary \textsc{qed} we start with a Dirac spinor representing the electron and positron. Since we've seen above that a chiral superfield only contains a Weyl spinor, we require two chiral superfields, $\Phi_\pm$, which we may interpret to be the electron and positron superfields. Our only two inputs are the electromagnetic coupling $e$ and the electron mass $m$. The latter suggests a superpotential
\begin{align}
W(\Phi_+,\Phi_-) =&\phantom{+} m \Phi_+\Phi_-.
\end{align}
Writing out the resulting Lagrangian in components:
\begin{align}
\mathcal L_\text{\textsc{sqed}} =&
\left[
    \frac 12 D^2
    - \frac 14 F_{\mu\nu}F^{\mu\nu}
    -i\lambda \sigma^\mu \partial_\mu \bar\lambda
\right]
\nonumber
\\
&+F_+^* F_+ 
+ \left|D_\mu \varphi_+\right|^2
+ i\bar\psi_+ D_\mu\bar\sigma^\mu\psi_+
\nonumber
\\
&+F_-^* F_-
+ \left|D_\mu \varphi_-\right|^2
+ i\bar\psi_- D_\mu\bar\sigma^\mu\psi_-
\nonumber
\\
&
-\frac{ie}{\sqrt{2}}\left(
\varphi_+ \bar\psi_+ \bar\lambda
-\varphi_- \bar\psi_- \bar\lambda
\right) + \text{h.c.}
\nonumber
\\
&
+ \frac{e}{2}D \left( |\varphi_+|^2 - |\varphi_-|^2 \right)
\nonumber
\\
& + m\left(\varphi_+ F_- + \varphi_-F_+ - \psi_+\psi_-\right) + \text{h.c.}
\end{align}
We can write this out explicitly by solving for the auxiliary fields $D$, $F_\pm$. The equations of motion are
\begin{align}
D &= -\frac e2 \left(|\varphi_+|^2-|\varphi_-|^2\right)
&
F_\pm =& -m \varphi_\mp^*.
\end{align}
Plugging this back into the Lagrangian gives
\begin{align}
\mathcal L_\text{SQED} &=
\sum_{i=\pm} \left( |D_\mu \varphi_i |^2 + i\bar\psi_i D_\mu \bar\sigma^\mu \psi_i \right)
-\frac 14 F_{\mu\nu}F^{\mu\nu} - i \lambda \sigma^\mu \partial_\mu \bar\lambda
\nonumber \\
& - m^2 \left(|\phi_+|^2+ |\phi_-|^2\right) - m\psi_+\psi_- - m\bar\psi_+\bar\psi_-
\nonumber\\
& - \frac{e^2}{8} 
\left(|\varphi_+|^2 - |\varphi_-|^2\right)^2 - \left[\frac{i e}{\sqrt{2}}\left( \varphi_+ \bar\psi_+ \bar\lambda - \varphi_-\bar\psi_-\bar\lambda \right) + \text{h.c.}\right].
\end{align}
The first line gives the kinetic terms for the electron $\psi_-$, positron $\psi_-$, selectron ($\phi_-$), spositron ($\phi_+$), photon $A_\mu$, and photino $\lambda$. The second line gives an equivalent mass to the chiral scalars and fermions. The last line gives vertices that come from the supersymmetrization of the kinetic terms: four-point scalar interactions from the $D$ terms and a three-point Yukawa-like vertex with the `chiral' scalars and photino. The relation between the gauge group and the four-point scalar interaction plays a central role in how the Higgs fits into \SUSY, as we show below.

%%% LECTURE 2

\subsection{The MSSM}

We now focus on the minimal supersymmetric extension of the Standard Model, the \MSSM. To go from the \SM to the \MSSM, it is sufficient to promote each \SM chiral fermion into a chiral superfield and each \SM gauge field into a vector superfield. Thus for each \SM fermion there is a new propagating scalar sfermion (squarks or sleptons) and for each \SM gauge field there is also a propagating gaugino, a fermion in the adjoint representation. As we showed above, off-shell \SUSY also implies non-propagating auxiliary fields.

The matter (\CSF) content of the \MSSM is shown in Table~\ref{tab:MSSM}. It is the same as the \SM except that we require two Higgs doublet chiral superfields. This is necessary for the cancellation of the SU(2)$_\text{L}^2\times$U(1)$_\text{Y}$ and SU(2)$_\text{L}$ Witten anomalies 
% "SU(2)_L^3 anomaly" is not correct, must be SU(2)L Witten anomaly. Something different.
coming from the Higgs fermions, or Higgsinos. An additional hint that this is necessary comes from the observation that the superpotential is a holomorphic function of the chiral superfields while the Standard Model up-type Yukawa coupling requires the conjugate of the Higgs, $\tilde H = i\sigma^2 H^*$. 

\begin{table}
\begin{center}
%\singlespacing
\begin{tabular}{cccc}
\toprule % requires: booktabs.sty.
\CSF & SU($3$)$_\text{c}$ & SU($2$)$_\text{L}$& U($1$)$_\text{Y}$\\
%\hline
\midrule % requires: booktabs.sty.
%% No more boxes, just us BF
%$Q$  &  {\tiny\yng(1)} & {\tiny\yng(1)} & $\phantom{+}\nicefrac{1}{6}$ \\
%$\bar U$  & $\overline{\tiny\yng(1)}$ & $\mathbbm{1}$ & $-\nicefrac{2}{3}$ \\
%$\bar D$  & $\overline{\tiny\yng(1)}$ & $\mathbbm{1}$ & $\phantom{+}\nicefrac{1}{3}$\\
%$L$  & $\mathbbm{1}$ & {\tiny\yng(1)} & $-\nicefrac{1}{2}$\\
%$\bar E$  & $\mathbbm{1}$ & $\mathbbm{1}$ & $-1$\\
%$H_d$  & $\mathbbm{1}$ & $\tiny\yng(1)$ & $\phantom{+}\nicefrac{1}{2}$\\
%$H_u$  & $\mathbbm{1}$ & $\tiny\yng(1)$ & $-\nicefrac{1}{2}$\\
%
$Q$  &  $\mathbf{3}$ & $\mathbf{2}$ & $\phantom{+}\nicefrac{1}{6}$ \\
$\bar U$  & $\overline{\mathbf{3}}$ & $\mathbf{1}$ & $-\nicefrac{2}{3}$ \\
$\bar D$  & $\overline{\mathbf{3}} $ & $\mathbf{1}$ & $\phantom{+}\nicefrac{1}{3}$\\
$L$  & $\mathbf{1}$ & $\mathbf{2}$ & $-\nicefrac{1}{2}$\\
$\bar E$  & $\mathbf{1}$ & $\mathbf{1}$ & $-1$\\
$H_d$  & $\mathbf{1}$ & $\mathbf{2}$ & $\phantom{+}\nicefrac{1}{2}$\\
$H_u$  & $\mathbf{1}$ & $\mathbf{1}$ & $-\nicefrac{1}{2}$\\
\bottomrule % requires: booktabs.sty.
\end{tabular}
\caption[Matter content of the MSSM]{Matter content of the \MSSM. Note that we have used $\mathbf{2}=\overline{\mathbf{2}}$
%$\tiny\yng(1) = \overline{\tiny\yng(1)}$ 
for SU(2)$_\text{L}$.}\label{tab:MSSM}
\end{center}
\end{table}

The most general renormalizable superpotential made with these fields can be split into two terms, $W = W^{(\text{good})} + W^{(\text{bad})}$,
\begin{align}
W^{(\text{good})} =&
y_u^{ij} Q^i H_u \bar U^j 
+ y_d^{ij} Q^i H_d \bar D 
+ y_e^{ij} L^i H_d \bar E^j + \mu H_u H_d\\
W^{(\text{bad})} =& 
\lambda^{ijk}_1 Q^i L^j \bar D^k
+ \lambda^{ijk}_2 L^i L^j \bar E^k 
+ \lambda^{i}_3 L^i H_u 
+ \lambda^{ijk}_4 \bar D^i \bar D^j \bar U^k.
\label{eq:MSSM:RPV:superpotential}
\end{align}
In $W^{(\text{good})}$ one can straight forwardly identify the Standard Model Yukawa couplings which give the \SM fermions their masses. Since these are packaged into the superpotential these terms also encode the additional scalar quartic interactions required by supersymmetry. The last term in $W^{(\text{good})}$ is a supersymmetric Higgs mass known as the $\mu$-term. By supersymmetry this term also gives a mass to the Higgsinos, which we require since we do not observe any very light chiral fermions with the quantum numbers of a Higgs. 

The $W^{(\text{bad})}$ terms, on the other hand, are phenomenologically undesirable. These are renormalizable interactions which violate baryon ($B$) and/or lepton ($L$) number and are thus constrained to have very small coefficients. Compare this to the \SM where $B$ and $L$ are accidental symmetries: all renormalizable interactions of \SM fields allowed by the \SM gauge group preserve $B$ and $L$. Violation of these symmetries only occurs at the non-renormalizable level and are suppressed by what can be a very high scale, e.g.\ $M_\text{GUT}$.

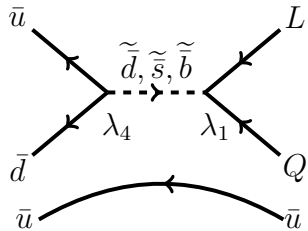
\begin{figure}
\begin{center}
\begin{tikzpicture}[line width=1.5 pt, scale=1.3]
	\draw[fermionbar] (-140:1)--(0,0);
	\draw[fermionbar] (140:1)--(0,0);
	\draw[scalar] (0,0)--(0:1) ;
	\node at (-140:1.2) {$\bar d$};
	\node at (140:1.2) {$\bar u$};
	\node at (.5,.3) {$\tilde{\bar d}, \tilde{\bar s}, \tilde{\bar b}$};	
	 \node at (.1,-.35) {$\lambda_4$};
	 \node at (1.1,-.35) {$\lambda_1$};
\begin{scope}[shift={(1,0)}]
	\draw[fermion] (-40:1)--(0,0);
	\draw[fermion] (40:1)--(0,0);
	\node at (-40:1.2) {$Q$};
	\node at (40:1.2) {$L$};	
\end{scope}
%\draw[fermionbar] (180:1) to [out=0,in=220] (40:1);
\draw[fermionbar] (-.7,-1.3) to [out = 30, in=150] (1.8,-1.3);
\node at (-.85, -1.3) {$\bar u$};
\node at (1.9, -1.3) {$\bar u$};
\end{tikzpicture}
\end{center}
\caption{Proton decay mediated by squarks. Arrows indicate \emph{helicity} and should not be confused with the `charge flow' arrows of Dirac spinors \cite{Dreiner:2008tw}. Tildes indicate superpartners while bars are used to write right-chiral antiparticles into left-chiral fields in the conjugate representation. %See Appendix~\ref{sec:SUSY:starsanddaggers} if one is confused by this ornament.
}
\label{fig:SUSY:p:decay}
\end{figure}

We see that in the \MSSM we must find ways to forbid, or otherwise strongly suppress, the terms in $W^{(\text{bad})}$. Otherwise one would be faced with dangerous rates for rare processes such as proton decay, $p^+ \to e^+ \pi^0$ or $\bar\nu \pi^+$ (or alternately with $\pi$ replaced with $K$) as shown in Fig.~\ref{fig:SUSY:p:decay}. %(or alternately with $\pi$ replaced with $K$): 
Observe that this is a tree level process and all of the couplings are completely unsuppressed. 

A simple way to forbid $W^{(\text{bad})}$ is to impose \textbf{matter parity}, which is a $\mathbbm{Z}_2$ symmetry with assignments:
\begin{center}
\begin{tabular}{lc}
Superfield & Matter parity\\
\hline
quark, lepton \CSF & $P_M=-1$\\
Higgs \CSF & $P_M=+1$\\
gauge \VSF & $P_M=+1$.
\end{tabular}
\end{center}
Under these assignments, all terms in $W^{(\text{good})}$ have $P_M=+1$ while all terms in $^{(\text{bad})}$ have $P_M=-1$. One can check that one may write matter parity in terms of baryon and lepton number as
\begin{align}
P_M = (-)^{3(B-L)}.
\end{align}
A common variation of this is to impose the above constraint using \textbf{$R$-parity},
\begin{align}
P_R = (-)^{3(B-L)+2s},
\end{align}
where $s$ is the spin of the field. Conservation of matter parity implies conservation of $R$-parity. This is because the $(-)^{2s}$ factor always cancels in any interaction term since Lorentz invariance requires that any such term has an even number of fermions. Observe that all \SM fields have $R$-parity $+1$ while all superpartner fields have $R$-parity $-1$. (This is similar to $T$-parity for Little Higgs models.) The diagrams assocaited with electroweak precision observables carry only \SM external states. Since $R$-parity requires pair-production of superpartners, this means that electroweak precision corrections cannot occur at tree-level and must come from loop diagrams.

It is important to understand that $R$-parity (or matter parity) is an additional symmetry that we impose on top of supersymmetry. $R$-parity has some important consequences:
\begin{enumerate}
\item The lightest $R$-parity odd particle is stable. This is known as the \textbf{lightest supersymmetric particle} or \LSP. If the \LSP is an electrically neutral color singlet---as we shall assume---it is a candidate for \WIMP-like \DM. 
\item Each superpartner (sparticle) other than the \LSP will decay. At the end of any such sequence of decays one is left with an odd number (usually one) of \textsc{lsp}s.
\item In collider experiments, the initial state has $P_R=+1$ so that only an even number of sparticles can be produced at a time (e.g.\ via pair production). At the end of the decay these end up as \textsc{lsp}s which manifest themselves as missing energy signals at colliders.
\end{enumerate}
For most of this document we postulate that the \MSSM has exact $R$-parity conservation---though this is something of an ad-hoc assumption. 

\subsection{Supersymmetry breaking}

Any scalar partners to the \SM leptons or quarks with exactly degenerate masses as their \SM partner would have been discovered long ago. 
Thus, the next piece required to construct a realistic \MSSM is a way to break supersymmetry and split the mass degeneracy between the \SM particles and their superpartners.
Since we want to keep the desirable ultraviolet behavior of supersymmetry, we assume that \SUSY is a fundamental symmetry of nature which is spontaneously broken. 

\SUSY is unbroken when the supercharges annihilate the vacuum, $Q|0\rangle = \overline Q|0\rangle = 0$. The \SUSY algebra, $\{Q,\overline Q\} = 2 \sigma^\mu P_\mu$ allows us to write the four-momentum operator as $P^\mu = \frac 14 \bar\sigma^\nu \{Q,\overline Q\}$ so that the Hamiltonian is
\begin{align}
H=P^0 = \frac 14 \left(
Q_1\overline Q_{\dot 1}
+
\overline Q_{\dot 1} Q_1
+
Q_2\overline Q_{\dot 2}
+
\overline Q_{\dot 2} Q_2
\right).
\end{align}
Observing that this expression is positive semi-definite, we see that
\begin{center}
\begin{tabular}{lcc}
if \SUSY is unbroken, & \quad & $\langle 0 | H | 0 \rangle =0$\\
if \SUSY is broken, & \quad & $\langle 0 | H | 0 \rangle >0$
\end{tabular}.
\end{center}
The vacuum energy can be read from the scalar potential,
\begin{align}
V[\phi] &= V_F[\phi] + V_D[\phi]\\
V_F[\phi] &= \sum_i \left|\frac{\partial W}{\partial\phi_i}\right|^2 = \sum_i |F_i|^2\\
V_D[\phi] &= \sum_a \frac 12 g^2 \left|\sum_i \phi^\dag_i T^a \phi_i \right|^2 = \sum_a \frac 12 g D^a D^a. \label{eq:SUSY:D-terms}
\end{align}
We see that \SUSY breaking corresponds to one of the auxiliary fields, $F_i$ or $D_i$, picking up a vacuum expectation value (\vev). We refer to the case $\langle F_i\rangle \neq 0$ as $F$-type \SUSY breaking and the case $\langle D\rangle \neq 0$ as $D$-type \SUSY breaking.

When an ordinary global symmetry is spontaneously broken due to a field picking up a \vev there exists a massless boson in the spectrum of the theory known as the Goldstone boson. In the same way, when \SUSY is broken spontaneously due to a auxiliary field picking up a \vev, there exists a massless \emph{fermion} in the theory known as the \textbf{Goldstino}\footnote{This is somewhat unfortunate nomenclature. One would expect the massless mode coming from spontaneously broken \SUSY to be called a Goldstone fermion whereas the `Goldstino' should refer to the supersymmetric partner of a Goldstone boson coming from the spontaneous breaking of an ordinary symmetry.}. The spin of this field is inherited by the spin of the \SUSY generators. Heuristically, the massless Goldstone modes correspond to acting on the \vev with the broken generators and promoting the transformation parameters to fields. Since the \SUSY transformation parameter is fermionic, the Goldstone field must also be fermionic.

For example, if $\langle F\rangle \neq 0$, then the transformation of the fermion $\psi$ under the broken (\SUSY) generator is
\begin{align}
\delta_\epsilon \psi = 2 \epsilon \langle F\rangle.
\end{align}
\SUSY acts as a shift in the fermion, analogously to the shift symmetry of a Goldstone boson under a spontaneously broken global internal symmetry. 
If there is more than one superfield with a non-zero $F$ term, then
\begin{align}
\delta_\epsilon \psi_i &= 2\epsilon \langle F_i\rangle\\
\psi_\text{Goldstone} &= \sum_i \frac{F_i}{\sqrt{\sum_i F_i^2}} \, \psi_i.
\end{align}
Note that we have used the convention that, when there is no ambiguity, $F$ refers to the \SUSY breaking background value, dropping the brackets $\langle \cdots \rangle$ to avoid clutter. One can further generalize this to include a linear combination of gauginos when there is also $D$-term \SUSY breaking.

When ordinary spontaneously broken internal symmetries are promoted to gauge symmetries, their Goldstone modes are `eaten' and become the longitudinal polarization of the gauge fields. Similarly, gauging supersymmetry corresponds to writing a theory of supergravity. The gravitino then becomes massive by eating the Goldstino from spontaneous \SUSY breaking.
%The Goldstino of spontaneously broken supersymmetry is then `eaten' by the gravitino to become massive.

\subsection{Sum rule for broken SUSY}

% See, e.g. 
%% http://www.weizmann.ac.il/particle/nir/uploads/file/chapter4.pdf

Even when it is spontaneously broken, \SUSY is a strong constraint on the parameters of a theory. One of the most important constraints is the \SUSY sum rule, which relates the traces of the mass matrices of particles of different spins.

First consider the mass terms for chiral fermions ($\psi$) and gauginos ($\lambda$):
\begin{align}
i\sqrt{2} g \left(T^a\right)^i_j
\left(\varphi_i \bar\lambda^a \bar\psi^j - \varphi^* \lambda \psi \right)
- \frac{\partial^2 W}{\partial \varphi_i \partial \varphi_j} \psi_i\psi_j + \text{h.c.}
\end{align}
We may write this succinctly as a mass matrix,
\begin{align}
\begin{pmatrix}
\psi_i & \lambda_a
\end{pmatrix}
\begin{pmatrix}
F_{ij} & \sqrt{2} D_{bi} \\
\sqrt{2}D_{aj} & 0 
\end{pmatrix}
\begin{pmatrix}
\psi_j\\
\lambda_b
\end{pmatrix},
\end{align}
where we use the shorthand notation
\begin{align}
F_{ij} &= \frac{\partial F_i}{\partial \varphi_j} = \frac{\partial^2 W}{\partial \varphi_i\partial \varphi_j}
&
D_{ai} &= \frac{\partial D_a}{\partial \varphi_i} = g \varphi_i^*T^a.
\end{align}
Call this fermion mass matrix $m^{(j=1/2)}$.
%
%\begin{align}
%m^{(j=1/2)} &= 
%\begin{pmatrix}
%F_{ij} & \sqrt{2} D_{bj}\\
%\sqrt{2} D_{aj} & 0
%\end{pmatrix}
%\end{align}
%
Next, the scalar mass matrix $\left(m^2\right)^{(j=0)}_{ij}$ is obtained by the Hessian of the scalar potential,
\begin{align}
\begin{pmatrix}
\frac{\partial^2 V}{\partial\varphi_i \partial \varphi^*_j}
&
\frac{\partial^2 V}{\partial\varphi_i\partial\varphi_j}
\\
\frac{\partial^2 V}{\partial \varphi_i^* \partial \varphi_j^*}
&
\frac{\partial^2 V}{\partial \varphi_i^*\partial \varphi_j}
\end{pmatrix}
=
\begin{pmatrix}
\bar F^{ij} F_{kj} + D_a^i D_{aj} + D_{aj}^{i}D_a&
\bar F^{ijk} F_k + D_a^j D_a^j\\
F_{ijk} \bar F^{k} + D_{ai} D_{aj}
&
F_{ik}\bar F^{jk} + D_{ai} D_a^j + D_{ai}^j D_a
\end{pmatrix}.
\label{eq:SUSY:scalar:mass}
\end{align}
Finally, the gauge boson matrix comes from the kinetic terms
\begin{align}
\sum_i g^2 |A^a_\mu T^{a\alpha}_{\phantom{a\alpha}\beta} \phi_{i\alpha}|^2
= 
|A_\mu^a D_a^i|^2,
\end{align}
and may thus be written
\begin{align}
(m^2)^{(j=1)}_{ab} &= D_a^iD_{bi} + D_{ai} D_b^i.
\end{align}

The traces of the squared mass matrices are, respectively,
\begin{align}
\text{Tr } m^{(j=1/2)} \left(m^{(j=1/2)}\right)^\dag = &
F_{ij} \bar F^{ij} + 4 | D_{ai}|^2
\\
\text{Tr } \left(m^{(j=0)}\right)^2 = &
2 F^{ij} \bar F_{ij} + 2D_a^i D_{ai} + 2D_a D_{ai}^i
\\
\text{Tr } \left(m^{(j=1)}\right)^2 = &
2 D_{ai} D_a^i.
\end{align}
For convenience, we may define the \textbf{supertrace}, a sum of the squared mass matrices weighted by the number of states,
\begin{align}
\text{STr } \left(m^{(j)}\right)^2 \equiv &
\sum_j \text{Tr } (2j+1)(-)^{2j} m^2\\
=& -2 F\bar F - u |D_{ai}|^2 + 2 F\bar F + 2 D_{a}^iD_{ai} + 2 D_a D_{ai}^i + 3\cdot 2 D_{ai}D_a^i\\
=& 2D_a(D_a)^i_i\\
=& 2 D_{\tilde a} \sum_i q_i^{(\tilde a)}
\end{align}
Note that $\langle D_a\rangle \neq 0$ only for U(1) factors, so $(D_{a})^i_i = \sum q_i$, the sum of all U(1) charges. We have written $\tilde a$ to index only the U(1) factors of the gauge group. Note, however, that usually
\begin{align}
\sum_i q_i^{(\tilde a)} = 0
\end{align}
due to anomaly cancellation. This leads to the very stringent constraint that
\begin{align}
\text{STr } m^2 = 0.
\label{eq:SUSY:supertrace}
\end{align}
Note that this is a tree-level result that assumes renormalizable interactions\footnote{Non-renormalizable terms in the K\"ahler potential, for example, modify how the superpotential terms contribute to the scalar potential since one has to rescale fields for them to be canonically normalized.}. 

\subsection{Soft breaking and the MSSM}
\label{sec:SUSY:MSSM:soft:breaking}

The sum rule (\ref{eq:SUSY:supertrace}) is a road block to \SUSY model building. To see why, consider the scalar mass matrix (\ref{eq:SUSY:scalar:mass}) applied to squarks. In order to preserve SU(3)$_c$, the squarks should not obtain a \vev. This implies that the $D$-terms vanish, $D_a^i = D_\text{color} = 0$, for squarks. Thus further means that quarks only get their masses from the superpotential. 

Similarly preserving U(1)$_\text{EM}$ implies that the $D$-terms corresponding to the electrically charged SU(2)$_L$ directions should also vanish: $D_\pm = D_{1,2} = 0$. This means that the only $D$-terms which are allowed to be non-trivial are $D_3$ and $D_Y$, corresponding to the third generator of $SU(2)_L$ and hypercharge. The scalar mass matrix for the up-type quarks is then
\begin{align}
m^2_{2/3} &=
\begin{pmatrix}
m_{2/3}m_{2/3}^\dag + \left(\frac 12 g D_3 + \frac 16 g' D_Y \right) \mathbbm{1} &
\Delta\\
\Delta^\dag & 
m_{2/3}m_{2/3}^\dag - \frac 23 g' D_Y\mathbbm{1}
\end{pmatrix}
\label{eq:SUSY:sum:rule:m23}
\\
m^2_{1/3} &=
\begin{pmatrix}
m_{1/3}m_{1/3}^\dag + \left(-\frac 12 g D_3 + \frac 16 g' D_Y \right) \mathbbm{1} &
\Delta'\\
\Delta'^\dag & 
m_{1/3}m_{1/3}^\dag + \frac 13 g' D_Y\mathbbm{1}
\end{pmatrix},
\label{eq:SUSY:sum:rule:m13}
\end{align}
where the $\Delta$ and $\Delta'$ are the appropriate expressions from (\ref{eq:SUSY:scalar:mass}) and $m_{2/3,1/3}$ correspond to the quadratic terms in the superpotential that contribute to the quark masses. 

Charge conservation requires the sum of $D$ terms to vanish, so that at least one $D$ term is less than or equal to zero. 
For example, suppose that
\begin{align}
\frac 12 g D_3 + \frac 16 g' D_Y \leq 0.
\label{eq:SUSY:sum:rule:suppose:D:term}
\end{align}
Let $\beta$ be the direction in field space corresponding to the up quark. Then $\beta$ is an eigenvector of the quark mass matrix $m_{2/3}$ with eigenvalue $m_u$. Then (\ref{eq:SUSY:sum:rule:suppose:D:term}) implies that
\begin{align}
\begin{pmatrix}
\beta^\dag & 0
\end{pmatrix}
m^2_{2/3}
\begin{pmatrix}
\beta \\
0
\end{pmatrix}
\leq m_u^2.
\end{align}
This implies that there exists a squark in the spectrum that has a tree-level mass less than the up quark. Such an object would have been discovered long ago and is ruled out. 

More generally, the observation that there is at least one negative $D$-term combined with the form of the squark matrices (\ref{eq:SUSY:sum:rule:m23}) and (\ref{eq:SUSY:sum:rule:m13}) implies that there must exist a squark with mass less than or equal to either $m_u$ or $m_d$. 
Thus even if \SUSY is broken, it appears that any supersymmetric version of the Standard Model is doomed to be ruled out at tree level.

In order to get around this restriction, one typically breaks \SUSY in a separate \textbf{supersymmetry breaking sector}  (\cancel{\SUSY}) that is not charged under the Standard Model gauge group. This \cancel{\SUSY} sector still obeys a sum rule of the form (\ref{eq:SUSY:supertrace}) but the spectrum is no longer constrained by observed \SM particles. In order for the \cancel{\SUSY} sector to lend masses to the \SM superpartners, one assumes the existence of a \textbf{messenger sector} which interacts with both the \SM and the \cancel{\SUSY} sectors. The messenger sector transmits the \SUSY-breaking auxiliary field \vev to the \SM sector and allows the \SM superpartners to become massive without violating the sum rule (\ref{eq:SUSY:supertrace}). Note that this also allows a large degree of agnosticism about the details of the \cancel{\SUSY} sector---as far as the phenomenology of the \MSSM is concerned, we only need to know about the \cancel{\SUSY} scale and the properties of the messenger sector. 

There are two standard types of assumptions for the messenger sector depending on how one assumes it couples to the \SM:
\begin{itemize}
\item \textbf{Gravity mediation}: here one assumes that the \SM and \cancel{\SUSY} breaking sectors only communicate gravitationally. The details of these interactions fall under the theory of local supersymmetry, or \textbf{supergravity} (\textsc{sugra}), but are typically not necessary for collider phenomenology. %
\item \textbf{Gauge mediation}: The messenger sector contains fields which are charged under the \SM gauge group.
\end{itemize}
An alternative way around the \cancel{\SUSY} sum rule is to construct a `single sector' model based on strong coupling~\cite{ArkaniHamed:1997fq, Luty:1998vr}. These turn out to be dual to 5D models of \SUSY breaking using tools that we introduce in Section~\ref{sec:XD}~\cite{Gabella:2007cp}.

Often we are only interested in the properties of the Standard Model particles and their superpartners. We can `integrate out' the details of the messenger sector and parameterize \SUSY breaking into non-renormalizable interactions. 
As an example, suppose that a superfield, $X$, breaks supersymmetry by picking up an $F$-term \vev: $\langle X\rangle = \cdots + \langle F \rangle \theta^2$. $X$ may also have a scalar \vev, but this does not break \SUSY. We then parameterize the types of non-renormalizable couplings that are generated when we integrate out the messenger sector.  We have four types of terms:
\begin{enumerate}
\item \textbf{Non-holomorphic scalar masses} are generated by higher order K\"ahler potential terms such as
\begin{align}
\int d^4\theta \; \frac{X^\dag X}{M^2}\, \Phi^\dag \Phi &=  \left(\frac FM \right)^2 \varphi^* \varphi + (\text{\SUSY preserving terms}).
\label{eq:SUSY:breaking:non:holomorphic:soft:scalar:mass}
\\
\int d^4\theta \; \frac{X+X^\dag}{M}\, \Phi^\dag \Phi  &=  \left(\frac{F^*}{M} \right) \int d^2 \theta\; \Phi^\dag \Phi + \text{h.c.} + (\text{\SUSY preserving terms}).
\label{eq:SUSY:breaking:non:holomorphic:soft:scalar:mass:2}
\end{align}
We have written the \SUSY-breaking part of (\ref{eq:SUSY:breaking:non:holomorphic:soft:scalar:mass:2}) suggestively to appear as a non-holomorphic superpotential term. %This, however, is a \SUSY-breaking term and so does not violate the holomorphy of the superpotential imposed by \SUSY.
Since $\Phi^\dag$ only contains $\bar\theta$s and not $\theta$, $\int d^2\theta\, \Phi^\dag \Phi = \varphi^*F_\varphi = \varphi^* W'[\varphi^*]$. For renormalizable superpotentials, this can give an $A$-term of the form (\ref{eq:SUSY:breaking:A:terms}) or a $b$-term of the form (\ref{eq:SUSY:breaking:holomorphic:soft:scalar:mass}) below; the latter with a slightly different scaling with $F$.

The mass scale $M$ is required by dimensional analysis and is naturally the scale of the mediator sector that has been integrated out. For gravity mediation $M\sim M_\text{Pl}$ while for gauge mediation $M\sim M_\text{mess}$, the mass of the messenger fields. Doing the Grassmann integral and picking the terms that depend on the \SUSY breaking order parameter $F$ gives a mass $m^2 = (F/M)^2$ to the scalar $\varphi$. Note that $F$ has dimension 2 so that this term has the correct mass dimension.
\item $\textbf{Holomorphic scalar masses}$ are generated by a similar higher order K\"ahler potential term,
\begin{align}
\int d^4\theta \; \frac{X^\dag X}{M^2}\, \left[\Phi^2+\left(\Phi^\dag\right)^2\right] = \left(\frac FM \right)^2 \left(\varphi^2 + \varphi^{*2}\right) + (\text{\SUSY preserving terms}).
\label{eq:SUSY:breaking:holomorphic:soft:scalar:mass}
\end{align}
These are often called $b$-terms. %
One may want to instead write these masses at lower order in $F$ by writing a superpotential term $W \supset X\Phi^2$. This, however, is a renormalizable interaction that does not separate the \cancel{\SUSY} sector from the visible sector---as one can see the mediator mass does not appear explicitly in such a term. Thus $W \supset X\Phi^2$ is subject to the \SUSY sum rule and is not the type of soft term we want for the \MSSM.
\item \textbf{Holomorphic cubic scalar interactions} are generated from the superpotential,
\begin{align}
\int d^2\theta \; \frac{X}{M} \Phi^3 + \text{h.c.} =
\frac{F}{M}\left(\varphi^3 + \varphi^{*3}\right) + (\text{\SUSY preserving terms}).\label{eq:SUSY:breaking:A:terms}
\end{align}
These are called $A$-terms  and are the same order as the scalar mass.
\item \textbf{Gaugino masses} are generated from corrections to the gauge kinetic term,
\begin{align}
\int d^2\theta \; \frac{X}{M} \mathcal{W}_\alpha \mathcal{W}^\alpha + \text{h.c.} =
\frac{F}{M} \lambda\lambda + \text{h.c.} + (\text{\SUSY preserving terms}).
\label{eq:SUSY:MSSM:gaugino:mass}
\end{align}
This is a gaugino mass on the same order as the scalar mass and the $A$-term.
\end{enumerate}
In principle one could also generate tadpole terms for visible sector fields, but we shall ignore this case and assume that all field are expanded about their minimum.
These four types of terms are known as \textbf{soft supersymmetry breaking} terms. The key point is that these do not reintroduce any quadratic \UV sensitivity in the masses of any scalars. This is clear since above the \SUSY breaking mediation scale $M$, the theory is supersymmetric and these divergences cancel. 

It is common to simply parameterize the soft breaking terms of the \MSSM in the Lagrangian:
\begin{align}
\mathcal L_\text{soft} =&
-\frac 12 \left( M_3 \tilde g\tilde g + M_2 \tilde W\tilde W + M_1 \tilde B\tilde B \right) +\text{h.c.}
\\&
- \left(a_u \tilde Q H_u \tilde{\bar u} + a_d \tilde Q H_d \tilde{\bar d} + a_e \tilde L H_d \tilde{\bar e} \right) + \text{h.c.}
\\&
-\tilde Q^\dag m_Q^2 \tilde Q -\tilde L^\dag m_L^2 \tilde L - \tilde u^\dag m_u^2 \tilde{\bar u} - \tilde{d}^\dag m_d^2 \tilde{\bar d} - \tilde{e}^\dag m_e^2 \tilde{\bar e}
- m_{H_u}^2 H_u^* H_u - m_{H_d}^2 H_d^*H_d
\\&
-\left(b H_u H_d +  \text{h.c.})\right).
\end{align}
This is simply a reparameterization of the types of soft terms described in (\ref{eq:SUSY:breaking:non:holomorphic:soft:scalar:mass} -- \ref{eq:SUSY:MSSM:gaugino:mass}), from which one can read off the scaling of each coefficient with respect to $F/M$.

Note that the trilinear soft terms, $a_{u,d,e}$, and the soft masses $m^2_{Q,L,u,d,e}$ are $3\times 3$ matrices in flavor space. The trilinear terms are in a one-to-one correspondence with the Yukawa matrices except that they represent a coupling between three scalars. In general, the soft masses cause the squarks and sleptons to have different mass eigenstates than the \SM fermions.

Phenomenologically, we assume that
\begin{align}
M_{1,2,3},\; a_{u,d,e} \quad &\sim\quad m_\text{\SUSY}\\
m^2_{Q,u,d,L,e,H_u,H_d},\; b \quad &\sim\quad m_\text{\SUSY}^2,
\end{align}
where $m_\text{\SUSY}$ is between a few hundreds of \GeV to a \TeV. This is the range in which generic \MSSM-like models provide a solution to the Hierarchy problem.

\begin{framed}
\noindent \footnotesize
\textbf{$R$-symmetry, gauginos, supersymmetry breaking.} Recall that when an $R$-symmetry exists, the different components of a superfield carry different $R$ charges. %From algebra with the \SUSY generators, one can see that the superspace coordinate $\theta$ carries unit $R$ charge so that the superpotential must have $[W]_R = 2$.
Because the $\mathcal O(\theta)$ component of $\mathcal W^\alpha$, $F_{\mu\nu}$, is real, it cannot carry an $R$ charge. This means that the lowest component, the gaugino $\lambda$, must have non-zero $R$-charge. Further, the gaugino mass term (\ref{eq:SUSY:MSSM:gaugino:mass}) breaks this symmetry. One will find that $R$-symmetry plays an important role in many non-perturbative results in \SUSY. Two important results related to \SUSY breaking and gaugino masses are \cite{Komargodski:2009jf, Nelson:1993nf}.
\end{framed}

\subsection{Electroweak symmetry breaking in the MSSM}

The most important feature of the Standard Model is electroweak symmetry breaking. Recall that this is due to a tachyonic Higgs mass at the origin being balanced by a positive quartic coupling leading to a non-zero vacuum expectation value.
In the \MSSM we have two Higgs doublets,
\begin{align}
H_u &= 
\begin{pmatrix}
H^+_u\\
H^0_u
\end{pmatrix}
&
H_d &=
\begin{pmatrix}
H^0_d\\
H^-_d
\end{pmatrix}.
\end{align}
We have already seen that supersymmetry relates the scalar quartic coupling to the other couplings of the theory. This then constrains the expected Higgs boson mass.

To preserve SU(3)$_\text{c}$ and U(1)$_\text{EM}$ we assume that no squarks or sleptons pick up \vevs. Then the quartic terms in the Higgs potential come from $D$-terms, (\ref{eq:SUSY:D-terms}):
\begin{align}
V_D 
&= \frac{g^2}{4}
\left(H_u^\dag \sigma^a H_u + H_d^\dag \sigma^a H_d\right)
\left(H_u^\dag \sigma^a H_u + H_d^\dag \sigma^a H_d\right)
+ \frac{g'^2}{4} \left( |H_u|^2 - |H_d|^2\right)^2
\nonumber
\\
&= \frac 12 g^2 |H_u^\dag H_d|^2 + \frac 18 (g^2 + g'^2) \left(|H_u|^2 - |H_d|^2\right)^2,
\label{eq:SUSY:MSSM:EWSB:D}
\end{align}
where we have simplified the SU(2)$_\text{L}$ terms using the relation $\sigma_{ij}^a\sigma_{k\ell}^a = 2\delta_{i\ell}\delta_{jk}-\delta_{ij}\delta_{k\ell}$.
We see immediately that the Higgs quartic $\lambda$ coupling goes like the squared electroweak couplings, $g^2$ and $g'^2$. %
This connection between the Higgs sector and the gauge parameters does not exist in the Standard Model

In addition to the $D$-term contribution, there is also the supersymmetric $F$-term contribution coming from the $\mu$-term in the superpotential. The quadratic contributions to the Higgs potential are,
\begin{align}
V_F = |\mu|^2 |H_u|^2 + |\mu|^2 |H_d|^2 + \cdots
\label{eq:SUSY:MSSM:EWSB:F}
\end{align}
We have dropped terms proportional to the Yukawa couplings since we assume the scalar partners of the \SM fermions do not acquire \vevs. 
On top of this, there are the soft supersymmetry breaking terms. These include soft masses for each Higgs doublet and a `holomorphic' $b$-term which is called $B_\mu$ (or sometimes $B\mu$),
\begin{align}
V_\text{soft} = m_{H_u}^2 |H_u|^2 + m_{H_d}^2 |H_d|^2 + \left(B_\mu H_u\cdot H_d + \text{h.c.}\right).
\label{eq:MSSM:Vsoft}
\end{align}
Note that the contraction of $H_u$ and $H_d$ in the $D$-term (\ref{eq:SUSY:MSSM:EWSB:D}) is different from that in the $B_\mu$ term (\ref{eq:SUSY:MSSM:EWSB:soft}). Specifically, $H_u \cdot H_d$ is contracted using the $\epsilon_{ab}$ tensor and gives $H_u^+H_d^- - H_u^0H_d^0$. Further, the $D$-term couplings are real since they are part of a real superfield. The $F$-term couplings are made real because they are the modulus of a complex parameter. The couplings of the soft terms, on the other hand, carry arbitrary sign and phase.

Combining all of these factors, the full Higgs potential is
\begin{align}
V_H = &
\phantom{+}
V_D + V_F + V_\text{soft}\\
=&
\phantom{+}
\frac 12 g^2 |H_u^\dag H_d|^2 + \frac 18 (g^2 + g'^2) \left(|H_u|^2 - |H_d|^2\right)^2\nonumber\\
&
+
\left(|\mu|^2 + m_{H_u}^2\right) |H_u|^2 + 
\left(|\mu|^2 + m_{H_u}^2\right) |H_d|^2 
+ \left(B_\mu H_u\cdot H_d + \text{h.c.}\right).
\label{eq:SUSY:MSSM:EWSB:soft}
\end{align}
To simplify this, we can assume that the charged components of the doublets pick up no \vev and write everything in terms of only the neutral components (we address the validity of this assumption below):
\begin{align}
V_H =&\phantom{+}
\frac 18 (g^2 + g'^2) \left(|H_u^0|^2 - |H_d^0|^2\right)^2 
%&
+\sum_{i=u,d} 
\left(|\mu|^2 + m_{H_i}^2\right) |H^0_i|^2
-2B_\mu \text{Re}(H^0_uH^0_d).
\end{align}
Observe that this potential has a direction in field space, $|H_u^0|^2 = |H_d^0|^2$ where the $D$-term quartic vanishes. This is called a \textbf{D-flat} direction and requires caution. In order to break electroweak symmetry, we must destabilize the origin of field space with a tachyonic mass term to force a linear combination of the neutral Higgses to pick up a \vev. In the \SM destabilization is balanced by the quartic coupling which forces the \vev to take a finite value. We see now in the \MSSM that one has to take special care to make sure that the destabilized direction does \textit{not} align with the $D$-flat direction or else the potential isn't bounded from below. In other words, we must impose a negative mass squared in one direction in the Higgs moduli space while making sure that there is a positive definite mass squared along the $D$-flat direction. This can be written as two conditions:
\begin{enumerate}
\item We require exactly one negative eigenvalue in the neutral Higgs mass matrix,
\begin{align}
\begin{vmatrix}
|\mu|^2 + m_{H_u}^2 & - B_\mu \\
- B_\mu & |\mu|^2 + m_{H_d}^2  
\end{vmatrix}
= 
\left(|\mu|^2 + m_{H_u}^2\right)
\left(|\mu|^2 + m_{H_d}^2\right)
- B_\mu^2
<
0.
\label{eq:SUSY:MSSM:EWSB:unstable}
\end{align}
\item The mass squared term is positive when $|H^0_u| = |H^0_d|$. For simplicity, suppose $B_\mu$, $\langle H^0_u\rangle$, and $\langle H^0_d\rangle$ are all real (see below). Then this imposes
\begin{align}
\left(|\mu|^2 + m_{H_u}^2\right)
+ \left(|\mu|^2 + m_{H_d}^2\right)
+2B_\mu >0.
\label{eq:SUSY:MSSM:EWSB:stable}
\end{align}
\end{enumerate}
The conditions (\ref{eq:SUSY:MSSM:EWSB:unstable}) and (\ref{eq:SUSY:MSSM:EWSB:stable}) are the requirements for electroweak symmetry breaking in the \MSSM.
%%
%\begin{framed}
%\noindent \footnotesize 
%Before moving on, let us justify some assumptions that we have made in this subsection:
%\begin{itemize}
%\item \textbf{The Higgs \vev doesn't break electric charge, $\langle H^\pm\rangle=0$.} Above we assumed that the charged Higgses do not pick up background values. One might worry whether or not such an assumption is reasonable. A simple way to do this is to use SU(2)$_L$ invariance to pick $\langle H_u^+\rangle = 0$. At the minimum of the potential, $\partial V/\partial H^+_u =0$, one can check that this generically gives $H^-_d=0$. An alternate possibility is that
%$$
%B_\mu + \frac{g^2}{2} H^{0\dag}_uH^{0\dag}_d = 0,
%$$
%though this is unfavorable to electroweak symmetry breaking since the $B_\mu$ lifts both flat directions.
%%
%\item \textbf{$B_\mu$, $\langle H^0_u\rangle$, and $\langle H^0_d\rangle$ are real and positive}. First note that by rephasing the fields we may absorb the phase of $B_\mu$ into the product $H^0_u H^0_d$. Then observe that for fixed moduli, $|H^0_u|$ and $|H^0_d|$,   $-B_\mu\text{Re}\left(2H_u^0 H_d^0\right)$ is minimized when $H^0_u$ and $H^0_d$ have opposite phases. Since $H^0_u$ and $H^0_d$ also have opposite hypercharges, we may use a U(1)$_Y$ transformation to rotate away this relative phase. Thus the Higgs sector does not break \CP and we may choose $B_\mu$ and the neutral Higgs \vevs to be real and positive.
%\end{itemize}
%\end{framed}

Note that a natural choice for the soft masses, $m_{H_u}^2 = m_{H_d}^2$, does not obey the restrictions (\ref{eq:SUSY:MSSM:EWSB:unstable}) and (\ref{eq:SUSY:MSSM:EWSB:stable}). One way to nevertheless enforce this relation is to impose it as a boundary condition at some high scale and allow the renormalization group flow to differentiate between them. This is actually quite reasonable, since the $\beta$-function for these soft masses include terms that go like the squared Yukawa coupling. The two soft masses flow differently due to the large difference in the top and bottom Yukawas. In fact, the up-type Higgs mass parameter shrinks in the \IR and it is natural to assume
\begin{align}
m_{H_u}^2 < m_{H_d}^2.
\end{align}
A convenient choice is $m_{H_u}^2 <0$ and $m_{H_d}^2 >0$. In this way the \MSSM naturally admits \textbf{radiative electroweak symmetry breaking} where the tachyonic direction at the origin is generated by quantum effects.

Since there are many parameters floating around, it use useful to summarize that the following all prefer electroweak symmetry breaking and no runaway directions:
\begin{itemize}
\item Relatively large $B_\mu$
\item Relatively small $\mu$
\item Negative $m_{H_u}^2$.
\end{itemize}
Be aware that these are only rough guidelines and are neither necessary nor sufficient.

It is standard to parameterize the \vevs of the two Higgses relative to the \SM Higgs \vev by introducing an angle, $\beta$,
\begin{align}
\langle H^0_u \rangle = \frac{v_u}{\sqrt{2}} = \frac{v}{\sqrt{2}} \sin\beta & &
\langle H^0_d \rangle = \frac{v_d}{\sqrt{2}} = \frac{v}{\sqrt{2}} \cos\beta.
\end{align}
Minimizing the potential, $\partial V/\partial H^0_u = \partial V/\partial H^0_d =0$, one obtains
\begin{align}
\sin 2\beta &= \frac{2B_\mu}{2|\mu|^2 + m_{H_u}^2 + m_{H_d}^2}
\label{eq:SUSY:MSSM:muBmu}
\\
\frac{M_Z^2}{2} &= -|\mu|^2 + \frac{m_{H_d}^2 - m_{H_u}^2 \tan^2\beta}{\tan^2\beta - 1}.
\label{eq:SUSY:MSSM:weird:cancellation}
\end{align}
The second relation is especially strange: it connects the supersymmetric $\mu$ term to the soft-breaking masses, even though these come from totally different sectors of the theory. In other words, unlike the quartic and gauge couplings which are tied together by supersymmetry, these parameters have no reason to have any particular relation with each other. Further, $M_Z^2$ is experimentally measured and much smaller than the typical expectation for either $\mu$ or $m_{H_{u,d}}^2$, so it appears that there's some cancellation going on. %This is known as the Little Hierarchy problem.

The Higgs sector of the \MSSM contains the usual \CP-even Higgs $h$, a heavier \CP-even Higgs, the Goldstones of electroweak symmetry breaking, an additional pair of charged Higgses $H^\pm$, and a \CP-odd Higgs $A$. With a little work, one can show that the \CP-even Higgs masses are
\begin{align}
m_h^2 = \frac 12 \left[
M_Z^2 + m_A^2 \pm \sqrt{\left(M_Z^2 + m_A^2\right)^2 - 4m_A^2 M_Z^2 \cos^2 2\beta}
\right],
\end{align}
where $m_A^2 = B_\mu/(\sin\beta \cos\beta)$. One can further show that this is bounded from above,
\begin{align}
m_{h} \leq M_Z \left| \cos 2\beta \right| \leq M_Z.
\label{eq:SUSY:mh:tree:bound}
\end{align}
Of course, we now know that $m_h \approx 125$ \GeV. In fact, even before the \LHC it was known from \LEP that $m_h \gtrsim 114$ \GeV. While at first glance (\ref{eq:SUSY:mh:tree:bound}) appears to be ruled out experimentally, this is only a tree-level bound. What this is really saying is that one requires large corrections to the quartic self-coupling to pull up the Higgs mass from its tree level value. Due to the size of $y_t$, the main effect comes from top and stop loops. 

To maximize the quartic coupling, we are pushed towards large values of $\tan \beta$ since this would put most of the Higgs \vev in $H_u$ and would make the light Higgs be primarily composed of $H_u$. Examining the $H_u^4$ coupling at loop level, consider diagrams of the form:
\begin{center}
	\begin{tikzpicture}[line width=1.5 pt, scale=1]
		\draw[scalar] (-1,2.3) -- (0,2);
		\draw[fermion] (0,2) -- (2,2);
		\draw[scalar] (2,2) -- (3,2.3);
		\draw[scalarbar] (-1,-.3) -- (0,0);
		\draw[fermionbar] (0,0) -- (2,0);
		\draw[scalarbar] (2,0) -- (3,-.3);
		\draw[fermion] (0,2) -- (0,0);
		\draw[fermionbar] (2,2) -- (2,0);
		\node at (-1.3,2.4) {$H_u^0$};
		\node at (3.3,2.4) {$H_u^0$};
		\node at (-1.3,-.4) {$H_u^0$};
		\node at (3.3,-.4) {$H_u^0$};
		\node at (-.5 ,1) {$t_R$};
		\node at (2.5 ,1) {$t_L$};
		\node at (1 ,2.4) {$t_L$};
		\node at (1 ,-.3) {$t_R$};
	\end{tikzpicture}
	\qquad\qquad
	\begin{tikzpicture}[line width=1.5]
    	\draw[scalar] (-1,2.3) -- (1,2);
    	\draw[scalar] (1,2) -- (3,2.3);
    	\draw[scalarbar] (-1,-.3) -- (1,0);
    	\draw[scalarbar] (1,0) -- (3,-.3);
	    \draw[scalar] (1,2) arc (90:270:1);
	    \draw[scalar] (1,0) arc (270:450:1);
	    \node at (-1.3,2.4) {$H_u^0$};
		\node at (3.3,2.4) {$H_u^0$};
		\node at (-1.3,-.4) {$H_u^0$};
		\node at (3.3,-.4) {$H_u^0$};
		\node at (-.6,1) {$\tilde t_{L,R}$};
		\node at (2.6,1) {$\tilde t_{L,R}$};
	\end{tikzpicture}
\end{center}
Assuming negligible $A$ terms, the result is
\begin{align}
\lambda(m_t) = \lambda_\text{SUSY} + \frac{2 N_c y_t^4}{16\pi^2} \ln \left(\frac{m_{\tilde t_{1}} m_{\tilde t_{2}}}{m_t^2}\right),
\end{align}
where $\lambda_\text{SUSY}$ comes from the $D$-term potential and $N_c$ is the number of colors. This equation tells us that in order to push the Higgs mass above the tree-level bound of $M_Z$, one must increase $m_{\tilde{t}}$. The correction is 
\begin{align}
\Delta m_h^2 = 
\frac{3}{4\pi^2} v^2 y_t^4 \sin^2 \beta
\ln \left(\frac{m_{\tilde t_1} m_{\tilde t_2}}{m_t^2}\right).
\end{align}
For further details, we refer to the treatment in \cite{Drees:2004jm} or the encyclopedic reference \cite{Djouadi:2005gj}.

\subsection{The little hierarchy problem of the MSSM}

It has been well known since \LEP that in order to push $m_h > 114$ \GeV in the \MSSM, one requires large stop masses, $m_{\tilde t} \sim 1 - 1.4$ \TeV.
Pushing the stop mass this heavy comes at a cost, unfortunately. The stops contribute not only to the Higgs quartic---which we need to push the Higgs mass up---but also to the soft mass $m_{H_u}^2$ from loops of the form
\begin{center}
\begin{tikzpicture}[line width=1.5 pt, scale=1.5]
  	\begin{scope}[shift={(2,0)}]
    \draw[scalarnoarrow] (-.2,-.5) -- (1.3,-.5);
    \draw[scalarnoarrow] (.55,-.1) circle (.4);
  	\end{scope}
  	\begin{scope}[shift={(5,0)}]
  	\node at (-1.4,0) {\Large{+}};
%    \draw[scalarnoarrow] (-.2,-.5) -- (1.3,-.5);
%    \draw[vector] (.55,-.1) circle (.4);
    \draw[fermion] (180:.4) arc (180:0:.4);
	\draw[fermion] (0:.4) arc (0:-180:.4);
	\draw[scalarnoarrow] (-.4,0) -- (-.9,0);
	\draw[scalarnoarrow] (.4,0) -- (.9,0);
  	\end{scope}
 \end{tikzpicture}
 \end{center}
 The larger one sets $m_{\tilde t}$, the larger the shift in $m_{H_u}^2$. Recall, however, the strange cancellation we noted in (\ref{eq:SUSY:MSSM:weird:cancellation}). This equation seems to want $m_{H_u}^2 \sim M_Z^2/2$. The loop corrections above contribute a shift of the form
 \begin{align}
 \Delta m_{H_u}^2 = \frac{3y_t^2}{4\pi^2} m_{\tilde t}^2 \ln \left(\frac{\Lambda_\text{UV}}{m_{\tilde t}}\right).
 \end{align}
For $m_{\tilde t} = 1.2$ \TeV and $\Lambda_\text{UV} = 10^{16}$ this balancing act between $m_{H_u}^2$ and $M_Z^2/2$ requires a fine tuning of %
$$\frac{M_Z^2/2}{\Delta m_{H_u}^2}\sim 0.1\%.$$
Physically what's happening is that the stop plays a key role in naturalness by canceling the sensitivity to the \UV scale. By pushing the stop to be heavier to increase the Higgs quartic, one reintroduces quadratic sensitivity up to the scale of the stop mass. This is known as the \textbf{little hierarchy problem} of the \MSSM.

\subsection{SUSY breaking versus flavor}
\label{sec:MSSM:SUSY:flavor}

The soft breaking Lagrangian introduces many new masses, phases, and mixing angles on top of those found in the Standard Model for a total of 124 parameters \cite{Haber:1997if}. Most of this huge parameter space, however, is already  excluded from flavor and \CP violating processes. 
Recall that in the \SM, there are no tree-level flavor-changing neutral currents (\FCNC) and loop-level contributions are suppressed by the \GIM mechanism. Lepton number violation is similarly strongly suppressed. 
In the limit where the Yukawa couplings vanish, $y \to 0$, the Standard Model has a U(3)$^5$ flavor symmetry where each of the five types of matter particles are equivalent. This flavor symmetry is presumably broken at some scale $\Lambda_F$ in such a way that the only imprint of this \UV physics at scales well below $\Lambda_F$ are the Yukawa matrices. This flavor scale can be very large so that effects of this flavor breaking go like $1/\Lambda_F$ and are plausibly very small.

In the \MSSM, one must further 
check that the flavor breaking dynamics has already `frozen out' at the \SUSY breaking scale so that the only non-trivial flavor structure in the \SUSY breaking parameters are the Yukawa matrices themselves. This means we would like the mediator scale $M$ to be below the flavor scale, $M \ll \Lambda_F$.
In gravity mediation, however, $\Lambda_\text{med} = M_\text{Pl}$, and we can no longer guarantee that the \SUSY breaking mediators are insulated from flavor violating dynamics. %Even if these parameters are flavor universal at the \SUSY breaking scale, one expects loop effects from the flavor breaking sector to be large and unsuppressed. 
This leads to strong constraints on the flavor structure of the \MSSM soft parameters.
%% Comment: There are other "gravity mediation" games you can play, e.g. putting this in an extra dimension with branes, but the point is that you really need something new.

For example, consider one of the most carefully studied \FCNC processes, kaon anti-kaon ($K$-$\bar K$) mixing. The quark content of the mesons are $K=d\bar s$ and $\bar K = \bar d s$. In the \SM this process is mediated by diagrams such as
\begin{center}
	\begin{tikzpicture}[line width=1.5 pt, scale=1]
		\draw[fermion] (-1,2.3) -- (0,2);
		\draw[fermion] (0,2) -- (2,2);
		\draw[fermion] (2,2) -- (3,2.3);
		\draw[fermionbar] (-1,-.3) -- (0,0);
		\draw[fermionbar] (0,0) -- (2,0);
		\draw[fermionbar] (2,0) -- (3,-.3);
		\draw[vector] (0,2) -- (0,0);
		\draw[vector] (2,2) -- (2,0);
		\node at (-1.25,2.4) {$d$};
		\node at (3.25,2.4) {$s$};
		\node at (-1.25,-.4) {$s$};
		\node at (3.25,-.4) {$d$};
		\node at (-.5 ,1) {$W$};
		\node at (2.5 ,1) {$W$};
		\node at (1 ,2.3) {$u_i$};
		\node at (1 ,-.3) {$u_j$};
	\end{tikzpicture}
\end{center}
Each vertex picks up a factor of the \CKM matrix. The \GIM observation is the fact that the unitarity of the \CKM matrix imposes an additional suppression. In the \MSSM, on the other hand, the squark soft masses introduce an additional source of flavor violation so that the quark and squark mass matrices are misaligned. This manifests itself as flavor-changing mass insertions, $\Delta m_{ds}^2$, on squark propagators when written in terms of the Standard Model mass eigenstate combinations:
\begin{center}
	\begin{tikzpicture}[line width=1.5 pt, scale=1]
		\draw[fermion] (-1,2.3) -- (0,2);
		\draw[scalar] (0,2) -- (1,2);
       \draw[scalarbar] (1,2) -- (2,2);
       \begin{scope}[shift={(1,2)}] % Mass Insertion
		    \clip (0,0) circle (.175cm);
		    \draw[fermionnoarrow] (-1,1) -- (1,-1);
		    \draw[fermionnoarrow] (1,1) -- (-1,-1);
	    \end{scope}
		\draw[fermionbar] (2,2) -- (3,2.3);
		\draw[fermionbar] (-1,-.3) -- (0,0);
		\draw[scalarbar] (0,0) -- (1,0);
       \begin{scope}[shift={(1,0)}] % Mass Insertion
		    \clip (0,0) circle (.175cm);
		    \draw[fermionnoarrow] (-1,1) -- (1,-1);
		    \draw[fermionnoarrow] (1,1) -- (-1,-1);
	    \end{scope}
		\draw[scalar] (1,0) -- (2,0);
		\draw[fermion] (2,0) -- (3,-.3);
		\draw[gluon, line width=.75 pt] (0,2) -- (0,0);
		\draw[gluon, line width=.75 pt] (2,2) -- (2,0);
%		\draw[line width=.75 pt] (0,2) -- (0,0);
%		\draw[line width=.75 pt] (2,2) -- (2,0);
		\draw[fermionbar] (0,2) -- (0,0);
		\draw[fermionbar] (2,2) -- (2,0);
		\node at (-1.25,2.4) {$d$};
		\node at (3.25,2.4) {$s$};
		\node at (-1.25,-.4) {$s$};
		\node at (3.25,-.4) {$d$};
		\node at (-.5 ,1) {$\tilde g$};
		\node at (2.5 ,1) {$\tilde g$};
		\node at (.5 ,2.4) {$\tilde d$};
		\node at (.5 ,-.3) {$\tilde s$};
		\node at (1.5 ,2.4) {$\tilde s$};
		\node at (1.5 ,-.3) {$\tilde d$};
	\end{tikzpicture}
\end{center}
Note that rather than $W$ bosons, this diagram is mediated by gluinos which carry much stronger coupling constants $\alpha_3 \gg \alpha_2$. Further, Since there are no factors of $V_\text{CKM}$, there is no \GIM suppression. The loop integral goes like $d^4k/k^{10} \sim 1/m_{\text{\SUSY}}^6$. Thus we can estimate this contribution to kaon mixing to be
\begin{align}
\mathcal M_{K\bar K}^{\text{\MSSM}} \sim \alpha_3^3 \left(\frac{\Delta m_{ds}^2}{m_{\text{\SUSY}}^2}\right)^2 \frac{1}{m_\text{\SUSY}^2}.
\end{align}
Comparing this to the experimental bound,
\begin{align}
\frac{\Delta m_{ds}^2}{m_\text{\SUSY}^2} 
\lesssim
4\cdot 10^{-3} \left(\frac{m_\text{\SUSY}}{500\text{ GeV}}\right).
\end{align}
There are similar constraints on \CP violating and lepton number violating processes (e.g.\ dipole moments and $\mu \to e \gamma$).
This is the \textbf{SUSY flavor problem}: a generic flavor structure for the \MSSM soft parameters is phenomenologically ruled out.
We are led to conclude that the off-diagonal flavor terms must be strongly suppressed to avoid experimental bounds.

One way to do this is to suppose an organizing principle in the \SUSY breaking parameters, \textbf{soft-breaking universality},
\begin{enumerate}
\item Soft breaking masses are all universal for all particles at some high scale. This means that $m_Q^2 \propto \mathbbm{1}$ in flavor space, and similarly for each \MSSM matter multiplet. 
\item If $a$-terms are not flavor-universal, then the Higgs \vev induces similar problematic mixings,
\begin{align}
\mathcal L_a = a^u_{ij} Q_i\bar U_j H_u
+ a^d_{ij} Q_i\bar D_j H_d
+ a^e_{ij} L_i\bar E_j H_d.
\end{align}
To avoid this, assume that $a^I_{ij}$ is proportional to the Yukawa matrix,
\begin{align}
a^I_{ij} = A^I y^I_{ij}.
\end{align}
This way, the rotation that diagonalizes the \SM fermions also diagonalizes their scalar partners. 
\item To avoid \CP violation, assume that all non-trivial phases beyond those in the Standard Model \CKM matrix vanish.
\end{enumerate}
These are phenomenological principles. Ultimately, one would like to explain why these properties should be true (or at least approximately so).

\subsection{Gauge mediated SUSY breaking}

One straightforward realization of soft-breaking universality is to have the messenger sector be flavor universal. A natural way to do this is \textbf{gauge mediation} since the \SM gauge fields are blind to flavor \cite{Dine:1982zb, Dine:1995ag, Dine:1994vc, Dine:1996xk}. See \cite{Giudice:1998bp} for a review.

\begin{center}
	\bigskip
	\begin{tikzpicture}[line width=1.5 pt, scale=.75]
 		\node at (0,0) [rectangle, draw=black, rounded corners, inner sep=.3cm]  (breaking) {$\cancel \SUSY$};
		\node at (5,0) [rectangle, draw=black, rounded corners, inner sep=.3cm] (messenger) {messenger};
		\node at (10,0) [rectangle, draw=black, rounded corners, inner sep=.3cm] (MSSM) {\MSSM};
		\draw [fermionnoarrow] (breaking) -- (messenger);
		\draw [vector] (messenger) -- (MSSM);
		\node at (0,-1) {$\langle F_X\rangle \neq 0$};
		\node at (5,-1) {$\Phi_i, \bar\Phi_i$};
		\node at (7.8,.75) {\footnotesize \SM gauge};
	 \end{tikzpicture}
	\bigskip	
\end{center}

The main idea is that the \SUSY breaking sector has some superfield (or collection of superfields) $X$ which pick up $F$-term \vevs, $\langle F_X\rangle \neq 0$. This generates mass splittings in the messenger sector superfields, $\Phi_i$ and $\bar\Phi_i$. These messengers obey the tree-level \SUSY sum rules discussed above but are not problematic since all of the components can be made heavy.
One then assumes that the messengers are charged under the \SM gauge group so that the \MSSM superfields will feel the effects of \SUSY breaking through loops that include the messenger fields. Note that anomaly cancellation of the \SM gauge group typically requires the messenger superfields to appear in vector-like pairs, $\Phi$ and $\bar\Phi$ with opposite \SM quantum numbers.

The messenger fields generate non-renormalizable operators that connect the \MSSM and the \SUSY breaking sector without introducing any flavor dependence for the soft masses. Further, because the messenger scale is adjustable, one can always stay in regime where it is parametrically smaller than the flavor scale $M \ll \Lambda_F$.  
Recall the estimates in Section~\ref{sec:SUSY:MSSM:soft:breaking} for the size of the \MSSM soft terms. For gauge mediation, $M$ is the mass of the messenger sector fields $\Phi_i$ and $\bar\Phi_i$ and $F$ is the \SUSY breaking \vev, $F_X$. Below $M$ we integrate out the messengers to generate the \MSSM soft parameters. %Assuming $\Lambda_F>M$, the soft parameters will be flavor universal.

The simplest realization of this is \textbf{minimal gauge mediation}. Here one assumes only one \SUSY breaking field $X$ and $N_m$ mediators, $\Phi_i$ and $\bar\Phi_i$, in the fundamental representation of an SU$(5)$ \GUT. The superpotential coupling between these sectors is
\begin{align}
%W = \lambda \bar\Phi X \Phi.
W = \bar\Phi X \Phi.
\end{align}
The contribution to the potential is
\begin{align}
%\left| \frac{\partial W}{\partial \Phi} \right|^2
%=&
%\left|\lambda \langle X \rangle\right|^2 |\varphi|^2
%+
%\left|\lambda \langle X \rangle\right|^2 |\bar\varphi|^2
%+
%\lambda \varphi\bar\varphi \langle F_X \rangle
\left| \frac{\partial W}{\partial \Phi} \right|^2
=&
\left|\langle X \rangle\right|^2 |\varphi|^2
+
\left|\langle X \rangle\right|^2 |\bar\varphi|^2
+
\varphi\bar\varphi \langle F_X \rangle
\end{align}
The messenger masses are
\begin{align}
%m_\psi &= \lambda X
%\\
%m_\varphi^2 &= \lambda^2 X^2 \pm \lambda F_X,
m_\psi &=  X
\\
m_\varphi^2 &=  X^2 \pm  F_X,
\end{align}
using the notation where the angle brackets $\langle\cdots \rangle$ are dropped when it is clear that we are referring to the \vev of a field. 
Observe that the messenger scale is set by the lowest component \vev of the \SUSY breaking parameter, $M=X$.
% Observe that this satisfies the \SUSY sum rule. 
%
In what follows we make the typical assumption that $F/M^2 \ll 1$.
%It is typical to assume $F/M^2 \ll 1$. % are we considering F/M^2 << 1? 
%\flip{Are we assuming $F \ll M^2$ in the `exact' formulas below?}
% Indeed, it doesn't make much sense to have F >> M^2.
%
Note that these masses satisfy the \SUSY sum rule.

Now let's consider the spectrum arising from this simple set up. The gauginos of the \SM gauge group pick up a mass contribution from diagrams of the form
\begin{center}
\begin{tikzpicture}[line width=1.5 pt, scale=1]
%    \draw[line width=.75 pt] (-2.75,0) -- (-1.25,0);
%    \draw[line width=.75 pt] (1.25,0) -- (2.75,0);
    \draw[fermion] (-2.75,0) -- (-1.25,0);
    \draw[fermionbar] (1.25,0) -- (2.75,0);

    \draw[vector, line width=1 pt] (-2.75,0) -- (-1.25,0);
	 \draw[vector, line width=1 pt] (2.75,0) -- (1.25,0);
	 \draw[scalar] (-1.25,0) arc (180:90:1.25);
	 \draw[scalar] (1.25,0) arc (0:90:1.25);
	 \draw[fermion] (-1.25,0) arc (180:270:1.25);
	 \draw[fermion] (1.25,0) arc (0:-90:1.25);
	  \begin{scope}[shift={(0,1.25)}] % Mass Insertion
		    \clip (0,0) circle (.175cm);
		    \draw[fermionnoarrow] (-1,1) -- (1,-1);
		    \draw[fermionnoarrow] (1,1) -- (-1,-1);
	    \end{scope}
	     \begin{scope}[shift={(0,-1.25)}] % Mass Insertion
		    \clip (0,0) circle (.175cm);
		    \draw[fermionnoarrow] (-1,1) -- (1,-1);
		    \draw[fermionnoarrow] (1,1) -- (-1,-1);
	    \end{scope}
	   \node at (0,1.75) {$\langle F_X\rangle$};
	   \node at (0,-1.65) {$\langle X\rangle$};
	   \node at (-45:1.7) {$\psi_{\bar\Phi}$};
	   \node at (225:1.7) {$\psi_{\Phi}$};
	   \node at (45:1.7) {$\bar\varphi$};
	   \node at (135:1.7) {$\varphi$};
	   \node at (-3,0) {$\lambda$};
	   \node at (3,0) {$\lambda$};
\end{tikzpicture}
\end{center}
The $\langle X\rangle$ insertion on the $\psi_\Phi$ line is required to flip the gaugino helicity (recall that arrows on fermion indicate helicity). The $F$ insertion on the $\varphi$ line is required to connect to \SUSY breaking so that this is indeed a mass contribution that is not accessible to the gauge boson. The $F$ \vev is also required to flip from a $\varphi$ to a $\bar\varphi$ so that the scalar of the chiral superfield picks up a sense of chirality as well. 
Using powerful methods based on holomporphy \cite{Giudice:1997ni, ArkaniHamed:1998kj}, the gaugino mass for the $i^{\text{th}}$ gauge factor is
\begin{align}
M_{\lambda_i} = \frac{F M}{M^2} \frac{g_i^2}{16\pi^2} N_m = \frac{\alpha_i}{4\pi} N_m \frac FM.
\end{align}
This expression---which one could have guessed from a back-of-the-envelope estimate---turns out to be exact to leading order in $F/M^2$. This is a reflection of the powerful renormalization theorems in supersymmetry, see e.g.\ \cite{Terning:2006uq}. One of the concrete predictions of minimal gauge mediation is the relation
\begin{align}
M_{\lambda_1} \; : \;
M_{\lambda_2} \; : \;
M_{\lambda_3} 
=
\alpha_1 \; : \;
\alpha_2 \; : \;
\alpha_3.
\label{eq:SUSY:MGM:gaugino:masses}
\end{align}
The heaviest superpartners are those which couple to the largest rank gauge group.

The scalar partners of the \SM matter particles do not directly couple to the messengers. Thus the masses for the squarks and sleptons must be generated at two loop level. There are many diagrams that include loops of both the messenger scalar and fermion:
\begin{center}
\begin{tikzpicture}[line width=1.5 pt, scale=.5]
	\draw[scalarnoarrow] (-3,0) -- (3,0);
	\draw[provector] (-1.5,0) -- (-1,2);
	\draw[antivector] (1.5,0) -- (1,2);
	\draw[scalarnoarrow] (0,2) circle (1);
	\node at (0,2) {$\varphi$};
\end{tikzpicture}
\quad\quad\quad
\begin{tikzpicture}[line width=1.5 pt, scale=.5]
	\draw[scalarnoarrow] (-3,0) -- (3,0);
	\draw[provector] (-1.5,0) -- (-1,2);
	\draw[antivector] (1.5,0) -- (1,2);
	\draw[fermionnoarrow] (0,2) circle (1);
	\node at (0,2) {$\psi$};
\end{tikzpicture}
\quad\quad\quad
\begin{tikzpicture}[line width=1.5 pt, scale=.5]
	\draw[scalarnoarrow] (-3,0) -- (3,0);
	\draw[scalarnoarrow] (-2,0) arc (135:45:2.8284);
	\draw[scalarnoarrow] (-2,0) arc (210:-30:2.309);
	% \draw[scalarnoarrow] (0,2) circle (1);
	\node at (0,2) {$\varphi$};
\end{tikzpicture}
\quad\quad\quad
\begin{tikzpicture}[line width=1.5 pt, scale=.5]
	\draw[scalarnoarrow] (-3,0) -- (3,0);
	\draw[provector] (-2,0) to [out=135,in=45] (0,1);
	\draw[antivector] (2,0) to [out=45,in=135] (0,1);
	\draw[scalarnoarrow] (0,2.1) circle (1);
	\node at (0,2.1) {$\varphi$};
\end{tikzpicture}

\vspace{1em}

\begin{tikzpicture}[line width=1.5 pt, scale=.5]
	\draw[scalarnoarrow] (-3,0) -- (3,0);
	\draw[provector] (0,0) to [bend left=45] (-1,3);
	\draw[antivector] (0,0) to [bend right=45] (1,3);
	\draw[scalarnoarrow] (0,3) circle (1);
	\node at (0,3) {$\varphi$};
\end{tikzpicture}
\quad\quad\quad
\begin{tikzpicture}[line width=1.5 pt, scale=.5]
	\draw[scalarnoarrow] (-3,0) -- (3,0);
	\draw[provector] (0,0) to [bend left=45] (-1,3);
	\draw[antivector] (0,0) to [bend right=45] (1,3);
	\draw[fermionnoarrow] (0,3) circle (1);
	\node at (0,3) {$\psi$};
\end{tikzpicture}
\quad\quad\quad
\begin{tikzpicture}[line width=1.5 pt, scale=.5]
	\draw[scalarnoarrow] (-3,0) -- (-2,0);
	\draw[scalarnoarrow] (3,0) -- (2,0);
	\draw[fermionnoarrow, line width =1pt ] (-2,0) -- (2,0);
	\draw[fermionnoarrow, line width =1pt ] (-2,0) -- (-2,2) -- (2,2) -- (2,0);
	\draw[vector, line width =1pt ] (-2,0) -- (-2,2);
	\draw[vector, line width =1pt ] (2,2) -- (2,0);
	\draw[scalarnoarrow] (-2,2) arc (180:0:2);
	\node at (0,3) {$\varphi$};
\end{tikzpicture}\quad\quad\quad
\begin{tikzpicture}[line width=1.5 pt, scale=.5]
	\draw[scalarnoarrow] (-3,0) -- (3,0);
	\draw[provector] (0,0) to [bend left=90] (0,2);
	\draw[antivector] (0,0) to [bend right=90] (0,2);
	\draw[scalarnoarrow] (0,3) circle (1);
	\node at (0,3) {$\varphi$};
\end{tikzpicture}
\end{center}
The loops either include a gauge boson or otherwise use the scalar quartic $D$-term interaction between messengers and sfermions. The result is that the soft scalar masses go like
\begin{align}
m_\text{soft}^2 \sim \left(\frac{g^2}{16\pi^2}\right)^2 N_m \frac{F^2}{M^2} C_i,
\end{align}
where $C_i$ is the relevant quadratic casimir.
Observe that $m_\text{soft}^2 \sim m_\lambda^2$ so that the sfermions which couple to the higher rank gauge factors pick up more mass. Including the various gauge charges and taking the limit $\alpha_3 \gg \alpha_2 \gg \alpha_1$ gives a prediction for the sfermion spectrum in minimal gauge mediation,
\begin{align}
	m_{\tilde q}^2 : m_{\tilde \ell}^2 : m_{\tilde E}^2 = \frac 43 \alpha_3^2 : \frac 34 \alpha_2^2 : \frac 35 \alpha_1^2.
\label{eq:SUSY:MGM:sfermion:masses}
\end{align}
Note that (\ref{eq:SUSY:MGM:gaugino:masses}) and (\ref{eq:SUSY:MGM:sfermion:masses}) are only predictions of minimal gauge mediation. A parameterization of the soft terms from a generic gauge mediation model is presented in \cite{Meade:2008wd, Carpenter:2008wi} under the banner of \textbf{general gauge mediation}.
Requiring that the superpartner masses are around the electroweak scale sets
\begin{align}
\frac{F}{M} \sim 100 \text{ TeV}.
\end{align}
Note that since the messengers interact with the \SM superfields only through gauge interactions, the holomorphic soft terms ($A$ and $B$ terms) are typically very small in gauge mediation.

One important phenomenological consequence of gauge mediation is that the lightest supersymmetric partner (\LSP) is not one of the \MSSM fields but rather the gravitino whose mass is \cite{Giudice:1998bp},
\begin{align}
%m_{3/2} \sim \frac F{M_\text{Pl}} \sim \left(\frac{\sqrt{F}}{100 \text{ TeV}}\right) 2.4 \text{ eV}.
m_{3/2} \sim \frac F{\sqrt{3} M_\text{Pl}} \sim \left(\frac{\sqrt{F}}{100 \text{ TeV}}\right)^2 2.4 \text{ eV}.
%% See Giudice Rattazzi eq (3.2)
\end{align}
Thus the gravitino is much lighter than the electroweak scale, but is also similarly weakly coupled. The relevant couplings at low energies are not gravitational, but rather through the Goldstino component of the gravitino. This coupling is proportional to the \SUSY breaking \vev $F$. 
Because of $R$-parity, any supersymmetric partner produced in the \MSSM will eventually decay into the next-to-lightest superpartner (\textsc{nlsp}). This \textsc{nlsp} must eventually decay into the gravitino \LSP since it is the only decay mode available.
When $\sqrt{F} \gtrsim 10^6\text{ GeV}$, the \textsc{nlsp} is so long lived that on collider scales it behaves effectively like the \LSP. On the other hand, if $\sqrt{F} \lesssim 10^6\text{ GeV}$, the \textsc{nlsp} decays within the detector. This gives a fairly unique signal with displaced photons and missing energy if the \textsc{nlsp} is the bino, $\tilde B$.

\subsection{The $\mu$--$B_\mu$ problem of gauge mediation}
% 9603238v1

Let's return to an issue we addressed earlier when discussing electroweak symmetry breaking. We wrote two relations (\ref{eq:SUSY:MSSM:muBmu} -- \ref{eq:SUSY:MSSM:weird:cancellation}) satisfied at the minimum of the Higgs potential. We noted the $\mu$-problem associated with (\ref{eq:SUSY:MSSM:weird:cancellation}): $\mu$ and $m_{H_{u,d}}^2$  seem to come from different sectors of the theory but must conspire to be roughly the same scale.
In principle, since $\mu$ is a supersymmetric dimensionful parameter (the only one in the \MSSM), it could take a value on the order of the Planck mass. We now present a solution to the $\mu$-problem, but we shall see that this solution will cause problems in gauge mediation due to the second relation, (\ref{eq:SUSY:MSSM:muBmu}).

One way to address this $\mu$-problem is to forbid it in the supersymmetric limit and then assume that it is generated through the \SUSY breaking sector.
For example, a global Peccei-Quinn (\textsc{pq}) symmetry,
\begin{align}
 H_u \to & e^{i\alpha}H_u\label{eq:MSSM:PQ:1}\\
 H_d \to & e^{i\alpha}H_d\label{eq:MSSM:PQ:2},
\end{align}
prohibits the $\mu$ term in the superpotential. Gravity, however, is believed to explicitly break global symmetries. Indeed, gravity mediation of \SUSY breaking will produce a $\mu$ term. Consider, for example, the higher order K\"ahler potential term that couples the \SUSY breaking superfield $X$ to the Higgses \cite{Giudice:1988yz},
\begin{align}
\int d^4\theta \frac{X^\dag H_u \cdot H_d}{M_\text{Pl}} + \text{h.c.}
\end{align}
When $\langle X\rangle \sim \theta^2 F$, one generates an effective $\mu$ term of order $\mu \sim F/M_\text{Pl}$.
%\begin{align}
%\mu \sim \frac{F}{M_\text{Pl}}.
%\end{align}
This neatly addresses the $\mu$-problem and ties the $\mu$ term to the \SUSY breaking masses.
The $B_\mu$ term that is generated comes from 
\begin{align}
\int d^4\theta \frac{X^\dag X H_u \cdot H_d}{M_\text{Pl}^2}
\end{align}
and thus is of the same order as $\mu^2$. This is consistent with the observation in (\ref{eq:SUSY:MSSM:muBmu}) that $B_\mu$, $\mu$, and the soft breaking terms seem to want to be the same order.
%
%
%% ASPEN UPDATE: Remove all this extra stuff, and just replace with a short sentence about gauge mediation being more difficult. 
%
We remark that this is no longer true in gauge mediation since $F \ll 10^{11}$ GeV, the $\mu$ and $B_\mu$ terms generated from gravitational breaking are far too small. This must be addressed separately in such theories. 

\subsection{Variations beyond the MSSM}

The \MSSM is under pressure from the \LHC. For a review of the status after Run I of the \LHC, see \cite{Craig:2013cxa}. There are two main issues:
\begin{enumerate}
\item The Higgs mass $m_h = 125$ GeV is hard to achieve in the \MSSM since it requires a large radiative correction to the tree level upper bound of $m_h = M_Z$.
\item There are no signs of superpartners. With the simplest assumption that $m_{\tilde q} \sim m_{\tilde g}$, the \LHC pushes the scale of colored superpartners to be over 1.2 \TeV. This appears to no longer be natural.
\end{enumerate}
In this section we present some model-building directions that the \LHC data may be suggesting.

\subsubsection{Additional $D$-term contributions}

One simple direction to increase the tree-level Higgs mass is to add extra $D$-terms to increase the Higgs quartic coupling \cite{Batra:2003nj, Maloney:2004rc, Medina:2009ey, Bellazzini:2009ix, Cheung:2012zq}. This requires charging the Higgs under an additional U(1)$_X$ gauge group which one must break above the weak scale. This technique is able to indeed push the tree-level Higgs mass up to the observed value, but one is constrained by changes to Higgs decay branching ratios, particularly $h\to b\bar b$ \cite{Blum:2012ii, Craig:2012bs}.

\subsubsection{The NMSSM}

At the cost of adding an additional singlet superfield $S$ to the \MSSM sector, one may solve the $\mu$ problem and also raise the Higgs mass by enhancing its quartic coupling \cite{Fayet:1974pd, Dine:1981rt, Nilles:1982dy, Ellis:1988er, Ellwanger:1993xa}. The Higgs sector superpotential for this ``next-to-minimal'' supersymmetric \SM (\NMSSM, see \cite{Ellwanger:2009dp, Maniatis:2009re} for reviews) is
\begin{align}W_\text{NMSSM} &=
y_u H_u Q \bar U + y_d H_d Q \bar D + y_e H_d \bar E + \lambda S H_u H_d + \frac 13 \kappa S^3.
\end{align}
The $\kappa$ term breaks the Peccei-Quinn symmetry, (\ref{eq:MSSM:PQ:1} -- \ref{eq:MSSM:PQ:2}), to a $\mathbb Z_3$. 
Since $S$ is a gauge singlet, the $D$-term potential is unchanged from (\ref{eq:SUSY:MSSM:EWSB:D}).
Note, however, that there is no longer a $\mu$ term in the superpotential, instead the $SH_uH_d$ coupling has taken its place. Thus the $F$-term potential differs from that of the \MSSM, (\ref{eq:SUSY:MSSM:EWSB:F}), and is instead
\begin{align}
V_{F,\text{\NMSSM}} &= \lambda^2 |S|^2 \left(|H_u|^2 + |H_d|^2\right) + \lambda^2 \left|H_u H_d\right|^2.
\end{align}
We observe that the combination $\lambda \langle S \rangle$ plays the role of an effective $\mu$ term and solves the $\mu$-problem.
%\begin{align}
%\mu_\text{eff} = \lambda S.
%\end{align}
%
Finally, there are additional soft terms allowed which augment $V_\text{soft}$ in (\ref{eq:MSSM:Vsoft}),
\begin{align}
\Delta V_\text{soft,\NMSSM}
&=
m_S^2 |S|^2 + \lambda A_\lambda (SH_uH_d + \text{h.c.}) + \frac 13 \kappa A_\kappa (S^3 + \text{h.c.}).
\end{align}
The resulting expression for the Higgs mass is approximately
\begin{align}
m_h^2 \approx M_Z^2 \cos^2 2\beta + \lambda^2 v^2 \sin^2 2\beta - \frac{\lambda^2v^2}{\kappa^2} (\lambda - \kappa \sin 2\beta)^2 + \frac{3m_t^4}{4\pi^2 v^2}
\left[
    \ln\left(\frac{m_{\tilde t}}{m_t}\right)
    + \frac{A_t^2}{m_{\tilde t}} \left( 1- \frac{A_t^2}{12 m_{\tilde t}}\right)
\right].
\end{align}
This can be larger than the value in the \MSSM depending on the value of $\lambda$. There are limits on the size of $\lambda$ coming from perturbativity, but lifting the Higgs mass to 125 \GeV is fine. The singlet $S$ contributes an additional complex scalar to the Higgs sector and an additional neutralino.

\subsubsection{Natural SUSY}

The simplest choices for the \MSSM parameters---those that treat all the flavors universally, as preferred by the flavor problem---are tightly constrained by the non-observation of new physics at the \LHC. Because the \LHC is a proton-proton collider, it is easy for it to produce colored superpartners such as squarks and gluinos. These, in turn, are expected to show up as events with many jets and missing energy as the heavy colored states decay into the \LSP. The fact that no significant excesses have been found pushes one to consider other parts of the large \MSSM parameter space.

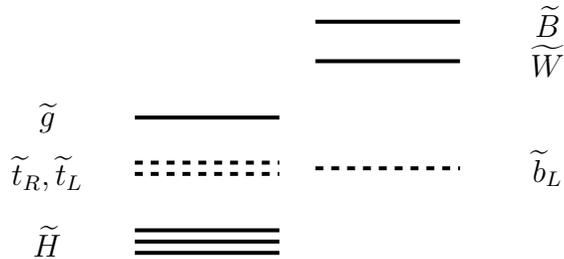
\begin{figure}
\begin{center}
%	\bigskip
	\begin{tikzpicture}[line width=1.5 pt, scale=.75]
		\node at (0,0) (l1a) {};
		\node at (3,0) (l1b) {};
       \node at (0,.2) (l2a) {};
		\node at (3,.2) (l2b) {};
		\node at (0,.4) (l3a) {};
		\node at (3,.4) (l3b) {};
		\draw [fermionnoarrow] (l1a) -- (l1b);
		\draw [fermionnoarrow] (l2a) -- (l2b);
		\draw [fermionnoarrow] (l3a) -- (l3b);
		\node at (0,1.4) (l4a) {};
		\node at (3,1.4) (l4b) {};
       \node at (0,1.6) (l5a) {};
		\node at (3,1.6) (l5b) {};
		\node at (3.2,1.5) (l6a) {};
		\node at (6.2,1.5) (l6b) {};				\draw [scalarnoarrow] (l4a) -- (l4b);		\draw [scalarnoarrow] (l5a) -- (l5b);
		\draw [scalarnoarrow] (l6a) -- (l6b);
		\node at (0,2.4) (l7a) {};
		\node at (3,2.4) (l7b) {};
       \draw [fermionnoarrow] (l7a) -- (l7b);
       \node at (3.2,3.4) (l8a) {};
		\node at (6.2,3.4) (l8b) {};
		\node at (3.2,3.6) (l8aa) {};
		\node at (6.2,3.6) (l8bb) {};
		\node at (3.2,4.1) (l9a) {};
		\node at (6.2,4.1) (l9b) {};
       \draw [fermionnoarrow] (l8a) -- (l8b);
       \draw [fermionnoarrow] (l9a) -- (l9b);
       \node [left of = l2a] (l2label) {$\tilde H$};
       \node [left of = l4a] (l4label) {$\tilde t_R,\tilde t_L$};
       \node [right of = l6b] (l6label) {$\tilde b_L$};
       \node [left of = l7a] (l7label) {$\tilde g$};
       \node [right of = l8b] (l8label) {$\tilde W$};
       \node [right of = l9b] (l9label) {$\tilde B$};
	 \end{tikzpicture}
%	\bigskip	
\end{center}
\caption{Heuristic picture of a natural \SUSY spectrum. All other superpartners are assumed to have masses well above the \TeV scale and decouple.}
\label{fig:SUSY:natural:spectrum}
\end{figure}

Instead of biasing our parameter preferences by simplicity, one may take a different approach and ask what is the minimal sparticle content required for naturalness? In other words, which superpartners are absolutely required to cancel quadratic divergences? Once these are identified, one may decouple the remaining sparticles and check the experimental constraints on the resulting spectrum. 
The ingredients of a `minimally' natural \MSSM spectrum are \cite{Dimopoulos:1995mi, Cohen:1996vb} (see \cite{Kats:2011qh, Brust:2011tb, Papucci:2011wy, Hall:2011aa} for a re-examination from the early \LHC run)
\begin{enumerate}
\item \textbf{Light stops}. The largest \SM contribution to the Higgs quadratic \UV sensitivity is the top loop. Naturalness thus requires that its superpartner, the stop, is also accessible to cancel these loops. Since the stop lives in both the $U_R$ and $Q_L$ superfields, this typically also suggests that the left-handed sbottom is also light.
\item \textbf{Light Higgsinos}. In order to preserve natural electroweak symmetry breaking, $\mu$ should be on the order of the electroweak scale. This is the same parameter that determines the Higgsino mass, so the Higgsinos should also be light.
\item \textbf{Not-too-heavy gluinos}. The stop is a scalar particle which is, itself, quadratically \UV sensitive at face value. The main contribution to the stop mass comes from gluon loops so that naturalness requires `not-too-heavy' ($\sim 1.5$ \TeV) gluinos to cancel these loops. In other words, the gluino feeds into the Higgs mass at two-loop order since it keeps the light-stop light enough to cancel the Higgs' one-loop \UV sensitivity.
\item \textbf{Light-ish electroweak-inos} (optional). Finally, if one insists on grand unification, the scale of the gluinos imposes a mass spectrum on the electroweak gauginos with $M_{\text{\textsc{ew}-ino}} < M_{\text{gluino}}$. As a rough estimate, Majorana gluinos should have mass $\lesssim 2m_t$ while Dirac gluinos should have mass $\lesssim 4m_t$. % 
\item \textbf{All other particles decoupled.} All of the other squarks and sleptons are assumed to be well above the \TeV scale and effectively inaccessible at Run-I of the \LHC. 
\end{enumerate}
These are shown in Fig.~\ref{fig:SUSY:natural:spectrum}.

The simplest models have a light stop $\tilde t_L$ which decays either to a top and neutralino/gravitino, $t + \tilde N$, or a bottom and a chargino, $b + \tilde C$.  Bounds on these decays depend on the $\tilde N$ ($\tilde C$) mass. 
The `stealthy' region near $m_{\tilde N} = 0$ and the `compressed' region near $m_{\tilde N} \approx m_t$ are especially difficult to probe kinematically.

\subsubsection{$R$-parity violation}

One of the main ways to search for `vanilla' \SUSY signatures is to trigger on the large amount of missing energy (\textsc{met} or $\cancel{E}_T$) expected from the neutral \LSP. % 
Underlying this assumption is $R$-parity, which forces the \LSP to to be stable. 

Recall that $R$-parity was something that we embraced because it killed the supersymmetric terms in the superpotential (\ref{eq:MSSM:RPV:superpotential}) that would violate lepton and baryon number and would be severely constrained by experiments, most notably proton decay. If, however, there were another way to suppress these dangerous operators, then perhaps we could avoid the experimental bounds while giving the \LSP a way to decay into non-supersymmetric particles. This would allow us to consider models with $R$-parity violation (\textsc{rpv}) with no missing energy signal \cite{Hall:1983id, Ross:1984yg, Barger:1989rk, Bhattacharyya:1997vv, Dreiner:1997uz}, see \cite{Barbier:2004ez} for a review. Such models would be immune to the usual \textsc{met}-based \SUSY search strategies.

The simplest way to do this is to turn on only the $\lambda_4 \bar U \bar D \bar D$ term. This violates baryon number but preserves lepton number so that protons remain stable. Motivated by naturalness, we may now allow the stop to be the \LSP since this is no longer a dark matter candidate. The \textsc{rpv} coupling would allow a decay $\tilde t \to \bar b \bar s$, which would be hidden in the large \textsc{qcd} di-jet background.

One still has to worry about the effects of this \textsc{rpv} coupling on the partners of the light squarks. Phenomenologically, the strictest bounds come from neutron--anti-neutron oscillation and dinucleon decay. Indeed, most of the flavor bounds on the \MSSM come from the first two generations of sparticles. One interesting model-building tool is to invoke \textbf{minimal flavor violation}, which posits that the flavor structure of the entire \MSSM is carried by the Yukawa matrices \cite{Csaki:2011ge}. This then implies that the coefficient of the $\bar U^i \bar D^j \bar D^k$ \textsc{rpv} coupling is proportional to a product of Yukawa elements depending on the generations $i$,$j$, and $k$. This gives a natural explanation for why the \textsc{rpv} couplings of the first two generation squarks are much smaller than the stop.

\section{Extra Dimensions}\label{sec:XD}

%\flip{Some things from my thesis as intro.}
The original proposal for extra dimensions by
Kaluza \cite{Kaluza:1921tu}, Klein \cite{Klein:1926tv}, and later Einstein \cite{Einstein:1938fk} were attempts to unify electromagnetism with gravitation.
Several decades later the development of string theory---originally as a dual theory to explain the Regge trajectories of hadronic physics---led physicists to revisit the idea of compact extra dimensions \cite{Scherk:1975fm, Cremmer:1975sj, Cremmer:1976ir}.
In early models, the non-observation of an additional spatial direction was explained by requiring the compactification radius to be too small for macroscopic objects. 

\begin{framed}
\noindent \footnotesize
\textbf{Further reading:} Two of the authors' favorite reviews on this subject are \cite{Csaki:2004ay} and \cite{Csaki:2005vy}. This lecture is meant to be largely complementary. Additional references include \cite{Rattazzi:2003ea, Ponton:2012bi, Cheng:2010pt, Gherghetta:2010cj, Sundrum:2011ic, Sundrum:2005jf}, which focus on different aspects.
\end{framed}

\subsection{Kaluza-Klein decomposition}
\label{sec:XD:KK:decomposition}

The simplest example to begin with is a real scalar field in 5D where the fifth dimension is compactified to a circle of radius $R$. The details of the compactification do not change the qualitative behavior of the theory at low energies. The Lagrangian is
\begin{align}
S = 
\int d^5x \; \frac 12 \partial_M \phi(x,y) \partial^M \phi(x,y)
=
\int d^5x \; \frac 12\left[ 
\partial_\mu \phi(x,y) \partial^\mu \phi(x,y)
-
\left(\partial_y \phi(x,y)\right)^2 
\right],
\end{align}
where $M=0,\cdots,5$ and $x_5 = y$. Since $y$ is compact, we may identify energy eigenstates by doing a Fourier decomposition in the extra dimension,
\begin{align}
\phi(x,y) = \frac{1}{\sqrt{2\pi R}} \sum_{n=-\infty}^\infty \phi^{(n)}(x)\, e^{i\frac{n}{R} y}.
\label{eq:XD:scalar:decomposition}
\end{align}
Since $\phi$ is real, $\left(\phi^{(n)}\right)^\dag = \phi^{(-n)}$. Plugging this expansion into the action allows us to use the orthogonality of the Fourier terms to perform the $dy$ integral. This leaves us with an expression for the action that is an integral over only the non-compact dimensions, but written in terms of the \KK modes $\phi^{(n)}(x)$,
\begin{align}
S &= \phantom{\frac 12} \int d^4x \sum_{mn} \left(\int dy \; \frac 1{2\pi R} e^{i\frac{(m+n)}{R}y}\right)
\frac 12 
\left[ 
\partial_\mu \phi^{(m)}(x)
\partial^\mu \phi^{(n)}(x)
+
\frac{mn}{R^2} \phi^{(m)}(x) \phi^{(n)}(x)
\right]
\\
& =
\frac 12 \int d^4 x 
\sum_n \left[ \partial_\mu \phi^{(-n)} \partial^\mu \phi^{(n)}
-
\frac{n^2}{R^2} \phi^{(-n)}\phi^{(n)}
\right]
\\
& =
\phantom{\frac 12}
\int d^4 x 
\sum_{n>0} 
\left[ 
\left(\partial_\mu \phi^{(n)}\right)^\dag \partial^\mu \phi^{(n)}
-
\frac{n^2}{R^2} \left|\phi^{(n)}\right|^2
\right].
\end{align}
From the 4D point of view, a single 5D scalar becomes a `Kaluza-Klein (\textsc{kk}) tower' of 4D particles, each with mass $n/R$. If there were more than one extra dimension, for example if one compactified on an $k$-dimensional torus with radii $R_5$, $R_6$, $\ldots$, then the \textsc{kk} tower would have $k$ indices and masses
\begin{align}
m_{n_5,n_6,\cdots,n_k}^2 = m_0^2 + \frac{n_5^2}{R_5^2} + 
\frac{n_6^2}{R_6^2} + \cdots +  \frac{n_k^2}{R_k^2},
\end{align}
where $m_0^2$ is the higher dimensional mass of the field.

\subsection{Gauge fields}

A more complicated example is a gauge field. We know that gauge fields are associated with vector particles, but in 5D the vector now carries five components, $A_M$. We perform the same \textsc{kk} decomposition for each component $M$,
\begin{align}
A_M(x,y) = \frac{1}{\sqrt{2\pi R}} \sum_n A_M^{(n)}(x)\, e^{i\frac nR y}.
\label{eq:XD:gauge:decomposition}
\end{align}
Note that this decomposes into a \KK tower of 4D vectors, $A_\mu^{(n)}$, and a  \KK tower of 4D scalars, $A_5^{(n)}$.
Similarly, the field strengths are antisymmetric with respect to indices $M$ and $N$ so that the action is decomposed according to
\begin{align}
S &= \int d^4x\, dy\; \left(-\frac 14 F_{MN}F^{MN} \right)
\\
 &= \int d^4x\, dy \; -\frac 14 F_{\mu\nu}F^{\mu\nu} + \frac 12
  \left(\partial_\mu A_5 - \partial_5 A_\mu\right)
   \left(\partial^\mu A_5 - \partial_5 A^\mu\right)
   \\
   &= \int d^4x \; \sum_n -\frac 14 F_{\mu\nu}^{(-n)} F^{(n)\mu\nu}
   + \frac 12
   \left(\partial_\mu A_5^{(-n)} - \partial_5 A_\mu^{(-n)}\right)
   \left(\partial^\mu A_5^{(n)} - \partial_5 A^{(n)\mu}\right).
\end{align}
This looks complicated because there is an odd mixing between the 4D vector, $A_\mu^{(n)}$, and the 4D scalar $A_5^{(n)}$. Fortunately, this mixing term can be removed by fixing to 5D axial gauge,
\begin{align}
A_\mu^{(n)} &\to A_\mu^{(n)} - \frac i{n/R} \partial_\mu A_5^{(n)}
&
A_5^{(n)} &\to 0,
\end{align}
for $n\neq 0$. Note that for $n=0$ there's no scalar--vector mixing anyway. The resulting action takes a much nicer form,
\begin{align}
S =& \int d^4x \; 
-\frac 14 \left(F_{\mu\nu}^{(0)}\right)^2
+ \frac 12 \left(\partial_\mu A_5^{(0)}\right)^2
+ 
\sum_{n\geq 1} 2\left(-\frac 14 F_{\mu\nu}^{(-n)} F^{(n)\mu\nu} + \frac 12 \frac{n^2}{R^2} A_\mu^{(-n)}A^{(n)\mu}\right).
\end{align}
The spectrum includes a tower of massive vector particles as well as a massless (zero mode) gauge boson and scalar. 

Recall the usual expression for the number of degrees of freedom in a massless 4D gauge boson:
\begin{align}
(\text{4 components in }A_\mu)
- (\text{longitudinal mode})
- (\text{gauge redundancy}).
\end{align}
When the gauge boson becomes massive, it picks up a longitudinal mode from eating a scalar by the Goldstone mechanism. This is precisely what has happened to our \KK gauge bosons, $A_\mu^{(n)}$: they pick up a mass by eating the scalar \KK modes, $A_5^{(n)}$.

In a theory with $(4+n)$ dimensions, the $(4+n)$-component vector $A_M$ decomposes into a massless gauge boson, $n$ massless scalars, a tower of massive \KK vectors $A_\mu$, and a tower of $(n-1)$ massive \KK scalars. 

One may similarly generalize to spin-2 particles such as the graviton. In $(4+n)$ dimensions these are represented by an antisymmetric $(4+n)\times(4+n)$ tensor,
\begin{align}
g_{MN} =
	\left(
	\begin{array}{cc:c}
	\multicolumn{2}{c:}{\multirow{2}{*}{\large{$g_{\mu\nu}$}}} & \multicolumn{1}{c}{\multirow{2}{*}{\large{$A_\mu$}}} \\
	 &&  \\
	\hdashline
	 &  & \varphi
%\multicolumn{2}{cc:}{$A_\mu$} & $\varphi$
	\end{array}
	\right).
\end{align}
The massless 4D zero modes include the usual 4D graviton, a vector, and a scalar. At the massive level, there is a \KK tower of gravitons with $(n-1)$ gauge fields and $[\frac 12 n(n+1) - n]$ scalars. Here we observe the graviton and vector eating the required degrees of freedom to become massive.

\subsection{Matching of couplings}

It is important to notice that the mass dimension   of couplings and fields depend on the number of spacetime dimensions. The action is dimensionless, $[S]=0$, since it is exponentiated in the partition function. Then, in $(4+n)$ dimensions, the kinetic term for a boson gives
\begin{align}
\left[ d^{(4+n)}x\; (\partial\phi)^2\right]
=
-(4+n)
+2
+2[\phi]
=0
\qquad
\Rightarrow
\qquad
[\phi] = 1 + \frac n2.
\end{align}
Note that this is consistent with the dimensions in the \KK expansion (\ref{eq:XD:scalar:decomposition}). The 5D scalar contains the 4D scalars with a prefactor $\sim R^{-1/2}$ that has mass dimension $1/2$.
Similarly, for fermions, $[\psi] = \frac 32 + \frac n2$. With this information, dimensions of the Lagrangian couplings can be read off straightforwardly. For example, the 5D gauge field lives in the covariant derivative,
\begin{align}
D_\mu = \partial_\mu - ig_5 A_\mu 
= \partial_\mu - i\frac{g_5}{\sqrt{2\pi R}}A^{(0)}_\mu + \cdots.
\end{align}
We see that $[g_5] = -1/2$ since $[\partial]=1$ and $[A_M]=3/2$. Further, we find an explicit relation between the 5D parameter $g_5$ and the observed 4D gauge coupling,
\begin{align}
g_4 = \frac{g_5}{\sqrt{2\pi R}}.
\end{align}
More generally, in $(4+n)$ dimensions the 4D coupling is related to the higher dimensional coupling by the volume of the extra dimensional space,
\begin{align}
g_4^2 = \frac{g_{(4+n)}^2}{\text{Vol}_n}.
\label{eq:XD:4D:5D:couplings}
\end{align}
One can read off the matching of the gravitational coupling by looking at the prefactor of the Ricci term in the action,
\begin{align}
S_{(4+n)} 
= - M_{(4+n)}^{2+n}
\int d^{4+n}x \; \sqrt{g}\, R_{(4+n)}
=
- %
M_{(4+n)}^{2+n} V_n
\int d^{4}x \; \sqrt{g_{(4)}}\, R_{(4)} + \cdots,
\end{align}
where we've written $g$ for the determinant of the metric.
From this we identify 4D Planck mass $M_\text{Pl}$ from the fundamental higher dimensional Planck mass, $M_{(4+n)}$,
\begin{align}
M_\text{Pl}^2 = M_{(4+n)}^{2+n} V_n.
\label{eq:XD:MPl:vs:Mstar}
\end{align}

The higher dimensional Planck mass is a good choice for a fundamental mass scale for the theory,
\begin{align}
M_* = M_{(4+n)}.
\end{align}
In a $(4+n)$ dimensional theory where the characteristic mass scale is $M_*$ and a compactification radius $R$. Then dimensional analysis tells us that the higher dimensional gauge couplings, which are dimensionful, characteristically scale like
\begin{align}
g_{(4+n)} \sim M^{-n/2}_*.
\end{align}
Relating this to the 4D couplings with (\ref{eq:XD:4D:5D:couplings}) and relating $M_*$ to the 4D Planck mass with (\ref{eq:XD:MPl:vs:Mstar}) gives 
\begin{align}
R \sim \frac{1}{M_\text{Pl}} g_4^{(n+2)/n}.
\label{eq:XD:no:go:UED}
\end{align}
Plugging in the observed \SM gauge couplings on the right hand side gives a compactification radius which is far too small to be relevant at colliders---the first \KK modes will be near the Planck scale.

\subsection{Branes and Large Extra Dimensions}

In the mid '90s, developments in string theory led to a new ingredient that renewed interest in extra dimensions that might be accessible at collider scales. The key idea is that \textbf{branes}, solitonic objects which form lower dimensional subspaces, can trap fields. In other words, not all fields have to propagate in all dimensions. 
This was introduced by Rubakov and Shaposhnikov \cite{Rubakov:1983bb}, who showed that instead of a very small radius of compactification, it may be that our observed universe is constrained to live in a (3+1)-dimensional subspace of a higher dimensional spacetime. %

\begin{framed}
\noindent \footnotesize \textbf{Terminology}.
Models that make use of branes to localize fields are known as \textbf{braneworld} models and are distinguished from models where all fields propagate in the extra dimensions, known as \textbf{universal extra dimensions}. In braneworld models, fields which are allowed to propagate in the full space are said to live in the \textbf{bulk}.
\end{framed}

Allowing the fields to be brane-localized buys us quite a lot. It allows us to separate particle physics from gravity.
One can, for example, force the \SM fields to be truly four-dimensional objects that are stuck to a (3+1)-dimensional brane. This avoids the bound on the size of the extra dimension in (\ref{eq:XD:no:go:UED}), since that relied on the \SM propagating in the bulk. 

\begin{figure}
\begin{center}
\includegraphics[height=4.5cm]{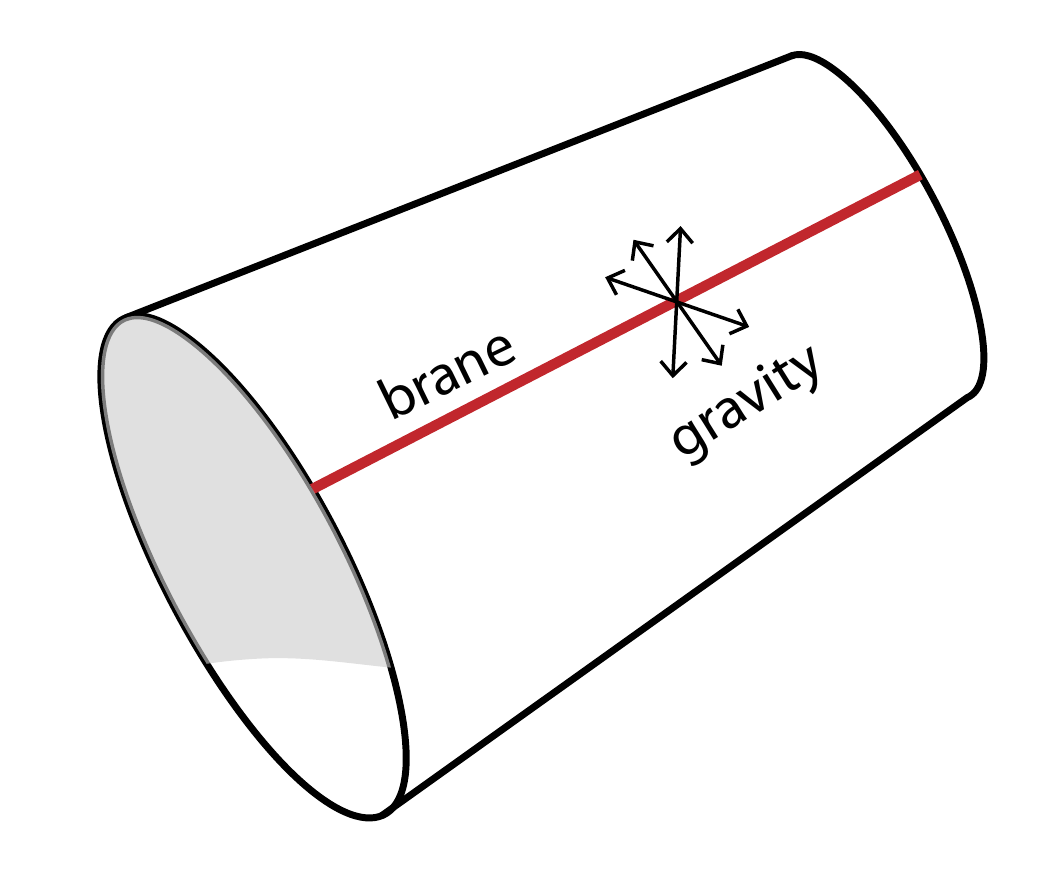}
\qquad \qquad \qquad
\includegraphics[height=4.5cm]{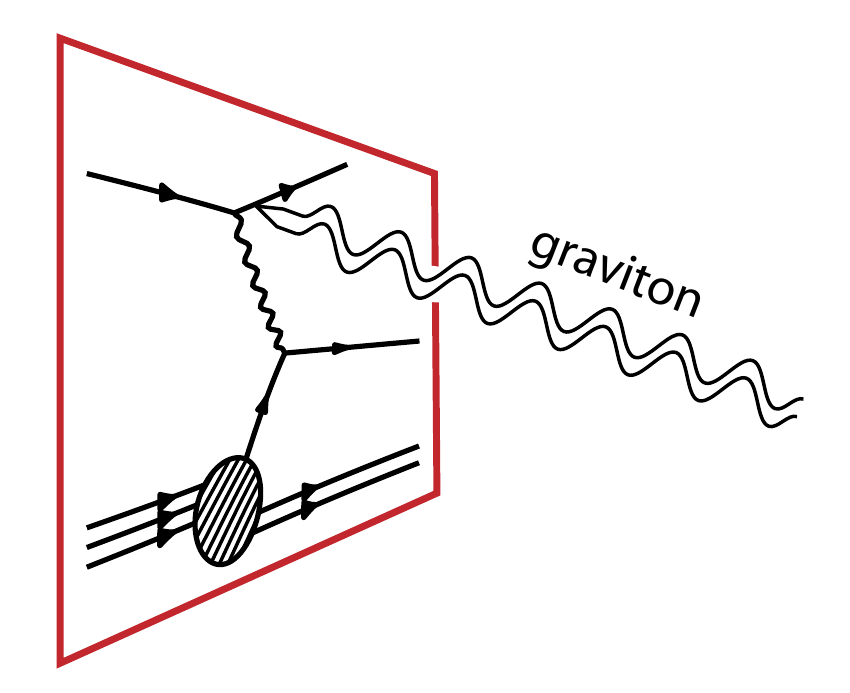}
\end{center}
\caption{Cartoon pictures of a (3+1) dimensional brane in a compact 5D space. (\textsc{left}) The brane (red line) as a subspace. Gravity propagates in the entire space `diluting' its field lines relative to forces localized on the brane. (\textsc{right}) \SM processes localized on the brane, now with an additional dimension drawn, emitting a graviton into the bulk.}
\label{fig:XD:branes}
\end{figure}

With this in mind, one could allow the volume of the extra dimensions to actually be quite large. This idea was explored by Arkani-Hamed, Dimopoulos, and Dvali in the \textsc{add} or \textbf{large extra dimension} scenario \cite{ArkaniHamed:1998nn}.
If this were feasible, then (\ref{eq:XD:MPl:vs:Mstar}) gives a new way to address the Hierarchy problem. The large volume factor allows the fundamental scale of nature to be much smaller than the observed Planck mass, $M_*  \ll M_\text{Pl}$. 
If, for example, $M_* \sim \text{1 \TeV}$, then there is no Hierarchy problem. Gravity appears to be weaker at short distances because its flux is diluted by the extra dimensions.
As one accesses scales smaller than $R$, however, one notices that gravity actually propagates in $(4+n)$ dimensions. 
A cartoon of the braneworld scenario is shown in Fig.~\ref{fig:XD:branes}.

How large can this extra dimension be? Doing a rough matching and using $\text{Vol}_n = r^n$ in (\ref{eq:XD:MPl:vs:Mstar}) gives
\begin{align}
R = \frac 1{M_*} \left(\frac{M_\text{Pl}}{M_*}\right)^{2/n}.
\label{eq:XD:R:Mpl:Mst}
\end{align}
Pushing the fundamental scale to $M_* \sim$ \TeV requires
\begin{align}
R = 10^{32/n}\text{ \TeV}^{-1} = 2\cdot 10^{-17}\; 10^{32/n}\text{ cm},
\end{align}
using \text{GeV}$^{-1} = 2\cdot 10^{-14}$ cm. We make the important caveat that this is specifically for the \textsc{ADD} model. 
Considering different numbers of extra dimensions,
\begin{itemize}
\item $n=1$. For a single extra dimension we have $R=10^{15}$ cm, which is roughly the size of the solar system and is quickly ruled out. 
\item $n=2$. Two extra dimensions brings us down to $R\approx 0.1$ cm, which is barely ruled out by gravitational Cavendish experiments.
\item $n=3$. Three extra dimensions pushes us down to $R<10^{-6}$ cm.
\end{itemize}
How much do we know about gravity at short distances? Surprisingly little, actually. Cavendish experiments (e.g.\ E\"ot-Wash\footnote{The name is a play on the E\"otv\"os experiment by University of Washington researchers.}) test the $r^{-2}$ law down to $10^{-4}$ m. These set a direct bound on the $n=2$ case that $R< 37 \text{ }\mu\text{m}$ and $M_* > 1.4$ TeV. For larger $n$ one is allowed to have $M_* = $ TeV.

One might have objected that one cannot say that $M_*$ is the fundamental scale while allowing $R$, itself a dimensionful quantity, float to take on any value. Indeed, in a completely natural theory, one expects $R\sim 1/M_*$ so that $R\sim \text{\TeV}^{-1}$. This is quite different from what we wrote in (\ref{eq:XD:R:Mpl:Mst}). Indeed, what we have done here is swapped the hierarchy in mass scales to a Hierarchy between $R$ and $M_*^{-1}$. In other words, we have reformulated the Hierarchy problem to a problem of \textbf{radius stabilization}. This is indeed very difficult to solve in \textsc{add}.

Nevertheless, we may explore the phenomenological consequences of an \textsc{add} type model at colliders and through astrophysical observations.
\begin{itemize}
\item The first thing to consider is the production of  \KK gravitons. 
\begin{center}
%\begin{tikzpicture}[line width=1.5 pt]
%\draw[vector, line width=.75] (0,0) -- (2,0);
%\draw[vector, line width=.75] (0,-.1) -- (2,-.1);
%\draw[fermion] (-1,.5) -- (0,0);
%\draw[fermion] (0,0) -- (0,-1.5);
%\draw[fermion] (0,-1.5) -- (-1,-2);
%\draw[vector, line width=.75] (0,-1.5) -- (2,-1.5);
%\draw[vector, line width=.75] (0,-1.4) -- (2,-1.4);
%\node at (-1.25 ,.5) {$f$};
%\node at (-1.25 ,-2) {$f$};
%\node at (2.25 ,-1.5) {$G$};
%\node at (2.25 ,0) {$G$};
%\end{tikzpicture}
%\qquad\qquad\qquad\qquad
\begin{tikzpicture}[line width=1.5 pt]
\draw[vector] (0,0) -- (2,0);
%\draw[vector, line width=.75] (0,0) -- (2,0);
%\draw[vector, line width=.75] (0,-.1) -- (2,-.1);
\draw[fermion] (-1,.5) -- (0,0);
\draw[fermion] (0,0) -- (0,-1.5);
\draw[fermion] (0,-1.5) -- (-1,-2);
\draw[vector, line width=.75] (0,-1.5) -- (2,-1.5);
\draw[vector, line width=.75] (0,-1.4) -- (2,-1.4);
\node at (-1.25 ,.5) {$f$};
\node at (-1.25 ,-2) {$f$};
\node at (2.25 ,-1.5) {$G$};
\node at (2.5 ,0) {$\gamma, g$};
\end{tikzpicture}
\end{center}
The \KK graviton couples too weakly to interact with the detector so it appears as missing energy. By itself, however, missing energy is difficult to disentangle from, say, neutrino production. Thus it's useful to have a handle for the hardness  of the event (more energetic than $Z\to \nu
\bar\nu$) so one can look for processes that emit a hard photon or gluon. Thus a reasonable search is a jet or photon with missing energy. It is worth noting that this is the same search used for searching for dark matter, which is also typically a massive particle which appears as missing energy.

\item Alternately, one may search for  $s$-channel virtual graviton exchange in processes like $e^+e^- \to f\bar f$. One expects a resonance at the \KK graviton mass.

\item Supernovae can cool due to the emission of gravitons. This is similar to the supernovae cooling bounds on axions. The strongest bounds on $n=2$ theories push $M_* \gtrsim 100$ TeV.

\item An additional byproduct of lowering the fundamental gravitational scale is that one may form microscopic black holes at energies kinematically accessible to the \LHC and cosmic rays. For $E_\text{CM} > M_*$ black holes  are formed with a radius
\begin{align}
R_S \sim 
\frac{1}{M_*} 
\left(\frac{M_{\text{BH}}}{M_*}\right)^{\frac{1}{n+1}}.
\end{align}
the cross section is roughly the geometric value, $\sigma_\text{BH} \sim \pi R_S^2$ and can be as large as 400 pb. These microscopic black holes decay via Hawking radiation,
\begin{align}
T_H \sim \frac{1}{R_S}
\end{align}
with this energy distributed equally to all degrees of freedom, for example 10\% going to leptons, 2\% going to photons, and 75\% going to many jets.
\end{itemize}

\subsection{Warped extra dimensions}

We've seen that the framework of large extra dimensions leads to interesting phenomenology, but the \textsc{add} realization leaves the size of the radius unexplained and is therefore not a complete solution to the Hierarchy problem. The Randall-Sundrum (\RS) proposal for a warped extra dimension offers a more interesting possibility \cite{Randall:1999ee}. The set up differs from \ADD in that the space between the two branes has a non-factorizable metric that depends on the extra space coordinate, $z$,
\begin{align}
ds^2 = \left(\frac Rz\right)^2 \left(\eta_{\mu\nu}dx^\mu dx^\nu - dz^2\right).
\label{eq:RS:metric}
\end{align}
This is the metric of anti-de Sitter space (\AdS) with curvature $k=1/R$. There are two branes located at $z=R$ (the \UV brane) and $z=R'>R$ (the \IR brane) that truncate the extra dimension; in this sense the \RS background is often described as a `slice of \AdS.' We see that $1/R$ is naturally a fundamental \UV scale of the theory. The metric (\ref{eq:RS:metric}) warps down the natural physical scale as a function of the position along the extra dimension. In particular, when $R'\gg R$ one finds that near $z=R'$, the scales are warped down to much smaller values. Note the different notation from the \ADD case: the size of the extra dimension is $R'-R\approx R'$, while $R$ should be identified with the radius of curvature.

\begin{figure}
\begin{center}
\includegraphics[height=7cm]{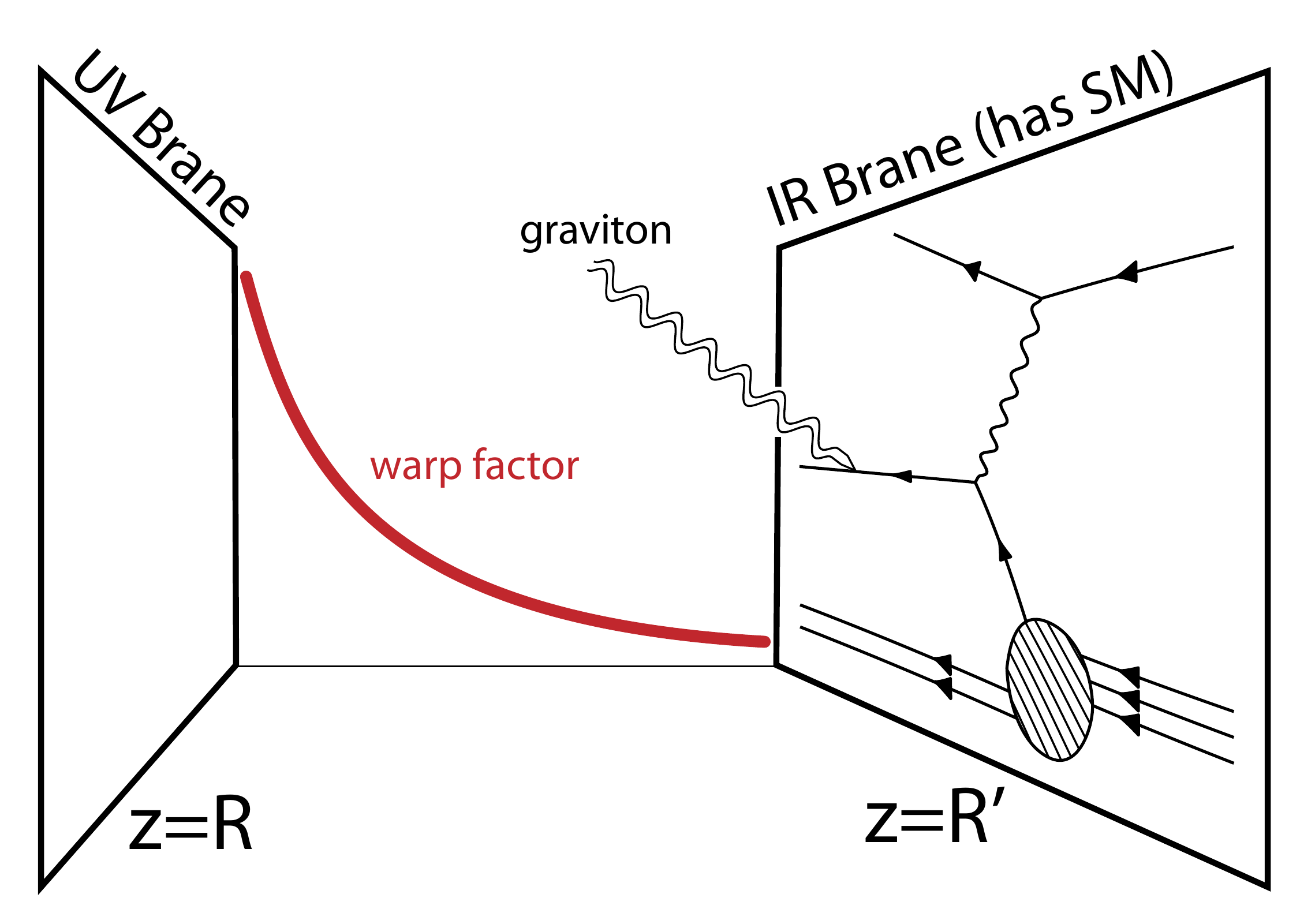}
\end{center}
\caption{Cartoon of the \RS scenario with a brane-localized \SM. The warp factor, $(R/z)^2$, causes energy scales to be scaled down towards the \IR brane.}
\label{fig:XD:RS:RS1}
\end{figure}

To see how this works, suppose that the Higgs is localized to live on the \IR brane at $z=R'$. The action on this brane depends on the 4D induced metric $\hat g_{\mu\nu}$ (note that $\sqrt{\hat g} = \sqrt{g}/\sqrt{g_{55}}$),
\begin{align}
S =
\int d^4x \sqrt{\hat g}
\left[
\partial_\mu H
\partial_\nu H
\hat g^{\mu\nu}
- 
\left(|H|^2 - \frac{v^2}{2}\right)^2
\right]%_{z=R'}. 
\end{align}
We assume that the Higgs \vev is on the order of the \UV scale, $v = 1/R$, since this is the fundamental 5D scale. Plugging in the metic gives
\begin{align}
S = &
\int d^4x \left(\frac{R}{R'}\right)^4
\left[
\partial_\mu H
\partial^\mu H
\left( \frac{R'}R\right)^2
- 
\left(|H|^2 - \frac{v^2}{2}\right)^2
\right]_{z=R'},
\end{align}
where indices are implicitly raised with respect to the Minkowski metric. Canonically normalizing the kinetic term via
\begin{align}
\hat H =  \frac{R}{R'}H,
\end{align}
allows us to write the action in the form,
\begin{align}
S = \int d^4x \; \left(\partial_\mu \hat H\right)^2 
- \lambda\left[|\hat H|^2 - \frac 12 \left( v \frac{R}{R'}\right)^2\right]^2,
\end{align}
where we see that the canonically normalized Higgs picks up a \vev that is warped down to the \TeV scale. One can further imagine that the cutoff for loops contributing to the Higgs mass are similarly warped down to, say, the \TeV scale. In this way, the warped extra dimension gives a new handle for generating hierarchies. Readers should be skeptical that we're not just hiding the Hierarchy problem in some fine tuning of the \IR scale $R'$ relative to the fundamental scale $R$. Indeed, the real solution to the Hierarchy problem requires a mechanism for radius stabilization, which we present below. Note that typically $R \approx M_\text{Pl}^{-1}$ and $R' \approx \text{TeV}^{-1}$ so that $R'$ is roughly the size of the extra dimension. A cartoon of this scenario is shown in Fig.~\ref{fig:XD:RS:RS1}.

In the remainder of this lecture we'll focus on the \RS background. In the appendices we present some additional technical results that may be useful for building \RS models.
Further details of the \RS gravitational background are discussed in Appendices \ref{app:XD:RS:Grav:BG} and \ref{app:XD:RS:Grav:BG:orbifold}. Details of bulk matter fields are discussed in Appendices  \ref{sec:XD:RS:bulk:fermion} and
\ref{sec:XD:RS:bulk:gauge}

\subsection{The Planck scale and hierarchy in RS}

We have seen how the \AdS curvature can warp mass scales to be much smaller than the fundamental 5D scale $1/R$. It is instructive to also check the observed Planck scale. With respect to the fundamental Planck scale $M_*$ (ostensibly $M_* \sim 1/R$), the gravitational action is
\begin{align}
S_g = M_*^3 \int_{R}^{R'} dz \int d^4x  \;
\sqrt{g_{(5)}} \mathcal R_{(5)},
\end{align}
where the quantities with subscripts are the determinant of the 5D metric and the 5D Ricci scalar, respectively. By performing the $dz$ integral one finds the effective 4D gravitational action,
\begin{align}
S_g &= M_*^3  \int_{R}^{R'} dz\; \left( \frac{R}{z}\right)^3 \int d^4 x \; \sqrt{g_{(4)}} \mathcal R_{(4)}
%\\
%&
=M_*^3 \frac 12 \left[1 - \left(\frac{R}{R'}\right)^2\right] \int d^4 x \; \sqrt{g_{(4)}} \mathcal R_{(4)}.
\end{align}
We can thus identify the effective 4D Planck mass by reading off the coefficient,
\begin{align}
M_\text{Pl}^2 =  \frac{M_*^3 R}{2} \left[1 - \left(\frac{R}{R'}\right)^2\right] 
	\sim 
	M_*^2, % cf eq 9 of hep-th/0012148
\end{align}
so that for a large extra dimension, $R' \gg R$, the 4D Planck mass is insensitive to $R'$ and is fixed by the 5D Planck mass, $M_* \sim 1/R$. This is precisely what we have set out to construct: assuming there is a dynamical reason for $R' \gg R$, we are able to warp down masses to the \TeV scale by forcing particles to localize on the \IR brane while simultaneously maintaining that 4D observers will measure a Planck mass that is much heavier.

An alternate way of saying this is that the Hierarchy problem is solved because the \SM Higgs is peaked towards the \IR brane while gravity is peaked towards the \UV brane. What we mean by the latter part of this statement is that the graviton zero mode has a bulk profile that is peaked towards the \UV brane. Recall that in flat space, zero modes have flat profiles since they carry no momentum in the extra dimension. In \RS, the warping of the space also warps the shape of the graviton zero mode towards the \UV brane; the weakness of gravity is explained by the smallness of the graviton zero mode profile where the Standard Model particles live. This should be compared to the case of a flat interval where the zero mode wave function decouples as the size of the extra dimension increases. In this case the coupling with the \IR brane indeed becomes weaker, but the graviton \KK modes become accessible and can spoil the appearance of 4D gravity. In \RS the zero mode doesn't decouple and one doesn't need to appeal to a dilution of the gravitational flux into the extra dimensions as in the \ADD model. See \cite{Csaki:2004ay} for more explicit calculations in this picture.

\subsection{Bulk scalar profiles in RS}
\label{sec:XD:bulk:scalar}

In the original \RS model, only gravity propagates in the bulk and has \KK modes. However, it is instructive to derive the \KK properties of a bulk scalar.
\begin{itemize}
\item This serves as a simple template for how to \KK reduce more complicated bulk fields, such as the graviton, in a warped background. See, e.g.\ \cite{Csaki:2004ay} for the analysis of the graviton.
\item We anticipate the `modern' incarnations of the \RS where gauge and matter fields are pulled into the bulk. The properties of these fields are detailed in Appendix~\ref{app:XD:RS} and follow this analysis of the bulk scalar.
\item As mentioned above, the solution to the Hierarchy problem depends on stabilizing the position of the \IR brane, $z=R'$, relative to the \UV brane, $z=R$. The standard technique for doing this requires a bulk scalar.
\end{itemize}

Start with a bulk complex scalar $\Phi(x,z)$ with a bulk mass parameter $m$. The bulk action is
\begin{align}
S =  \int_R^{R'} dz \int d^4x \sqrt{g} \,
\left[
    \left(\partial_M \Phi\right)^*
    \partial^M\Phi
    -m^2 \Phi^* \Phi
\right].
\end{align}
In principle one may have additional brane-localized interactions proportional to $\delta(z-R')$ or $\delta(z-R)$. We use $M,N$ to index 5D coordinates while $\mu,\nu$ only run over 4D coordinates. 
Varying with respect to $\Phi^*$ yields an equation of motion
\begin{align}
-\partial_M\left(\sqrt{g} g^{MN} \partial_N \Phi\right) - \sqrt{g} m^2 \Phi = 0.
\end{align}
In writing this we have dropped an overall surface term that we picked up when integrating by parts. Specializing to the \RS metric, this amounts to picking boundary conditions such that
\begin{align}
\left.\Phi^*(z) \partial_z\Phi(z)\right|_{R,R'} = 0,
\end{align}
with the appropriate modifications if there are brane-localized terms. We see that we have a choice of Dirichlet and Neumann boundary conditions.
We now plug in the Kaluza-Klein decomposition in terms of yet-unknown basis functions $f^{(n)}(z)$ which encode the profile of the $n^{\text{th}}$ mode in the extra dimension:
\begin{align}
\Phi(x,z) = \frac{1}{\sqrt{R}} \sum_n^{\infty} \phi^{(n)}(x) f^{(n)}(z).
\end{align}
The factor of $1/\sqrt{R}$ is pulled out for explicit dimensional analysis; that is, the profile $f^{(n)}(z)$ is defined to be dimensionless. 
By assumption, the $\phi^{(n)}$ are eigenstates of $\eta^{\mu\nu}\partial_\mu\partial_\nu$ with eigenvalue $-m_{(n)}^2$, the $\KK$ mass. We are thus left with a differential equation for $f^{(n)}(z)$,
\begin{align}
\left[
    \left(\frac{R}{z}\right)^3 m_{(n)}^2
    - \frac 3z \left(\frac{R}{z}\right)^3 \partial_z
    + \left(\frac{R}{z}\right)^3 \partial_z^2
    - \left(\frac{R}{z}\right)^5 m^2
    \right]
    \frac{f^{(n)}(z)}{\sqrt{R}} = 0.
    \label{eq:XD:RS:scalar:KK:eom}
\end{align}
This is a Sturm-Liouville equation with real eigenvalues and real, orthonormal eigenfunctions,
\begin{align}
\int_R^{R'} \frac{dz}{R}\, \left(\frac{R}{z}\right)^3 f^{(n)}(z)f^{(m)}(z) = \delta^{mn}.
\label{eq:XD:RS:scalar:orthonormal}
\end{align}
Just as we saw in Section~\ref{sec:XD:KK:decomposition} for a flat extra dimension, this orthonormality relation diagonalizes the \KK kinetic terms. One may now solve (\ref{eq:XD:RS:scalar:KK:eom}) 
by observing that through suitable redefinitions this is simply a Bessel equation.
The result is a general solution for $n>0$ of the form
\begin{align}
f^{(n)}(z) = c_1 z^2 J_{\alpha} (m_{(n)}z)
+ c_2 z^2 Y_{\alpha} (m_{(n)}z),
\end{align}
where $J, Y$ are the familiar Bessel functions and $\alpha=\sqrt{4+m^2R^2}$. The integration constants $c_{1,2}$ and the spectrum of \KK masses $m_{(n)}^2$ can be found using boundary conditions on each brane and the orthonormality relation (\ref{eq:XD:RS:scalar:orthonormal}). 
The states have a discrete spectrum with spacing of approximately $R'^{-1} \sim \text{\TeV}$ with profiles peaked towards the \IR brane.

\begin{framed}
\noindent \footnotesize \textbf{Sturm-Liouville theory}.
The orthonormality relations for bulk fields in a warped background are results of the Sturm-Liouville form of the equations of motion for these fields. The generic form of such an equation is
\begin{align}
	\partial_z\left[p(x)\partial_z f^{(n)}(z)\right]
	+ q(z)f^{(n)}(z) = - \lambda w(z) f^{(n)}(z) \ , 
	\label{eq:xd:rs:sturm:liouville:box}
\end{align}
where the \emph{weight} $w(z)>0$ and the eigenvalue $\lambda$ is identified with the squared \KK mass, $m_n^2$. The regular solutions of such an equation satisfy the orthonormality relation
\begin{align}
	\int dz \, f^{(n)}(z) f^{(m)}(z) w(z) = \delta_{mn} \ , 
	\label{eq:xd:rs:sturm:liouville:orthogonality}
\end{align}
where the integration is over the relevant interval. In the case of the \RS model, this range is $(R,R')$ and one must use the dimensionful profiles, $f^{(n)}(z) \to f^{(n)}(z)/\sqrt{R}$ so that the integral is dimensionless. As shown in Appendix~\ref{app:XD:RS}, the equations of motion for bulk fields of non-trivial spin will lead to Sturm-Liouville equations with different weight functions. The orthonormality condition for the \KK profiles of these fields then differ in the power of $(R/z)$ weighting the overlap integral. We note that the equations of motion for bulk fields of any spin can be massaged into Bessel equations, which are themselves a special case of a Sturm-Liouville equation.
\end{framed}

For $n=0$ the zero mode profile is
\begin{align}
f^{(0)}(z) &= c_1 z^{2 - \sqrt{4+m^2R^2}}
+ c_2 z^{2 + \sqrt{4+m^2R^2}}.
\label{eq:XD:RS:scalar:zero:mode}
\end{align}
We use this result in the Goldberger-Wise mechanism discussed below, but let us remark that the zero mode is neither consistent with Neumann nor Dirichlet boundary conditions and requires brane localized terms to generate boundary conditions that permit a zero mode.

The same general procedure can be used to find the profiles of higher spin bulk fields. In Appendices~\ref{sec:XD:RS:bulk:fermion} and
\ref{sec:XD:RS:bulk:gauge} we work through the additional subtleties coming from fermions and gauge bosons. A Standard Model field is associated with the zero mode of a 5D field, where the \SM mass is a correction from electroweak symmetry breaking on the zero mass from the \KK decomposition. Note that the meaning of the 5D profile is that a 4D particle, even though it is localized and pointlike in the four Minkowski dimensions, is an extended plane wave in the fifth dimension. The boundary conditions imposed by the branes mean that this system is essentially identical to a waveguide in electrodynamics\footnote{This should have been no surprise given the appearance of Bessel functions.}.

\subsection{Radius stabilization}
\label{XD:RS:Radius:stabilization}

We've now shifted the Hierarchy problem to a question of why the \IR scale $R'$ is so much larger than the \UV scale $R$. In fact, one should think about $R'$ as the expectation value of a dynamical degree of freedom, $R'=\langle r(x,z)\rangle$, called the \textbf{radion}.
This is identified with the 4D scalar arising from the dimensional decomposition of the 5D metric. This isn't surprising since the metric is, of course, the quantity which measures distances.
% % 
%
Thus far in our description of the \RS framework, the radion is a \textbf{modulus}---it has no potential and could take any value. This is problematic since excitations of this field would be massless and lead to long-range modifications to gravity. It is thus important to find a mechanism that dynamically fixes $R' \sim \text{\TeV}^{-1}$ to (1) provide a complete solution to the Hierarchy problem and (2) avoid constraints from modifications to gravity. 
\begin{framed}
\noindent\footnotesize
\textbf{Don't be fooled by coordinate choices}.
The original \RS literature used variables such that the metric explicitly contained an exponential warping $ds^2 = e^{-2ky}dx^2-dy^2$ so that an $\mathcal O(10)$ value of $k\pi R'$ leads to large hierarchies. Do not confuse this variable choice with a solution to the Hierarchy problem---it just shifts the fine tuning into a parameter to which the theory is exponentially sensitive. %
The reason why the exponential hierarchy is actually physical in \RS (with a dynamically stabilized radius) is that fields propagating in the space are redshifted as they `fall' towards the \IR brane in the gravitational well of the \AdS background.
\end{framed}

A standard solution in the \RS model is the Goldberger-Wise mechanism\footnote{
While the Goldberger-Wise mechanism is just one simple option to stabilize the size of the extra dimension, it is close to what actually happens in string compactifications that tacitly \UV complete the \RS scenario \cite{Brummer:2005sh}. 
} \cite{Goldberger:1999uk, Goldberger:1999un}, where radion kinetic
and potential energy terms conspire against one another to select vacuum with finite $R'$. 
To do this, we introduce a massive bulk scalar field $\Phi(x,z)$ of the type in Section~\ref{sec:XD:bulk:scalar}. We introduce brane-localized potentials for this field which force it to obtain a different \vev at each brane, $\varphi_\text\UV \neq \varphi_\text{\IR}$,
\begin{align}
\Delta \mathcal L &= -\lambda \delta(z-R) \left(\Phi^2 - \varphi^2_\text\UV\right)^2 - \lambda \delta(z-R') \left(\Phi^2 - \varphi^2_\text\IR\right)^2
&
\lambda & \to \infty.
\end{align}
This causes the scalar to pick up a $z$-dependent \vev that interpolates between $\varphi_\UV$ and $\varphi_\IR$,
\begin{align}
\langle\Phi(x,z)\rangle &= \varphi(z)
&
\varphi(R) = \varphi_\text\UV \qquad\quad
\varphi(R') = \varphi_\text\IR.
\label{eq:XD:RS:Goldberger:Wise:BC}
\end{align}
The general form of $\varphi(z)$ is precisely the zero mode profile in (\ref{eq:XD:RS:scalar:zero:mode}) since the $\vev$ carries zero momentum in the Minkowski directions.
One may now consider the terms in the action of $\Phi(x,z)$ (evaluated on the \vev $\varphi(z)$) as contributions to the potential for the radion via $R' = \langle r(x,z)\rangle$. The kinetic term for $\Phi(x,z)$ contributes a potential to $r(x,z)$ that goes like $\varphi'(z)^2$. 
\begin{enumerate}
\item This gradient energy is minimized when $\varphi(z)$ has a large distance to interpolate between $\varphi_{\text\UV,\text\IR}$ since larger $R'$ allows a smaller slope. 

\item On the other hand, the bulk mass for $\Phi(x,z)$ gives an energy per unit length in the $z$-direction when $\varphi(z)\neq 0$. Thus the energy from this term is minimized when $R'$ is small. 
\end{enumerate}
By balancing these two effects, one is able to dynamically fix a value for $R'$. A pedagogical derivation of this presented in \cite{Rattazzi:2003ea}. The main idea is that for small values of the bulk $\Phi(x,z)$ mass, $m^2 \ll R^{-2}$, one may write the $\Phi(x,z)$ \vev as
\begin{align}
\varphi = c_1 z^{-\varepsilon} + c_2 z^{4+\varepsilon},
\end{align}
where $\varepsilon = \alpha - 2 = \sqrt{4+m^2R^2}-2 \approx m^2R^2/4$ is small. The coefficients $c_{1,2}$ are determined by the boundary conditions (\ref{eq:XD:RS:Goldberger:Wise:BC}). The potential takes the form
\begin{align}
V[R'] = \varepsilon \frac{\varphi_\text\UV^2}{R}
+
\frac{R^3}{R'^4}\left[
(4+2\varepsilon)
\left(\varphi_\text\IR - \varphi_\text\UV \left(\frac R{R'}\right)^\varepsilon\right)^2
-\varepsilon \varphi_\text\IR R'^{-4}
\right]
+ \mathcal O\left(\frac{R^4}{R'^8}\right),
\end{align}
where judicious checkers of dimensions will recall that the dimension of the 5D scalar is $[\varphi(x,z)] = \frac{3}{2}$. The minimum of this potential is
\begin{align}
R' = R\left(\frac{\varphi_\text\UV}{\varphi_\text\IR}\right)^{1/\varepsilon}.
\end{align}
We can generate the Planck-weak hierarchy with $1/\epsilon \sim 20$ and $\varphi_\text\UV/\varphi_\text\IR \sim 10$. A key point here is that we may write the radius in terms of a characteristic energy scale, $R' \sim 1/\mu$, and the potential for $\mu$ carries terms that go like $\mu^4$ times a polynomial in $\mu^\epsilon$. This is reminiscent of dimensional transmutation and, indeed, we explain below that the \RS scenario can be understood as a dual description of strongly coupled 4D dynamics.

The above description of the Goldberger-Wise mechanism neglects the effect of the background $\Phi$ field on the \RS geometry. For example, one may wonder if the \RS metric is even compatible with the $\Phi$ \vev. In order to account for this gravitational backreaction, one must solve the $\Phi$ equation of motion combined with the Einstein equation as a function of the metric (discussed in Appendix~\ref{app:XD:RS:Grav:BG}) in the presence of the $\Phi$ \vev. This set of coupled second order differential equations is generically very difficult to solve. Fortunately, there exists a `superpotential\footnote{
The trick was inspired by similar calculations in supergravity but otherwise is only related to \SUSY in the sense that the `superpotential' here also allows one to write first order equations of motion\cite{DeWolfe:1999cp}.
}' trick that one may apply to solve the system exactly. This method is described and demonstrated pedagogically for the Goldberger-Wise field in \cite{Csaki:2004ay, Ponton:2012bi}. One finds that it is indeed possible to maintain the \RS background in the presence of the bulk field necessary to stabilize the radius.

\subsection{Holographic interpretation}
\label{fig:xd:holography}

\begin{framed}
\noindent \footnotesize\textbf{Gauge/gravity duality} is a way to understand the physics of a warped extra dimension as the dual to a strongly coupled 4D theory. Our goal here is to develop the intuition to use and understand the \AdS/\CFT dictionary as an interpretational tool. The most rigorous explicit derivations of this duality are often presented in the language of string theory. %
This idea is presented pedagogically in the language of 4D quantum field theory (rather than string theory) in \cite{ArkaniHamed:2000ds, Sundrum:2011ic, Gherghetta:2006ha, Gherghetta:2010cj, Rattazzi:2000hs, Skenderis:2002wp, McGreevy:2009xe}.
Those interested in presentations that connect to supergravity and string theory may explore \cite{
Nastase:2007kj,
Ramallo:2013bua,
DHoker:2002aw,
Aharony:1999ti,
Polchinski:2010hw,
Blumenhagen:2005mu},
listed roughly in order of increasing formal theory sophistication starting from very little assumed background.
We also point out \cite{Strassler:2005qs} which is an excellent presentation of dualities between 4D supersymmetric gauge theories that are analogous to the gauge/gravity correspondence.
\end{framed}

\begin{figure}
\begin{center}

\begin{tikzpicture}[scale=.6,every node/.style={minimum size=1cm},on grid, rotate=0]
		
% Adapted from TikzExamples

    \begin{scope}[
            yshift=-83,every node/.append style={
            yslant=0.5,xslant=-1},yslant=0.5,xslant=-1
            ]
            
        %% BOTTOM LAYER
        % opacity to prevent graphical interference
        \fill[white,fill opacity=0.9] (0,0) rectangle (5,5);
        \begin{scope}[shift={(1.945,1.945)}]
            \draw[black, thick] (0,0) rectangle (1.111,1.111);
            \draw[step=.370333, gray] (0,0) grid (1.111,1.111);
        \end{scope}
        \draw[blue, dashed] (0,0) rectangle (5,5);%marking borders
    \end{scope}
    
    %% NEXT TO BOTTOM	
    \begin{scope}[
    	yshift=0,every node/.append style={
    	    yslant=0.5,xslant=-1},yslant=0.5,xslant=-1
    	             ]
        \fill[white,fill opacity=.8] (0,0) rectangle (5,5);
        \draw[blue,dashed] (0,0) rectangle (5,5);
        \begin{scope}[shift={(1.667,1.667)}]
            \draw[black, thick] (0,0) rectangle (1.666,1.666);
            \draw[step=.555333, gray] (0,0) grid (1.666,1.666);
        \end{scope}
    \end{scope}
    
    %% NEXT TO TOP
    \begin{scope}[
    	yshift=90,every node/.append style={
    	yslant=0.5,xslant=-1},yslant=0.5,xslant=-1
    	             ]
    	\fill[white,fill opacity=.7] (0,0) rectangle (5,5);
    	\begin{scope}[shift={(.84,.84)}]
    	    \draw[black, step=1.106] (0,0) grid (3.318,3.318);
        	\draw[black,very thick] (0,0) rectangle (3.318,3.318);
    	\end{scope}
    	\draw[blue,dashed] (0,0) rectangle (5,5);
    \end{scope}

    %% TOP
    \begin{scope}[
    	yshift=170,every node/.append style={
    	    yslant=0.5,xslant=-1},yslant=0.5,xslant=-1
    	  ]
        \fill[white,fill opacity=0.8] (0,0) rectangle (5,5);
        \draw[step=5/3, black] (0,0) grid (5,5);
        \draw[black, very thick] (0,0) rectangle (5,5);
    \end{scope}

%  http://tex.stackexchange.com/questions/33607/easy-curves-in-tikz

%    \draw[fill=blue] (0,-1) circle (.08);
%    \draw[fill=blue] (-1.1,-.43) circle (.08);
%    \draw[fill=blue] (1.1,-.43) circle (.08);
%    \draw[fill=blue] (1.65,2.48) circle (.08);
%    \draw[fill=blue] (3.3,5.65) circle (.08);
%    \draw[fill=blue] (4.98,8.46) circle (.08);
%    \draw[fill=blue] (5.7,9) circle (.08);
    
    \node[name=a1] at (1.1,-.43) {};
    \node[name=a2] at (1.65,2.48) {};
    \node[name=a3] at (3.3,5.65) {};
    \node[name=a4] at (4.98,8.46) {};

    \node[name=b1] at (-1.1,-.43) {};
    \node[name=b2] at (-1.65,2.48) {};
    \node[name=b3] at (-3.3,5.65) {};
    \node[name=b4] at (-4.98,8.46) {};
    
    \draw [red, line width=2, opacity=.8] plot [smooth, tension=1] coordinates { (a1) (a2) (a3) (a4)};
    \draw [red, line width=2, opacity=.8] plot [smooth, tension=1] coordinates { (b1) (b2) (b3) (b4)};
    
    \draw [black, line width = 2,->] (-9, -1) -- (-9, 9.9);
    \node at (-9.5, 9) {\large $z$}; 
    \draw [black, line width = 2,->] (-8, 9.5) -- (-8, -1.5);
	 \node at (-7.5, -.5) {\large $\mu$}; 
	 
	 \node[rotate=90] at (-9.5, 4) {\large extra dimension};
	 \node[rotate=-90] at (-7.5, 5) {\large renormalization scale};
	
\end{tikzpicture}
\end{center}
\caption{Cartoon of the \AdS/\CFT correspondence. The isometries of the extra dimensional space enforce the conformal symmetry of the 4D theory. Moving in the $z$ direction corresponds to a renormalization group transformation (rescaling) of the 4D theory.}
\label{fig:xd:holographic}
\end{figure}
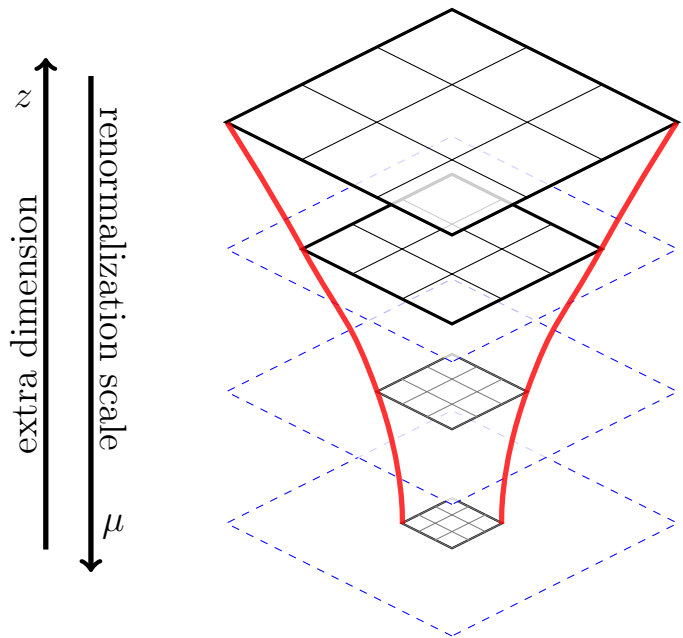

We now introduce a way to re-interpret the observables of \RS scenario in terms of the dynamics of a purely four-dimensional theory in its non-perturbative regime. The idea is that the symmetries of the bulk \AdS space enforce the symmetries of a conformal theory in 4D---this latter theory approximates a strongly coupled theory near a fixed point. Combined with the observation that a shift in $z$ causes an overall rescaling of the \AdS metric (\ref{eq:RS:metric}), we can identify slices of constant $z$ as scale transformations of the 4D [approximately] conformal theory. In this way, the 5D \AdS theory `geometrizes' the renormalization group flow of the 4D theory. One then interprets the physics on the \UV brane as a 4D conformal theory that sets the boundary conditions for the 5D fields. Slices of constant $z$ describe the \RG evolution of this theory at lower energies, $\mu \sim 1/z$.
Because the higher-dimensional theory encodes information about the behavior of a lower-dimensional theory on its boundary, this identification is known as the \textbf{holographic} interpretation of warped extra dimensions. This interpretation is sketched in  Fig.~\ref{fig:xd:holographic}.

\subsubsection{Plausibility check from an experimentalist's perspective}

As a very rough check of why this would be plausible, consider the types of spectra one expects from an extra dimensional theory versus a strongly coupled 4D theory. In other words, consider the first thing that an experimentalists might want to check about either theory. The theory with an extra dimension predicts a tower of Kaluza-Klein excitations for each particle. The strongly coupled gauge theory predicts a similar tower of bound states such as the various meson resonances in \QCD. From the experimentalist's point of view, these two theories are qualitatively very similar.

\subsubsection{Sketch of a more formal description}

We can better motivate the holographic interpretation by appealing to more formal arguments.
One of the most powerful developments in theoretical physics over the past two decades is the \AdS/\CFT correspondence---more generally, the \textbf{holographic principle} or the \textbf{gauge/gravity correspondence} \cite{Maldacena:1997re, Gubser:1998bc, Witten:1998qj, ArkaniHamed:2000ds}. 
The conjecture states that type \textsc{iib} string theory on $\text{\textsc{a}d\textsc{s}}_5\times S^5$ is equivalent to 4D $\mathcal N=4$ superconformal SU$(N)$ theory on Minkowski space in the large $N$ limit: 
\begin{align}
\text{\textsc{a}d\textsc{s}}_5\times S^5  
\qquad\qquad\Longleftrightarrow\qquad\qquad 
\mathcal N=4 \text{ super Yang-Mills.}
\label{eq:XD:AdS:CFT:duality}
\end{align}
The essence of this duality is the observation that a stack of $N$ so-called $D3$-branes in string theory can be interpreted at low energies in two ways:
\begin{enumerate}
\item A solitonic configuration of closed strings which manifests itself as an extended black hole-like object for which $\text{\textsc{a}d\textsc{s}}_5\times S^5$ is a solution.
\item Dirichlet boundary conditions for open strings which admit a non-Abelian $U(N)$ gauge symmetry associated with the $N$ coincident $D3$-branes.
\end{enumerate}
These correspond to the left- and right-hand side of (\ref{eq:XD:AdS:CFT:duality}) and form the basis of the \AdS/\CFT correspondence.

%% See: 1310.4319
The key for us is that the $\text{\textsc{a}d\textsc{s}}_5\times S^5$ extra dimension `geometrizes' the renormalization group flow of the strongly coupled theory by relating the position in the extra dimension $z$ with the \RG scale $\mu$. 
An operator $\mathcal O_i$ in the 4D theory has a source $j_i(x,\mu)$ that satisfies an \RG equation
\begin{align}
\mu \frac{\partial}{\partial \mu} j_i(x,\mu) &= \beta_i(j_j(x,\mu),\mu).
\end{align}
The gauge/gravity correspondence identifies this source as the value of a bulk field $j_i(x,\mu)\Leftrightarrow \Phi_i(x,z)$ at the \UV boundary of the $\text{\textsc{a}d\textsc{s}}_5$ extra dimension. The profile of $\Phi_i$ in the extra dimension is associated with the \RG flow of $j_i(x,\mu)$. Each Minkowski slice of $\text{\textsc{a}d\textsc{s}}_5$ represents a picture of the 4D theory probed at a different energy scale $\mu \sim 1/z$. %%

More concretely, the duality gives a prescription by which the correlation functions of one theory are identified with correlation functions of the other.
The parameters of these two theories are related by
\begin{align}
\frac{R^4}{\ell^4} = 4\pi g^2 N,
\end{align}
where $R$ is the \AdS curvature, $\ell$ is the string length, and $g$ is the Yang Mills coupling. Here we see why \AdS/\CFT is such a powerful tool. In the limit of small string coupling $\alpha' \sim \ell^2$ where string theory can be described by classical supergravity, the dual gauge theory is strongly coupled and very `quantum'. The correlation functions of that theory are non-perturbative and difficult to calculate, whereas the dual description is weakly coupled. The duality gives a handle to calculate observables in theories outside the regime where our usual tools are applicable.

\subsubsection{What it means to geometrize the RG flow}

For our purposes, it is only important that we understand the warped extra dimension as the renormalization group flow of a strongly coupled 4D gauge theory. 
To see how this \RG flow is `geometrized,' we consider the internal symmetries of the two theories.
\begin{itemize}
\item
The isometry of the $S^5$ space is $\text{SO}(6) \cong \text{SU}(4)$. This is precisely the $R$-symmetry group of the $\mathcal N = 4$ gauge theory. 
\item The isometry of the AdS$_5$ space is SO$(4,2)$, which exactly matches the spacetime symmetries of a 4D conformal theory. 
\end{itemize}
Since \RS only has a slice of the \AdS space without the $S^5$, we expect it to be dual to a conformal theory without supersymmetry. Steps towards formalizing the holographic interpretation of Randall-Sundrum are reviewed in \cite{Strassler:2005qs}.

Armed with this background, we can develop a working understanding of how to interpret \RS models as a picture of a strong, four-dimensional dynamics. Observe that in the conformal coordinates that we've chosen, the metric has a manifest scale symmetry
\begin{align}
z &\to \alpha z & x \to \alpha x.
\end{align}
Consider 4D cross sections perpendicular to the $z$ direction. Moving this cross section to another position $z\to \alpha z$ is equivalent to a rescaling of the 4D length scales. Increasing $z$ thus corresponds to a decrease in 4D energy scales. In this way, the \AdS space gives us a holographic handle on the renormalization group behavior of the strongly coupled theory. 

\subsubsection{What it means to take a slice of Anti-de Sitter}

The \RS scenario differs from $\text{\textsc{a}d\textsc{s}}_5$ due to the presence of the \UV and \IR branes which truncate the extra dimension. Since flows along the extra dimension correspond to scale transformations, the branes represent scales at which conformal symmetry is broken. The \UV brane corresponds to an explicit \UV cutoff for the 4D conformal theory. The \IR brane sets the scale of the \KK modes. We heuristically identified these with bound states of the strongly coupled theory, and so we can identify the \IR brane as a scale where conformal symmetry is spontaneously broken, the theory confines, and one finds a spectrum of bound states. Recall that the bound state profiles are localized toward the \IR brane; this is an indication that these bound states only exist as one approaches the confinement scale. The picture of the \RS `slice of \AdS' is thus of a theory which is nearly conformal in the \UV that runs slowly under \RG flow down to the \IR scale where it produces bound state resonances.

The \SM, and in particular the Higgs, exist on the \IR brane and are thus identified with composite states of the strongly coupled theory. In the extra dimensional picture, we argued that the Higgs mass is natural because the \UV cutoff was warped down to the \TeV scale. In the dual theory, the solution to the Hierarchy problem is compositeness (much like in technicolor): the  scalar mass is natural because above the confinement scale the scalar disappears and one accesses its strongly coupled constituents. By comparison, a state stuck on the \UV brane is identified with an elementary (non-composite) field that couples to the \CFT.

\subsubsection{The meaning of 5D calculations}
\label{sec:xd:RS:meaning:of:5d}

At the level presented, it may seem like the \AdS/\CFT correspondence is a magic wand for describing strong coupling perturbatively---and indeed, if you have started to believe this, it behooves you to always know the limits of your favorite tools. A 5D calculation includes entire towers of 4D strongly coupled bound states---in what sense are are we doing a perturbation expansion? First of all, we underscore that the \AdS/\CFT correspondence assumes the 't Hooft large $N$ limit, where $N$ is the rank of the gauge group \cite{tHooft:1973jz}. Further, whether in four or five dimensions, a scattering calculation assumes a gap in the particle spectrum. This gap in the 5D mass is translated into a gap in the scaling dimension $\Delta$ of the 4D \CFT operators. Thus one of the implicit assumptions of a holographic calculation is that the spectrum of the \CFT has a gap in scaling dimensions.  More practically, a scattering process in 5D include 4D fields with large \KK masses. We can say definite things about these large \KK mass states, but only as long as these questions include a sum over the entire tower.

\subsection{The RS Radion is a Dilaton}
\label{sec:XD:radion:is:dilaton}

We have already met the \textbf{radion} as the dynamical field whose \vev sets the distance between the \UV and \IR brane. Excitations of the radion about this \vev correspond to fluctuations in the position of the \IR brane. From its origin as a part of the 5D dynamical metric, it couples to the trace of the energy-momentum tensor,
\begin{align}
\frac{r}{\Lambda_\text\IR} T^\mu_\mu.
\end{align}
Observe that this is very similar to the coupling of the \SM Higgs except that it is scaled by a factor of $\frac{v}{\Lambda_\text\IR}$ and there are additional couplings due to the trace anomaly---for example, a coupling to gluons of the form \cite{Csaki:1999mp, Goldberger:1999un}
\begin{align}
\left[
\frac{r}{\Lambda_{\text\IR}}
- \frac 12 \frac{r}{\Lambda_{\text\IR}} F_{1/2}(m_t)\right]
\frac{\alpha_s}{8\pi}(G_{\mu\nu}^a)^2,
\end{align}
where $F_{1/2}(m_t) = -8m_t^2/m_h^2+\cdots$ is a triangle diagram function, see e.g.\ (2.17) of \cite{Gunion:1989we}.

Why should the radion coupling be so similar to the Higgs?
Before one stabilizes the radion \vev (e.g. as in Section~\ref{XD:RS:Radius:stabilization}), the radion is a modulus and has a flat potential. In the holographic 4D dual, the radion corresponds to the Goldstone boson from the spontaneous breaking of conformal symmetry by the confining dynamics at the \IR scale. In other words, in the 4D theory, the radion is a \textbf{dilaton}. 
This is the reason why it is so similar to the \SM Higgs: the Higgs is \textit{also} a dilaton in a simple limit of the Standard Model.

In the \SM the only dimensionful parameter is that of the Higgs mass,
\begin{align}
V(H) = \lambda\left(H^\dag H - \frac{v^2}{2}\right)^2.
\end{align}
In the limit when $\lambda \to 0$, the Standard Model thus enjoys an approximate scale invariance. If we maintain $v\neq 0$ while taking $\lambda \to 0$, that is, we leave the Higgs \vev on, then:
\begin{itemize}
\item Electroweak breaking SU$(2)\times$U$(1)\to$U(1) gives the usual three Goldstone bosons eaten by the $W^\pm$ and $Z$
\item The breaking of scale invariance gives an additional Goldstone boson, which is precisely the Higgs.
\end{itemize}
Indeed, the Higgs couples to the sources of scale invariance breaking: the masses of the fundamental \SM particles,
\begin{align}
\frac{h}{v}\left(m_f \bar\Psi \Psi + M_W^2W_\mu W^\mu + \cdots\right).
\end{align}

This observation leads to an interesting possibility: could one construct a complete model with no elementary scalar Higgs, but where a condensate breaks electroweak symmetry and scale invariance? Then the dilaton of this theory may have the properties of the \SM Higgs. If one can reproduce the observed Higgs mass then it could be very difficult to tell the scenario apart from the \SM \cite{Bellazzini:2012vz}.

\subsection{Realistic Randall-Sundrum Models}
\label{sec:XD:realistic:RS}

While the original \RS model is sometimes used as a template by \LHC experiments to put bounds on \KK gravitons, most theorists usually refer to \RS to mean a more modern variant than the model presented thus far. In the so-called `realistic' version of Randall-Sundrum, all of the Standard Model fields are allowed to propagate in the bulk \cite{Davoudiasl:1999tf, Grossman:1999ra, Gherghetta:2000qt}. Doing this allows one to use other features of the \RS framework to address other model building issues. For example, pulling the gauge fields into the bulk can help for grand unification, but this typically leads to unacceptably large corrections to the Peskin-Takeuchi $S$-parameter. One way to control this is to also allow the fermions to live in the bulk. We explain below that the bulk fermions open up a powerful new way to use the \RS background to generate the hierarchies in the Yukawa matrix.

Solving the Hierarchy problem requires the Higgs to either be stuck on the \IR brane or otherwise have a bulk profile that is highly peaked towards it. 
Allowing the fermions and gauge fields to propagate in the bulk introduces a tower of \KK modes for each state. These tend to be peaked towards the \IR brane and, as we learned above, are identified with bound states of the strongly coupled holographic dual. 
The Standard Model matter and gauge content are identified with the zero modes of the bulk fields. These carry zero \KK mass and pick up small non-zero masses from their interaction with the Higgs. 
When boundary conditions permit them, zero mode profiles can have different types of behavior:
\begin{itemize}
\item Fermion zero modes\footnote{One immediate concern with bulk fermions is that in 5D the basic spinor representation is Dirac rather than Weyl. Thus one does not automatically obtain a chiral spectrum of the type observed in the \SM. While heavy \KK states indeed appear as Dirac fermions, one may pick boundary conditions for the bulk fermion field that project out the `wrong chirality' zero-mode state. See Appendix~\ref{sec:xd:rs:chiral:bc}.} are either exponentially peaked toward the \IR brane or the \UV brane. The parameter controlling this behavior is the bulk mass\footnote{Observe that this is a manifestation of our identification of bulk masses and scaling dimension in Sec.~\ref{sec:xd:RS:meaning:of:5d}.}, see (\ref{eq:XD:RS:fermion:profile}).
\item Gauge boson zero modes are flat in the extra dimension, though electroweak symmetry breaking on the \IR brane distorts this a bit, see (\ref{eq:XD:RS:Z:profile}).
\end{itemize}
The holographic interpretation of a Standard Model field with a bulk profile is that the \SM state is \textbf{partially composite}. That is to say that it is an admixture of elementary and composite states. This is analogous to the mixing between the $\rho$ meson and the photon in \QCD. States whose profiles are peaked towards the \UV brane are mostly elementary, states peaked toward the \IR brane are mostly composite, and states with flat profiles are an equal admixture.

\begin{figure}
\begin{center}
\includegraphics[height=5.6cm]{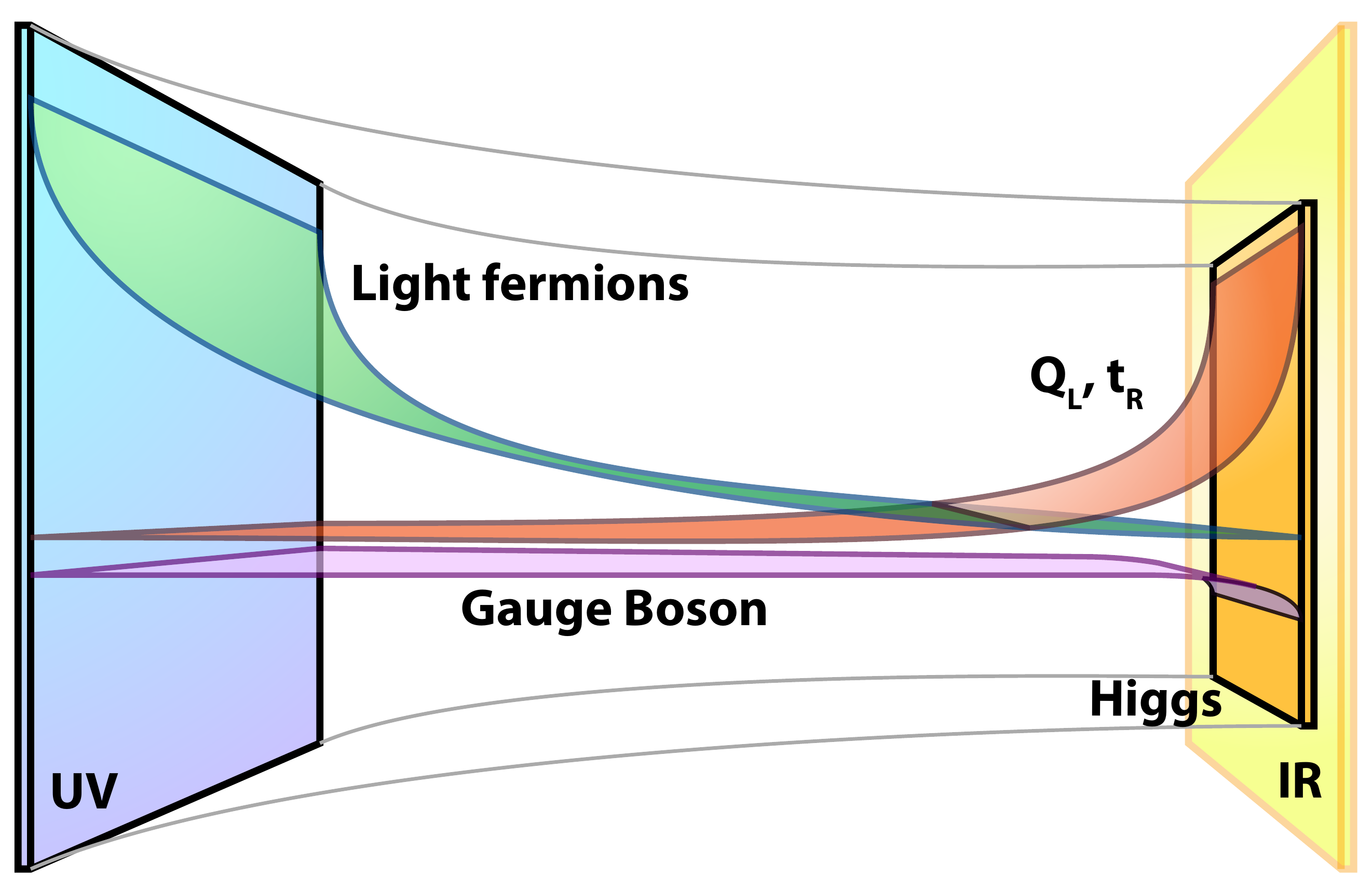}
\end{center}
\caption{A cartoon of the zero mode profiles of various \SM particles in the `realistic' \RS scenario.}
\label{fig:xd:rs:realistic}
\end{figure}

The effective 4D coupling between states depends on the overlap integral of their extra dimensional profiles. This gives a way to understand the hierarchies in the Yukawa matrices, since these are couplings to the Higgs, which is mostly localized on the \IR brane \cite{
Grossman:1999ra, Gherghetta:2000qt,
Huber:2000ie,
Huber:2003tu,
Burdman:2002gr,
Agashe:2004cp,
Agashe:2004ay}.
This is a realization of the split fermion scenario\footnote{Note that the use of an extra dimension to explain flavor hierarchies does not require warping.} \cite{ArkaniHamed:1999dc, Kaplan:2000av, Kaplan:2001ga, Grossman:2002pb}.
The zero-mode fermions that couple to the Higgs, on the other hand, can be peaked on either brane. We can see that even with $\mathcal O(1)$ 5D couplings, if the zero-mode fermions are peaked away from the Higgs, the $dz$ overlap integral of their profiles will produce an exponentially small prefactor. We can thus identify heavier quarks as those whose bulk mass parameters cause them to lean towards the Higgs, while the lighter quarks are those whose bulk mass parameters cause them to lean away from the Higgs. Because the 5D couplings can be arbitrary $\mathcal O(1)$ numbers, this is often called \textbf{flavor anarchy}. This scenario is sketched in Fig.~\ref{fig:xd:rs:realistic}.

This framework tells us how to search for `realistic' \RS models. Unlike the original \RS model, whose main experimental signature were \KK gravitons decaying to \SM states like leptons, the profiles of our \SM fields tell us what we expect realistic \RS to produce. The most abundantly produced new states are those with strong coupling, say the \KK gluon. Like all of the \RS \KK states, this is peaked towards the \IR brane. The \SM field which couples the most to this state are the right-handed tops. This is because we want the tops to have a large Yukawa coupling, and the left-handed top cannot be too peaked on the \IR brane or else the bottom quark---part of the same electroweak doublet---would become heavy. These \KK gluons are expected to have a mass $\gtrsim$ 3 \TeV, so we expect these tops to be very boosted. This suggests experimental techniques like jet substructure (see \cite{Altheimer:2012mn, Shelton:2013an, Salam:2009jx} for reviews).

There are additional features that one may add to the \RS scenario to make it even more realistic. From the picture above, the electroweak gauge \KK modes lean towards the \IR brane where the Higgs can cause large mixing with the \SM $W$ and $Z$. This causes large corrections to the Peskin-Takeuchi $T$ parameter which seems to push up the compactification scale, causing a reintroduction of tuning. A second issue is that the third generation \SM fermions also have a large overlap with the Higgs and can induce a large $Z\bar b b$ coupling through the neutral Goldstone. This coupling is well measured and would also require some tuning in the couplings. It turns out, however, that imposing custodial symmetry in the bulk can address both of these problems \cite{Agashe:2003zs, Agashe:2006at}. The symmetry is typically gauged and broken on the \IR brane so that it is holographically identified with a global symmetry of the 4D theory---just as in the \SM. This introduces several new states in the theory, many of which are required to have boundary conditions that prevent zero modes.

\subsection{A sketch of RS flavor}
\label{sec:xd:rs:flavor:sketch}

Let us assume that the Higgs is effectively \IR brane-localized. The effective 4D Yukawa coupling between a left-handed quark doublet and a right-handed quark singlet is given by the $\mathcal O(1)$ anarchic (non hierarchicial) 5D Yukawa coupling multiplied by the zero-mode fermion profiles evaluated on the \IR brane, $\epsilon$,
\begin{align}
y^u_ij \sim \mathcal O(1)_{ij} \times \epsilon_i^{Q} \epsilon_j^{u_R}.
\end{align}
Here we have implicitly treated the Higgs boson profile as a $\delta$-function on the \IR brane and integrated over the profiles.
In the 4D mass eigenstate basis, $y_t\sim 1$, we can write $\epsilon^{u_R}_3 \sim \epsilon^Q_3 \sim 1$. For a choice of these parameters, one may then use the bottom mass to determine the value of $\epsilon^{d_R}_3$. This, in turn, may be used in conjunction with the \CKM matrix,
\begin{align}
%V_{\text{\CKM} i \lesseq j} \sim \mathcal O\left(
%\frac{\epsilon_i^{Q}}{\epsilon_j^Q}
%\right),
V_{\text{CKM}, i\leq j} \sim \mathcal O\left(
\frac{\epsilon_i^{Q}}{\epsilon_j^Q}
\right),
\end{align}
to determine the $\epsilon$s of lower generations and so forth. One automatically obtains a hierarchical pattern of mixing. 

Neutrino zero modes, on the other hand, must be highly peaked on the \UV brane. In fact, these are typically even more peaked on the \UV brane than the Higgs is peaked on the \IR brane. In other words, one should no longer treat the Higgs as purely brane localized\footnote{This itself causes some conceptual issues since the interactions of a purely brane Higgs is incompatible with the boundary conditions required to make the fermion zero modes chiral\cite{Csaki:2003sh}.} and rather as a profile which is exponentially small on the \UV brane. In this limit, one can treat the right-handed neutrinos as each having a $\delta$-function profile on the \UV brane. Even with $\mathcal O(1)$ anarchic Yukawa couplings, the smallness of the Higgs profile then suppresses the neutrino mass to automatically be small. Further, since each neutrino Yukawa coupling has the same Higgs mass, one finds larger mixing than in the quark sector, as phenomenologically observed.

\subsection{Example: the coupling of the $Z$ in RS}

As a sample calculation, consider the coupling of the $Z$ boson in \RS. We first derive the effective 4D (\SM) coupling of the $Z$ in terms of the 5D parameters and then calculate the \FCNC induced by the zero mode $Z$. In the \SM the $Z$ is, of course, flavor universal and flavor-changing coupling. Indeed, at zeroth order, \RS also prevents such a \FCNC since the gauge boson zero mode profile is flat and therefore universal. We will see, however, that the correction to the $Z$ profile induces a small \FCNC term.

Let us first state some results that are derived in the appendix. The localization of the normalized zero mode fermion profile is controlled by the dimensionless parameter $c$, 
\begin{align}
%    \chi^{(0)}_{c} (x,z) &=\frac{1}{\sqrt{R'}} \left(\frac{z}{R}\right)^2 \left(\frac{z}{R'}\right)^{-c} f_c \;\chi^{(0)}_c(x)
%    \quad\quad\quad \text{and }\quad\quad\quad  \psi_c^{(0)} (x,z) = \chi_{-c}^{(0)} (x,z), \label{BSG:eq:fermion:profile}
%    \\
    \Psi_c^{(0)}(x,z) = \frac{1}{\sqrt{R'}} 
\left(\frac zR \right)^2
\left(\frac{z}{R'}\right)^{-c}
%\sqrt{
%    \frac{1-2c}{1-(R/R')^{1-2c}}
%    } 
f_c
P_L \Psi_c^{(0)}(x),
\end{align}
where $c/R$ is the fermion bulk mass and $P_L$ is the left-chiral projection operator. Right chiral states differ by $P_L\to P_R$ and $c\to -c$.
We have also used the \RS flavor function characterizing the fermion profile on the \IR brane (larger $f$ means larger overlap with the Higgs),
\begin{align}
f_c &=\sqrt{\frac{1-2c}{1-(R/R')^{1-2c}}}.\label{BSG:eq:flavor:function}
\end{align}
Each \SM fermion has a different bulk mass $c$ which according to the size of its \SM Yukawa coupling. For simplicity of notation, we will simultaneously use $c$ as the bulk mass parameter and as a flavor index rather than $c_i$.
Further, the profile for the zero mode $Z$ boson is
%\begin{align}
%h^{(0)}(z) = \frac{1}{\sqrt{R\log R'/R}}
%\left[
%\right]
%\end{align}
\begin{align}
h_Z^{(0)}(z) = \frac{1}{\sqrt{R\log R'/R}}\left[
1 -
\frac{M_Z^2}{4} \left(z^2 - 2z^2 \log \frac zR\right)
\right],
\label{eq:XD:RS:Z:profile}
\end{align}
Starting in the canonical 5D basis where the bulk masses ($c$ parameters) are diagonal, the zero mode fermion coupling to the zero mode $Z$ is
\begin{align}
g_{\text{4D}} Z^{(0)}_\mu(x) \Bar \Psi_c^{(0)}(x) \gamma^\mu \Psi_c^{(0)}(x) + \cdots
=
\int dz \left(\frac Rz\right)^5 g_\text{5D} Z_M^{(0)}(x,z) \bar\Psi^{(0)}_c(x,z) \Gamma^M\Psi^{(0)}(x,z),
\end{align}
where $\Gamma^M = \frac zR \gamma^M$, the prefactor coming from the vielbein. Plugging in the profiles gives
\begin{align}
g_{\text{4D}}^{cc} %Z^{(0)}_\mu(x) \Bar \Psi_c^{(0)}(x) \gamma^\mu \Psi_c^{(0)}(x)
&= 
g_\text{5D}
\int_R^{R'} dz\;
\frac{1}{R'}
\left(\frac zR\right)^{-2c}
f_c^2 
\frac{1}{\sqrt{R \log R'/R}}
\left[1 + \frac{M_Z}{4}\left(z^2 - 2z^2 \log \frac zR\right)\right],
\end{align}
where the $cc$ superscripts index fermion flavor. We write $g_\text{4D}^{cc} = g_\text\SM + g_\text\FCNC^{cc}$ in anticipation that the term in the bracket proportional to $M_Z$ is non-universal and will contribute a \FCNC. The leading term, on the other hand, gives the usual \SM coupling. Performing the $dz$ integral for that term gives
\begin{align}
g_\text{\SM} = \frac{g_5 f_c^2 (R')^{2c}}{R'\sqrt{R \log R'/R}} \frac{R'}{1-2c}\left[
1-\left(\frac{R}{R'}\right)^{1-2c}
\right]
=
\frac{g_5}{\sqrt{R\log R'/R}}.
\end{align}
This is indeed flavor-universal since it is independent of $c$ so that upon diagonalization of the zero mode mass matrix with respect to the Yukawa matrices, this contribution remains unchanged.

On the other hand, the term proportional to $M_Z$ gives a non-universal contribution. Performing a change of variables to $y=z/R$ and performing the $dy$ integral gives
\begin{align}
g_\text\FCNC^{cc} = -g_5 \frac{(M_ZR')^2 \log R'/R}{2(3-2c)}f_c^2,
\end{align}
where we've dropped a subleading term that doesn't have the $\log R'/R$ enhancement. Consider, for example, the coupling between a muon and an electron through the zero mode $Z$. The unitary transformation that diagonalizes the Yukawa mass matrix goes like $f_i/f_j$ so that
\begin{align}
g^{Z_0\mu e}_\text\FCNC 
= 
\left( U^\dag g^{ee} U \right)_{\mu e}
\sim
-\frac{f_e}{f_\mu} 
\left(
    \frac{f_\mu^2}{3-2c_\mu} 
    - \frac{f_e^2}{3-2c_e}
    \right) 
    (M_ZR')^2 
    \frac 12 \log \frac{R'}{R} g_{\text{\SM}}.
\end{align}
We can drop the second term since flavor anarchy requires $f_e^2 \ll f_\mu^2$. The result is
\begin{align}
g^{Z_0\mu e}_\text\FCNC 
=- g_\text\SM
\frac{(M_ZR')^2}{2(3-2c_\mu} \log \frac{R'}{R} f_\mu f_e.
\end{align}
The observation that the coupling is suppressed by $(M_zR')^2$ is sometimes called the `\textbf{RS GIM mechanism}.' Note that in order to do a full calculation, one must also include the non-universal contribution from \KK $Z$ bosons. These couplings do not have a $(M_zR')^2$ suppression, but \FCNC diagrams with these \KK modes are  suppressed by the $Z^{(n)}$ mass.

\section{The Higgs from Strong Dynamics}

\begin{framed}
\noindent \footnotesize
\textbf{Further reading:} The original phenomenological Lagrangian papers lay the foundation for the general treatment of Goldstone bosons \cite{Coleman:1969sm, Callan:1969sn}. See \S 19.6 of \cite{Weinberg:1996kr} for a slightly more pedagogical treatment that maintains much of the rigor of \cite{Coleman:1969sm, Callan:1969sn}, or Donoghue, \textit{et al.}\ for a discussion tied closely to \QCD \cite{Donoghue:DynamicsSM}. Very readable discussions can be found in \cite{Georgi:2009vn, EFTBook}. For a rather comprehensive review that emphasizes the role of `gauge' symmetries, see \cite{Bando:1987br}. For  the composite Higgs see \cite{Contino:2010rs, PreSUSY2015} or the 2012 \textsc{ictp} ``School on Strongly Coupled Physics Beyond the Standard Model''~\cite{ICTP2012School} for a modern sets of lectures and \cite{Bellazzini:2014yua, Csaki:2015hcd} for a phenomenological reviews. See \cite{Perelstein:2005ka, Schmaltz:2005ky} for reviews of the little Higgs scenario. Finally, a recent comprehensive review can be found in~\cite{Panico:2015jxa}.
\end{framed} 

For our last topic we explore models where strong dynamics at a scale $\Lambda \sim 10$ \TeV produces a light, composite Higgs. The solution to the Hierarchy problem is that there is no elementary scalar---beyond $\Lambda$ one becomes sensitive to the underlying `partons' that make up the Higgs.
Through the holographic principle, we have already discussed many broad features of this paradigm in the context of warped extra dimensions above.

One key question to address is the lightness of the Higgs mass. If $\Lambda \sim$ 10 \TeV, how is it that the Higgs appears at 125 \GeV? By comparison, the strong coupling scale for quantum chromodynamics is $\Lambda_\text\QCD \sim \mathcal O(300 \text{ \MeV})$ while most \QCD states, such as the $\rho$ meson and proton are at least as heavy as this\footnote{A better comparison is $\Lambda = 4\pi f_\pi \sim \mathcal O(\text{\GeV})$, where $f_\pi$ is the pion decay constant. `Typical' \QCD states such as the $\rho$ meson have masses of at least this value, $m_\rho \sim \Lambda$. We explain the distinction in Section~\ref{sec:comp:NDA}.}. Those who are sharp with their meson spectroscopy will quickly observe that there is a counter-example in \QCD: the pions are all \emph{lighter} than $\Lambda_\text\QCD$, albeit by only an $\mathcal O(1)$ factor. 

The reason that the pions can be appreciably lighter than the other \QCD states is the well-known story of \textbf{chiral perturbation theory}, a subset of the more general \textbf{nonlinear $\Sigma$ model} (\NLSM) construction. The pions are the Goldstone bosons of the spontaneously broken SU(2)$_L\times$SU(2)$_R$ flavor symmetry coming from chiral rotations of the up and down quarks. Small explicit breaking of this symmetry generates a mass for the pions so that they are \textbf{pseudo-Goldstone} modes. In the composite models that we consider in this section, we assume a similar structure where the Higgs is a pseudo-Goldstone boson of some symmetry for which $\Lambda \approx 4\pi f$ with breaking scale $f\approx 1$ \TeV. We show that the generic composite Higgs set up still requires some tuning between the electroweak scale $v$ and the symmetry breaking scale $f$. One way to generate this `little hierarchy' is through the mechanism of collective symmetry breaking.%, which we introduce in the context of the little Higgs models.
	We close this section by drawing connections to models of an extra dimension and by providing a phenomenological taxonomy of composite Higgs models to help clarify nomenclature.

\subsection{Pions as Goldstone bosons}

Before exploring composite Higgs models in earnest, it is useful to review strong electroweak symmetry breaking in \QCD since this gives a concrete example of the effective theory of Goldstone bosons. It is also useful because electroweak symmetry breaking in \QCD formed the motivation for technicolor models that have since fallen out of favor---it is useful to see why this is, and how composite Higgs models are different from a revival of technicolor.

First, consider the Lagrangian for pure \QCD: a theory of vector-like quarks and gluons, where `vector-like' mean the left- and right-handed quarks come in conjugate representations,
\begin{align}
\mathcal L_\text\QCD = - \frac 14 G^a_{\mu\nu} G^{a\mu\nu} + \bar q (i\slashed{D} - m) q.
\end{align}
This is a theory which becomes strongly coupled and confines at low energies, leading to a spectrum of composite states. This makes it a good template for our own explorations into compositeness. We can already guess that at low energies the effective theory is described by Goldstone bosons, the pions. In anticipation, we examine the global symmetries of the theory.

We focus only on the three lightest quarks with masses $m_i \ll \Lambda_\text\QCD$.
In the chiral limit, $m\to 0$, the physical quarks are Weyl spinors and have an enhanced U(3)$_\text{L}\times$U(3)$_\text{R}$ global flavor symmetry acting separately on the left- and right-handed quarks,
\begin{align}
q_L^i &\to (U_{L})^{i}_{\phantom{i}j} q^j_L\\
q_R^i &\to (U_{R})^{i}_{\phantom{i}j} q^j_R.
\end{align}
%Moving back to Dirac spinors, one may write the conserved currents in the form $(j_{L,R})_\mu^a = \bar q \gamma^\mu T^a P_{L,R} q$.
One may write the currents for this global symmetry. For compactness we move back to Dirac spinors and write in terms of the vector ($U_\text{L}=U_\text{R}$) and axial ($U_\text{L}=U_\text{R}^\dag$) transformations:
\begin{align}
(j_V^a)^{\mu} &= \bar q \gamma^\mu T^a q
%\label{eq:comp:vect:j}\\
&
(j_A^a)^{\mu} &= \bar q \gamma^\mu \gamma_5 T^a q
\label{eq:comp:axial:j}\\
(j_V)^\mu &= \bar q \gamma^\mu q
%\\
&
(j_A)^\mu &= \bar q \gamma^\mu \gamma_5 q,
\end{align}
where the $T^a$ are the generators of SU(3).
We can identify $j_V$ with baryon number, which is conserved in \QCD, and we note that $j_A$ is anomalous so that it is not a good symmetry and we don't expect to see it at low energies\footnote{What happens to this symmetry at low energies is rather subtle and was known as the `U(1) problem.' There is a lot more to the story than simply saying that the axial U(1) is anomalous and so does not appear at low energies. One can construct a current out of $j_A$ and a Chern-Simons (topological) current that is anomaly-free and spontaneously broken. This current indeed has a Goldstone pole. However, Kogut and Susskind showed  that this current is not gauge invariant. There are actually two Goldstone bosons that cancel in any gauge invariant operator~\cite{Kogut:1973ab}.}. The vectorial SU(3), with current $j^a_V$, is precisely the symmetry of Gell-Mann's eightfold way and can be used to classify the light hadrons. What do we make of the axial SU(3), $j^a_A$?

Phenomenologically we can observe that the axial SU(3) is not a symmetry of the low energy spectrum, otherwise we would expect a parity doubling of all the `eightfold way' multiplets. There is one way out: this symmetry must be spontaneously broken. What could possibly enact this breaking in a theory with no Higgs boson? It turns out that \QCD itself can do the job!  We assume that the axial SU(3)$_\text{A}$ is broken spontaneously by a quark--anti-quark condensate,
\begin{align}
\langle \bar q q\rangle = \langle \bar q_L^i q_{Ri} + \text{h.c.} \rangle \neq 0
\end{align}
in such a way that the vector SU(3)$_\text{V}$ is preserved. This is the unique combination that preserves Lorentz invariance and breaks SU(3)$_\text{A}$. By dimensional analysis, this `chiral condensate' takes the form
%\begin{align}
$
\langle \bar q^i q_j \rangle \sim \delta_{ij}\Lambda_\text\QCD^3.
$
%\end{align}
Given that \QCD is strongly interacting in the \IR, the existence of this non-trivial vacuum condensate should not be surprising and is indeed supported by lattice calculations. However, the exact mechanism by which this condensate forms is non-perturbative and not fully understood. This also gives a robust prediction: we should have eight pseudoscalar Goldstone bosons as light excitations. These are precisely the pions, kaons, and $\eta$. Because SU(3)$_\text A$ is only a symmetry in the chiral $m\to 0$ limit, these are not exactly Goldstone bosons as the symmetry is explicitly broken by the quark masses and electromagnetism. However, because this explicit breaking is small relative to $\Lambda_\text\QCD$, these excitations are still very light $m_\pi \ll \Lambda_\text\QCD$ and are often referred to as \textbf{pseudo-Goldstone bosons} (sometimes pseudo-Nambu--Goldstone bosons, p\textsc{ngb}, in the literature).

Note that the electroweak group sits inside the \QCD flavor symmetry\footnote{It has to be true that the electroweak gauge group sits in the full \QCD global symmetry group in order for some of the quarks to have non-trivial electroweak charges.},
\begin{align}
\text{SU(3)}_\text{L} \times \text{SU(3)}_\text{R} \times \text{U(1)}_\text{B} \supset
\text{SU(2)}_\text{L} \times \text{U(1)}_\text{Y}.
\label{eq:comp:QCD:flavor:and:EW}
\end{align}
We can see this since an SU(3)$_\text L$ fundamental contains $(u_L, d_L, s_L)$, where the first two components form the usual SU(2)$_\text L$ first generation quark doublet. In this way, SU(2)$_\text L$ is simply the upper left $2\times 2$ component of the SU(3)$_\text L$ generators. Similarly, hypercharge is a combination of the diagonal generators,
\begin{align}
Y = T_{R3} + \frac B2.
\label{eq:comp:Y:in:TR:B}
\end{align}
We say that the electroweak group is \textbf{weakly gauged} with respect to low energy \QCD. By this we mean that the gauge couplings are perturbative in all energy scales of interest. This weak gauging is a small explicit breaking of the \QCD flavor symmetries and accounts for the mass splitting between the $\pi^0$ and $\pi^\pm$.

\subsection{A farewell to technicolor}
\label{sec:comp:no:technicolor}

Because of (\ref{eq:comp:QCD:flavor:and:EW}), the spontaneous breaking of SU(3)$_\text{A}$ by the chiral condensate $\langle \bar q q\rangle$ also breaks electroweak symmetry. This is an important observation: even if there were no Higgs boson, electroweak symmetry would still be broken and $W$ and $Z$ bosons would still be massive, albeit with a much smaller mass. This mass comes from `eating' part of the appropriately charged pseudo-Goldstone bosons. We will see this in slightly more detail below. Readers unfamiliar with this story are encouraged to follow the treatment in \cite{Quigg:2009xr}.

The observation that strong dynamics can---and indeed, \emph{does} in \QCD---break electroweak symmetry led to the development of \textbf{technicolor} theories where the \SM is extended by a confining sector \cite{Jackiw:1973tr, Cornwall:1973ts, Weinstein:1973gj, Weinberg:1979bn, Susskind:1978ms}. By the holographic interpretation of extra dimensions, this type of electroweak symmetry breaking is analogous to the \RS scenario where a brane-localized Higgs picks up a \vev. The large hierarchy between the Planck and electroweak scales is then understood to be a result of dimensional transmutation. The simplest constructions of these models, however, suffer from several issues. These include the requirement for an additional mechanism to generate fermion masses \cite{Dimopoulos:1979es, Eichten:1979ah} and generically large deviations in flavor and electroweak precision observables \cite{Dashen:1969eg, Dashen:1970et, Eichten:1979ah}. However, the nail in the coffin for most of these models is observation of the Higgs boson at 125 \GeV, much lighter than the compositeness scale.  Such a state---even if it is not the \emph{Standard Model} Higgs---is very difficult to explain in the context of these models.

As such, even though the models we consider here encode strong dynamics, they are completely different from the pre-Higgs technicolor strong dynamics of the past. To repeat: composite Higgs is \emph{not} technicolor. In technicolor, the strong dynamics generates a techni-condensate $\langle \bar Q Q\rangle$ of techni-quarks which spontaneously breaks SU(2)$_\text{L}\times$U(1)$_\text{Y}$. In the composite Higgs models we consider, there is a spontaneous breaking of some symmetry which produces Goldstone bosons but does \emph{not} itself break electroweak symmetry. Instead, one of the Goldstone bosons develops a non-trivial potential and is identified with the Higgs doublet.

\subsection{Chiral perturbation theory}
\label{sec:comp:chipt}

In this section we review the main framework for describing Goldstone bosons of chiral symmetry breaking, known as \textbf{chiral perturbation theory}. Many of the results highlight general principles that appear in any theory of Goldstone bosons, known as \textbf{nonlinear sigma models}. A completely general treatment of spontaneously broken global symmetries is captured in the the so-called Callan-Coleman-Wess-Zumino (\CCWZ) construction, which we present in Appendix~\ref{sec:comp:CCWZ}.

The importance of having a Lagrangian theory of Goldstone bosons is clear from the success of \SM predictions before the Higgs discovery. Na\"ively, one might wonder how we knew so much about the Standard Model before the Higgs discovery---isn't the Higgs a very central piece to the theory? As we saw above, the key feature is actually electroweak symmetry breaking: whether or not there is a Higgs, one always has the Goldstone bosons which are eaten by the $W^\pm$ and $Z$ to become massive. It is this nonlinear sigma model that pre-Higgs experiments had studied so carefully. The discovery of the Higgs is a statement that the nonlinear sigma model is \UV completed into a linear sigma model.

\subsubsection{Framework} \label{sec:composite:framework}

We begin with the concrete example of low-energy \QCD that we described above. Given that the chiral condensate $\langle \bar q q \rangle$ breaks SU(3)$_\text A$, we proceed to write down the effective theory describing the interaction of the resulting Goldstone bosons. Let us write $U_0$ to refer to the direction in field space associated with the chiral condensate, $U_0 \sim \langle\bar q q \rangle$. This transforms as a \textbf{bifundamental} with respect to SU(3)$_\text{L}\times$SU(3)$_\text{R}$,%%, that is it transforms in the $(\mathbf{3}_L, \bar\mathbf{3}_R)$ representation,
\begin{align}
U(x) \to U_\text L U(x) U_\text R^\dag,
\end{align}
where $U_L$ and $U_R$ are the transformation matrices under the SU(3)$_\text{L}$ and SU(3)$_\text{R}$ respectively. The observation that SU(3)$_\text A$ is broken corresponds to $U_0 = \mathbbm{1}$.
Note that this indeed preserves the SU(3)$_\text{V}$ transformations $U_\text{L} = U_\text{R}$.

\begin{figure}
\begin{center}
\includegraphics[width = .6\textwidth]{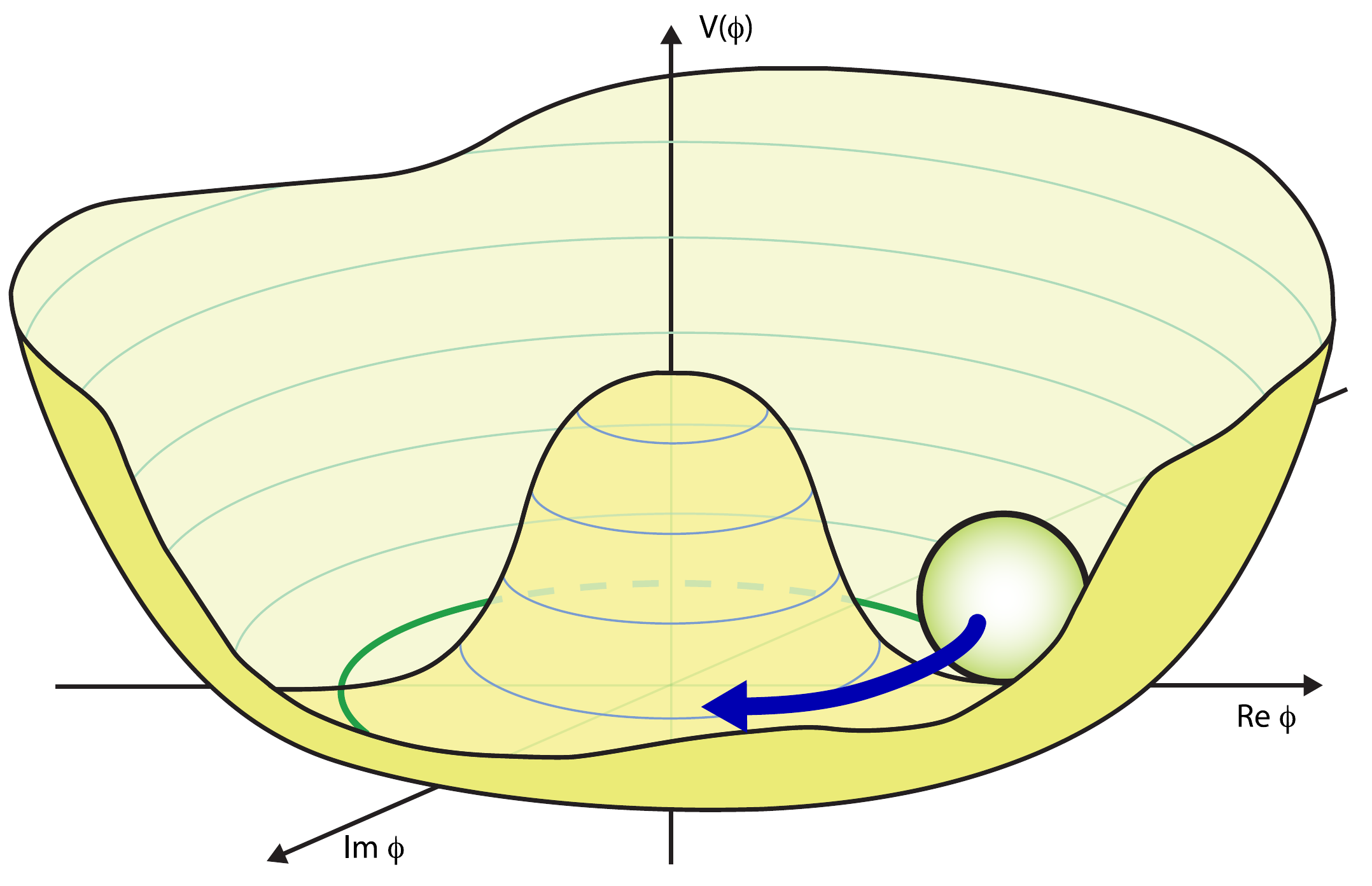}
\end{center}
\caption{Cartoon of the Goldstone excitation for a `Mexican hat' potential. Image from \cite{HiggsVectorImage}.}
\label{fig:comp:mexicanhat}
\end{figure}

We now consider the fluctuations $U(x)$ about $U_0$---these are what we identify with the Goldstone bosons. Recall the picture of spontaneous symmetry breaking through the `Mexican hat' potential in Fig.~\ref{fig:comp:mexicanhat}. The action of an unbroken symmetry does not affect the \vev (represented by the ball), while broken symmetries shift the \vev along the vacuum manifold.  This gives an intuitive picture of how to identify the Goldstone modes:
\begin{enumerate}
\item Identify a convenient \vev, $U_0$
\item Act on that \vev with the broken group elements
\item Promote the transformation parameter to a field, identify these with the Goldstones.
\end{enumerate}
For the chiral Lagrangian, our broken symmetries are those for which $U_L = U_R^\dag$. Writing $U_L = \exp(i \epsilon^a T^a)$, we act on $ U_0  = \mathbbm{1}$,
\begin{align}
e^{i \epsilon^a T^a} 
\begin{pmatrix}
1 &&\\
&1&\\
&&1
\end{pmatrix}
e^{i \epsilon^a T^a} 
=
e^{2 i \epsilon^a T^a}.
\end{align}
We now promote the transformation parameter $\epsilon^a$ to Goldstone fields, $\epsilon^a \sim \pi^a(x)$. Since $\epsilon^a$ is dimensionless, in order for $\pi^a$ to have canonical scaling dimension we should rescale by the \textbf{decay constant}\footnote{The name comes from identifying the appearance of this factor in the matrix element for pion decays, e.g.\ $\langle 0| \bar u \gamma^\mu \gamma_5 d | \pi^-\rangle \equiv i f p^\mu$.} $f$.  We may understand the physical meaning of $f$ if we recall Fig.~\ref{fig:comp:mexicanhat}, since we want $\epsilon$ to be an angle that parameterizes the position along the vacuum circle: the Goldstone is a periodic variable with period $2\pi f$, so that $f$ is identified with the value of the symmetry breaking \vev. The angle $\epsilon$ is then $\pi(x)/f$. %where one should be careful not to confuse $\pi(x) \in (0,2\pi)$ with $\pi = 3.14$. 
We thus promote $\epsilon^a \to \pi(x)/f$ so that we may define the field $U(x)$,
\begin{align}
U(x) = e^{i \frac{\pi^a(x)}f T^a} \, U_0 \, e^{i \frac{\pi^a(x)}{f} T^a} = e^{2i \frac{\pi^a(x)}f T^a}.
\label{eq:comp:chiral:pt:U:field}
\end{align}
We now have an object $U(x)$ which packages the Goldstone fields, $\pi^a(x)$. Note that $U(x)$ transforms linearly under the full SU(3)$_\text L \times$SU(3)$_\text R$ group, $U(x) \to U_L U(x) U_R^\dag$, but the fields that actually describe the low energy spectrum are related in a non-trivial way to $U(x)$.

\subsubsection{How pions transform}

We can determine the transformation of the pions $\pi^a$ by using the transformation of the linear field $U(x)$. Under the SU(3)$_\text V$ (unbroken) symmetry, 
$U_\text L = U_\text R = U_\text V$, we have
\begin{align}
U(x) \to U_\text V U(x) U_\text V^\dag = U_\text V \left(1 + 2i\frac{\pi^a(x)}{f}T^a + \cdots\right) U_\text V^\dag,
\end{align}
where we can see from the first term that $\pi^a(x)T^a \to U_V \pi^a(x) T^a U_V^\dag$. In other words, $\pi^a(x)$ transforms \emph{linearly} under the unbroken symmetry. Note that the higher order terms also obey this by trivially inserting factors of $U_V^\dag U_V = \mathbbm{1}$. Indeed, we expected this result because we know that Gell-Mann's eightfold way is precisely a realization of SU(3)$_\text V$, so our pions must transform as octets. 

Things are not as simple for the broken symmetry, $U_\text L = U_\text R^\dag=U_\text A$. In this case the transformation is
\begin{align}
U(x) \to U_\text A U(x) U_\text A \equiv e^{2i\frac{\pi'^{a}(x)}f T^a}.
\end{align}
In this case the pion does \emph{not} transform in a nice, linear way\footnote{This may seem confusing since $U(x)$ transforms as a bifundamental under SU(3)$_\text{L}\times\text{SU}(3)_\text{R}$. However, components of $U(x)$ are not independent due to the nonlinear constraints of being unitary and having unit determinant.}. Unlike the above case, there is no sense in which this looks like $\pi^a(x)T^a \to U_\text A \pi^a(x)T^a U_\text A^\dag$.  The best we can do is say that we have moved $U_0$ to a new point on the vacuum manifold, which we parameterize by an angle $2\pi'^a(x)/f$. The transformation $\pi^a(x) \to \pi'a(x)$ is nonlinear. To leading order,
\begin{align}
1 + 2i \frac{\pi'^a(x)}f T^a
&= \left(
1+ic^a T^a
\right)
\left(
1 + 2i \frac{\pi^a(x)}f T^a
\right)
\left(
1+ic^a T^a
\right)
\end{align}
so that
\begin{align}
\pi'^a(x) T^a = \pi^a(x) T^a + f c^a T^a.
\end{align}
In other words, to leading order the pion shifts $\pi^a \to fc^a$. This shift symmetry in the nonlinear realization is precisely why the pion is massless; the only non-trivial pion Lagrangian terms must carry derivatives.

\begin{framed}
\noindent \footnotesize
\textbf{Coset space description}. In anticipation of the more general \CCWZ construction, let us restate the above arguments in a more compact way. The symmetry breaking pattern is the coset SU(3)$_\text L\times$SU(3)$_\text R/$SU(3)$_\text V$. Using the notation above, this means that group elements of the full symmetry $U_{L,R}$ can be written as a product of  elements of the unbroken group $U_\text V$ and the [left] coset $U_\text A \in$ SU(3)$_\text L\times$SU(3)$_\text R/$SU(3)$_\text V$,
\begin{align}
U_\text{L} &= U_\text{A} U_\text{V}
&
U_\text{R} &= U_\text{A}^\dag U_\text{V}.
\label{eq:comp:chiral:coset}
\end{align}
One can check that this matches the above cases when one sets $U_\text{A}=\mathbbm{1}$ or $U_\text{V}=\mathbbm{1}$. The general transformation of the linear packaging of the pions, $U(x) = \exp\left(2i\pi^a(x)T^a/f\right)$, is
\begin{align}
U(x) \to U_\text A \left( U_\text V U(x) U_\text{V}^\dag \right) U_\text{A}.
\end{align}
From here it is clear that SU(3)$_\text V$ is realized linearly while SU(3)$_\text A$ is realized non-linearly.
\end{framed}

\begin{framed}
\noindent \footnotesize
\textbf{$\text{SU(3)}_\text{A}$ is not a subgroup of $\text{SU(3)}_\text{L}\times \text{SU(3)}_\text{R}$}. While one can divide the algebra of $\text{SU(3)}_\text L\times \text{SU(3)}_\text R$ into axial and vector generators, one should note that there is no such thing as an `axial subgroup' of $\text{SU(3)}_\text L\times \text{SU(3)}_\text R$. One can check that the commutation relations of axial generators include vector generators so that the SU(3)$_\text{A}$ algebra doesn't close by itself.
\end{framed}

\subsubsection{Lagrangian description}
\label{sec:comp:xpt:lagrangian}

Thus far we have found a convenient way to package the Goldstone fields $\pi^a(x)$ into a linear realization of the full SU(3)$_\text L \times$SU(3)$_\text R$ symmetry. We would like to write down a Lagrangian describing the dynamics of the Goldstones. 
Our strategy will be to write the lowest order terms in $U(x)$ that are SU(3)$_\text L \times$SU(3)$_\text R$ invariant and then expand $U(x)$ in Goldstone excitations about $U_0$.
One can see that many invariants, such as $U(x)^\dag U(x)$, are independent of the Goldstones. In fact, only derivative terms contain the Goldstone fields. This is consistent with our argument that Goldstones must have derivative couplings. The lowest order non-trivial term is 
\begin{align}
\mathcal L = \frac{f^2}4 \text{Tr}\left[ \left(\partial^\mu U^\dag(x)\right)\partial_\mu U(x)\right]
\label{eq:comp:chiral:pert}
\end{align}
The pre-factor is fixed by expanding $U(x) = 1 + 2i\frac{\pi^a(x)}f T^a + \cdots$ and ensuring that the kinetic term for $\pi^a(x)$ is canonically normalized. We have used the normalization that $\text{Tr}\, T^aT^b = \frac 12 \delta^{ab}$. The higher order terms in the expansion of $U$ yield a series of non-renormalizable pion--pion interactions.

Next we weakly gauge the electroweak group. Recall that this sits in SU(3)$_\text L\times$SU(3)$_\text R\times$U(1)$_\text B$. The left- and right-chiral quarks are fundamentals under SU(3)$_\text L$ and SU(3)$_\text{R}$ respectively and have baryon number $1/3$. This information, combined with knowing how SU(2)$_\text L$ sits in SU(3)$_\text L$ and (\ref{eq:comp:Y:in:TR:B}), determines the quantum numbers of the linear field $U(x)$, which transforms as a $\bar{\mathbf{3}} \times \mathbf{3} \times 0$ under SU(3)$_\text L\times$SU(3)$_\text R\times$U(1)$_\text B$. To `turn on' the electroweak gauge interactions, we simply promote derivatives to covariant derivatives $\partial_\mu \to D_\mu$ where
\begin{align}
D_\mu U(x)^i_{\phantom{i} j}
&=
\partial_\mu U(x)^i_{\phantom{i} j}
- ig W_\mu^a \frac 12\left( \tau^a\right)^{i}_{\phantom{i} k}U(x)^k_{\phantom{k} j}
+ i g' B_\mu \frac 12 
U(x)^i_{\phantom{i} k} 
\left( {T^3_R}\right)^{k}_{\phantom k j} 
.
\end{align}
We have written the SU(2)$_\text{L}$ generators as
\begin{align}
\frac 12 \tau^a
= 
\frac 12
	\left(
	\begin{array}{cc:c}
	\multicolumn{2}{c:}{\multirow{2}{*}{\large{$\tau^a$}}} & \multicolumn{1}{c}{\multirow{2}{*}{\large{$0$}}} \\
	 &&  \\
	\hdashline
	0 & 0 & 0
	\end{array}
	\right)
	\subset \text{SU(3)}_\text{L}.
\end{align}
Promoting $\partial_\mu \to D_\mu$ in (\ref{eq:comp:chiral:pert}) yields
\begin{align}
%\mathcal L = \frac{f^2}{4}
%\text{Tr}
%\left|
%\left(
%\frac{2i}{f} \partial_\mu \pi^a(x) T^a
%- \frac{ig}{2}W^a_\mu(x) \tau^a
%\right)U(x)
%\right|^2 + \cdots,
\mathcal L = \frac{f^2}{4}
\text{Tr}
\left|
\left(
\partial_\mu 
- \frac{ig}{2}W^a_\mu(x) \tau^a
\right)U(x)
\right|^2 + \cdots,
\label{eq:comp:EWSB:in:QCD:L}
\end{align}
where we leave the similar term with $B_\mu(x)$ implicit.

\subsubsection{Electroweak symmetry breaking}

One may check that (\ref{eq:comp:EWSB:in:QCD:L})  has terms that are linear in $W(x)$ such as $\frac {g}2 f W_\mu^+(x)\partial^\mu \pi^-(x) + \text{h.c.}$ This is precisely a mixing term between the $\pi^+(x)$ and the $W_\mu^+(x)$. In other words, the $W$ has eaten the Goldstone boson to pick up a longitudinal polarization. This is precisely electroweak symmetry breaking at work. Note that similar terms mixing the $W_\mu^3$ and $B_\mu$ with the $\pi^0$. As usual, the masses of the heavy gauge bosons come from the gauge fields acting on the $U_0$ `\vev' part of $U(x)$, the resulting spectrum is
\begin{align}
\Delta \mathcal L = \frac{g^2f^2}4 W^+W^- + \frac{g^2+g'^2}{4}f^2 \frac{Z^2}{2}.
\end{align}
The characteristic mass scale is 100 \MeV, much smaller than the actual $W$ and $Z$ since most of the mass contribution to those fields comes from the Higgs \vev. Diagrammatically, we can imagine the mixing as follows:
%%% This seems to mis-behave when I use externalize
\begin{align}
\vcenter{\hbox{
\begin{tikzpicture}[line width=1.5 pt, scale=.75]
	\draw[vector] (0,0) -- (1,0);
	\begin{scope}[shift = {(1,0)}]
%	\draw[line width=1.5pt, color=red] (-1,-.3) circle (1.75);
	\clip (.5,0) circle (.5);
	\foreach \x in {-1.9,-1.8,...,1.3}
	\draw[line width=.5 pt, color=black] (\x,-1.3) -- (\x+.6,1.3);
\end{scope}
    \draw (1.5,0) circle (.5);
	\draw[vector] (2,0) -- (3,0);
\end{tikzpicture}
}}
=
\vcenter{\hbox{
\begin{tikzpicture}[line width=1.5 pt, scale=.75]
\draw[vector] (0,0)--(1.5,0);
\end{tikzpicture}
}}
+
\vcenter{\hbox{
\begin{tikzpicture}[line width=1.5 pt, scale=.75]
\draw[vector] (0,0)--(1,0);
\draw (1.5,0) circle (.5);
\node at (1.5,0) {$\Pi$};
\draw[vector] (2,0)--(3,0);
\end{tikzpicture}
}}
+
\vcenter{\hbox{
\begin{tikzpicture}[line width=1.5 pt, scale=.75]
\draw[vector] (0,0)--(1,0);
\draw (1.5,0) circle (.5);
\node at (1.5,0) {$\Pi$};
\draw[vector] (2,0)--(3,0);
\draw (3.5,0) circle (.5);
\node at (3.5,0) {$\Pi$};
\draw[vector] (4,0)--(5,0);
\end{tikzpicture}
}}
+
\cdots
\label{eq:comp:chiral:W:corrected:prop}
\end{align}
We have parameterized the strong dynamics in terms of a momentum-dependent form factor $\Pi(q^2)$. What the $W$ boson is really coupling to is the $\text{SU}(2)_\text L$ current formed from the quarks,
\begin{align}
\vcenter{\hbox{
\begin{tikzpicture}[line width=1.5 pt, scale=1]
\draw[vector] (0,0)--(1,0);
\draw (1.5,0) circle (.5);
\node at (1.5,0) {$\Pi$};
\draw[vector] (2,0)--(3,0);
\node at (-.2,0) {$\mu$};
\node at (3.2,0) {$\nu$};
\end{tikzpicture}
}}
=
\vcenter{\hbox{
\begin{tikzpicture}[line width=1.5 pt, scale=1]
\draw[vector] (-.3,0)--(.7,0);
\draw[fermionnoarrow, line width=1.2] (1.2,-.325) -- (.7,0);
\draw[fermionnoarrow, line width=1.2] (.7,0) -- (1.2,.325);
\draw[fermionnoarrow, line width=1.2] (1.8,-.325) -- (2.3,0);
\draw[fermionnoarrow, line width=1.2] (2.3,0) -- (1.8,.325);
\draw[fill=white, line width=1.2] (1.5,0) circle (.5);
\node at (1.5,0) {\footnotesize QCD};
\draw[vector] (2.3,0)--(3.3,0);
\node at (-.5,0) {$\mu$};
\node at (3.5,0) {$\nu$};
\end{tikzpicture}
}}
\end{align}
where the $W$ bosons are coupling to quarks which then interact strongly with one another.
In other words,
\begin{align}
i\Pi_{\mu\nu}(q) = \langle J^+_\mu(q) J^-_\nu(-q)\rangle.
\label{eq:comp:form:factor:JJ}
\end{align}
The \QCD corrected $W$ propagator $\Delta_{\mu\nu}(q)$ from resumming the diagrams in (\ref{eq:comp:chiral:W:corrected:prop}) is
\begin{align}
\Delta_{\mu\nu}(q) &= \frac{-i}{q^2 - g^2 \Pi(q^2)/2}
&
\Pi_{\mu\nu}(q) &= \left(\eta_{\mu\nu} - \frac{q_\mu q_\nu}{q^2} \right) \Pi(q^2).
\label{eq:comp:form:factor:prop}
\end{align}
The observation that a charged pion has been `eaten' to make the $W$ massive is the statement that $\Pi_{\mu\nu}(q^2)$ has a zero-momentum pole. Indeed, $\langle 0 | J_\mu^+ | \pi^-(p)\rangle = i f_\pi p_\mu/\sqrt{2}$. The \QCD blobs in (\ref{eq:comp:chiral:W:corrected:prop}) also encode, however, the effects of heavier resonances and has poles at the masses of these states. In the `large $N$' limit (large number of colors) one may write the current-current correlation function as a sum of resonances \cite{tHooft:1973jz, Witten:1979kh, Coleman:Aspects},
\begin{align}
\left(\eta_{\mu\nu} - \frac{q_\mu q_\nu}{q^2}\right)\Pi(q^2)
&=
 \left(q^2\eta_{\mu\nu} - {q_\mu q_\nu}\right)
\sum_n \frac{f^2_n}{q^2 -m_n^2},
\label{eq:comp:form:factor:poles}
\end{align}
where the Goldstone pole appears for $m_0=0$.

\subsubsection{Electromagnetic mass splitting}

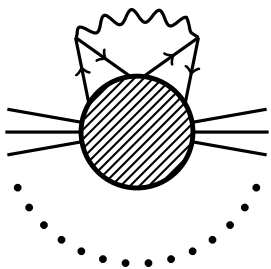
\begin{figure}
\begin{center}
	\begin{tikzpicture}[line width=1.5 pt, scale=2.5]
		
		\draw[fermionbar, line width=1.2 pt] (57:.6) -- (85:.29);
		\draw[fermion, line width=1.2 pt] (57:.6) -- (29:.29);

		\draw[fermionbar, line width=1.2 pt] (123:.6) -- (151:.29);
		\draw[fermion, line width=1.2 pt] (123:.6) -- (95:.29);
		
		\draw[vector, line width=1.2pt] (123:.6) arc (123:57:.6);
		
		\draw[fermionnoarrow, line width=1.2pt] (0,0) -- (170:.7);
		\draw[fermionnoarrow, line width=1.2pt] (0,0) -- (180:.7);
		\draw[fermionnoarrow, line width=1.2pt] (0,0) -- (190:.7);
		
		\draw[fermionnoarrow, line width=1.2pt] (0,0) -- (10:.7);
		\draw[fermionnoarrow, line width=1.2pt] (0,0) -- (0:.7);
		\draw[fermionnoarrow, line width=1.2pt] (0,0) -- (-10:.7);

       \begin{scope}		
		\foreach \x in {205,215,...,335}
		    \draw[fill=black] (\x:.7) circle (.01);
       \end{scope}
      
       % 		\draw[fill=white] (0,0) circle (.29cm);
		
		\draw[fill=black] (0,0) circle (.3cm);
		\draw[fill=white] (0,0) circle (.295cm);
		\begin{scope}
	    	\clip (0,0) circle (.3cm);
	    	\foreach \x in {-.95,-.9,-.85,...,.3}
				\draw[line width=.8 pt] (\x,-.3) -- (\x+.6,.3);
	  	\end{scope}
	 \end{tikzpicture}
\end{center}
	 \caption{`Cat diagram' adapted from \cite{Kaplan:1983fs}. Despite the silly appearance, the key point is that the photon couples to the electric current $J_\mu = e \bar\Psi\gamma_\mu \Psi$ (`ears') formed from interactions with fundamental quarks in the strongly coupled sector. The `whiskers' are the pseudo-Goldstone external states when expanding the $U(x)$ field in (\ref{eq:comp:cat:operator}). The contribution to the charged meson masses come from the `two whisker' diagram.}
	 \label{fig:comp:cat}
\end{figure}

In addition to the spontaneous chiral symmetry breaking by strong dynamics, the SU(3)$_\text L\times$SU(3)$_\text R$ group is also broken \emph{explicitly} from the gauging of $\text U(1)_\text{EM}\subset \text{SU}(3)_\text V$. The neutral Goldstones (pions, kaons, and the $\eta$) are unaffected by this. The charged Goldstones, on the other hand, pick up masses from photon loop diagrams of the form in Fig.~\ref{fig:comp:cat}. These diagrams contribute to an operator that gives a shift in the [pseudo-]Goldstone mass,
\begin{align}
\Delta \mathcal L \sim e^2 \text{Tr}\left[Q U(x)^\dag Q U(x)\right],
\label{eq:comp:cat:operator}
\end{align}
where $Q=\frac 13 \text{diag}(2,-1,-1)$ is the matrix of quark electric charges. Since the electromagnetic force does not distinguish between the down and strange quarks, this diagram gives an equal shift to both the charged pions (e.g.\ $u\bar d$) and kaons (e.g.\ $u\bar s$). Since the up and anti-down/strange quark have the same charge, the bound state is more energetic than the neutral mesons and we expect the shift in the mass-squared to be positive \cite{Kaplan:1983fs, Georgi:2007zza}. Note that the contribution to the charged pion mass is quadratically sensitive to the chiral symmetry breaking scale, though it is also suppressed by the smallness of $\alpha_\text{EM}$.

\subsubsection{Explicit breaking from quark spectrum}

One can add quark masses that constitute a small ($m_q \ll \Lambda_\text\QCD$) explicit breaking of the global symmetry and generate small masses to the pseudo-Goldstone bosons. One can write this as a spurion $M=\text{diag}(m_u,m_d,m_s)$ which has the same quantum numbers as $U(x)$. One can add these terms to the effective Lagrangian by forming the appropriate global symmetry group invariant. In particular, we add to the Lagrangian 
\begin{align}
\Delta \mathcal L \sim \text{Tr}\left[M U(x)\right]  \sim \text{Tr}\left[M \left(\frac{\pi^a(x)}{f}T^a\right)^2\right]+\cdots
\end{align}
In the limit where $m_u=m_d$ and ignoring the electromagnetic splitting above, one may identify the masses for the pions, kaons, and $\eta$ (different components of $\pi^a$) to derive the Gell-Mann--Okubo relation,
\begin{align}
m_\eta^2 + m_\pi^2 = 4m_K^2.
\end{align}

\subsubsection{NDA: When the theory breaks down}\label{sec:comp:NDA}

Finally, let us note that the effective Lagrangian for pions is non-renormalizable, so we should say something about the cutoff for this theory. At tree-level, the two-to-two scattering of pions with characteristic momentum $p$ goes like $p^2/f^2$ from (\ref{eq:comp:chiral:pert}). Using na\"ive dimensional analysis (\textsc{nda}) \cite{Georgi:1986kr, Manohar:1983md, Georgi:1992dw, Gavela:2016bzc}, we see that the loop contributions go like
\begin{align}
\vcenter{\hbox{
\begin{tikzpicture}[line width=1.5 pt, scale=.5]
	\draw [scalarnoarrow] (-2,1) -- (-1,0);
	\draw [scalarnoarrow] (-1,0) -- (-2,-1);
	\draw [scalarnoarrow] (-1,0) arc (180:0:1);
	\draw [scalarnoarrow] (-1,0) arc (180:360:1);
	\draw [scalarnoarrow] (2,-1) -- (1,0);
	\draw [scalarnoarrow] (1,0) -- (2,1);
	\node at (1,0) [draw=black, circle, fill=black, inner sep=0.5 mm] {};
	\node at (-1,0) [draw=black, circle, fill=black, inner sep=0.5 mm] {};
\end{tikzpicture}
}}
\quad
\sim
\quad
\int \dbar^4k  \left(\frac{p^2k^2}{f^4}\right) \frac{1}{k^4} 
\quad\sim\quad
 \frac{\Lambda^2p^2}{16 \pi^2 f^4}.
 \quad\sim\quad
 \vcenter{\hbox{
\begin{tikzpicture}[line width=1.5 pt, scale=.5]
	\draw [scalarnoarrow] (-1,1) -- (0,0);
	\draw [scalarnoarrow] (0,0) -- (-1,-1);
	\draw [scalarnoarrow] (1,-1) -- (0,0);
	\draw [scalarnoarrow] (0,0) -- (1,1);
	\node at (0,0) [draw=black, circle, fill=black, inner sep=0.5 mm] {};
\end{tikzpicture}
}}
\times \frac{\Lambda^2}{16\pi^2 f^2},
\end{align}
where we have used the shift symmetry (the full SU(3)$^2$ group structure) to tell us that at the numerator of the integrand carries at least two powers of the external momenta. Validity of our loop expansion thus requires that $\Lambda \sim 4\pi f \sim \text\GeV$, and this is indeed the scale at which additional \QCD states appear. Note that this cutoff, based on perturbativity of the $1/f$ couplings in the chiral Lagrangian, is slightly different from $\Lambda_\text{QCD} \sim  \mathcal O(300 \text{ \MeV})$, which is the scale where $\alpha_s$ becomes non-perturbative.

Indeed, this \UV behavior of the theory of Goldstones is one of the reasons why we expected either the Higgs or something new to be manifest at the \LHC: the \SM without a Higgs is simply a nonlinear sigma model. By the Goldstone equivalence theorem, the scattering cross section for longitudinal $W$ boson scattering grows linearly with the center of mass energy. In order to maintain unitarity, one requires that either there is a Higgs boson (a linearization of the nonlinear sigma model) or that the theory becomes strongly coupled so that higher order terms can cancel the unphysical behavior.

\subsubsection{NDA: Characteristic couplings}
\label{sec:comp:nda:lagrangian}

To show the power of \textsc{nda}, let us explore the qualitative behavior of a strongly coupled theory without doing any calculations.
The rules of \textsc{nda} boil down to (1) a factor of $1/f$ for each particle involved and (2) powers of a heavy scale, $m_\rho$, to make up the remaining dimensions~\cite{Georgi:1992dw}. 
Rule 1 comes from the fact that the Goldstone fields\footnote{%
This rule of applies even for non-Goldstone fields. For example, baryons appear a factor of $1/f\sqrt{\Lambda}$, where the $\sqrt{\Lambda}$ makes up for the difference between the scalar and fermion mass dimensions~\cite{Georgi:1992dw}.
} $\pi(x)$ appear in the Lagrangian pre-packaged as $U(x) = \exp(i \pi(x)/f)$. Rule 2 is straight dimensional analysis with respect to a mass scale which one can take to be a \UV scale $\Lambda$, or more phenomenologically, the mass of the lowest non-Goldstone resonances, $m_\rho$. We use the $\rho$ meson as an example of such a state. 
Let us parameterize the separation between the heavy mass scale $m_\rho$ and the compositeness scale $f$ by $g_\rho = m_\rho/f$. We may interpret $g_\rho$ as the `natural' coupling size of the heavy state $\rho$ to the strong sector.

For a strong sector field $\phi$, define the dimensionless combinations
\begin{align}
	x&= \frac{\phi}{f} =g_\rho \frac{\phi}{m_\rho}
	&
	y&= \frac{\partial}{m_\rho}.
\end{align}
We'd like to build an \textsc{nda} Lagrangian to estimate the size of couplings. We start by writing some dimensionless function $\tilde{\mathcal L}(x,y)$. In order to obtain the correct mass dimension of a Lagrangian, we further define $\mathcal L_0(x,y) = m_\rho^4 \tilde{\mathcal L}(x,y)$. This function is assumed to contain a kinetic term,
\begin{align}
	\mathcal L_0(x,y) &\supset m_\rho^4 x^2 y^2 = g_\rho^2 \mathcal O(\partial^2,\phi^2).
	\label{eq:NDA:prefactor:1}
\end{align}
We see that we have to rescale by $g_\rho^{-2}$ to obtain a canonically normalized Lagrangian, 
\begin{align}
	\mathcal L = \frac{1}{g_\rho^2} \mathcal L_0(x,y) = \frac{m_\rho^4}{g_\rho^2} \tilde{\mathcal L}(x,y)
	= m_\rho^2 f^2 \tilde{\mathcal L}(x,y).
	\label{eq:NDA:prefactor:2}
\end{align}

%As an example that is useful below, 
Let us use this to determine the expected size of a quartic coupling of strong sector fields. This comes from the $\mathcal O(x^4)$ term in the expansion of $\tilde{\mathcal L}(x,y)$ so that
\begin{align}
	\mathcal L \supset \frac{m_\rho^2}{g_\rho^2}g_\rho^4 \frac{\phi^4}{m_\rho^4} = g_\rho^2 \phi^4.
	\label{eq:comp:NDA:quartic}
\end{align}
Thus we expect the quartic coupling of the $\phi$ to go like $g_\rho^2 \sim m_\rho^2/f^2$, justifying the intepretation of $g_\rho$ as a strong sector coupling scale.

\subsection{Composite, pseudo-Goldstone Higgs}

The main idea for composite pseudo-Goldstone Higgs models is that the Higgs mass parameter is protected against quadratic quantum corrections up to the compositeness scale because it is a pseudo-Goldstone boson. 
Above the scale of compositeness, it is simply not an elementary scalar.
This should be contrasted with the solutions to the Hierarchy problem already discussed:
\begin{itemize}
\item \textsc{Supersymmetry}: due to the extended spacetime symmetry, there is a cancellation of the quadratic corrections through the introduction of different-spin partners.
\item \textsc{Technicolor/Higgs-less}: there is no elementary Higgs and electroweak symmetry breaking proceeds through a Fermi condensate. This is now excluded.
\item \textsc{Warped extra dimensions}: the Higgs itself is a composite state so that above the compositeness scale it no longer behaves like a fundamental scalar. However, there is no explanation for why the Higgs is lighter than the confinement scale. 
\end{itemize}
Note, in particular, that the composite Higgs scenario that we're interested in is distinct from technicolor: the pseudo-Goldstone nature of the Higgs is an explanation for why the Higgs mass is so much lighter than the other bound states in the strongly coupled sector. 

Goldstone bosons, however, behave very differently from the Standard Model Higgs. We saw that Goldstone bosons have derivative couplings owing to their shift symmetry. The Higgs, on the other hand, has Yukawa couplings and the all important electroweak symmetry-breaking potential. Our goal in this section is to see how to construct a theory of Goldstones which can produce a Higgs particle that has all of the required couplings of the \SM Higgs. 

We shall closely follow the discussion in \cite{Contino:2010rs} and refer the reader there for further details and references.

\subsubsection{The framework}\label{sec:comp:framework}

\begin{figure}
\begin{center}
\begin{tikzpicture}[line width=1.5 pt, scale=.8]
\draw (0,0) circle (3);
\draw (1,-.3) circle (1.75);
\draw (-1,-.3) circle (1.75);
\node at (0,2) {\Large $G$};
\node at (1.75,-.3) { $H_\text{global}$};
\node at (-1.75,-.3) { $H_\text{gauge}$};
%\node at (0,-.3) {\Large $H$};
\node at (0,.2) {$H$};
\draw[line width=1pt] (0,-.8) circle (.5);
\node at (0,-.8) {\footnotesize EW};
\end{tikzpicture}
\qquad
\qquad
\qquad
\begin{tikzpicture}[line width=1.5 pt, scale=.8]
\draw (0,0) circle (3);
\draw[line width=1pt, dashed] (1,-.3) circle (1.75);
\draw[line width=1pt, red] (45:1) circle (1.75);
\draw[line width=1pt] (-1,-.3) circle (1.75);
\draw[line width=1pt] (0,-.8) circle (.5);
%\node at (0,-.8) {EW};
%\node[scale=.75, rotate=-30] at (.09,-.65) {\footnotesize \textcolor{blue}{EM}};
\begin{scope}
	\clip (45:1) circle (1.75);
	\draw[line width=1.5pt, color=red] (-1,-.3) circle (1.75);
	\clip (0,-.8) circle (.5);
%	\draw[line width=1pt, fill=blue] (0,-.8) circle (.5);
	\foreach \x in {-1.9,-1.8,...,1.3}
	\draw[line width=.5 pt, color=blue] (\x,-1.3) -- (\x+.6,1.3);
\end{scope}
%\node[scale=.75, rotate=-30] at (.09,-.65) {\footnotesize \textcolor{white}{EM}};
%
\path[->, color=blue, line width=1.5] (1.25,.25) edge [out = 160, in = 70] (.4,-.3);
\node at (1.8,.25) {\small \textcolor{blue}{EM}};
\node at (120:2.5) {$G$};
%\node at (1.75,-.3) { $H_\text{global}$};
\node at (-1.75,-.3) { $H_\text{gauge}$};
\node at (.75,1.9) {\textcolor{red}{$H_\text{oblique}$}};
%\node at (0,.2) {$H$};
\fill[fill=white] (-.4,.5) circle (.35);
\node at (-.4,.55) {\textcolor{red}{$H'$}};
\draw[<->, color = red] (5:3.3) arc (5:40:3.3);
\node at (20:4.9) {\textcolor{red}{$\displaystyle \xi = \left(\frac{v}{f}\right)^2$}};
\end{tikzpicture}
\end{center}
\caption{Pattern of symmetry breaking. (\textsc{left}, tree level) Strong dynamics breaks $G\to H_\text{global}$ spontaneously, while $H_\text{gauge} \subset G$ is explicitly broken through gauging. The unbroken group $H = H_\text{gauge}\cap H_\text{global}$ contains the \SM electroweak group, $\text{SU}(2)_\text{L}\times\text{U}(1)_\text{Y}$. 
(\textsc{right}, loop level) Vacuum misalignment from \SM interactions shifts the unbroken group $H\to H'$ and breaks the electroweak group to $\text{U}(1)_\text{EM}$. The degree of misalignment is parametrized by $\xi$, the squared ratio of the \EWSB \vev to the $G\to H$ \vev. Adapted from \cite{SafariThesis}.}
\label{fig:comp:breaking:pattern}
\end{figure}
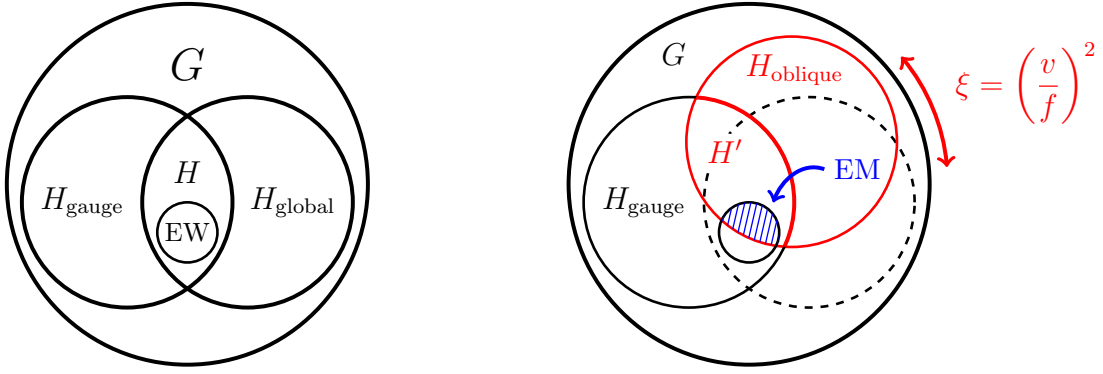

Start with a large global symmetry group $G$, analogous to the `large' $\text{SU}(3)_\text L\times\text{SU}(3)_\text R$ global symmetry of low energy \QCD. We will break this symmetry in two ways:
\begin{enumerate}
\item We assume that the strong dynamics spontaneously breaks $G$ to a subgroup $H_\text{global}$. This is analogous to chiral symmetry breaking in \QCD, $\text{SU}(3)_\text L\times\text{SU}(3)_\text R \to \text{SU}(3)_\text{V}$.
\item In addition to this, we will explicitly break $G$ by weakly gauging a subgroup $H_\text{gauge}$ which contains the \SM electroweak group $\text{SU}(2)_\text{L} \times \text{U}(1)_\text{Y}$. This is analogous to the gauging of $\text{U}(1)_\text{EM}$.
\end{enumerate}
We assume that the \SM electroweak group is a subgroup of $H = H_\text{gauge} \cup H_\text{global}$ so that it is gauged and preserved by the strong dynamics. This is shown on the left of Fig.~\ref{fig:comp:breaking:pattern}. This results in 
$\text{dim}\,H_\text{gauge}$ 
transverse gauge bosons and  
$\left(\text{dim}\,G-\text{dim}\,H_\text{global}\right)$ 
Goldstone bosons. The breaking 
$G\to H_\text{global}$ 
also breaks some of the gauge group so that there are a total of 
$\left( \text{dim}\,H_\text{gauge} - 
H\right)
$ %\text{dim}\,H_\text{gauge}\cup H_\text{global}\right)
massive gauge bosons and 
%$\left(\text{dim}\,G - \text{dim}\,H\right)$ 
$\left(\text{dim}\,G - \text{dim}\,H_\text{global}\right)
-
\left(\text{dim}\,H_\text{gauge} - \text{dim}\,H\right)$ 
`uneaten' massless Goldstones. 

Now we address the white elephant of the Higgs interactions---can we bequeath to our Goldstone bosons the necessary non-derivative interactions to make one of them a realistic Higgs candidate? This is indeed possible through \textbf{vacuum misalignment}, which we illustrate on the right of Fig.~\ref{fig:comp:breaking:pattern}. 
The gauging of $H_\text{gauge}$ gives loop-level corrections to the dynamical symmetry breaking pattern since this is an explicit breaking of the global symmetry. This is analogous to how the gauged $\text{U}(1)_\text{EM}$ splits the masses of the charged and neutral pions through a photon loop. Loops of \SM gauge bosons can generate an electroweak symmetry breaking potential for the Higgs. We illustrate this below.

One key point here is that since the Higgs potential is generated dynamically through \SM gauge interactions, the electroweak scale $v$ is distinct from the $G\to H$ symmetry breaking scale $f$. The `angle' 
\begin{align}
\xi = \left( \frac vf \right)^2
\label{eq:comp:comphiggs:angle}
\end{align}
parameterizes this separation of scales and quantifies the degree of vacuum misalignment.
Note that this is a separation of scales which does \emph{not} exist in technicolor and is the key to parameterizing how the Higgs remains light relative to the heavier resonances despite not being a `true' Goldstone boson. The limits $\xi \to 0$ and $\xi \to 1$ correspond to the \SM (heavy states completely decoupled) and technicolor, respectively. We note that this parameter is also a source of tuning in realistic composite Higgs models. Once the pseudo-Goldstone Higgs state is given non-derivative interactions, these interactions generically introduce quadratic divergences at loop level which would lead to an expected $\mathcal O(1\%)$ tuning.
To avoid this, one needs to introduce a smart way of dealing with these explicit breaking terms called \textbf{collective symmetry breaking} which we discuss below. First, however, we focus on the effects of gauge bosons on the Higgs potential.

We have the following constraints for picking a symmetry breaking pattern:
\begin{enumerate}
\item The \SM  electroweak group is a subgroup of the unbroken group, $\text{SU}(2)_\text L \times \text{U}(1)_\text Y \subset H$. In fact, it is better to have the full custodial $\text{SU}(2)_\text L\times \text{SU}(2)_\text R \cong \text{SO}(4)$ group embedded in $H$ since this will protect against large contributions to the $\rho$-parameter.
\item There is at least one pseudo-Goldstone boson with the quantum numbers of the \SM Higgs. To protect the $\rho$-parameter, it is better to have a $(\mathbf{2},\mathbf{2})$ under the custodial group.
\end{enumerate}

At this point we have said nothing about the \SM fermions. These, too, will have to couple to the strong sector to generate Yukawa couplings with the Higgs. We show below that a reasonable way to do this is to allow the \SM fermions to be \textbf{partially composite}, a scheme that we had already seen in the holographic interpretation of the \RS scenario. Indeed, extra dimensions provide a natural language to construct composite Higgs models.

\subsubsection{Minimal Composite Higgs: set up}
\label{sec:comp:MCH}

We now consider an explicit example, the \textbf{minimal composite Higgs model}, which was explored in \cite{Agashe:2004rs, Contino:2006qr} using the intuition from the \RS framework. Following the guidelines set above, we would like to choose choose $H_\text{global} = \text{SO}(4)$, the custodial group which is the minimal choice to protect the $\rho$-parameter. However, the $\text{SO}(4) = \text{SU}(2)_L\times\text{SU}(2)_R$ charge assignments don't give the correct $\text U(1)_\text Y$ charges, as is well known in left-right symmetric models. Thus our `minimal' choice for $H_\text{global}$ requires an additional $\text U(1)_\text X$ so that one may include hypercharge in the unbroken group, $H$,
\begin{align}
Y = (T^{R})^3 + X.
\label{eq:comp:hypercharge:X}
\end{align}
We then choose $G=\text{SO}(5)\times \text U(1)_\text X$ and introduce a linear field $\Sigma$ that is an $\text{SO}(5)$ fundamental and uncharged under $\text U(1)_\text X$. Note that we can ignore the $\text U(1)_\text X$ charge in our spontaneous symmetry breaking analysis since it's really just `coming along for the ride' at this point. $\Sigma$ acquires a \vev to break $\text{SO}(5)\to \text{SO}(4)$,
\begin{align}
\langle\Sigma\rangle = (0,0,0,0,1)^T.
\end{align}
This is analogous to the \QCD chiral condensate. We can now follow the intuition we developed with chiral perturbation theory. The Goldstone bosons of this breaking are given by transforming this \vev by the broken generators. A useful parameterization of the four broken generators is
\begin{align}
T^{\hat a}_{ij} = \frac{i}{\sqrt{2}} \left(
\delta^k_i \delta^5_j - \delta^k_j\delta^5_i
\right),
\end{align}
where $\hat a \in \{1,\cdots 4\}$. We refer to the unbroken generators with an undecorated index: $T^a$. 
The SO(5) group element that acts non-trivially on the \textsc{vev},
$\exp({i h^{\hat a}T^{\hat a}/f})$,
can be written in terms of sines and cosines by separately summing the odd and even terms of the exponential.
The linear field $\Sigma$ can then be decomposed into the Goldstone pieces $h^{\hat a}(x)$ and a radial component $h(x) = \sqrt{h^{\hat a}(x) h^{\hat a}(x)}$,
\begin{align}
\Sigma = e^{i h^{\hat a}(x)T^{\hat a}/f} \langle \Sigma \rangle =  \frac{\sin(h/f)}{h} \left(h^1,h^2,h^3,h^4, h\cot (h/f) \right).
\label{eq:comp:linear:field:Sigma}
\end{align}
With this parameterization, the \SM Higgs doublet is
\begin{align}
H = \frac{1}{\sqrt{2}}\begin{pmatrix}
 h^1 + i h^2 \\
 h^3 + i h^4
 \end{pmatrix}.
\end{align}

\subsubsection{Gauge couplings}
\label{sec:comp:mch:gauge}

\begin{framed}
\noindent \footnotesize
\textbf{Why $\xi$ is an angle}. As a quick exercise, let's see why we said (\ref{eq:comp:comphiggs:angle}) should be identified with an angle. Starting from the kinetic term for $\Sigma$ in (\ref{eq:comp:linear:field:Sigma}) with electroweak covariant derivative, one finds that the $W$ and $Z$ mass terms are
\begin{align}
	\mathcal L &\supset 
	\frac{g^2 f^2}{4}\
	\sin\frac{\langle h\rangle}{f}
	\left( |W|^2 +\frac{1}{2\cos^2\theta_W} Z^2 \right).
\end{align}
Using the relation $M_W = \cos\theta_W\, m_Z$, this tells us that
\begin{align}
	\sin\frac{\langle h\rangle}{f} &= \frac{v}{f},
\end{align}
so that we now see the relation between $\xi = v^2/f^2$ and the `angle' $\langle h\rangle/f$. Note that the \vev $\langle h\rangle \neq 246$~GeV.
\end{framed}

We would like to write down a Lagrangian for this theory and parameterize the effects of the strong sector on the \SM couplings. A useful trick for this is to pretend that the global $\text{SO}(5)\times \text U(1)_\text X$ symmetry is gauged and then `demote' the additional gauge fields to spurions---i.e.~turn them off. We can then parameterize the quadratic part of the Lagrangian for the full set of SO(5) [partially spurious] gauge bosons, $V_\mu = A^a_\mu T^a + A^{\hat a}_\mu T^{\hat a}$, and the U(1)$_\text X$ gauge boson, $X$, by writing down the leading $\text{SO}(5)\times \text U(1)_\text X$-invariant operators:
\begin{align}
\Delta \mathcal L = \frac 12 \left(\eta^{\mu\nu} + \frac{q^\mu q^\nu}{q^2}\right)
\left[
\Pi_X(q^2) X_\mu X_\nu + 
\Pi_0(q^2) \text{Tr} (A_\mu A_\nu)
+ 
\Pi_1(q^2) \text{Tr} (\Sigma A_\mu A_\nu\Sigma^T)
\right].
\end{align}
Where the form factors are completely analogous to (\ref{eq:comp:form:factor:JJ}) and (\ref{eq:comp:form:factor:prop}). 
Contained in this expression are the kinetic and mass terms of the \SM electroweak gauge bosons. To extract them, we must expand the form factors $\Pi(q^2)$ in momenta and identify the $\mathcal O(q^0)$ terms as mass terms and the $\mathcal O(q^2)$ terms as kinetic terms. 
Since the $\Pi_X$ and $\Pi_0$ terms include gauge fields in the unbroken directions, they should vanish at $q^2=0$, otherewise masses would be generated for those directions.
The $\Pi_1$ term, however, selects out the broken direction upon inserting the $\Sigma \to \Sigma_0$ and thus contains the Goldstone pole, (\ref{eq:comp:form:factor:poles}). We thus find
\begin{align}
\Pi_0(0) = \Pi_X(0) &= 0 
&
\Pi_1(0) = f^2.
\end{align}
Assuming that the Higgs obtains a \vev, one may rotate it into a convenient location $(h^1,\cdots,h^4) = (0,0,v/\sqrt{2},0)$ corresponding to the usual \SM Higgs \vev parameterization. We now assume that $H_\text{gauge}$ is the \SM electroweak group and drop all spurious gauge bosons. Using (\ref{eq:comp:linear:field:Sigma}), the strong sector contribution to the Lagrangian of these gauge bosons to $\mathcal O(q^2)$ is
\begin{align}
\Delta \mathcal L_{q^0}
&=
\left(\eta^{\mu\nu} + \frac{q^\mu q^\nu}{q^2}\right)\frac 12 \left(\frac{f^2}4 \sin^2\frac{\langle h\rangle}{f} \right)\left(B_\mu B_\nu + W^3_\mu W^3_\nu - 2W^3_\mu B_\nu + 2W^+_\mu W^-_\nu \right)
\label{eq:comp:Delta:L:q0}
\\
\Delta \mathcal L_{q^2}
&=
\frac{q^2}{2} \left[
\Pi'_0(0) W^a_\mu W^a_\nu 
+ \left(\Pi'_0(0) + \Pi_X'(0)\right)B_\mu B_\nu 
\right],
\end{align}
where we have used the choice of $\text{SO}(5)$ generators in the appendix of \cite{Contino:2010mh}.
$\Delta \mathcal L_{q^2}$ gives contributions to the kinetic terms of the gauge bosons. Observe that these are not canonically normalized, but instead can be thought of as shifts in the gauge coupling,
\begin{align}
\Delta\left(\frac 1{g^2}\right) &= -\Pi'_0(0)
&
\Delta\left(\frac 1{g'^2}\right) &= -\left(\Pi'_0(0) + \Pi'_X(0)\right).
\end{align}
Thus if the SU(2)$_\text L$ gauge bosons have a `pure' gauge coupling $g_0$ when one turns off the strongly coupled sector, the full observed SU(2)$_\text L$ gauge coupling is
\begin{align}
\frac{1}{g_\text{SM}^2} = \frac{1}{g_0^2} - \Pi_0'(0),
\end{align}
and similarly for $g'_\text{SM}$.

$\Delta \mathcal L_{q^0}$ corresponds to contributions the masses of the heavy electroweak gauge bosons. Taking into account the need to canonically normalize with respect to $\Delta \mathcal L_{q^2}$, we obtain the usual $W^\pm$ and $Z$ masses by identifying the \SM Higgs $\vev$ as $v = f \sin(\langle h\rangle/f)$. We see the appearance of the misalignment angle,
\begin{align}
\xi = \sin^2 \frac{\langle h\rangle}{f} \equiv \frac{v^2}{f^2}.
\end{align}
Finally, by restoring $\langle h\rangle \to h(x)$ in (\ref{eq:comp:Delta:L:q0}) we may determine the composite Higgs couplings to the gauge bosons\footnote{%
This is a trivial use of the Higgs low-energy theorem: the low-momentum Higgs couplings are equivalent to promoting the \vev to $h(x)$ like \cite{Ellis:1975ap, Shifman:1978zn} This theorem can be used, for example, to calculate the Higgs coupling to photons by evaluating the mass dependence of the running of the \textsc{qed} gauge coupling. The application of the theorem to composite Higgs models is explored in \cite{Gillioz:2012se}.%
}. The key is the expansion
\begin{align}
f^2 \sin^2 \frac{h(x)}f &=
v^2 + 2v\sqrt{1-\xi} h(x) + (1+2\xi)h(x)^2 + \cdots.
\end{align}
From this we can make the prediction that the $\text{SO}(5)/\text{SO}(4)$ composite Higgs couplings to the massive electroweak gauge bosons $V=W^\pm,Z$ deviate from their \SM values,
\begin{align}
g_{VVh} &=  \sqrt{1-\xi} g^\text{SM}_{VVh} 
&
g_{VVhh} &=  (1-2\xi) g^\text{SM}_{VVhh}. 
\label{eq:comp:mch:VVh:VVhh}
\end{align}
At this point, these couplings introduce gauge boson loops which are quadratically divergent. These loops go like
\begin{align}
\vcenter{\hbox{
\begin{tikzpicture}[line width=1.5pt, scale=1.5]
    \draw[scalarnoarrow] (-.2,-.5) -- (1.3,-.5);
    \draw[vector] (.55,-.1) circle (.4);
\end{tikzpicture}
}}
\sim\;
\frac{g^2}{16\pi^2} \Lambda^2
\sim\; g_\text{SM}^2 (1-\xi) f^2,
\label{eq:comp:reintroduction:of:quadratic}
\end{align}
where we have used the dimensional analysis limit $\Lambda = 4\pi f$. We see that having explained the lightness of the Higgs by appealing to the Goldstone shift symmetry, reintroducing the Higgs couplings to the gauge bosons breaks this shift symmetry and wants to push the Higgs mass back up towards the symmetry breaking scale. %
In order to avoid this, one additional ingredient called collective breaking (along with light gauge and top partners) is necessary. We present this in Section~\ref{sec:comp:collective}.

\subsubsection{Partial compositeness}

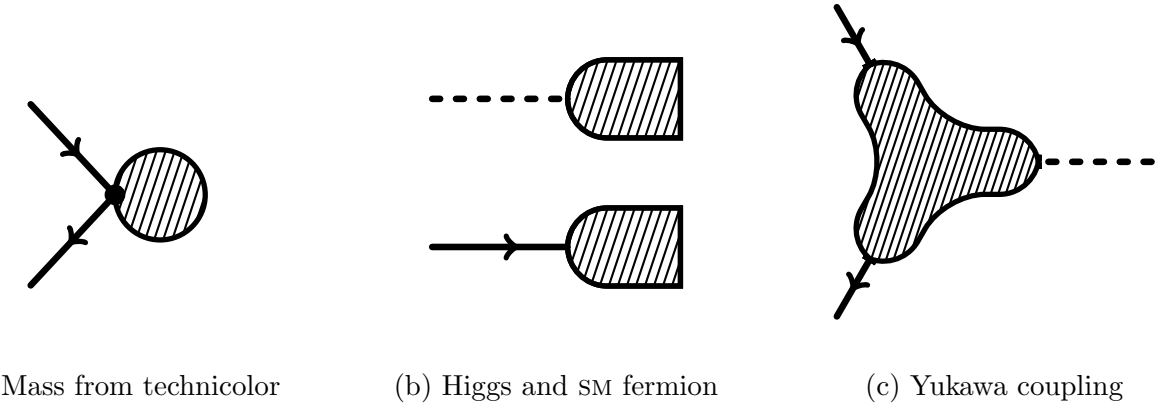
\begin{figure}
        \centering

\begin{subfigure}[b]{0.3\textwidth}
	
	\begin{center}
\begin{tikzpicture}[line width=1.5,scale=.6]
	\pgfmathsetmacro{\offset}{2.5}
	\pgfmathsetmacro{\rounding}{15}

    %\draw[fermionbar, line width=2.5, line cap=round]
%    \draw[line width=2.5, line cap=round, loosely dashed] (1,0)--(3,0);
    
    \draw[fermionbar, line width=2.5, line cap=round] (-1,0) -- (145:3.5);
    
    \draw[fermion, line width=2.5, line cap=round] (-1,0) -- (-145:3.5);

%	\begin{scope}
%	    	\clip (0,0) circle (3);
%	    	\foreach \x in {-.95,-.9,-.85,...,.3}
%				\draw[line width=.8 pt] (\x,-.3) -- (\x+.6,.3);
%	  	\end{scope}
	  	
	\begin{scope}
	    \clip (0,0) circle (1);
	   \foreach \x in {-5,-4.8,...,3}
		\draw[line width=.8 pt] (\x,-3) -- (\x+2,3);
	\end{scope}

    \draw[
    line width=2, 
    draw=black,
    rounded corners=\rounding](0,0) circle (1);
    
    \draw[
    line width=1, 
    fill=black,
    rounded corners=\rounding](-1,0) circle (.2);

% http://tex.stackexchange.com/questions/130838/tikz-segments-with-5pt-dotted-ends

\end{tikzpicture}

\end{center}

\vspace{1em}

\caption{Mass from technicolor}
\label{fig:comp:TC:Yukawa}
\end{subfigure}
\quad
\begin{subfigure}[b]{0.3\textwidth}
	
	\begin{center}
\begin{tikzpicture}[line width=1.5,scale=.6]
	\pgfmathsetmacro{\offset}{2.5}
	\pgfmathsetmacro{\rounding}{15}

	\draw[line width=2.5, line cap=round, loosely dashed] (-6,0) -- (-2.5,0);
		
	 \draw[
    line width=2, 
    fill=white,
    rounded corners=\rounding]
    (120:1) 
    -- ($(120:1)-(0:\offset)$)
    -- ($(120:1)-(0:\offset)-(0,2*sin(60)$) 
    [sharp corners]
    -- (-120:1)
%    -- ($(-120:1)+(120:1)$)
    -- cycle;
	
	\begin{scope}
	    \clip[rounded corners=\rounding] (120:1) 
    -- ($(120:1)-(0:\offset)$)
    -- ($(120:1)-(0:\offset)-(0,2*sin(60)$) 
    [sharp corners]
    -- (-120:1)
    -- cycle;
	   \foreach \x in {-5,-4.8,...,3}
		\draw[line width=.8 pt] (\x,-3) -- (\x+2,3);
	\end{scope}
	
    \draw[
    line width=2, 
    draw=black,
    rounded corners=\rounding]
    (120:1) 
    -- ($(120:1)-(0:\offset)$)
    -- ($(120:1)-(0:\offset)-(0,2*sin(60)$) 
    [sharp corners]
    -- (-120:1)
%    -- ($(-120:1)+(120:1)$)
    -- cycle;

% http://tex.stackexchange.com/questions/130838/tikz-segments-with-5pt-dotted-ends

\end{tikzpicture}

\vspace{2em}

\begin{tikzpicture}[line width=1.5,scale=.6]
	\pgfmathsetmacro{\offset}{2.5}
	\pgfmathsetmacro{\rounding}{15}

	\draw[fermion, line width=2.5, line cap=round] (-6,0) -- (-2.5,0);
		
	 \draw[
    line width=2, 
    fill=white,
    rounded corners=\rounding]
    (120:1) 
    -- ($(120:1)-(0:\offset)$)
    -- ($(120:1)-(0:\offset)-(0,2*sin(60)$) 
    [sharp corners]
    -- (-120:1)
%    -- ($(-120:1)+(120:1)$)
    -- cycle;
	
	\begin{scope}
	    \clip[rounded corners=\rounding] (120:1) 
    -- ($(120:1)-(0:\offset)$)
    -- ($(120:1)-(0:\offset)-(0,2*sin(60)$) 
    [sharp corners]
    -- (-120:1)
    -- cycle;
	   \foreach \x in {-5,-4.8,...,3}
		\draw[line width=.8 pt] (\x,-3) -- (\x+2,3);
	\end{scope}
	
    \draw[
    line width=2, 
    draw=black,
    rounded corners=\rounding]
    (120:1) 
    -- ($(120:1)-(0:\offset)$)
    -- ($(120:1)-(0:\offset)-(0,2*sin(60)$) 
    [sharp corners]
    -- (-120:1)
%    -- ($(-120:1)+(120:1)$)
    -- cycle;

% http://tex.stackexchange.com/questions/130838/tikz-segments-with-5pt-dotted-ends

\end{tikzpicture}
\end{center}

\vspace{1em}

\caption{Higgs and \SM fermion}
\label{fig:comp:hig:sm:comp}
\end{subfigure}
\quad
%
%
%add desired spacing between images, e. g. ~, \quad, \qquad etc.
\begin{subfigure}[b]{0.3\textwidth}
\begin{center}
\begin{tikzpicture}[line width=1.5,scale=.5]
	\pgfmathsetmacro{\offset}{2.5}
	\pgfmathsetmacro{\rounding}{15}
	
	\begin{scope}
	    \clip[rounded corners=\rounding] (60:1) 
    -- ($(60:1)+(0:\offset)$)
    -- ($(60:1)+(0:\offset)-(0,2*sin(60)$) 
    -- (-60:1)
    -- ($(-60:1)+(240:\offset)$)
    -- ($(180:1)+(240:\offset)$)
    -- (180:1)
    -- ($(180:1)+(120:\offset)$)
    -- ($(60:1)+(120:\offset)$)
    -- cycle;
	   \foreach \x in {-5,-4.8,...,3}
		\draw[line width=.8 pt] (\x,-3) -- (\x+2,3);
	\end{scope}
	
    \draw[
    line width=2, 
    draw=black,
    rounded corners=\rounding]
    (60:1) 
    -- ($(60:1)+(0:\offset)$)
    -- ($(60:1)+(0:\offset)-(0,2*sin(60)$) 
    -- (-60:1)
    -- ($(-60:1)+(240:\offset)$)
    -- ($(180:1)+(240:\offset)$)
    -- (180:1)
    -- ($(180:1)+(120:\offset)$)
    -- ($(60:1)+(120:\offset)$)
    -- cycle;

% http://tex.stackexchange.com/questions/130838/tikz-segments-with-5pt-dotted-ends
    \draw[line width=2.5, line cap=round, loosely dashed] ($(0:1)+(0:0.9*\offset)$) -- ($(0:1)+(0:2*\offset)$);
    
    \draw[fermionbar, line width=2.5, line cap=round] ($(120:1)+(120:0.8*\offset)$) -- ($(120:1)+(120:1.5*\offset)$);
    
    \draw[fermion, line width=2.5, line cap=round] ($(-120:1)+(-120:0.8*\offset)$) -- ($(-120:1)+(-120:1.5*\offset)$);

\end{tikzpicture}
\end{center}
         \caption{Yukawa coupling}
        \label{fig:comp:yukawa:comp}
        \end{subfigure}
        \caption{Fermion couplings to the composite sector, represented by shaded blobs.
        %Partial compositeness. The confining sector is represented by shaded blobs. 
        (a): Bilinear coupling of fermions to the composite sector (\ref{eq:comp:technicolor:bilinear:fermion:coupling}) lead to %`technicolor' types of Yukawa couplings.
        	fermion masses from the condensate of techniquarks.
        (b): Partial compositeness scenario. In addition to the Higgs being part of the strong sector, the elementary \SM fermions mix linearly with strong sector operators with the same quantum numbers. (c): Yukawa interactions are generated through the strong sector dynamics. Adapted from \cite{Contino:2010mh}. }\label{fig:comp:partial:compositeness}
\end{figure}

Having introduced the Higgs couplings to the gauge bosons, we can move on to finding a way to incorporate the Yukawa couplings into composite Higgs models. The way this is done in technicolor is to introduce a four-Fermi interaction that is bilinear in \SM fields, e.g.\ %$(\bar Q_L u_R)(\bar\psi_\text{TC}\bar\psi_\text{TC})$
\begin{align}
\Delta\mathcal L \sim (\bar Q_L u_R)(\bar\psi_\text{TC}\psi_\text{TC})
\label{eq:comp:technicolor:bilinear:fermion:coupling}
\end{align}
 where the $(\bar\psi_\text{TC}\psi_\text{TC})$  are bilinears of the techni-quarks. The resulting fermion mass is shown in Fig.~\ref{fig:comp:TC:Yukawa}. This strategy typically runs afoul of constraints on \CP violation and flavor-changing neutral currents since one can imagine the composite sector similarly generating a four-fermion operator between \SM states unless elaborate flavor symmetry schemes are assumed.

Instead of connecting the strong sector to a \SM fermion bilinear, we can consider a \emph{linear} connection. This is known as \textbf{partial compositeness}~\cite{Kaplan:1991dc, Gherghetta:2000qt} and is shown in Fig.~\ref{fig:comp:hig:sm:comp}. %
We assume that instead of (\ref{eq:comp:technicolor:bilinear:fermion:coupling}), the elementary fermions mix with a fermionic composite operator,
\begin{align}
\Delta \mathcal L \sim \bar Q_L \mathcal O_{Q_L},
\label{eq:comp:partial:comp:sim}
\end{align}
where $\mathcal O_{Q_L}$ is a strong sector operator that is interpolated into a composite quark doublet. We assume similar mixing terms for each of the other \SM fermions. In order to preserve the \SM quantum numbers we must assume that the the \SM gauge group is a weakly gauged subgroup of the strongly coupled sector's flavor symmetries.
Note that the gauge bosons are also partially composite\footnote{In this framework the longitudinal modes of the massive \SM gauge bosons pick up this partial compositeness from the Higgs. It is also possible to have a scenario where the transverse modes are partially composite, see \cite{Craig:2011ev,Csaki:2011xn} for explicit realizations.}, as we saw in (\ref{eq:comp:chiral:W:corrected:prop}).
The resulting Yukawa interactions are shown in Fig.~\ref{fig:comp:yukawa:comp}. 

The degree of mixing is now a freedom in our description. Let us parameterize the elementary--composite mixing by `angles' $\epsilon$,
\begin{align}
| \text{observed particle}\rangle \sim |\text{elementary} \rangle + \epsilon |\text{composite} \rangle.
\end{align}
We can use this degree of compositeness to control flavor violation. Since the strongest flavor constraints are for the first two generations, we assume that the first two generations have very small mixing with the composite sector. This suppresses dangerous flavor-violating four-fermion operators. On the other hand, we may assume that the third particles are more composite than the first two generations,
\begin{align}
\epsilon_3 \gg \epsilon_{1,2}.
\end{align}
Since the degree of compositeness also controls the interaction with the Higgs, this means that the third generation particles have a larger Yukawa coupling and, upon electroweak symmetry breaking, have heavier masses. The astute reader will note that this is exactly the same as the flavor structure of the `realistic' Randall-Sundrum models in Sec.~\ref{sec:XD:realistic:RS}-\ref{sec:xd:rs:flavor:sketch}. The observation that light fermions can automatically avoid flavor bounds is precisely what we called the `\RS \GIM  mechanism.' This is no surprise since the holographic interpretation of the \RS model is indeed one where the Higgs is composite.

Let us briefly see how this works with an explicit example. Let us write out (\ref{eq:comp:partial:comp:sim}) with a coupling $\lambda_{Q_L}$ and cutoff $\Lambda=4\pi f$:
\begin{align}
	\Delta \mathcal L &\supset \frac{\lambda_{Q_L}}{\Lambda^{\text{dim}\mathcal O_{Q_L}-5/2}} \bar Q_L \mathcal O_{Q_L} + \text{h.c.},
	\label{eq:comp:partial:comp:mass1}
\end{align}
where the power of $\Lambda$ is chosen so that $\Delta \mathcal L$ has mass dimension four. We assume that at low energies, the operator $\mathcal O_{Q_L}$ dimensionally transmutes into a fermion $\Psi_{Q_L}$ with canonical mass dimension. We say that $\Psi_{Q_L}$ is an interpolating field for the composite operator $\mathcal O_{Q_L}$. It is treated as a local field in the same way that one may treat the proton as a local interpolating field for a $uud$ composite operator in \QCD. We further assume that $\Psi_{Q_L}$ comes with conjugate $\Psi_{Q_L}^c$ (interpolating a conjugate operator $\mathcal O_{Q_L}^c$) so that it may form a vectorlike mass,
\begin{align}
	\Delta \mathcal L &\supset g_* \Lambda^2 \bar\Psi_{Q_L} \Psi^c_{Q_L}+\text{h.c.},
	\label{eq:comp:partial:comp:mass2}
\end{align}
which comes from a coupling $g_* 
\Lambda^{4-2\text{dim}\mathcal O_{Q_L}} 
\bar{\mathcal O}_{Q_L} \mathcal O_{Q_L}^c$ in the \UV; the powers of $\Lambda$ sort themselves out as the operators dimensionally transmute into the interpolating fields. The coupling $g_*$ is a characteristic coupling of the strong sector discussed below in (\ref{eq:comp:taxonomy:partners}). Together, (\ref{eq:comp:partial:comp:mass1}) and (\ref{eq:comp:partial:comp:mass2}) give the mass matrix
\begin{align}
	\Delta\mathcal L = -\Lambda^2
	\begin{pmatrix}
		\bar Q_L &
		\bar \Psi_{Q_L} &
		\bar \Psi_{Q_L}^c
	\end{pmatrix}
	\begin{pmatrix}
		0 & 0 & \lambda_{Q_L} \\
		0 & 0 & g_* \\
		\lambda_{Q_L} & g_* & 0
	\end{pmatrix}
	\begin{pmatrix}
		Q_L \\
		\Psi_{Q_L} \\
		\Psi_{Q_L}^c
	\end{pmatrix}.
\end{align}
The light eigenstate is:
\begin{align}
\left| Q_L^\text{physical}\right\rangle &= 
\cos \theta_{Q_L} \left| Q_L \right\rangle
+
\sin \theta_{Q_L} \left| \Psi_{Q_L} \right\rangle,
\end{align}
with mixing angle
\begin{align}
	\sin \theta_{Q_L} = \frac{\lambda_{Q_L}}{\sqrt{g_*^2 + \lambda_{Q_L}^2}}.
\end{align}
We typically expect $\lambda_{Q_L} \ll g_*$ so that this simplifies to $\theta_{Q_L} \approx \lambda_{Q_L}/g_*$.

\subsubsection{Breaking electroweak symmetry}

Having addressed the Higgs couplings to both the \SM gauge bosons and fermions, we move on to the Higgs self-couplings. Until now we have simply assumed that the strong sector generates an electroweak-symmetry breaking potential. We now check that this assumption is plausible by arguing that loops involving the third generation quarks generate such a potential; this is similar to the Nambu--Jona-Lasinio model~\cite{Nambu:1961fr, Nambu:1961tp}.

The \SM fermions do not form complete SO(5) multiplets. In fact, they cannot even be embedded into SO(5), as we noted in (\ref{eq:comp:hypercharge:X}). The \SM fermions can be embedded in the global group $G=\text{SO}(5)\times U(1)_\text X$, but certainly do not fill out complete representations. We thus follow the same strategy that we used for the gauge bosons in Sec.~\ref{sec:comp:mch:gauge}. Let us promote the \SM fermions to full SO(5) spinor representations,
\begin{align}
\Psi_Q &= 
\begin{pmatrix}
Q_L\\
\hdashline
\chi_{Q}
\end{pmatrix}
&
\Psi_u &=
\begin{pmatrix}
\psi_u\\
\hdashline
u_R\\
\chi_u
\end{pmatrix}
&
\Psi_d &=
\begin{pmatrix}
\psi_d\\
\hdashline
\chi_d\\
d_R
\end{pmatrix},
\label{eq:comp:fermion:SO5}
\end{align}
where the dashed line separates the $\text{SU}(2)_\text L\times \text{SU}(2)_\text R$ parts of $\text{SO}(4)\subset \text{SO}(5)$. The $\psi$ and $\chi$ fields are spurions. Recall from Sec.~\ref{sec:comp:fermions:in:5D} that the fundamental spinor representation for SO(5) is a Dirac spinor which decomposes into two Weyl spinors. Do not confuse these Weyl spinors (\ref{eq:comp:fermion:SO5}) with Poincar\'e representations---these are representations of the global SO(5) internal group. In other words, the entire $\Psi$ multiplet are Weyl spinors with respect to Poincar\'e symmetry but are Dirac spinors with respct to the internal SO(5) symmetry. The upper half of the Dirac $\Psi$ spinors are charged under $\text{SU}(2)_\text L$ while the lower half is charged under $\text{SU}(2)_\text R$. This imposes a $U(1)_\text X$ charge of $1/6$ on the $\Psi$ fields to give the correct hypercharge assignments on the \SM fields. 

Now let us parameterize the strong sector dynamics in the couplings of the SO(5) fermions $\Psi$ and the linear field $\Sigma$ in (\ref{eq:comp:linear:field:Sigma}) that encodes the composite Higgs. Since the $\Sigma$ is an SO(5) vector, it can appear in a fermion bilinear as $\Sigma_i \Gamma^i$, where the $\Gamma$ are the 5D Euclidean space representation of the Clifford algebra. The effective \SM fermion bilinear terms are
\begin{align}
\mathcal L =
\sum_{r=Q,u,d}\bar\Psi_r \slashed{p} \left[\Pi_{r0} + \Pi_{r1} (\Gamma\cdot\Sigma)\right]\Psi_r
+
\sum_{r=u,d}\bar\Psi_Q \slashed{p} \left[M_{r0} + M_{r1} (\Gamma\cdot\Sigma)\right]\Psi_r + \text{h.c.}
\end{align}
where, as before, the form factors $\Pi$ and $M$ are momentum-dependent. We shall focus on only the $Q_L$ and $t_R$ pieces since they have the largest coupling to the strong sector. 
\begin{framed}
\noindent \footnotesize
\textbf{Keeping track of conjugate fields}. One should be careful with the conjugate fields in the above expression. For the Lorentz group in four and five dimensions, SO(3,1) and SO(4,1), we use the Dirac conjugate $\bar\Psi \equiv \Psi^\dag \gamma^0$ to form Lorentz invariants. Recall that this is because objects like $\Psi^\dag \Psi$ are not necessarily invariant because representations of the Lorentz group are not unitary---boosts acting on the spinor representation do not satisfy $U^\dag U = 1$. This is due to the relative sign between the time-like and space-like directions in the Minkowski metric. The Dirac conjugate is a way around this. For the case of the $G=SO(5)$ \emph{internal} symmetry, however, there is no issue of non-unitarity. Hence no additional $\Gamma^0$ (acting on the internal SO(5) space) is necessary in the Lagrangian. To be clear, we can write out the spacetime $\gamma$ and internal $\Gamma$ matrices explicitly:
\begin{align}
\bar\Psi = \Psi^\dag \gamma^0 \neq \Psi^\dag \gamma^0 \Gamma^0.
\end{align}%
\end{framed}
The matrix $\Gamma\cdot \Sigma$ takes the form 
\begin{align}
\Gamma\cdot \Sigma =
\frac{1}{h}
\begin{pmatrix}
h \cos(h/f) & \slashed{h} \sin(h/f)
\\
\slashed{\bar h} \sin(h/f) & - h\cos(h/f)
\end{pmatrix},
\end{align}
where $\slashed{h}$ and $\slashed{\bar h}$ are appropriate contractions with Pauli matrices. With the above caveat that there is no $\Gamma^0$ acting on the SO(5) conjugate, we may write out the Lagrangian for $Q_L$ and $t_R$ by dropping the spurious components of the $\Psi$ fields,
\begin{align}
\mathcal L = 
\bar Q_L \slashed{p}
\left[
\Pi_{Q0} + \Pi_{Q1} \cos\frac{h}{f}
\right] Q_L
+
\bar t_R \slashed{p}
\left[
\Pi_{t0} + \Pi_{t1} \cos\frac{h}{f}
\right] t_R
+
\bar Q_L M_{u1} \left[h \sin \frac{h}{f} H^c \right] t_R,
\label{eq:comp:fermion:couplings}
\end{align}
where $H^c = i\sigma^2 H$ is the usual conjugate Higgs doublet in the \SM\footnote{Here we have used the SO(5) basis in \cite{Contino:2010rs}.}. Observe that upon canonical normalization, the top mass can be read off the Yukawa term,
\begin{align}
m_t^2 = \left(\frac{v}{f}\right)^2
\frac{M_{t1}^2}{
\left(\Pi_{Q0} + \Pi_{Q1}\right)
\left(\Pi_{t0} - \Pi_{t1}\right)
},
\end{align}
where the form factors are evaluated at zero momentum. One may write similar expressions for the other fermions. Observe from (\ref{eq:comp:fermion:couplings}) that the fermions which are more composite (e.g.\ the top quark) will also experience deviations from their \SM couplings depending on $\xi$, analogously to the deviation of the Higgs--gauge boson couplings, (\ref{eq:comp:mch:VVh:VVhh}). 

In order to determine whether electroweak symmetry is broken, we can now plug this information into the Coleman-Weinberg potential for the Higgs, also known as the [quantum] effective potential. This is the potential term in the effective action after taking into account quantum corrections from integrating out the top quarks. In other words, it is the potential that determines the vacuum expectation value of fields. The result is
\begin{align}
V_\text{CW} &= 
-6 \int \dbar^4 p \;
\log \left( \Pi_{Q0} + \Pi_{Q1} \cos\frac hf \right) \nonumber \\
&\phantom{-6 \int \dbar^4 p \;}
+ 
\log \left[ 
p^2\left( \Pi_{Q0} + \Pi_{Q1} \cos\frac hf \right)
\left(\Pi_{t0} - \Pi_{t1} \cos \frac hf\right)
M_{t1}^2 \sin^2\frac vf
\right].
\end{align}
Expanding this to first order and keeping the leading order terms in the Higgs gives
\begin{align}
V_\text{CW}(h) = \alpha \cos \frac hf - \beta \sin^2 \frac hf,
\label{eq:comp:CW:Higgs:pot}
\end{align}
where $\alpha$ and $\beta$ are integrals over functions of the form factors where $\beta$ is typically of the order the top Yukawa. If $\alpha \leq 2\beta$, then the Higgs acquires a \vev parameterized by
\begin{align}
\xi \equiv \sin^2 \frac{\langle h\rangle}{f} = 1-\left(\frac{\alpha}{2\beta}\right)^2.
\end{align}
This means that a small $\xi$ typically requires a cancellation between $\alpha$ and $\beta$. Since these come from different sources, this is generically a tuning in the theory.

One can also ask if it was necessary to rely on the top quark. For example, we know that the gauge sector also breaks the Goldstone shift symmetry so that loops of gauge bosons can generate quadratic and quartic terms in the Higgs potential. However, for a vector-like strong sector, gauge loops contribute with the wrong sign to the $\beta$ term  and pushes to align---rather than misalign---the vacuum \cite{Witten:1983ut, Vafa:1983tf}. 

There are, however, alternate mechanisms to enact electroweak symmetry breaking for a composite Higgs. For example,
\begin{itemize}
\item mixing the composite Higgs with an elementary state \cite{Kaplan:1983fs},
\item making use of an explicitly broken global symmetry \cite{Kaplan:1983sm}
\item enlarging the $H_\text{gauge}$ so that it cannot be completely embedded into $H_\text{global}$ \cite{Banks:1984gj,Georgi:1984ef,Georgi:1984af}.
\end{itemize}

%% SILH pheno section used to be here, now moved to bottom

%% Former title: Little Higgs 
\subsection{Collective symmetry breaking}\label{sec:comp:collective}

% {sec:comp:framework}, {eq:comp:reintroduction:of:quadratic}

%We have seen in Section~\ref{sec:comp:framework} and equation~(\ref{eq:comp:reintroduction:of:quadratic}) that although the Goldstone nature of the composite Higgs protects against quadratic tuning, the introduction of the non-derivative interactions required to produce an electroweak-symmetry breaking potential generically reintroduces quadratic sensitivity to the compositeness scale.

The general composite Higgs is a useful framework for working with the Higgs as a psuedo-Goldstone boson. However, we saw in Section~\ref{sec:comp:framework} and equation~(\ref{eq:comp:reintroduction:of:quadratic}) that this is not enough to avoid tuning. The source is clear: a pure Goldstone Higgs is protected from quadratic corrections to its mass because of its shift symmetry. This very same shift symmetry prevents the required Higgs couplings to gauge bosons, fermions, and itself. One must break this shift symmetry in order to endow the Higgs with these couplings; this generically reintroduces a dependence on the cutoff, $\Lambda = 4\pi f$. 

This may make it seem like a no-go theorem for any realistic model of a pseudo-Goldstone Higgs. However, there is a nice way out of this apparent boondoggle called \textbf{collective symmetry breaking} that was originally introduced in `little Higgs' models~\cite{Georgi:1974yw, Georgi:1975tz, ArkaniHamed:2001nc} (see \cite{Perelstein:2005ka, Schmaltz:2005ky} for reviews) and is now an a key ingredient in composite Higgs models\footnote{In Section~\ref{sec:comp:twin:higgs} we present an alternate protection mechanism based on a $\mathbb{Z}_2$ symmetry.}. The idea is that one can separate the scales $v$ and $f$ by introducing new particles which cancel the quadratic divergences at one-loop order. Unlike supersymmetry, these partner particles carry the same spin as the Standard Model particles whose virtual contributions are to be canceled. Further, this cancellation only occurs for one-loop diagrams: higher loop diagrams are expected to contribute quadratically at their na\"ive dimensional analysis size, but these are suppressed relative to the leading term.

The general principle that allows this cancellation is that the shift symmetry is redundantly protected. A process is only sensitive to explicit symmetry breaking---as necessary for \SM-like Higgs couplings---if this explicit breaking is communicated by at least two different sectors of the theory. More concretely, the symmetry is only explicitly broken if multiple couplings are non-zero in the theory so that any diagram that encodes this explicit breaking must include insertions from at least two different couplings. This softens the cutoff sensitivity of various operators by requiring additional field insertions that decrease the degree of divergence of loop diagrams.

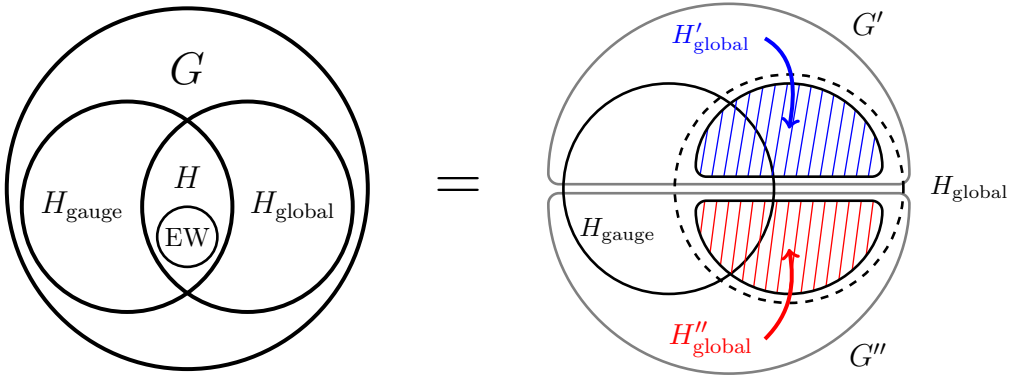
\begin{figure}
\begin{center}

\begin{tikzpicture}[line width=1.5 pt, scale=.8]
%\draw (0,0) circle (3);

\begin{scope}[shift={(-9,0)}]
\draw (0,0) circle (3);
\draw (1,-.3) circle (1.75);
\draw (-1,-.3) circle (1.75);
\node at (0,2) {\Large $G$};
\node at (1.75,-.3) { $H_\text{global}$};
\node at (-1.75,-.3) { $H_\text{gauge}$};
%\node at (0,-.3) {\Large $H$};
\node at (0,.2) {$H$};
\draw[line width=1pt] (0,-.8) circle (.5);
\node at (0,-.8) {\footnotesize EW};
\end{scope}

\begin{scope}[shift={(-4.5,0)}]
\node at (0,0) {\huge =};
\end{scope}

\begin{scope}[shift={(0,0.075)}]
\draw[rounded corners, line width=1, color=gray] (3,0) arc (0:180:3) -- cycle;
\end{scope}

\begin{scope}[shift={(0,-0.075)}]
\draw[rounded corners, line width=1, color=gray] (-3,0) arc (180:360:3) -- cycle;
\end{scope}

\draw[line width=1] (-1,0) circle (1.75);
\draw[dashed, line width=1] (1,0) circle (1.9);

\begin{scope}[shift={(1.,0)}]
\begin{scope}[shift={(0,.2)}]%[rotate=0]% [rotate=-45]
    \begin{scope}
    	\clip [rounded corners] (1.55,0) arc (0:180:1.55) -- cycle;
    	\foreach \x in {-1.9,-1.7,...,1.3}
	    \draw[line width=.5 pt, color=blue] (\x,-1.3) -- (\x+.6,2.3);
    \end{scope}
    \draw[rounded corners, line width=1] (1.55,0) arc (0:180:1.55) -- cycle;
\end{scope}
\begin{scope}[shift={(0,-.2)}]%[rotate=0]%[rotate=45]
    \begin{scope}
    	\clip [rounded corners] (-1.55,0) arc (180:360:1.55) -- cycle;
    	\foreach \x in {-1.9,-1.7,...,1.3}
	    \draw[line width=.5 pt, color=red] (\x,-2.3) -- (\x+.6,2.3);
    \end{scope}
    \draw[rounded corners, line width=1] (-1.55,0) arc (180:360:1.55) -- cycle;
\end{scope}
\end{scope}
%\draw[line width=2pt] (0,-.4) circle (.4);
\node at (4,0) {\footnotesize $H_\text{global}$};
\node at (-1.85,-.7) {\footnotesize $H_\text{gauge}$};
\node at (50:3.6) {$G'$};
\node at (-50:3.6) {$G''$};
\path[->, color=blue, line width=1.5] (.6,2.5) edge [out = -30, in = 80] (1,1);
\node at (-.3,2.5) {\footnotesize \textcolor{blue}{$H_\text{global}'$}};
\path[->, color=red, line width=1.5] (.6,-2.5) edge [out = 45, in = 280] (1,-1);
\node at (-.3,-2.5) {\small \textcolor{red}{$H_\text{global}''$}};
\end{tikzpicture}
\end{center}
\caption{Anatomy of collective symmetry breaking, following the conventions in Fig.~\ref{fig:comp:breaking:pattern}.}
\label{fig:comp:collective:breaking}
\end{figure}

\subsubsection{Collective breaking in action}

We now demonstrate collective symmetry breaking 
%the framework of a simple little Higgs 
in a model based on the `anatomy' in Fig.~\ref{fig:comp:collective:breaking}. The reader may find it useful to refer to the explicit example of a simple little Higgs model in Section~\ref{sec:comp:little:higgs:explicit} below. Instead of a simple global group $G$, suppose that $G = G'\times G''$. Each of these factors breaks spontaneously to subgroups $H'_\text{global}$ and $H''_\text{global}$, respectively. The spontaneous symmetry breaking pattern is thus
\begin{align}
G = G'\times G'' \to H'_\text{global} \times H''_\text{global}.
\end{align}
This gives us two linear fields $\Sigma'$ and $\Sigma''$ analogous to (\ref{eq:comp:linear:field:Sigma}) so that there are two separate sets of Goldstone bosons.

\begin{figure}
\begin{center}
\begin{subfigure}[b]{0.3\textwidth}
\begin{center}
\begin{tikzpicture}[line width=1.5 pt, scale=.8]
\begin{scope}[rotate=90]
\begin{scope}
    	\clip [rounded corners] (2.1,0) arc (0:180:2.1) -- cycle;
    	\foreach \x in {-2.9,-2.7,...,3.3}
	    \draw[line width=.5 pt, color=gray, dashed] (\x,-2.3) -- (\x-.6,3.3);
    \end{scope}
    \draw[rounded corners, line width=1.5] (2.1,0) arc (0:180:2.1) -- cycle;
\end{scope}

\begin{scope}[shift={(0,.2)}]%[rotate=0]% [rotate=-45]
    \begin{scope}
    	\clip [rounded corners] (1.55,0) arc (0:180:1.55) -- cycle;
    	\foreach \x in {-1.9,-1.7,...,1.3}
	    \draw[line width=.5 pt, color=blue] (\x,-1.3) -- (\x+.6,2.3);
    \end{scope}
    \draw[rounded corners, line width=1] (1.55,0) arc (0:180:1.55) -- cycle;
\end{scope}
\begin{scope}[shift={(0,-.2)}]%[rotate=0]%[rotate=45]
    \begin{scope}
    	\clip [rounded corners] (-1.55,0) arc (180:360:1.55) -- cycle;
    	\foreach \x in {-1.9,-1.7,...,1.3}
	    \draw[line width=.5 pt, color=red] (\x,-2.3) -- (\x+.6,2.3);
    \end{scope}
    \draw[rounded corners, line width=1] (-1.55,0) arc (180:360:1.55) -- cycle;
\end{scope}
\end{tikzpicture}
\end{center}
\caption{$q'$, $q'' \neq 0$}
\label{fig:comp:little:eg:vector}
\end{subfigure}%
\quad
\begin{subfigure}[b]{0.3\textwidth}
\begin{center}
\begin{tikzpicture}[line width=1.5 pt, scale=.8]
\begin{scope}[rotate=180]
\begin{scope}
    	\clip [rounded corners] (2.1,0) arc (0:180:2.1) -- cycle;
    	\foreach \x in {-2.9,-2.7,...,3.3}
	    \draw[line width=.5 pt, color=gray, dashed] (\x,-2.3) -- (\x-1.6,3.3);
    \end{scope}
    \draw[rounded corners, line width=1.5] (2.1,0) arc (0:180:2.1) -- cycle;
\end{scope}

\begin{scope}[shift={(0,.2)}]%[rotate=0]% [rotate=-45]
    \begin{scope}
    	\clip [rounded corners] (1.55,0) arc (0:180:1.55) -- cycle;
    	\foreach \x in {-1.9,-1.7,...,1.3}
	    \draw[line width=.5 pt, color=blue] (\x,-1.3) -- (\x+.6,2.3);
    \end{scope}
    \draw[rounded corners, line width=1] (1.55,0) arc (0:180:1.55) -- cycle;
\end{scope}
\begin{scope}[shift={(0,-.2)}]%[rotate=0]%[rotate=45]
    \begin{scope}
    	\clip [rounded corners] (-1.55,0) arc (180:360:1.55) -- cycle;
    	\foreach \x in {-1.9,-1.7,...,1.3}
	    \draw[line width=.5 pt, color=red] (\x,-2.3) -- (\x+.6,2.3);
    \end{scope}
    \draw[rounded corners, line width=1] (-1.55,0) arc (180:360:1.55) -- cycle;
\end{scope}
\end{tikzpicture}
\end{center}
\caption{$q'=0$, $q''\neq 0$}
\label{fig:comp:little:eg:left}
\end{subfigure}
\quad
\begin{subfigure}[b]{0.3\textwidth}
\begin{center}
\begin{tikzpicture}[line width=1.5 pt, scale=.8]
\begin{scope}[rotate=0]
    \begin{scope}
    	\clip [rounded corners] (2.1,0) arc (0:180:2.1) -- cycle;
    	\foreach \x in {-2.9,-2.7,...,3.3}
	    \draw[line width=.5 pt, color=gray, dashed] (\x,-2.3) -- (\x-1.6,3.3);
    \end{scope}
    \draw[rounded corners, line width=1.5] (2.1,0) arc (0:180:2.1) -- cycle;
\end{scope}

\begin{scope}[shift={(0,.2)}]%[rotate=0]% [rotate=-45]
    \begin{scope}
    	\clip [rounded corners] (1.55,0) arc (0:180:1.55) -- cycle;
    	\foreach \x in {-1.9,-1.7,...,1.3}
	    \draw[line width=.5 pt, color=blue] (\x,-1.3) -- (\x+.6,2.3);
    \end{scope}
    \draw[rounded corners, line width=1] (1.55,0) arc (0:180:1.55) -- cycle;
\end{scope}
\begin{scope}[shift={(0,-.2)}]%[rotate=0]%[rotate=45]
    \begin{scope}
    	\clip [rounded corners] (-1.55,0) arc (180:360:1.55) -- cycle;
    	\foreach \x in {-1.9,-1.7,...,1.3}
	    \draw[line width=.5 pt, color=red] (\x,-2.3) -- (\x+.6,2.3);
    \end{scope}
    \draw[rounded corners, line width=1] (-1.55,0) arc (180:360:1.55) -- cycle;
\end{scope}
\end{tikzpicture}
\end{center}
\caption{$q'\neq 0$, $q''= 0$}
\label{fig:comp:little:eg:right}
\end{subfigure}
\end{center}
\caption{Collective symmetry breaking. Upper (blue) and lower (red) blobs represent $H'$ and $H''$ in Fig.~\ref{fig:comp:collective:breaking}. The thick black line represents the gauged symmetry $H_\text{gauge}$ under which $\Sigma'$ has charge $q'$ and $\Sigma''$ has charge $q''$.  When either $q'$ or $q''$ vanishes, the unbroken group is $H'_\text{global}\times H''_\text{global}$.}\label{fig:comp:little:eg}
\end{figure}
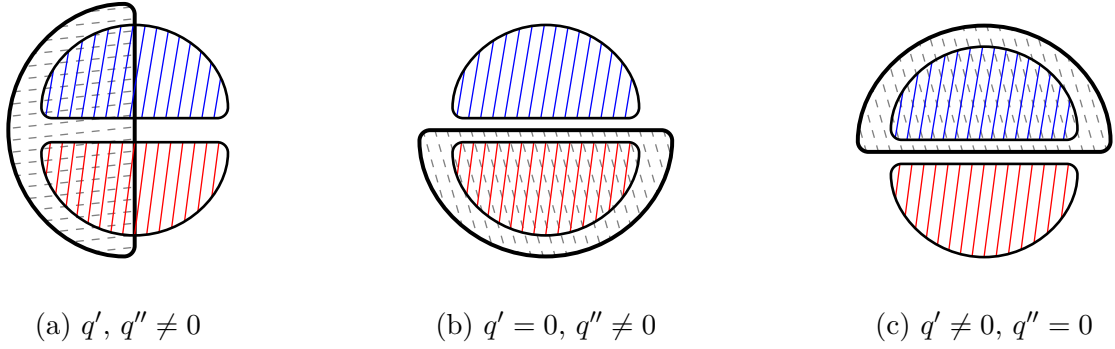

We explicitly break $G$ by gauging $H_\text{gauge} \subset G$. Suppose that both $H'_\text{global}$ and $H''_\text{global}$ are subgroups of $H_\text{gauge}$ in such a way that both $\Sigma'$ and $\Sigma''$ are charged under $H_\text{gauge}$ with nonzero charges $q'$ and $q''$ respectively.  A piece of each subgroup is gauged, as shown in Fig.~\ref{fig:comp:little:eg:vector}. $H'_\text{global}\times H''_\text{global}$ is then explicitly broken to a smaller subgroup, for example a vectorial subgroup identified by the gauging, $H$. 

On the other hand, when either $q'$ or $q''$ is set to zero, only one of the global subgroups is gauged, as shown in Fig.~\ref{fig:comp:little:eg:left} and \ref{fig:comp:little:eg:right}. In either of these cases, the resulting global symmetry group is still $H'_\text{global} \times H''_\text{global}$. In other words, one requires both $q'$ and $q''$ to explicitly break $H_\text{global}=H'_\text{global} \times H''_\text{global}$.

When one of the global subgroups is uncharged under the gauged subgroup, say $H''_\text{global}$, those Goldstone bosons pick up no mass from the gauge sector. For the other global subgroup which is charged under the gauge group, say $H'_\text{global}$, there are two possibilities:
\begin{enumerate}
\item If $H_\text{gauge} \subseteq H'_\text{global}$, then loops of the gauge bosons will feed into the mass of the pseudo-Goldstone bosons. In the absence of collective symmetry breaking, this gives a contribution that is quadratic in the cutoff.
\item If, on the other hand\footnote{It is sufficient to consider some subgroup $\tilde G' \subseteq G'$ that contains $H_\text{global}'$ as a proper subgroup}, $G'\subset H_\text{gauge}$, then the would-be Goldstone bosons from $G'\to H'_\text{global}$ are eaten by the $(G/H_\text{global}') \cap H_\text{gauge}$ gauge bosons. There is no quadratic sensitivity to the cutoff. 
\end{enumerate}
In the second case, the Higgs mechanism removed the $\Lambda^2$ contribution to the pseudo-Goldstone mass, but it also got rid of the pseudo-Goldstones themselves.

This leads us to consider the case when both $G'$ and $G''$ (not just their $H_\text{global}$ subgroups ) are charged under the gauged symmetry. For simplicity, suppose $G'=G''=H_\text{gauge}$ so that one gauges the vectorial combination. In this case, both the $\Sigma'$ and $\Sigma''$ fields carrying our Goldstone bosons are charged under the gauge group. The gauge fields become massive by the Higgs mechanism, but there are twice as many Goldstone bosons than it can eat\footnote{\label{foot:comp:bear:analogy}This is a manifestation of general outdoors advice: if you (a Goldstone boson) are being chased by a hungry bear (a gauge boson), it is not necessary for you survival that you can outrun it (have zero coupling). It is sufficient that you are with friends whom you can outrun. Collective breaking is, in part, the requirement that you have more slow friends than hungry bears.}. Indeed, the `axial' combination of $G'$ and $G''$ furnishes a set of Goldstone bosons that remain uneaten and are sensitive to explicit breaking effects so that they are formally pseudo-Goldstones. Any contribution to the pseudo-Goldstone mass, however, must be proportional to $(gq')(gq'')$, where $g$ is the gauge coupling. In other words, it requires interactions from both $\Sigma'$ and $\Sigma''$. The resulting mass term is suppressed since this requires factors of the $\Sigma'$ and $\Sigma''$ \textsc{vev}s to soak up additional boson legs. We now demonstrate this with an explicit example.

\begin{framed}
\noindent \footnotesize
\textbf{Why can't you just rotate to a different basis?} Based on Fig.~\ref{fig:comp:little:eg}, one might wonder if we can repartition $G = G'\times G''$ so that the $H'_\text{global}$ and $H''_\text{global}$ subgroups are always both gauged. Alternately, perhaps one can repartition $G$ so that only one subgroup is ever gauged. This cannot be done, even when $q'=q''$. The reason is precisely what we pointed out above Sec.~\ref{sec:comp:xpt:lagrangian}: the axial combination of two groups is not itself a group since its algebra doesn't close. 
\end{framed}

\subsubsection{Explicit example: $\left(\text{SU(3)}\to\text{SU(2)}\right)^2$}
\label{sec:comp:little:higgs:explicit}

Let us see how this fits together in a simple little Higgs model---though we emphasize that collective symmetry breaking is a generic feature of all realistic composite Higgs models, not just those of little Higgs type. We classify composite Higgs models in Section~\ref{sec:comp:classification} to clarify any ambiguity.
Consider the case where $G'=G''=\text{SU}(3)$ and $H'_\text{global}=H''_\text{global}=\text{SU}(2)$.  We thus have two fields which are linear representations of $\text{SU}(3)$ and carry the Goldstone bosons,
\begin{align}
\Sigma' = \text{exp}
\left[
\frac i{f'}
\left(
	\begin{array}{cc:c}
	\multicolumn{2}{c:}{
	    \multirow{2}{*}{
	        \large{$0_{2\times 2}$}
	      }
	 } & \multicolumn{1}{c}{\multirow{2}{*}{{$H'$}}} \\
	 &&  \\
	\hdashline
%	 &  & 0
	 \multicolumn{2}{c:}{
	     H'^\dag
    } & 0
%\multicolumn{2}{cc:}{$A_\mu$} & $\varphi$
	\end{array}
\right)
\right]
\begin{pmatrix}
 0 \\
 0 \\
 \hdashline
 f'
\end{pmatrix}
=
\begin{pmatrix}
 0 \\
 0 \\
 \hdashline
 f'
\end{pmatrix}
+
i
\begin{pmatrix}
\multirow{2}{*}{{$H'$}}\\ \\
 \hdashline
 0
\end{pmatrix}
-\frac{1}{2f}
\begin{pmatrix}
\multirow{2}{*}{{$0$}}\\ \\
 \hdashline
 H'^\dag H'
\end{pmatrix},
\label{eq:comp:little:sigma:expansion}
\end{align}
and similarly for $\Sigma''$. For simplicity let us set $f'=f''\equiv f$. The kinetic terms for the $\Sigma$ fields are
\begin{align}
\mathcal L = |D_\mu \Sigma'|^2 + |D_\mu \Sigma''|^2 = \cdots + (gq')^2 \left|V^a_\mu T'^a  \Sigma'\right|^2 + (gq'')^2 \left|V^a_\mu T''^a  \Sigma''\right|^2,
\end{align}
where $T'^{a} = T''^{a}$ are the generators of the gauged group. To see the contribution to the Higgs mass, one can Wick contract the two gauge bosons in these terms---this is precisely the analog of the `cat diagram' in Fig.~\ref{fig:comp:cat}. This contraction ties together the gauge boson indices so that the resulting term goes like
\begin{align}
\left[\text{loop factor}\right]
(gq')^2 \Sigma'^\dag T'^a T'^a \Sigma' = 
\left[\text{loop factor}\right]
\frac{(gq')^2}2 C_2 \Sigma'^\dag \,\mathbf{1}_\text{gauge}\, \Sigma',
\label{eq:comp:little:pgb:mass:vector:op}
\end{align}
and similarly for $\Sigma''$.
Here the loop factor contains the quadratic dependence on the cutoff, $\left[\text{loop factor}\right] \sim \Lambda^2/16\pi^2$, and the factor $\mathbf{1}_\text{gauge}$ is the identity matrix in the appropriate gauged subgroup. Here we have used $T^a T^a = C_2 \mathbf{1}$, where $C_2$ is the quadratic Casimir operator of the representation\footnote{$C_2(\text{fundamental}) = (N^2-1)/2N$ for SU($N$).}. Now let's explicitly demonstrate how collective breaking works. 
% We now gauge the vector subgroup of $G'\cap G''$, that is we set $q'=q''=1$.
\begin{itemize}
\item If only the SU(2)$=H'_\text{global}=H''_\text{global}$ parts of $G'$ and $G''$ were gauged, then there would be two separate sets of pseudo-Goldstone bosons $H'$ and $H''$. We plug in the expansion of $\Sigma'$ (\ref{eq:comp:little:sigma:expansion}) into (\ref{eq:comp:little:pgb:mass:vector:op}) and note that in this case,
\begin{align}
\mathbf{1}_\text{gauge} = 
\begin{pmatrix}
1 &0&0\\
0& 1 & 0\\
0 & 0& 0
\end{pmatrix}.
\end{align}
This picks up the Goldstones in the second term on the right-hand side of (\ref{eq:comp:little:sigma:expansion}) so that there is indeed a Goldstone mass term  proportional to $\Lambda^2 = (4\pi f)^2$ for each set of Goldstones. 
\item On the other hand, in the case where $G'$ and $G''$ are both gauged with $q'=q''$, the matrix $\mathbf{1}_\text{gauge}$ becomes a true identity operator,
\begin{align}
\mathbf{1}_\text{gauge} = 
\begin{pmatrix}
1 &0&0\\
0& 1 & 0\\
0 & 0& 1
\end{pmatrix}.
\end{align}
Now the global symmetry breaking \textsc{vev}s $\langle \Sigma'\rangle$ and $\langle \Sigma''\rangle$ break part of the gauge symmetry and the Higgs mechanism tells us that there are gauge bosons that eat would-be Goldstones. Indeed, the first term on the right-hand side of (\ref{eq:comp:little:sigma:expansion})---which is no longer projected out by $\mathbf{1}_\text{gauge}$---encodes the mass picked up by the gauge bosons. Observe, however, what has happened to the $\Lambda^2$ mass contribution in the previous scenario: it is now canceled by the cross term between the first and third terms on the right hand side of (\ref{eq:comp:little:sigma:expansion}). In other words, the terms which gave the quadratic sensitivity to the cutoff have vanished. 
\end{itemize}
If we were only considering a single SU(3)$\to$SU(2) global symmetry breaking, then we would still be out of luck since the massive gauge bosons would have eaten all of our Goldstone bosons---so even though we got rid of the $\Lambda^2$ sensitivity of the pseudo-Goldstone masses, we also would have gotten rid of the pseudo-Goldstones themselves. With foresight, however, we have followed the advice of footnote~\ref{foot:comp:bear:analogy}: we have more Goldstones than our gauge bosons can possibly eat.

A useful way to parameterize our Goldstones is to follow the convention in (\ref{eq:comp:chiral:coset}): 
\begin{align}
\Sigma' =& 
\text{exp}
\left[
\frac i{f}
\left(
	\begin{array}{cc:c}
	\multicolumn{2}{c:}{
	    \multirow{2}{*}{
	        \large{$0_{2\times 2}$}
	      }
	 } & \multicolumn{1}{c}{\multirow{2}{*}{{$V$}}} \\
	 &&  \\
	\hdashline
%	 &  & 0
	 \multicolumn{2}{c:}{
	     V^\dag
    } & 0
%\multicolumn{2}{cc:}{$A_\mu$} & $\varphi$
	\end{array}
\right)
\right]
\text{exp}
\left[
\frac i{f}
\left(
	\begin{array}{cc:c}
	\multicolumn{2}{c:}{
	    \multirow{2}{*}{
	        \large{$0_{2\times 2}$}
	      }
	 } & \multicolumn{1}{c}{\multirow{2}{*}{{$H$}}} \\
	 &&  \\
	\hdashline
%	 &  & 0
	 \multicolumn{2}{c:}{
	     H^\dag
    } & 0
%\multicolumn{2}{cc:}{$A_\mu$} & $\varphi$
	\end{array}
\right)
\right]
\begin{pmatrix}
 0 \\
 0 \\
 \hdashline
 f
\end{pmatrix}
\label{eq:comp:little:sigma:p}
\\
\Sigma'' =& 
\text{exp}
\left[
\frac i{f}
\left(
	\begin{array}{cc:c}
	\multicolumn{2}{c:}{
	    \multirow{2}{*}{
	        \large{$0_{2\times 2}$}
	      }
	 } & \multicolumn{1}{c}{\multirow{2}{*}{{$V$}}} \\
	 &&  \\
	\hdashline
%	 &  & 0
	 \multicolumn{2}{c:}{
	     V^\dag
    } & 0
%\multicolumn{2}{cc:}{$A_\mu$} & $\varphi$
	\end{array}
\right)
\right]
\text{exp}
\left[
\frac{-i}{f}
\left(
	\begin{array}{cc:c}
	\multicolumn{2}{c:}{
	    \multirow{2}{*}{
	        \large{$0_{2\times 2}$}
	      }
	 } & \multicolumn{1}{c}{\multirow{2}{*}{{$H$}}} \\
	 &&  \\
	\hdashline
%	 &  & 0
	 \multicolumn{2}{c:}{
	     H^\dag
    } & 0
%\multicolumn{2}{cc:}{$A_\mu$} & $\varphi$
	\end{array}
\right)
\right]
\begin{pmatrix}
 0 \\
 0 \\
 \hdashline
 f
\end{pmatrix},
\label{eq:comp:little:sigma:pp}
\end{align}
where we have identified the Higgs as the axial combination of global shifts, while the vector combination of Goldstones, $V$, is eaten by the gauge bosons to become massive. 

Now the $H$ pseudo-Goldstones only pick up mass from diagrams that involve both the $(gq')$ and the $(gq'')$ couplings. In other words, it requires a combination of the $\Sigma'$ and the $\Sigma''$ fields. The leading order contribution comes from diagrams of the form
\begin{align}
\vcenter{
\hbox{
\begin{tikzpicture}[line width=1.5, scale=.8]
% [dashed, dash pattern=on 10 off 5]
%, postaction={decorate},
%        decoration={markings,mark=at position .55 with {\arrow{>}}}
%[dashed, dash pattern=on 10 off 5, postaction={decorate},
%        decoration={markings,mark=at position .55 with {\arrow{>}}}
%]
\draw[scalar] (-2,2.5) -- (0,2);
\draw[scalar] (0,2) -- (2,2.5);
\draw[scalarbar] (-2,-0.5) -- (0,0);
\draw[scalarbar] (0,0) -- (2,-.5);
\draw[vector] (0,2) arc (90:270:1);
\draw[vector] (0,0) arc (-90:90:1);
\node at (-2.5,2.5) {$\Sigma'$};
\node at (2.5,2.5) {$\Sigma'$};
\node at (-2.5,-.5) {$\Sigma''$};
\node at (2.5,-.5) {$\Sigma''$};
\end{tikzpicture}
}
}
\sim
\frac{g^4}{16\pi^2} \log \Lambda^2 \; \left|\Sigma'^\dag \Sigma''\right|^2.
\end{align}
Since $\Sigma'^\dag\Sigma'' = f^2 - 2H^\dag H + \cdots$, we see that the leading term in the Higgs mass is only logarithmically sensitive to $\Lambda$ because it required one power each of the $\Sigma'$ and $\Sigma''$ \vevs. 
The Higgs mass sets the electroweak scale to be on the order of $f/(4\pi)$. This is a factor of $(4\pi)$ suppressed compared to the global symmetry breaking scale $f$---generating the hierarchy in $\xi$ that we wanted---and also a further factor of $(4\pi)$ from the cutoff $\Lambda = 4\pi f$. In this sense, collective symmetry breaking shows us what we can buy for factors of $(4\pi)$ and why those factors are important in na\"ive dimensional analysis.

\subsubsection{Top partners}\label{sec:comp:top:partner}

As before, the largest contribution to the Higgs mass comes from the top quark. In the simple scenario above, we have extended our gauge group\footnote{For simplicity we ignore the U(1)$_Y$ factor, it is straightforward to assign charges appropriately.} from SU(2)$_\text L$ to $H_\text{gauge}=$SU(3) so we'll need to also extend the usual top doublet to include a %`little partner,'
partner $T_L$
\begin{align}
Q=
\begin{pmatrix}
t_L\\
b_L
\end{pmatrix}
\to
Q= 
\begin{pmatrix}
t_L\\
b_L\\
T_L
\end{pmatrix}
.
\end{align}
%We must also include a right-handed SU(3) singlet $T'_{R}$ as a partner for the $t_L$, in parallel to the usual right-handed $t_{R}'$ partner of the \SM $t_L$. The Yukawa terms for the top quarks are,
We must also include a right-handed SU(3) singlet $T'_{R}$ as a partner for the $T_L$, in parallel to the usual right-handed $t'_{R}$ partner of the \SM $t_L$. The prime on the $t'_R$---what is normally called $t_R$ in the \SM---is for future convenience. The Yukawa terms for the top quarks are,
\begin{align}
\mathcal L_\text{top} = 
\lambda' \Sigma'^\dag Q t'^\dag_{R}
+
\lambda'' \Sigma''^\dag Q T'^\dag_R
+ \text{h.c.}
\label{eq:comp:little:yukawa}
\end{align}
where the fermions are written in terms of Weyl spinors. 
Other terms, such as ${\Sigma'}^\dag Q {T'_R}^\dag$ or ${\Sigma''}^\dag Q {t'_R}^\dag$, can typically be prohibited by invoking chiral symmetries.
Observe that the $\lambda'$ term is invariant under $G'$ if $Q$ is a fundamental under $G'$. Similarly, the $\lambda''$ term is invariant under $G''$ if $Q$ is a fundamental under $G''$. This is indeed consistent since $Q$ is a fundamental under $H_\text{gauge}$ which is the diagonal subgroup of $G'\times G''$. 
This shows us how collective symmetry breaking is embedded in the Yukawa sector. When only one of the $\lambda$ terms is nonzero, $\mathcal L_\text{top}$ is $G'\times G''$ invariant. However, when both are turned on, the global symmetry is broken down to the diagonal subgroup. 

This is collective breaking is similar to the breaking of the global U(3)$_Q\times \text{U}(3)_U \times \text{U}(3)_D$ flavor symmetry to $\text{U}(3)$ by the up- and down-type Yukawas in the Standard Model. If $y_u=0$ and $y_d\neq 0$, then the flavor symmetry would be enhanced to $\text{U}(3)^2$ since the right-handed up-type quarks could be rotated independently of the other fields. 

We can now plug in the expansion (\ref{eq:comp:little:sigma:p} -- \ref{eq:comp:little:sigma:pp}) into the Yukawa terms (\ref{eq:comp:little:yukawa}), ignoring the $V$ terms since we now know those are eaten by the gauge bosons. Expanding the resulting product gives
\begin{align}
\mathcal L_\text{top}
=&
%iH^\dag Q (\lambda'' T_R^\dag -\lambda' t_R^\dag) + \left(f-\frac{H^\dag H}{2f}\right)T_L\left(\lambda't_R^\dag + \lambda'' T_R^\dag\right).
iH^\dag Q (\lambda'' T_R^\dag -\lambda' t_R'^\dag) + \left(f-\frac{H^\dag H}{2f}\right)T_L\left(\lambda't_R'^\dag + \lambda'' T_R^\dag\right).
\end{align}
From this we can write out the right-handed top eigenstates
\begin{align}
T_R &= \frac{\lambda' t_R' + \lambda'' T_R'}{\sqrt{\lambda'^2 + \lambda''^2}}
&
t_R = i\frac{\lambda'' T_R' - \lambda' T_R'}{\sqrt{\lambda'^2 + \lambda''^2}}
\end{align}
and the resulting top Yukawa, top partner mass, and top partner coupling to $H^\dag H$,
\begin{align}
\mathcal L_\text{top}
=
\lambda_t H^\dag Q t_R^\dag
+ \lambda_t f T_L T_R^\dag - \frac{\lambda_t}{2f}H^\dag H T_L T_R^\dag,
\end{align}
where we see that all of the couplings are simply related to the \SM top Yukawa, $\lambda_t = \sqrt{\lambda'^2+ \lambda''^2}$. These relations ensure the cancellation between diagrams that give a $\Lambda^2$ contribution to the Higgs mass,
%
%\begin{align}
%\vcenter{
%    \hbox{
%        \begin{tikzpicture}[line width=1.5pt, scale=1.5]
%        \draw[scalarnoarrow] (-1,0) -- (-.5,0);
%        \draw[fermion] (.5,0) arc (0:180:.5);
%        \draw[fermion] (-.5,0) arc (180:360:.5);
%        \draw[scalarnoarrow] (1,0) -- (.5,0);
%        \node at (150:.8) {$t$};
%        \node at (-20:.7) {$\lambda_t$};
%        \node at (200:.7) {$\lambda_t$};
%        \node at (-1.2,0) {$h$};
%        \node at (1.2,0) {$h$};
%        \end{tikzpicture}
%    }
%}
%+
%\vcenter{
%    \hbox{
%        \begin{tikzpicture}[line width=1.5pt, scale=1.5]
%        \draw[scalarnoarrow] (-1,0) -- (1,0);
%        \draw[fermion] (0,0) arc (-90:90:.5);
%        \draw[fermion] (0,1) arc (90:270:.5);
%        \begin{scope}[shift={(0,1)}]
%		    \clip (0,0) circle (.1);
%		    \draw[fermionnoarrow] (-1,1) -- (1,-1);
%		    \draw[fermionnoarrow] (1,1) -- (-1,-1);
%	     \end{scope}
%        \node at (160:.8) {$T$};
%        \node at (.5,1.1) {$\lambda_t f$};
%        \node at (0,-.3) {$-\lambda_t/f$};
%        \node at (-1.2,0) {$h$};
%        \node at (1.2,0) {$h$};
%        \end{tikzpicture}
%    }
%}
%=
%\mathcal O(\log\,\Lambda).
%\end{align}
\begin{align}
\begin{tikzpicture}[line width=1.5 pt, scale=1.3,
baseline=(current  bounding  box.center) % for align
]
	\coordinate (vll) at (.25,0);
	\coordinate (vl) at (1,0);	
	\coordinate (vr) at (2,0);
	\coordinate (vrr) at (2.75,0);
	\draw[dashed, dash pattern=on 10 off 5] (vl)--(vll);
	\draw[fermion] (vl) arc (180:0:.5);
	\draw[fermion] (vr) arc (0:-180:.5);
	\draw[dashed, dash pattern=on 10 off 5] (vr) --(vrr);
	\node at ($(-.25,0)+(vll)$) {$\displaystyle h$};
	\node at ($(.2,0)+(vrr)$) {$\displaystyle h$};
	\node at ($(vl)+(-.2,.25)$) {$\displaystyle t$};
	\node at ($(vl)+(-.2,-.25)$) {$\displaystyle \lambda_t$};
	\node at ($(vr)+(.2,-.25)$) {$\displaystyle \lambda_t$};
\end{tikzpicture}
\;
	+
\begin{tikzpicture}[line width=1.5 pt, scale=1.3,
baseline=(current  bounding  box.center) % for align
]
	\coordinate (vl) at (-1,-.2);
	\coordinate (vc) at (0,0);
	\coordinate (vt) at (0,1);
	\coordinate (vr) at (1,-.2);
	\draw[fermion] (vc) arc (-90:90:.5);
	\draw[fermion] (vt) arc (90:270:.5);
	\draw[dashed, dash pattern=on 10 off 5] (vl)--(vc);
	\draw[dashed, dash pattern=on 10 off 5] (vc)--(vr);
	\begin{scope}[shift={(vt)}]
		\clip (0,0) circle (.175);
		\draw (135:1) -- (-45:1);
		\draw (45:1) -- (225:1);
	\end{scope}
	\node at ($(-.25,0)+(vl)$) {$\displaystyle h$};
	\node at ($(.25,0)+(vr)$) {$\displaystyle h$};
	\node at ($(vt) + (.45,.25)$) {$\displaystyle \lambda_t f$};
	\node at ($(vc) + (-.1,-.45)$) {$\displaystyle -\lambda_t/f$};
	\node at ($(vc) + (-.65,.2)$) {$\displaystyle T$};
\end{tikzpicture}
	=
	\mathcal O(\log\,\Lambda).	
	\label{eq:comp:top:partner:cancellation}
\end{align}
Note the symmetry factor of $1/2$ in the $h^2 T_LT_R^\dag$ Feynman rule. For simplicity we also drop an overall $\sqrt{2}$ in the normalization of the $h$ field which is irrelevant for the $\Lambda^2$ cancellation. We see that indeed collective symmetry breaking can protect against the reintroduction of quadratic sensitivity to the cutoff by the Yukawa interactions.

Just as in the case of natural \SUSY, an important signature of this class of models is to look for the `partner top' particles which are responsible for the softening of the cutoff dependence of Higgs mass from the top sector. One can search for these objects at the \LHC through either pair production,
\begin{align}
q \bar q/gg \; \to \; T\bar T,
\end{align}
or through single production in association with a \SM quark,
\begin{align}
bq \; &\to\; T q' 
&
qq' \; &\to \; Tb.
\end{align}
The top partner decays are fixed by the Goldstone equivalence theorem.
The partner top decays approximately 50\% of the time to $bW$, with the remaining decay products split evenly between $tZ$ and $th$~\cite{Perelstein:2003wd}. The lower bound on the top partner mass from vector-like heavy top (also referred to as fourth generation) searches is $\gtrsim 700$ GeV~\cite{Chatrchyan:2013uxa}. 

One can continue to calculate the Coleman-Weinberg potential in %the little Higgs 
this scenario to check for electroweak symmetry breaking and further study the phenomenology of these models.
As discussed below (\ref{eq:comp:Higgs:taxonomy:fit}), in addition to playing an important role generating the Higgs potential, these top partners are of phenomenological significance since they are expected to be lighter than the other strong sector resonances. By virtue of filling out representations of the global group, some of these top partners are expected to have exotic electromagnetic charges, such as $Q_\text{EM}=5/3$; these states are considered to be `smoking gun' signals of a composite Higgs scenario.
We refer the reader to the excellent reviews \cite{Perelstein:2005ka, Schmaltz:2005ky} for a pedagogical introduction in the context of the little Higgs.
See \cite{DeSimone:2012fs, Aguilar-Saavedra:2013qpa, Buchkremer:2013bha} for a more general discussion of experimental bounds on top partners.

\subsection{Deconstruction and moose models}
\label{sec:deconstruction}

We now briefly mention some connections with extra dimensional models and introduce a diagrammatical language that is sometimes used to describe the symmetry breaking pattern in composite models.

In Section~\ref{fig:xd:holography} we introduced the holographic principle as a connection between strongly coupled 4D theories and weakly coupled theories on a curved spacetime with an extra spatial dimension. This turns out to be a natural tool to get a handle for some of the strong dynamics encoded into the form factors. Indeed, the minimal composite Higgs model described above was developed using these insights \cite{Agashe:2004rs}.

There is, however, another way to connect 5D models to 4D models. 5D models have dimensionful couplings and are manifestly non-renormalizable. One proposal for a \UV completion is to discretize (`latticize') the extra dimension \cite{Hill:2000mu, ArkaniHamed:2001nc, ArkaniHamed:2001ca}. In this picture, the extra dimension is split into $N$ discrete sites which should no longer be thought of as discrete spacetimes, but rather as nodes in a `theory space' that describe a gauge symmetry structure on a single 4D spacetime. The bulk gauge symmetry $G$ latticized into a 4D gauged $G$ on each of the $N$ nodes,
\begin{center}
\begin{tikzpicture}[line width=1.5]
%\draw (0,0) circle (.5);
%\node at (0,0) {$G$};
\draw (2,0) circle (.5);
\node at (2,0) {$G$};
\draw (4,0) circle (.5);
\node at (4,0) {$G$};
\draw (6,0) circle (.5);
\node at (6,0) {$G$};
\draw[fill=black] (7,0) circle (.02); 
\draw[fill=black] (7.5,0) circle (.02); 
\draw[fill=black] (8,0) circle (.02); 
\draw (9,0) circle (.5);
\node at (9,0) {$G$};
\end{tikzpicture}
\end{center}
At this level the nodes are just $N$ separate gauge groups; after all, this is precisely what we mean by a local symmetry (see~\cite{Thaler:2005kr, Kahn:2012as} for a discussion in depth). We next introduce a set of $(N-1)$ scalar \textbf{link fields} $\Phi_i$ which are in the bifundamental representation with respect to the $N^\text{th}$ and $(N+1)^\text{th}$ gauge groups: $(\mathbf{N}_i,\bar{\mathbf{N}}_{i+1})$. We may draw these link fields as lines between the nodes,
\begin{center}
\begin{tikzpicture}[line width=1.5]
%\draw (0,0) circle (.5);
%\node at (0,0) {$G$};
\draw (2,0) circle (.5);
\node at (2,0) {$G$};
\draw (4,0) circle (.5);
\node at (4,0) {$G$};
\draw (6,0) circle (.5);
\node at (6,0) {$G$};
\draw[fill=black] (8,0) circle (.02); 
\draw[fill=black] (8.5,0) circle (.02); 
\draw[fill=black] (9,0) circle (.02); 
\draw (11,0) circle (.5);
\node at (11,0) {$G$};
\draw[fermion] (2.5,0) -- (3.5,0);
\draw[fermion] (4.5,0) -- (5.5,0);
\draw[fermion] (6.5,0) -- (7.5,0);
\draw[fermion] (9.5,0) -- (10.5,0);
\node at (3,.4) {$\Phi_1$};
\node at (5,.4) {$\Phi_2$};
\node at (7,.4) {$\Phi_3$};
\node at (10,.4) {$\Phi_{N-1}$};
\end{tikzpicture}
\end{center}
The arrow on the link field keeps track of the representation with respect to a group:
\begin{itemize}
\item Arrows leaving a node are fundamental with respect to that group.
\item Arrows entering a node are anti-fundamental with respect to that group.
\end{itemize}
Now suppose each of these link fields acquires a \vev proportional to $\mathbf{1}$ in their respective $G_i\times G_{i+1}$ internal spaces. Each link field would spontaneously break the symmetry $G_i\times G_{i+1} \to G_{\text{diag}}$. The symmetries are broken down to $G$. One can diagonalize the mass matrix for the gauge boson---a problem that is mathematically identical to solving the waves in a system of $N-1$ springs in series\cite{Georgi:2004iy}---to find that the spectrum looks like a tower of Kaluza-Klein modes. 
In fact, the link fields can be identified with the \KK modes of the fifth component of the bulk gauge field $A_5$. This construction also shows explicitly that the Kaluza-Klein gauge fields in 5D acquire their masses from eating the \KK modes of the $A_5$, which are here manifestly would-be Goldstone bosons.
By coupling matter appropriately, one constructs a \UV complete 4D model of a product of gauge groups that gives the same `low' energy physics as an extra dimension. 
We refer the reader to the original literature for details
\cite{Hill:2000mu, ArkaniHamed:2001nc, ArkaniHamed:2001ca} or \cite{Cheng:2010pt} for a brief summary.

Rather than just way to \UV complete extra dimensions, deconstructions are also a useful tool for motivating models of chiral symmetry breaking. In fact, they are a manifestation of a more general tool for composite models called \textbf{moose diagrams}\footnote{These diagrams are also called \textbf{quiver diagrams} by string theorists \cite{Douglas:1996sw}.}~\cite{Georgi:1985hf, Georgi:1986dw}. One can use this diagrammatic language to construct little Higgs models; indeed, this was the original inspiration for the development of collective symmetry breaking paradigm in Section~\ref{sec:comp:collective}. The topology of these diagrams encodes information about spectrum of Goldstone modes \cite{Gregoire:2002ra}. From the dimensional deconstruction of an extra dimension, it's clear that all of the Goldstones are eaten by the \KK modes of gauge bosons. More general connections between nodes, however, allow more Goldstones to survive hungry gauge bosons.

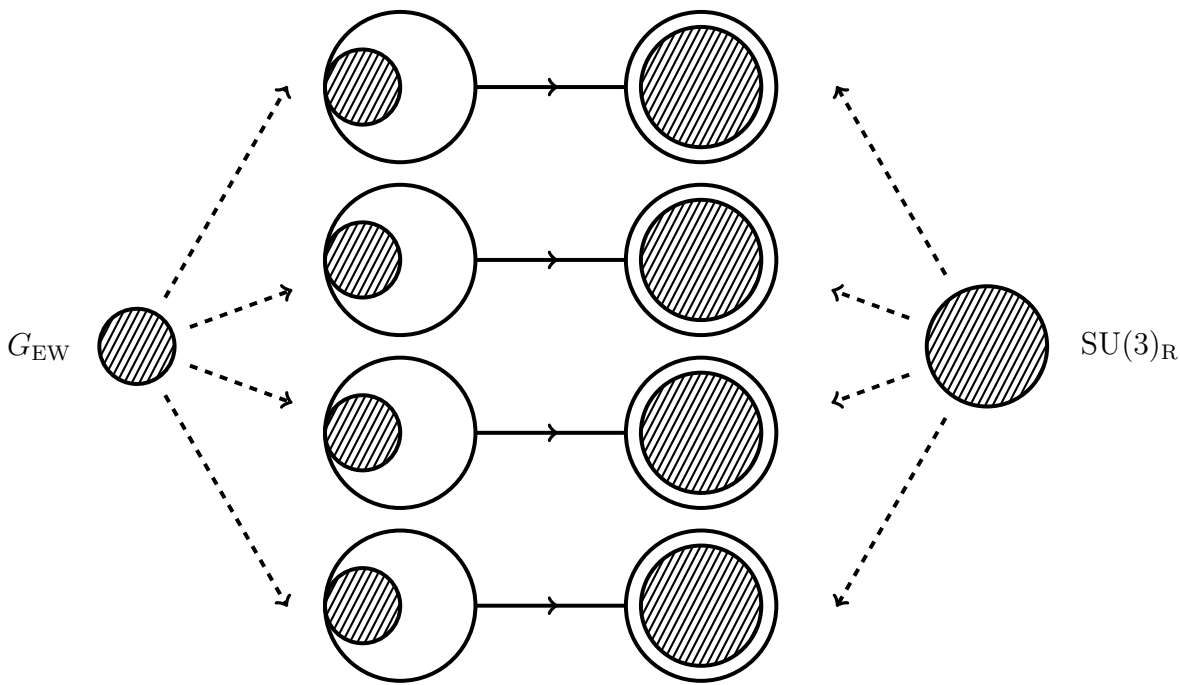
\begin{figure}
\begin{center}
\begin{tikzpicture}[line width=1.5]
	\begin{scope}
	    \begin{scope}
	    	\clip (-.5,0) circle (.5);
	    	\foreach \x in {-2.5,-2.4,...,.3}
				\draw[line width=.8 pt] (\x,-1.9) -- (\x+1.5,1);
	  	\end{scope}
    	\draw[line width = 1.5] (0,0) circle (1);
    	\draw[line width = 1.5] (-.5,0) circle (.5);
    	
    	\draw[line width = 1.5] (4,0) circle (1);
    	\begin{scope}
	    	\clip (4,0) circle (.8);
	    	\begin{scope}[shift={(4,0)}]
	    	\clip (0,0) circle (.8);
	    	\foreach \x in {-2.5,-2.4,...,.3}
				\draw[line width=.8 pt] (\x,-1.9) -- (\x+1.5,1);
	  	\end{scope}
	  	\end{scope}
    	\draw[line width = 1.5] (4,0) circle (.8);
    	
    	\draw[fermion] (1,0) -- (3,0);
	 \end{scope}
	\begin{scope}[shift={(0,-2.3)}]
	    \begin{scope}
	    	\clip (-.5,0) circle (.5);
	    	\foreach \x in {-2.5,-2.4,...,.3}
				\draw[line width=.8 pt] (\x,-1.9) -- (\x+1.5,1);
	  	\end{scope}
    	\draw[line width = 1.5] (0,0) circle (1);
    	\draw[line width = 1.5] (-.5,0) circle (.5);
    	
    	\draw[line width = 1.5] (4,0) circle (1);
    	\begin{scope}
	    	\clip (4,0) circle (.8);
	    	\begin{scope}[shift={(4,0)}]
	    	\clip (0,0) circle (.8);
	    	\foreach \x in {-2.5,-2.4,...,.3}
				\draw[line width=.8 pt] (\x,-1.9) -- (\x+1.5,1);
	  	\end{scope}
	  	\end{scope}
    	\draw[line width = 1.5] (4,0) circle (.8);
    	
    	\draw[fermion] (1,0) -- (3,0);
	 \end{scope}
	 \begin{scope}[shift={(0,-4.6)}]
	    \begin{scope}
	    	\clip (-.5,0) circle (.5);
	    	\foreach \x in {-2.5,-2.4,...,.3}
				\draw[line width=.8 pt] (\x,-1.9) -- (\x+1.5,1);
	  	\end{scope}
    	\draw[line width = 1.5] (0,0) circle (1);
    	\draw[line width = 1.5] (-.5,0) circle (.5);
    	
    	\draw[line width = 1.5] (4,0) circle (1);
    	\begin{scope}
	    	\clip (4,0) circle (.8);
	    	\begin{scope}[shift={(4,0)}]
	    	\clip (0,0) circle (.8);
	    	\foreach \x in {-2.5,-2.4,...,.3}
				\draw[line width=.8 pt] (\x,-1.9) -- (\x+1.5,1);
	  	\end{scope}
	  	\end{scope}
    	\draw[line width = 1.5] (4,0) circle (.8);
    	
    	\draw[fermion] (1,0) -- (3,0);
	 \end{scope}
	 \begin{scope}[shift={(0,-6.9)}]
	    \begin{scope}
	    	\clip (-.5,0) circle (.5);
	    	\foreach \x in {-2.5,-2.4,...,.3}
				\draw[line width=.8 pt] (\x,-1.9) -- (\x+1.5,1);
	  	\end{scope}
    	\draw[line width = 1.5] (0,0) circle (1);
    	\draw[line width = 1.5] (-.5,0) circle (.5);
    	
    	\draw[line width = 1.5] (4,0) circle (1);
    	\begin{scope}
	    	\clip (4,0) circle (.8);
	    	\begin{scope}[shift={(4,0)}]
	    	\clip (0,0) circle (.8);
	    	\foreach \x in {-2.5,-2.4,...,.3}
				\draw[line width=.8 pt] (\x,-1.9) -- (\x+1.5,1);
	  	\end{scope}
	  	\end{scope}
    	\draw[line width = 1.5] (4,0) circle (.8);
    	
    	\draw[fermion] (1,0) -- (3,0);
	 \end{scope}
	 \begin{scope}[shift={(-3.5, -3.45)}]
	     \begin{scope}
	    	\clip (0,0) circle (.5);
	    	\foreach \x in {-2.5,-2.4,...,.3}
				\draw[line width=.8 pt] (\x,-1.9) -- (\x+1.5,1);
	  	\end{scope}
	  	\draw[line width = 1.5] (0,0) circle (.5);
	  	\draw[dashed, line width=1.5, ->] (60:.75) -> (60:4);
	  	\draw[dashed, line width=1.5, ->] (-60:.75) -> (-60:4);
	  	\draw[dashed, line width=1.5, ->] (20:.75) -> (20:2.2);
	  	\draw[dashed, line width=1.5, ->] (-20:.75) -> (-20:2.2);
	 \end{scope}
    \begin{scope}[shift={(7.8, -3.45)}]
	     \begin{scope}
	    	\clip (0,0) circle (.8);
	    	\foreach \x in {-2.5,-2.4,...,.3}
				\draw[line width=.8 pt] (\x,-1.9) -- (\x+1.5,1);
	  	\end{scope}
	  	\draw[line width = 1.5] (0,0) circle (.8);
	  	\draw[dashed, line width=1.5, ->] (120:1.1) -> (120:4);
	  	\draw[dashed, line width=1.5, ->] (-120:1.1) -> (-120:4);
	  	\draw[dashed, line width=1.5, ->] (160:1.1) -> (160:2.2);
	  	\draw[dashed, line width=1.5, ->] (-160:1.1) -> (-160:2.2);
	 \end{scope}
	\node at (-4.8,-3.45) {$G_\text{EW}$};
	\node at (9.7,-3.45) {$\text{SU}(3)_\text{R}$};
    \end{tikzpicture}

\end{center}
\caption{Full symmetry structure of the minimal moose little Higgs model. Shaded blobs represent gauged subgroups. We explicitly show that only the `diagonal' subgroups are gauged.}
\label{fig:comp:minimal:moose:full}
\end{figure}

As an example, we present the `minimal moose' little Higgs model from \cite{ArkaniHamed:2002qx}. The basic building block is the coset for chiral symmetry breaking, $\text{SU}(3)_\text L \times \text{SU}(3)_\text R / \text{SU}(3)_\text V$. We gauge the electroweak subgroup $G_\text{EW}$ of $\text{SU}(3)_\text L$ and the entire $\text{SU}(3)_\text R$, which we represent schematically with shaded blobs:
\begin{center}
	\begin{tikzpicture}[line width=1.5]
	\begin{scope}
	    \begin{scope}
	    	\clip (-.5,0) circle (.5);
	    	\foreach \x in {-2.5,-2.4,...,.3}
				\draw[line width=.8 pt] (\x,-1.9) -- (\x+1.5,1);
	  	\end{scope}
    	\draw[line width = 1.5] (0,0) circle (1);
    	\draw[line width = 1.5] (-.5,0) circle (.5);
    	
    	\draw[line width = 1.5] (4,0) circle (1);
    	\begin{scope}
	    	\clip (4,0) circle (.8);
	    	\begin{scope}[shift={(4,0)}]
	    	\clip (0,0) circle (.8);
	    	\foreach \x in {-2.5,-2.4,...,.3}
				\draw[line width=.8 pt] (\x,-1.9) -- (\x+1.5,1);
	  	\end{scope}
	  	\end{scope}
    	\draw[line width = 1.5] (4,0) circle (.8);
    	
    	\draw[fermion] (1,0) -- (3,0);
    	\node at (-1.6,0) {$G_\text{EW}$};
    	\node at (0.1,1.4) {$\text{SU}(3)_\text L$};
    	\node at (4.1,1.4) {$\text{SU}(3)_\text R$};
    	\node at (5.9,0) {$\text{SU}(3)_\text R$};
    	\node at (2,.5) {$\Sigma$};
	\end{scope}
	\end{tikzpicture}
\end{center}
The minimal moose model actually requires four copies of this basic structure. As before, we only gauge the vectorial $G_\text{EW}$ of each of the $\text{SU}(3)_\text L$ factors and similarly for the $\text{SU}(3)_\text R$ factors. In other words, the theory only has two gauge couplings. This is shown schematically in Fig.~\ref{fig:comp:minimal:moose:full}.
We note that typically one only draws nodes for the gauge groups so that the usual moose diagram for this model is:
\begin{center}
\begin{tikzpicture}[line width=1.5]
%\draw (0,0) circle (.5);
%\node at (0,0) {$G$};
\draw[fermion] (2,0) to [out=60, in=120] (6,0);
%\draw[fermion] (2,0) to [out=25 in=180] (6,0);
\draw[fermion, yscale=-1] (2,0) to [out=-20, in=200] (6,0);
\draw[fermion] (2,0) to [out=-20, in=200] (6,0);
\draw[fermion] (2,0) to [out=-60, in=240] (6,0);
\draw[fill=white] (2,0) circle (.5);
\node at (2,0) {$G_\text{EW}$};
\draw[fill=white] (6,0) circle (.75);
\node at (6,0) {$\text{SU(3)}_\text{R}$};
\end{tikzpicture}
\end{center}
See \S 4.1 of \cite{Schmaltz:2005ky} for a review of this particular model.
A full discussion of these moose-based little Higgs models is outside of the scope of these lectures. In addition to the reviews mentioned above \cite{Perelstein:2005ka, Schmaltz:2005ky}, see \cite{Schmaltz:2010ac} for the self-described `bestest' little Higgs model and \cite{Vecchi:2013bja, DeSimone:2012fs} for a discussion of the status of composite Higgs models after the first run of the \LHC.

\subsection{A taxonomy of composite Higgs models}
\label{sec:comp:classification}

Having surveyed the main features of composite Higgs models, let us classify the landscape of such theories. This section is meant to clarify the distinctions between what is colloquially called a `composite Higgs' versus a `little Higgs' or a `holographic composite Higgs' versus a `dilatonic Higgs.' We closely follow the discussion in Sections~2 -- 3 of~\cite{Bellazzini:2014yua}, to which we refer the reader for further details and references.
 
As a warm up and review, recall the Standard Model Higgs potential
\begin{align}
	V(h) &= -\mu^2 |H|^2 + \lambda |H|^4 
	\quad\longrightarrow \quad
	-\frac{1}{2} \mu^2 h^2 + \frac{\lambda}{4}h^4.
	\label{eq:comp:SM:Higgs:V}
\end{align}
Minimizing the potential and matching to experiment yields
\begin{align}
	v^2 = \langle h \rangle &= \frac{\mu^2}{\lambda} = 246\text{ \GeV}
	&
	m_h^2 &= 2\mu^2 = \left(125 \text{ \GeV}\right)^2,
	\label{eq:comp:SM:Higgs:fit}
\end{align}
where $v$ has long been known from the masses and couplings of the electroweak gauge bosons, but $m_h^2$ is new data from 2012. This new information tells us that $\mu = 89$ \GeV and, from the expression for $v$, that $\lambda = 0.13$. 

Let us now map this onto a convenient parameterization of the Higgs potential in composite Higgs models. 
\begin{align}
V(h) &= \frac{g_\text{SM}^2 M^2}{16\pi^2}
\left(
-a h^2 + \frac{b}{2f^2} h^4
\right).
	\label{eq:comp:taxonomy}
\end{align}
One can compare this to (\ref{eq:comp:CW:Higgs:pot}). Here $g_\text{SM}$ is a characteristic Standard Model coupling, such as $g_\text{SM}^2 = N_c y_t^2$. Implicit in this parameterization is the expectation that the Higgs potential is radiatively generated, giving a $g_\text{SM}^2/16\pi^2$ prefactor. With this normalization, tree-level contributions appear as coefficients $a,b$ that go like $16\pi^2/g_\text{SM}^2$.
The mass scale $M$ is typically that of the new states (e.g.\ top partners) that cut off the quadratic divergence introduced by the explicit breaking of the Goldstone shift symmetry, as discussed in Section~\ref{sec:comp:mch:gauge}. It is useful to parameterize this in terms of the coupling of these new states to the strong sector $g_*$,
\begin{align}
M &= g_* f.
	\label{eq:comp:taxonomy:partners}
\end{align}
These states are typically lighter than the cutoff, $4\pi f$, to help with the little hierarchy problem. We expect the lighter mass comes from a weaker coupling to the strong sector, $g_*$, motivating the definition (\ref{eq:comp:taxonomy:partners}).
This coupling is sometimes written as $g_* = g_\rho$ in the literature, making the analogy to the coupling of the spin-1 $\rho$ meson in \QCD, see Section~\ref{sec:comp:nda:lagrangian}. 
(\ref{eq:comp:taxonomy:partners}) defines $g_*$ as a ratio of mass scales, but when one includes this state in chiral perturbation theory (using the \CCWZ formalism introduced in Appendix~\ref{sec:comp:CCWZ}), this ratio is manifestly the value of the $\rho \pi\pi$ coupling. 
In this sense, $g_\rho$ is the `gauge coupling' of the $\rho$ as a massive gauge boson, $g_\rho = m_\rho/f$.
%
%%%

%
The experimental information that the \SM quartic is $\lambda = 0.13$ is strongly suggestive of a loop induced coupling. Using the \textsc{nda} scaling of a strong sector quartic  (\ref{eq:comp:NDA:quartic}) and a proportionality factor from an explicit global symmetry breaking \SM loop, $g_\text{SM}^2/16\pi^2$, we estimate
\begin{align}
	\lambda_\text{loop} \approx 2 \frac{1}{16\pi^2}\,g_\text{SM}^2 g_*^2
	%\left(\frac Mf\right)^2
	\approx
	0.15 \times \left(\frac{g_\text{SM}}{\sqrt{N_c} y_t}\right)^2
	\left(\frac{g_*}{2}\right)^2.
	\label{eq:comp:lambda:vs:gstar}
\end{align}
Here the factor of 2 comes from two top partner polarizations and the scaling with respect to $g_* = M/f$ comes from \textsc{nda}~\cite{Georgi:1992dw}.
Thus the coupling of the new state is $g_* \ll 4\pi$ and is expected to be \emph{weakly coupled}. This is a more quantitative version of the statement that the discovery of the 125 \GeV Higgs signaled the death of technicolor, as we explained qualitatively in Section~\ref{sec:comp:no:technicolor}. The other implication of this weak coupling is that the new particles that cancel the quadratic sensitivity of the Higgs potential have masses well below the strong coupling scale,  $M \ll \Lambda = 4\pi f$; where we recall the \textsc{nda} cutoff from Section~\ref{sec:comp:NDA}.

Comparing (\ref{eq:comp:taxonomy}) to (\ref{eq:comp:SM:Higgs:V} -- \ref{eq:comp:SM:Higgs:fit}) gives
\begin{align}
	v^2 &= \frac{a}{b} f^2 = (246\text{ \GeV})^2
	&
	m_h^2 &= 
	2 \frac{g_\text{SM}^2}{16\pi^2}M^2 a
	=
	4v^2 \frac{g_\text{SM}^2 g_*^2}{16\pi^2} b
	= \left(125 \text{ \GeV}\right)^2.
	\label{eq:comp:Higgs:taxonomy:fit}
\end{align}

We can restate the discussion below (\ref{eq:comp:lambda:vs:gstar}) in terms of (\ref{eq:comp:Higgs:taxonomy:fit}). Prior to the Higgs discovery, one could have tuned $\xi=v^2/f^2$ by, say, increasing the parameter $b$. With the discovery of a 125~\GeV Higgs boson, one can no longer do this since increasing $b$ also increases $m_h^2$. Indeed, this is why prior to the Higgs discovery people said that composite Higgs models predict a heavier Higgs mass of $m_h \sim 300$~\GeV. One way to evade making $m_h^2 > m_t^2$ is to observe that most of the contributions to $b$ comes from the fermionic top partner resonances. We've been characterizing all of the heavy particle couplings as $g_*$, but in principle the top partners could have a different coupling, $g_T$, in which case $g_*\to g_T$ in (\ref{eq:comp:Higgs:taxonomy:fit}). If this top partner coupling is smaller than the general resonance coupling $g_T < g_*$, while also satisfying $g_T \gtrsim 1$ to push the mass up, then one can keep $m_h^2 < m_t^2$ while pushing up $b$ to achieve tuning in $\xi$. This is why one may expect the `light' top partners described in Section~\ref{sec:comp:top:partner} to have masses lighter than the other strong sector resonances $m_T \approx g_T f < M$.

%%%

In the remainder of this section we examine five classes of composite Higgs models and classify them according to their natural expectations for $a$, $b$, and $g_*$. These are summarized in Table~\ref{tab:comp:models}. %We close with a brief discussion of phenomenology.

\begin{table}
\begin{center}
\begin{tabular}{l|ccc|l}
\toprule % requires: booktabs.sty.
\textsc{model} 
& $\mathcal O(a)$ 
& $\mathcal O(b)$
& $\mathcal O(g_*)$
& \textsc{comments}
\\
\midrule 
Bona-fide composite Higgs
%& $\mathcal O(1)$
%& $\mathcal O(1)$
& 1
& 1
& $4\pi$
& Requires tuning of both $a$ and $b$.
\\
Little Higgs
%& $\mathcal O(1)$
%& $\mathcal O\left(\frac{16\pi^2}{g_*^2}\right)$
& 1
& $\frac{16\pi^2}{g_*^2}$
& $\ll 4\pi$
& Tree level quartic, $h$ too heavy.
\\
Holographic Higgs
%& $\mathcal O(1)$
%& $\mathcal O(1)$
& 1
& 1
& $\ll 4\pi$
& $\sim$ little Higgs with loop-level quartic.
\\
Twin Higgs
& 1
& $1-\frac{16\pi^2}{g_*^2}$
& $g_\text{SM}$
& $\mathbbm{Z}_2$ rather than collective breaking.
\\
Dilatonic Higgs
& \multicolumn{3}{c|}{\textsc{see text}}
& Related to \RS radion Higgs.
\\
\bottomrule % requires: booktabs.sty.
\end{tabular}
\caption{Taxonomy of composite Higgs models according to the couplings in (\ref{eq:comp:taxonomy}) and (\ref{eq:comp:taxonomy:partners}); based on~\cite{Bellazzini:2014yua}. Models must be tuned when phenomenology requires values of the couplings that are very different from the expected magnitudes shown here.}\label{tab:comp:models}
\end{center}
\end{table}

\subsubsection{Bona-Fide Composite Higgs}

The `bona-fide composite Higgs' models in the first row of Table~\ref{tab:comp:models} are the simplest realizations of the Higgs as pseudo-Nambu--Goldstone boson idea: a strongly coupled sector has a global symmetry which is spontaneously broken and yields a Goldstone with the quantum numbers of the Higgs.
The Higgs potential is assumed to be radiatively generated by explicit breaking terms so that in the parameterization (\ref{eq:comp:taxonomy}), $a\sim b\sim \mathcal O(1)$. From the left-side equation of (\ref{eq:comp:Higgs:taxonomy:fit}), a parametric separation between $v$ and $f$ requires $a$ to be tuned small by an amount $\xi$ in (\ref{eq:comp:comphiggs:angle}). 

Even with this, however, this is a second tuning required on $b$ since the new states are expected to couple to the strong sector with strong couplings, $g_* \sim 4\pi$. Thus one finds that the quartic coupling is too large in (\ref{eq:comp:lambda:vs:gstar}) compared to $\lambda = 0.13$. In other words, one predicts a Higgs mass that is heavier than observed in (\ref{eq:comp:Higgs:taxonomy:fit}). This is mapped onto a tuning of $b$.

\subsubsection{Little Higgs}

In little Higgs models, collective symmetry breaking naturally gives a hierarchy 
\begin{align}
	\xi = \frac{v^2}{f^2} \sim \frac{g_*^2}{16\pi^2} \ll 1.
\end{align}
The quartic coupling appears at tree-level, $\lambda\sim g_\text{SM}$. This is shown as $b\sim 16\pi^2/g_*^2$ in Table~\ref{tab:comp:models}.
Prior to the Higgs discovery, this set up was seen to be a feature: one explains the separation between $v$ and $f$. However, (\ref{eq:comp:Higgs:taxonomy:fit}) shows that this predicts a Higgs mass that is on the order of 500 \GeV for $g_\text{SM}\sim 1$. 

\subsubsection{Holographic Higgs}

These models are motivated by \AdS/\CFT duals of warped extra dimensional models, as we discussed in Section~\ref{fig:xd:holography}.
Like the `bona-fide composite Higgs,' the entire potential for these models are radiatively generated. This thus suffers the same $\mathcal O(\xi=v^2/f^2)$ to obtain the correct electroweak symmetry breaking scale. Unlike the `bona-fide composite Higgs,' however, $g_*$ is still weak and thus no additional tuning is required to keep the Higgs light. The four-dimensional effective theory (or deconstruction) of this scenario is what is most commonly meant when referring to a [modern] `composite Higgs' model; see e.g.~\cite{Agashe:2004rs}.

The holographic Higgs also has a version of collective symmetry breaking that is a result of locality in 5D~\cite{Thaler:2005kr}. Unlike the little Higgs models above, however, holographic Higgs models have radiative quartics. These models have the minimal amount of tuning: just $\xi$, which is a tuning of a few percent.

\subsubsection{Twin Higgs and neutral naturalness}\label{sec:comp:twin:higgs}

Twin Higgs models~\cite{Chacko:2005pe, Barbieri:2005ri} have received a lot of interest after the non-discovery of any top-partners at Run~I of the \LHC. Rather than protecting the pseudo-Goldstone Higgs from quadratic corrections with collective symmetry breaking, these models impose a $\mathbbm{Z}_2$ symmetry that protects the Higgs potential. The key phenomenological feature of this framework is that the partner particles that enact this protection are uncharged under the Standard Model. Since the top partners aren't colored, one no longer expects a large production cross section at the \LHC and one avoids the Run~1 bounds. These models are thus often referred to under the banner of `neutral naturalness' and are considered a last bastion for naturalness against collider bounds.

	We illustrate the twin mechanism with the toy example presented in~\cite{Chacko:2005pe}; the interested reader is encouraged to read the succinct paper in its entirety. Suppose a theory has a global $G=$SU(4) symmetry and a field $H$ in the fundamental representation with a symmetry-breaking potential,
	\begin{align}
		V(H) = - \mu^2 |H|^2 + \lambda |H|^4.
	\end{align}
	The field develops a \vev $\langle |H| \rangle = m/\sqrt{2\lambda} \equiv f$ and breaks SU(4) $\to$ SU(3). Now let us gauge a subgroup SU(2)$_\text{A}\times$SU(2)$_\text{B}$ of the global symmetry. We decompose $H$ into a doublet under each gauge group, $H_A$ and $H_B$. We may identify $A$ with the Standard Model SU(2)$_\text{L}$. As we saw in Section~\ref{sec:comp:mch:gauge}, this gauging generates mass terms for the would-be Goldstone bosons,
	\begin{align}
		V \supset \frac{9\Lambda^2}{64\pi^2}
		\left(g_A^2 |H_A|^2 + g_B^2 |H_B|^2\right).
	\end{align}
	Next impose a $\mathbb{Z}_2$ `twin' symmetry which swaps $A\leftrightarrow B$. This imposes $g_A = g_B$ so that the quadratic potential becomes,
	\begin{align}
		V \supset \frac{9g^2\Lambda^2}{64\pi^2}
		|H|^2 + \cdots,
	\end{align}
	which respects the original SU(4) symmetry of the theory and thus does not contribute to the mass of the Goldstone bosons. The higher order terms still introduce logarithmically divergent terms that break this SU(4) symmetry.

	We can see the `twin' cancellation in the top couplings:
	\begin{align}
		\mathcal L \supset 
		- y_t H_A \bar t_L^{(A)} t_R^{(A)}
		- y_t H_B \bar t_L^{(B)} t_R^{(B)}.
		\label{eq:comp:twin:top:yukawas}
	\end{align}
	The SU(4)$\to$SU(3) breaking imposes $\langle h_a\rangle^2 + \langle h_b\rangle^2 = f^2$. Expanding the SU(4) fundamental $H$ analogously to (\ref{eq:comp:linear:field:Sigma}), one may expand to $\mathcal O(h^2/f^2)$,
	\begin{align}
		H_A &\to h,
		&
		H_B \to f - \frac{h^2}{2f}.
	\end{align}
	Inserting this into (\ref{eq:comp:twin:top:yukawas}) yields a cancellation that is diagramatically identical to (\ref{eq:comp:top:partner:cancellation}) with the important difference that the $t^{(B)}$ and $\bar t^{(B)}$ are charged under a `twin' \QCD, but not ordinary \QCD.

Having demonstrated the basic principle, we refer the reader to the original literature for a demonstration of a complete model. In our phenomenological taxonomy of composite Higgs models, we have written $b \sim \mathcal O(1 - 16\pi^2/g_*^2)$ reflecting that the original twin Higgs models included a tree-level quartic put in by hand to generate the $v\ll f$ hierarchy, though this is not an intrinsic feature of these models. As we have discussed, the observed $\lambda=0.13$ disfavors the inclusion of this tree-level term.

\subsubsection{Dilatonic Higgs}

Rather than being a pseudo-Goldstone of an internal global symmetry, this scenario assumes that the Higgs is a \textbf{dilaton} coming from the spontaneous breaking of scale invariance~\cite{Goldberger:2008zz, Fan:2008jk, Vecchi:2010gj, Chacko:2012sy, Bellazzini:2012vz, Bellazzini:2013fga, Coradeschi:2013gda}. We have already explored this scenario in Section~\ref{sec:XD:radion:is:dilaton}, where we identified the \textbf{radion} in a warped extra dimension as a state which is holographically dual to the dilaton. This is distinct from the `holographic Higgs' scenario where the Higgs is the Goldstone of an internal global symmetry.

In this scenario the \vev that breaks scale invariance, $f$, sets the scale of the potential and is unrelated to the electroweak \vev: for example, the order parameter for the breaking of scale invariance talks to all massive particles, whereas the electroweak \vev ought to only talk to electroweak doublets. Thus the parameterization in (\ref{eq:comp:taxonomy}) is not relevant for comparison with the other composite Higgs models discussed: the dilaton is a completely different type of beast. For a dilaton to play the role of a Higgs, one assumes that the \UV theory also has a small explicit breaking of scale invariance to allow a \vev $f$ that breaks electroweak symmetry spontaneously.

Scale transformations act on spacetime as $x\to x' = e^{-\alpha} x$ and on mass-dimension-1 objects as $\phi \to \phi' = e^\alpha \phi$. For a refresher, see \cite{Coleman:Aspects, La:1996pq, Coleman:1970je}. Suppose that scale invariance is spontaneously broken with some order parameter, $f$. Applying the rule of thumb from Section~\ref{sec:composite:framework}, we identify the dilaton, $\sigma$, with the parameter of scale transformations:
\begin{align}
	f \to \chi f \equiv e^{\sigma/f} f.
\end{align}
Unlike the case for internal symmetries, the dilaton, $\sigma$, points in the same direction in field space as the order parameter, $f$. Thus the potential for the dilaton can be read off the potential for the \vev, $V(f)$. For internal symmetries, $V(f)$ is trivial since without explicit breaking since these the Goldstones only couple derivatively. For scale symmetries, $V(f)$ is allowed to have a quartic term, $V(f)\supset \lambda f^4$. This term is identified with a cosmological constant. The na\"ive dimensional analysis size of its coefficient is $\lambda \sim 16\pi^2$ coming from vacuum bubbles---this is essentially the cosmological constant problem. In order for scale symmetry to have been spontaneously broken, $V(f)$ must realize its minimum value at $f \neq 0$. This is not possible for purely quartic potential, so that one still requires a small explicit breaking of scale invariance that would give a $V(f)$ with a nontrivial vacuum. From our intuition with the Higgs, one might want to write a breaking term like $V(f) \supset -\mu^2 f^2$, but this would be a large breaking of scale invariance. Instead, we consider operators that are close to marginal,
\begin{align}
	V(f) \sim \lambda f^4 + a f^4 \log f  \sim \lambda f^4 + a f^{4-\epsilon} \ .
\end{align}
In order for this potential to spontaneously break scale invariance, both terms must be of roughly the same size. Since we expect $\lambda$ to be large, this means that the `small explicit breaking' must actually also be large. This is a contradiction to our initial assumption that the theory is approximately scale invariant so that the dilaton is a pseudo-Goldstone boson: with the large explicit breaking of scale invariance, one also expects the dilaton to be a heavy field. In essence this is the source of tuning in dilatonic Higgs models: one needs $\lambda$ to be small so that the explicit breaking required for $f\neq 0$ can also be small which, in turn, allows the dilaton to be light. A detailed discussion of this tuning is presented in Section~5 of~\cite{Bellazzini:2012vz} and \cite{Coradeschi:2013gda}. One way out of this apparent contradiction is supersymmetry, which gives a symmetry reason to protect against the cosmological constant: we saw from the \SUSY algebra that $\langle 0 | H | 0 \rangle = 0$ when supersymmetry is unbroken. Five-dimensional variants have been constructed in~\cite{Bellazzini:2013fga, Coradeschi:2013gda}.

As a qualitative example of the above considerations, one may ask whether \QCD could furnish a light dilaton since it is an approximately scale-invariant theory for which the chiral condensate is a spontaneous breaking of scale invariance. While $\alpha_s$ is small in the \UV, we know that it grows in the \IR. As the couplings get larger, so does the $\beta$ function, which is a concrete manifestation of scale invariance being broken by a larger amount. Thus the would-be dilaton in \QCD is not a light state in the effective theory.

\subsection{Parameterization of phenomenology}

We see that the composite Higgs can be probed through its deviations from the \SM Higgs couplings. A phenomenological parameterization of the space of light, composite Higgs models is presented in \cite{Giudice:2007fh} as the `strongly interacting light Higgs' (\textsc{silh}, pronounced ``silch'') effective Lagrangian and extended in \cite{Contino:2013kra}. We briefly review the main results and refer to \cite{Giudice:2007fh, Contino:2013kra} for a detailed discussion including matching to specific composite Higgs models. We now examine a convenient parameterization of the phenomenological Lagrangian,
\begin{align}
\mathcal L_\text{\textsc{silh}} =& \mathcal L_H^{\text{SM}}
+ 
\frac{\bar c_H}{2v^2} \partial^\mu (H^\dag H) \partial_\mu (H^\dag H) 
+ 
\frac{\bar c_T}{2v^2} \left( H^\dag \overleftrightarrow{D} H\right)^2
-
\frac{\bar c_6}{v^2} \lambda (H^\dag H)^3
\nonumber\\
&
+\left(\frac{\bar{c}_u}{2v^2} y_u H^\dag H \bar Q_L \tilde H u_R + \cdots\right)
+ 
\frac{i \bar c_W g}{2M_W^2} \left( H^\dag \sigma^i \overleftrightarrow H\right)\left(D^\nu W_{\nu\mu}\right)^i + \cdots,
\label{eq:comp:SILH:bar}
\end{align}
where $H^\dag \overleftrightarrow D_\mu H= H^\dag D_\mu H - D_\mu H^\dag H$ and the $\cdots$ represent similar terms for the other fermions and gauge bosons.  The expected sizes of these coefficients are
\begin{align}
\bar c_H, \bar c_T, \bar c_6, \bar c_\psi &\sim \frac{v^2}{f^2}
&
\bar c_{W,B} &\sim \frac{M_W^2}{g_*^2 f^2},
\end{align}
where $g_*$ is defined in (\ref{eq:comp:taxonomy:partners}).
Following \cite{Contino:2013kra}, the operators in (\ref{eq:comp:SILH:bar}) are normalized with respect to the Higgs \vev $v$ rather than the scale $f$ in \cite{Giudice:2007fh}; this is why the expected values of the barred couplings $\bar c_i$ differ by factors of $v/f$ from the couplings $c_i \sim 1$ in equation (15) of \cite{Giudice:2007fh}.

The phenomenological Lagrangian (\ref{eq:comp:SILH:bar}) can be constructed systematically from the non-linear sigma model including symmetry breaking terms which we assume are parameterized by the \SM couplings that break those symmetries: the Higgs quartic coupling $\lambda$ (breaking the pseudo-Goldstone Higgs shift symmetry) and the Yukawas (breaking shift and flavor symmetries). See Appendix~B of \cite{Panico:2011pw} for a detailed discussion in terms of na\"ive dimensional analysis. Following the \textsc{nda} of Section~\ref{sec:comp:nda:lagrangian}, the general strategy is to write 
\begin{align}
\mathcal L_\text{\textsc{silh}}= \frac{M^4}{g_*^2} 
\tilde{\mathcal L}\left(%\Sigma
U,\frac{\partial}{M}\right),
\end{align}
%where $U=\Sigma/\langle\Sigma\rangle$ is the linear field containing the Goldstones (\ref{eq:comp:linear:field:Sigma}). 
%
where $U=U(\pi/f)$ is the dimensionless linear field containing the Goldstones (\ref{eq:comp:chiral:pt:U:field}) and the partial derivative carries a factor of $M^{-1}$, the scale of new states~(\ref{eq:comp:taxonomy:partners}), to make it dimensionless. 
%
%The expansion of $U$ in Goldstones, $\pi$, automatically comes with factors of $f^{-1}$ to keep each term dimensionless. 
%
Each Goldstone field $\pi$ comes with a factor of $1/f$.
The \textsc{nda} prefactor $M^4 g_*^{-2} = M^2 f^2$ is derived in (\ref{eq:NDA:prefactor:1}--\ref{eq:NDA:prefactor:2}). 
Recall that this is precisely the $\Lambda^2 f^2$ prefactor in the \textsc{nda} literature except that we replace $\Lambda$ with a scale which exists in the effective theory, $M < \Lambda$ \cite{Manohar:1983md, Georgi:1992dw}. 
This inequality is equivalent to $g_* < 4\pi$ and parameterizes the regime in which the \NLSM is weakly coupled. We then take the dimensionless function $\tilde{\mathcal L}$ to be a derivative expansion analogous to (\ref{eq:comp:chiral:pert}); the \textsc{silh} interactions appear at higher order from the term
\begin{align}
{\mathcal L}_\text{\textsc{silh}} =
\frac{M^4}{g_*^2}\left(\cdots + \frac 1{3} 
%\left(\frac{1}{f^2 m_\rho}\pi(x) \overleftrightarrow\partial \pi(x)\right)^2
\left(\frac{\pi(x)}{f} \frac{\overleftrightarrow\partial}{M} \frac{\pi(x)}{f}\right)^2
%\left(\pi(x) \overleftrightarrow\partial_\mu \pi(x)\right)
+\cdots\right)
\end{align}
where $\pi(x)$ is identified with the Higgs doublet $H(x)$ and the partial derivatives $\partial_\mu$ are promoted to \SM gauge covariant derivatives. Gauge field strengths are included with factors of $m_\rho^{-2}$ since $F_{\mu\nu} \sim [D_\mu,D_\nu]$. 
The $\bar c_H$ and $\bar c_T$ terms encode the $\mathcal O(H^4,\partial^2)$ interactions after shifting the Higgs by a factor proportional to $(H^\dag H)H/f^2$ (see \cite{Giudice:2007fh}).
The $\bar c_6$, $\bar c_u$, $\bar c_W$ (and analogous terms) break the shift symmetries of the \NLSM and carry explicit factors of the \SM couplings that break those symmetries: the Higgs quartic interaction, the Yukawas, or \SM gauge couplings, respectively. %For example, 
Electroweak precision observables\footnote{See \cite{Skiba:2010xn, Wells:2005vk} for pedagogical reviews and \cite{Cacciapaglia:2006pk,Han:2004az,Barbieri:2004qk} for details.} set bounds on composite Higgs models %\cite{Giudice:2007fh, Barbieri:2004qk}
\cite{Contino:2013kra, Baak:2012kk}; at $2\sigma$:
\begin{align}
% -1.1 \times 10^{-3} &< \bar c_T < 1.3 \times 10^{-3} 
%% %\cite{Giudice:2007fh, Barbieri:2004qk}
-1.5 \times 10^{-3} &< \bar c_T < 2.2 \times 10^{-3}
%% Contino:2013kra, eq 2.11
\\
% m_\rho &\gtrsim \sqrt{c_W + c_B} \, 2.5 \text{ \TeV},
-1.4 \times 10^{-3} &< \bar c_W + \bar c_B < 1.9 \times 10^{-3}
,
\end{align}
coming from the $\hat T$ and $\hat S$ parameters respectively. The former condition reflects the requirement of custodial symmetry \cite{Sikivie:1980hm} (see \cite{Willenbrock:2004hu} for an introduction) which is assumed in the latter bound. The $\hat S$ parameter bound comes from the exchange of new spin-1 states of characteristic mass $M$ (analogous to the $\rho$ meson) and goes parametrically like $\hat S \sim M_W/M^2$. This pushes $M\gtrsim 2.5$~\TeV.
The observation of the 125 \GeV Higgs and the opportunity to measure its couplings offers additional data to fit the phenomenological Lagrangian. For example, the $\bar c_H$ and $c_f$ ($f$ running over the \SM fermions) are related to each other via the couplings of the Higgs to $W$ bosons \cite{Barbieri:2007bh}. The Higgs mass sets (at $3\sigma$) %\cite{Azatov:2012bz}
\begin{align}
\bar c_H  \leq 0.16.
\label{eq:comp:cH:bound}
\end{align}
This and other bounds on
composite Higgs models coming from Higgs observables are reviewed in \cite{Azatov:2012bz, Azatov:2012qz} using a slightly different effective theory parameterization introduced in \cite{Contino:2010mh}. In that notation, (\ref{eq:comp:cH:bound}) comes from $a^2 \geq 0.84$. Further phenomenological bounds and their relations to specific models can be found in \cite{Giudice:2007fh, Contino:2013kra, Bellazzini:2012tv, Bellazzini:2014yua}.
At $2\sigma$, Higgs data constraints the minimal composite Higgs model to satisfy \cite{Falkowski:2013dza}
\begin{align}
\frac{v}{f} \lesssim 0.5.
\end{align}
 The bounds on composite Higgs models coming from Higgs observables are reviewed in \cite{Azatov:2012bz, Azatov:2012qz, Contino:2013kra}. %using a slightly different effective theory parameterization introduced in \cite{Contino:2010mh}. 
 Further phenomenological bounds and their relations to specific models can be found in \cite{Giudice:2007fh, Bellazzini:2012tv, Bellazzini:2014yua}.

\section{Closing Thoughts}

We briefly review interconnections between some of the salient ideas in these lectures,  acknowledge topics omitted, and point to directions of further study.
%
%\subsection{Multiple guises of strong dynamics}
%
One of the themes in the latter part of these lectures were weakly coupled descriptions of strong dynamics and we close by highlighting this common thread.

\subsection{Covariant Derivatives}

Each of the scenarios that we explored carries its own sense of covariant derivative. The most explicit example is in a warped extra dimension, where spacetime is explicitly curved. The holographic principle made use of this geometry: the isometries of \AdS match the conformal symmetries of the strongly coupled theory near a fixed point. The system is so constrained by these symmetries that the behavior of 5D fields could be identified with the renormalization group flow of 4D operators.
Even in supersymmetry---where superspace can be thought of a `fermionic' extra dimension---we introduced a \SUSY covariant derivative. 
%
%Even though superspace is flat, there is a covariant derivative came from it being torsion-free \cite{gieres1988geometry}. 
%
The practical significance of this covariant derivative is to define chiral superfields, the irreducible representation of $\mathcal N=1$ \SUSY that we use for matter fields.
%
% And also how to write effective Lagrangians in Kahler terms
%
Finally, the nonlinear realizations we used for composite Higgs models also has a geometric structure coming from the coset space. This is seen in the \CCWZ formalism reviewed in Appendix~\ref{sec:comp:CCWZ}, where one identifies covariant derivative and gauge field for the coset space that are necessary to construct invariant Lagrangians.

\subsection{Nonlinear realizations}

The simplest handle on strong dynamics is to work in an effective theory of pseudo-Goldstone bosons given by the pattern of global symmetry breaking in the strong sector.  
In composite Higgs models, one addresses the Hierarchy problem by assuming that the Higgs is a pseudo-Goldstone boson associated with the dynamics of a strongly coupled sector that break global symmetries at a scale $f$. We saw that generically the \SM interactions required for a Higgs boson tend to push its mass back up towards the compositeness scale, $\Lambda \sim 4\pi f$. One way to push the Higgs mass back down is to invoke collective symmetry breaking, which can often be described succinctly using `moose' diagrams.

\subsection{Holographic and deconstructed extra dimensions}

We saw that the holographic principle is an alternative way to describe the dynamics of a strongly coupled sector through the use of a higher dimensional theory with a non-trivial geometry. From the extra dimensional perspective, the Higgs mass is natural because it is localized towards the \IR brane where the Planck scale is warped down to \TeV scale. Holographically, this is interpreted as the Higgs being a  composite state. 
Indeed, the minimal composite Higgs model \cite{Agashe:2004rs} was constructed using holography as a guiding principle. 

Further, the little Higgs models that first highlighted collective symmetry breaking were motivated by deconstructions of an extra dimension.
%
%It was in some sense a coincidental footnote that the moose constructions for little Higgs models were originally motivated by the deconstruction of an extra dimension. 
This picture allowed us to clearly see that the Goldstone modes of the spontaneously broken global symmetries can be identified with the scalar component of a 5D gauge field. In the deconstruction, the gauge bosons from each copy of the gauge group eat these Goldstone modes to become the spectrum of heavy \KK modes. In this sense, little Higgs and holographic composite Higgs constructions are similar to gauge-Higgs unification scenarios in 5D where the Higgs is the zero mode of a bulk gauge field, see \cite{Gogoladze:2013rqa} and references therein. One way to interpret the lightness of the Higgs mass is via locality in the deconstructed extra dimension: the symmetries are only broken on the boundaries and one needs a loop that stretches between the boundaries to generate a Higgs potential. This implies that the loop cannot be shrunk to zero and that the Higgs potential is finite since it can have no short-distance divergences. The natural cutoff is set by the size of the extra dimension.

Deconstruction itself, however, is rooted in the idea of a \textbf{hidden local symmetry} in nonlinear models. See \cite{Bando:1987br} for a comprehensive review. A 5D version of the little Higgs in \AdS was presented in \cite{Thaler:2005en}. Shortly after, \cite{Thaler:2005kr} connected the holographic composite Higgs to a little Higgs theory, relating the \CCWZ formalism of
Appendix~\ref{sec:comp:CCWZ} to the hidden local symmetry  construction.

\subsection{Natural SUSY and partial compositeness}

We began these lectures with what appeared to be a completely different subject: supersymmetry. We saw that the natural setting for \SUSY is superspace, which is superficially an `extra quantum dimension' that is both Grassmannian and spinorial. One way to see how \SUSY solves the Hierarchy problem is to observe that it requires the existence of superpartners (differing by half integer spin) that cancel the loop contributions of  particles to superpotential parameters such as the Higgs mass. We saw a similar cancellation when invoking collective symmetry breaking (or the twin Higgs mechanism) in composite Higgs theories with the notable difference that the  partner particles had the same spin as their \SM counterparts.

\begin{figure}
\begin{center}
	\begin{tikzpicture}%[show background grid] %% Use grid for positioning, then turn off
	 \node[circle, draw, line width=1mm, minimum size = 1.2cm] (GSM1) at (0,0) {\Large $G_\text{SM}$};
	 \node[circle, draw, line width=1mm, minimum size = 1.2cm, fill, color=green!50!gray] (SUSY) at (-4,0) {\large \textcolor{white}{$\cancel{\text{SUSY}}$}};
	 \node[circle, draw, line width=1mm, minimum size = 1.2cm] (GSM2) at (4,0) {\Large $G_\text{SM}$};
	 \draw[color=green!50!gray, line width=1mm] (SUSY) -- (GSM1);
	 \draw[line width=1mm] (GSM1) -- (GSM2);
	 \node[draw, anchor=north, rounded corners, line width=.2ex]
	 	(lightfermions) at (0,-1.5) {\normalsize Light fermions};
	  \node[draw, anchor=north, rounded corners, line width=.2ex] 
	  	(gen3) at (4,-1.5) {\normalsize 3$^\text{rd}$ gen, Higgs};
	  \draw[line width=.2ex] (lightfermions) -- (GSM1);
	  \draw[line width=.2ex] (gen3) -- (GSM2);
	\end{tikzpicture}
	\end{center}
\caption{Schematic moose diagram for natural \SUSY.}
\label{fig:close:natural:SUSY}
\end{figure}
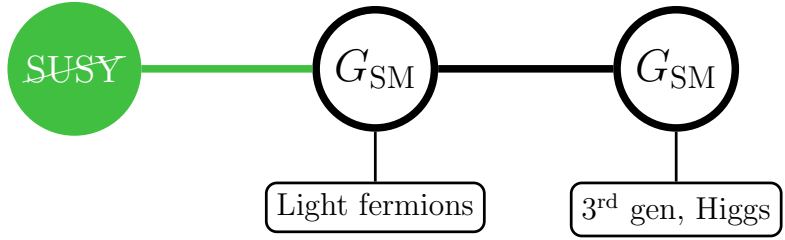

\SUSY, however, must be broken. These effects feed into the large parameter space of the minimal supersymmetric Standard Model and are required (in the \MSSM) for electroweak symmetry breaking. The \LHC puts tight bounds on the simplest \MSSM spectra and leads us to consider ways to hide \SUSY. One of these solutions is `natural \SUSY' where one only maintains the minimal spectrum of superpartners required for the naturalness of the Higgs mass. Among the predictions of natural \SUSY is a light stop and heavy first and second generation quarks. This type of spectrum, however, is automatic when supersymmetrizing the \RS model with anarchic flavor\footnote{One should note that because 5D spinors are Dirac, $\mathcal N=1$ \SUSY in 5D corresponds to $\mathcal N=2$ \SUSY in 4D. $\mathcal N=2$ was used in \cite{Fox:2002bu} to generate Dirac gaugino masses, which can help soften the two-loop quadratic corrections to the Higgs mass. See \cite{Csaki:2013fla} for a recent analysis of prospects.}. When \SUSY is broken on the \UV brane\footnote{This is one of the ways to interpret anomaly mediation of \SUSY breaking \cite{Randall:1998uk}.}, 5D \emph{super}fields which are localized near the \UV brane are more sensitive to the splitting between the \SM and superpartner masses~\cite{Gherghetta:2003he, Goh:2003yr, Gabella:2007cp, Gherghetta:2011wc}.  Invoking what we know about the anarchic flavor 5D mass spectrum (i.e.\ localization of the fermion profiles), we come to the conclusions in Table~\ref{tab:concl:natural:SUSY}.
\begin{table}
\begin{center}
\begin{tabular}{llll}
\hline
\textsc{Field} & 
\textsc{5D localization} &
\textsc{4D interpretation} &
\textsc{Superpartner}
\\
Higgs 
& \IR localized 
& Composite state 
& $\approx$ degenerate, mixes with $\tilde B$, $\tilde W$
\\
Top
& Peaked toward \IR
& Mostly composite
& Slightly heavier than top
\\
Light quarks
& Peaked toward \UV
& Mostly elementary
& Large \SUSY breaking masses\\
\hline
\end{tabular}
\end{center}
\caption{Holographic picture of natural \SUSY spectra. Superfields localized near the \IR brane have a large overlap with the Higgs so that the \SM component of the superfield picks up a large mass. Superfields localized near the \UV brane have a large overlap with \SUSY breaking so that the `superpartner' component of the superfield picks up a large mass. Thus light \SM fermions have heavy superpartners and vice versa.}
\label{tab:concl:natural:SUSY}
\end{table}
Holographically this is interpreted as supersymmetry being an accidental symmetry in the \IR. That is, the strong sector flows to a fixed point that is supersymmetric, even though the theory at the \UV is not manifestly supersymmetric. As a particle becomes more composite, it becomes more degenerate in mass with its superpartner. A schematic moose diagram is shown in Fig.~\ref{fig:close:natural:SUSY}; note that one of the sacrifices of this realization of natural \SUSY is conventional unification, see e.g.~\cite{Weiner:2001pv}.

\subsection{Naturalness and top partners}

The three classes of physics beyond the Standard Model that we have explored all generically predict new particles accessible at high energy colliders. For supersymmetry and extra dimensions, these particles were a manifestation of the extended spacetime symmetry under which the \SM particles must transform. For a composite Higgs, this reflected a larger global symmetry breaking pattern and included additional fermions that appear necessary to generate an \SM-like Higgs potential. At a technical level, we needed new particles to run in Higgs loops to soften the quadratic sensitivity to the cutoff. Since the top quark has the largest coupling to the Higgs, a generic prediction for naturalness are light (i.e.\ accessible at the \LHC) states to cancel the top loop. While these particles may have different spin, the examples we've explored focused on the case where they have the same \SM quantum numbers as the top. The color charge of these new particles make them easy to produce at the \LHC so that their non-observation is particularly disconcerting. One model building direction out of this puzzle is to consider models where the top partner is not color charged. 
%Such models have been constructed and invoke the tools introduced in these lectures---often in combinations, such as `double protected' \SUSY little Higgs models~\cite{Birkedal:2004xi, Berezhiani:2005pb, Roy:2005hg, Csaki:2005fc, Falkowski:2006qq, Chang:2006ra,}---under the titles of `twin Higgs' or `folded \SUSY' \cite{Craig:2013fga, Craig:2013cxa, Burdman:2006tz, Chang:2006ra,  Falkowski:2006qq, Chacko:2005un, Chacko:2005vw, Barbieri:2005ri, Chacko:2005pe, Chacko:2005ra}.
%
We saw this in the twin Higgs model in Section~\ref{sec:comp:twin:higgs}. A supersymmetric cousin on these models go under the name of folded \SUSY~\cite{Burdman:2006tz}, where the top partners are uncolored but still carry electroweak charges. 
Non-supersymmetric variants include the quirky little Higgs~\cite{Cai:2008au} and the orbifold Higgs~\cite{Craig:2014roa}.

\subsection{Seiberg duality}

In the composite Higgs models, same-spin partners cancel the leading \SM particle contributions to the quadratically sensitive terms in the Higgs mass. 
We saw that this is not coincidental, but is in fact imposed by the structure of collective symmetry breaking. In the same way, the protection against quadratic divergences in \SUSY is most clearly understood from the tremendous constraints put on the theory by supersymmetry. Among other things, these constraints impose the holomorphy of the superpotential which, in turn, prevents the perturbative renormalization of any of the superpotential terms. A derivation of this important result is beyond the scope of these lectures, but can be found---along with further implications of \SUSY---in the reviews already mentioned. 

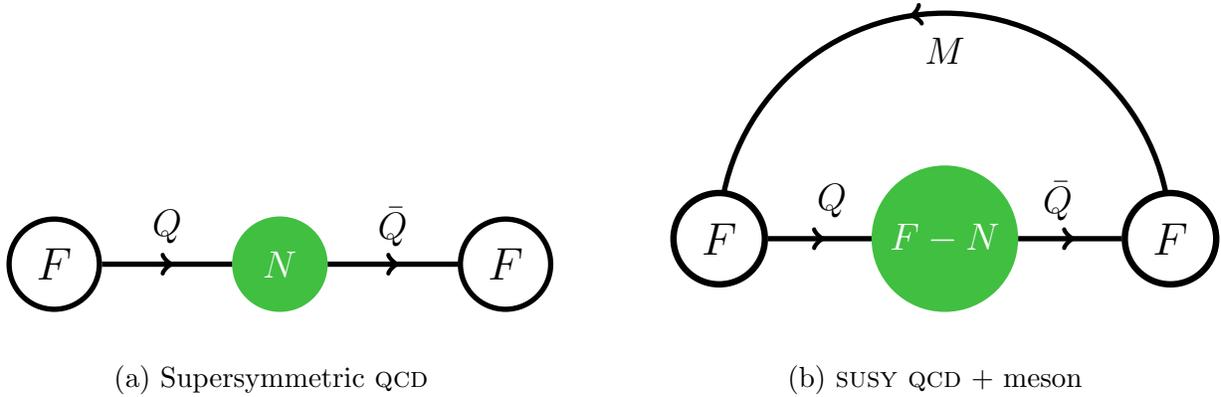
\begin{figure}
\begin{center}
\begin{subfigure}[b]{.4\textwidth}
\begin{center}
	\begin{tikzpicture}%[show background grid] %% Use grid for positioning, then turn off
	 \node[circle, draw, line width=2, minimum size = 1.2cm] (F1) at (-3,0) {\Large $F$};
	 \node[circle, draw, line width=2, minimum size = 1.2cm] (F2) at (3,0) {\Large $F$};
	 \node[circle, draw, line width=2, minimum size = 1.2cm, fill, color=green!50!gray] (N) at (0,0) {\large \textcolor{white}{$N$}};
    \draw[fermion, line width=2] (F1) -- (N);
    \draw[fermion, line width=2] (N) -- (F2);
    \node at (-1.5,.5) {\large $Q$};
    \node at (1.5,.5) {\large $\bar Q$};
	\end{tikzpicture}
	\end{center}
\caption{Supersymmetric \QCD}
\label{fig:close:elec:moose}
\end{subfigure}
\qquad\qquad
\begin{subfigure}[b]{.4\textwidth}
\begin{center}
		\begin{tikzpicture}%[show background grid] %% Use grid for positioning, then turn off
	 \node[circle, draw, line width=1mm, minimum size = 1.2cm] (F1) at (-3,0) {\Large $F$};
	 \node[circle, draw, line width=1mm, minimum size = 1.2cm] (F2) at (3,0) {\Large $F$};
	 \node[circle, draw, line width=1mm, minimum size = 1.2cm, fill, color=green!50!gray] (N) at (0,0) {\large \textcolor{white}{$F-N$}};
    \draw[fermion, line width=2] (F1) -- (N);
    \draw[fermion, line width=2] (N) -- (F2);
    \draw[fermion, line width=2] (F2) arc (0:180:3);
    \node[circle, draw, line width=2, minimum size = 1.2cm, fill=white] at (F1) {\Large $F$};
	 \node[circle, draw, line width=2, minimum size = 1.2cm, fill=white] at (F2) {\Large $F$};
	 \node at (-1.5,.5) {\large $Q$};
    \node at (1.5,.5) {\large $\bar Q$};
    \node at (0,2.5) {\large $M$};
	\end{tikzpicture}
	\end{center}
	\caption{\SUSY \QCD + meson}
\label{fig:close:mag:moose}
\end{subfigure}
\end{center}
\caption{Moose diagrams for a pair of Seiberg duals. Green nodes are gauged symmetries while white notes are global symmetries. Note that the lines now represent superfields.}
\label{fig:close:duality}
\end{figure}

Supersymmetry turns out to also be a powerful constraint on the behavior of gauge theories. In fact, they allow one to map out the entire phase structure of the supersymmetric generalization of \QCD, \textsc{sqcd}. This, in itself, is a topic of depth and elegance which is covered very well in \cite{Peskin:1997qi, Seiberg:1994pq, Intriligator:2007cp, Strassler:2003qg, Terning:2006uq}. One key outcome of this exploration in the 1990s was the observation that two distinct supersymmetric non-Abelian gauge theories, shown in Fig.~\ref{fig:close:duality}, flow to the same \IR theory.
%
% IR theory vs fixed point
%
One theory, \textsc{sqcd}, is a standard SU($N$) supersymmetric gauge theory with $F$ flavors such that
\begin{align}
%\frac{3}{2}N < F < 3N.
F>N+1.
\label{eq:closing:conformal:window}
\end{align}
%
%In the case where $N+1 < F < \frac 32 N$, this theory is asymptotically free and 
In the case where $N+1 < F < 3N$, this theory is asymptotically free and 
becomes strongly coupled
in the \IR. 
The other theory is an SU($F-N$) gauge theory with $F$ flavors and an additional color singlet `meson' which is a bifundamental under the SU($F$)$\times$SU($F$) flavor symmetry. This theory has a superpotential,
\begin{align}
W \sim \bar Q M Q,
\end{align}
which can be understood as a loop in the moose diagram since all indices are contracted.
When $N+1 < F < \frac 32 N$, the dual theory is \IR free and is perturbative. On the other hand, when $\frac 32 N \leq F < 3N$, the dual theory is also asymptotically free and flows to the same Banks-Zaks--like non-trivial fixed point that is the same endpoint of the original `electric' theory's \RG flow.
%
% "Fixed point" suggests a conformal theory, but this is NOT
% It's in the free magnetic phase

The fact that two \emph{a priori} unrelated theories flow to the same 
\IR theory suggest a compelling interpretation: the initial asymptotically free theory becomes strongly interacting at low energies and can be equivalently described by its dual theory.
In the case where $N+1 < F < \frac 32 N$, the dual theory is \IR free and perturbative precisely in the regime where the original theory is not.
This is an `electromagnetic' duality in the sense of exchanging strongly and weakly coupled descriptions of the same physics, similar to the \AdS/\CFT correspondence. %This is also an `\IR duality' in the sense that the two theories flow into one another deep in the \RG flow. 

This \textbf{Seiberg duality} is a powerful handle on strongly coupled physics via a weakly coupled 4D dual description. One popular application was to simplify the construction of models with dynamical \SUSY breaking, see \cite{Intriligator:2007cp, Shirman:2009mt} for reviews. In some sense this is completely analogous to using chiral perturbation theory to describe low-energy \QCD. However, unlike \QCD, the low energy (`magnetic') theory is not composed of gauge singlets. In fact, there is an \emph{emergent} SU($F-N$) gauge symmetry that appears to have nothing to do with the original SU($N$) gauge symmetry of the `electric' theory. One recent interpretation, however, is that this magnetic gauge group can be identified with the `{hidden local symmetry}' in nonlinear models \cite{Komargodski:2010mc}, which we previously mentioned in the context of deconstruction and moose models. In this construction, the $\rho$ meson in \QCD (the lightest spin-1 meson) is identified as the massive gauge boson of a spontaneously broken gauge symmetry present in the nonlinear Lagrangian. 

One can also relate Seiberg duality to the \AdS/\CFT correspondence through explicit string realizations. Note that (\ref{eq:closing:conformal:window}) is typically a different regime from the large $N$ limit invoked in \AdS/\CFT. From a purely field theoretical point of view, the \AdS/\CFT correspondence can be understood as a \textbf{duality cascade} where a \SUSY gauge theory has a renormalization trajectory that zig-zags between a series of fixed points. This is reviewed pedagogically for a field theory audience in \cite{Strassler:2005qs}.

\subsection{Multiple guises of strong dynamics}

In this final section we have touched on multiple ways in which we can address strong dynamics in field theory: nonlinear realizations based on the symmetry breaking structure, holographic extra dimensions, and Seiberg duality in \SUSY. The lesson to take away from this overview is that one should be flexible to think about strong dynamics in different languages. Often the intuition from one understanding of strong dynamics can shed light on constructions based on a different description. 

One example is the use of Seiberg duality to describe a [partially-]composite electroweak sector based on the `fat Higgs' model \cite{Harnik:2003rs}. The idea is to take super-\QCD with $F=N+2$ flavors so that the magnetic gauge group can be linked with SU$(2)_\text{L}$. The realization of this idea in \cite{Craig:2011ev} described this in terms of moose diagrams where the magnetic gauge group is `color-flavor locked' with an externally gauged SU(2)$_\text L$. This mixes the magnetic gauge bosons with the external gauge bosons so that the observed $W$ and $Z$ are partially composite. Independently, a similar model was presented in \cite{Csaki:2011xn} where the nature of this mixing was explained in terms of the intuition from a warped extra dimension. In particular, one hope is  that one can directly identify the magnetic SU(2) with the electroweak SU(2)$_\text L$. This, however, is not possible since---as we know from composite model building---at the compositeness scale the n\"aive dimensional analysis expectation is that the composite vector boson couples strongly: $g\sim 4\pi/\sqrt{N}$. In other words, if the electroweak gauge bosons are strongly coupled bound states, then one would expect a large residual interaction with other strongly coupled bound states. In the \RS language, a composite $W$ and $Z$ would have \IR brane localized profiles and this would typically predict very strong couplings. This would require a very large running to squeeze the profile on the \IR brane. In the Seiberg dual picture, this requires a very large number of flavors if one maintains that the $W$ and $Z$ are purely composite but have the observed \SM couplings, leading one to prefer partial compositeness of these particles.
This general framework was later used to construct a model of natural \SUSY in which follows the general deconstruction/moose in Fig.~\ref{fig:close:natural:SUSY} \cite{Csaki:2012fh}.

\subsection{Omissions}

We have necessarily been limited in scope. Even among the topics discussed, we have omitted an exploration of \SUSY gauge theories (leading up to Seiberg duality), variants of the `realistic' \RS models (as well as `universal extra dimension' models), the virtues of different cosets for composite Higgs model building, and an overview of product space (moose-y) little Higgs models. Many explicit calculations were left out and are left to the dilligent reader as exercises, and we only made cursory nods to the phenomenology of these models. 
In addition to the three major topics covered in these lectures, there are various other extensions to the Standard Model that we have not discussed. Our preference focused on models that address the Higgs hierarchy problem, and as such we have omitted discussions of many important topics such as grand unification, dark matter, flavor, strong \CP, cosmology (of which the cosmological constant is the most extreme fine tuning problem), or any of the phenomenology of interpreting possible experimental signals from colliders/telescopes/underground experiments/etc. 
We have only presented very cursory comparisons to current data; we refer the reader to the appropriate experiments' results pages and conference proceedings for the latest bounds. See also~\cite{Halkiadakis:2014qda} for an overview of the Run~I searches for new physics.
All of these topics---and perhaps many others---are, in some combination, key parts of a model builder's toolbox in the \LHC era.

\section*{Acknowledgements}

%\textsc{c.c.}\ thanks the organizers of the 2013 European School of High-Energy Physics for the invitation to give these lectures.
%\textsc{p.t.}\ is grateful to Mohammad Abdullah, Jack Collins, Anthony DiFranzo, Javi Serra, Yuri Shirman, and Daniel Stolarski for useful comments and discussions. 
%
%%
%%
%\textsc{c.c.}\ is supported in part by the \textsc{nsf} grant \textsc{phy}-1316222.
%\textsc{p.t.}~is supported in part by the \textsc{nsf} grant \textsc{phy}-1316792 and a \textsc{uci} Chancellor's \textsc{advance} fellowship.
 
\textsc{c.c.}\ thanks the organizers of the 2013 European School of High-Energy Physics for the invitation to give these lectures.
\textsc{p.t.}\ is grateful to Brando Bellazzini, Nathaniel Craig, Tony Gherghetta, Roni Harnik, Javi Serra, Yuri Shirman, and and Tim M.P.~Tait for enlightening discussions about topics in these lectures. We appreciated feedback from Mohammad Abdullah, Jack Collins, Anthony DiFranzo, Benjamin Lillard, Mario Martone, and Javi Serra who read parts of this manuscript. 
We thank Gilad Perez and Martijn Mulders for their patience with this manuscript.
We thank the Aspen Center for Physics (\textsc{nsf} grant \#1066293) for its hospitality during a period where part of this work was completed.
\textsc{c.c.}\ is supported in part by the \textsc{nsf} grant \textsc{phy}-1316222.
\textsc{p.t.}~is supported in part by the \textsc{nsf} grant \textsc{phy}-1316792 and a \textsc{uci} Chancellor's \textsc{advance} fellowship.

%%% Appendices

\appendix

\section{Details of the Randall-Sundrum Scenario}
\label{app:XD:RS}

\subsection{The RS gravitational background}
\label{app:XD:RS:Grav:BG}

We have assumed the metric (\ref{eq:RS:metric}). In this appendix we derive it from the assumption of a non-factorizable metric of the form
\begin{align}
ds^2 &= e^{-A(z)}\left( \eta_{\mu\nu} dx^\mu dx^\nu - dz^2\right)
\end{align}
and check the conditions for which a flat 4D background exists. This generic form of the metric is useful since it is an overall rescaling of the flat metric, that is, it is conformally flat. We can thus use a convenient relation between the Einstein tensors $G_{MN} = R_{MN} - \frac 12 g_{MN} R$ of two conformally equivalent metrics $g_{MN} = e^{-A(x)} \tilde g_{MN}$ in $d$ dimensions \cite{wald1984general},
\begin{align}
G_{MN} = \tilde G_{MN}
+ \frac{d-2}{2} \left[
\frac 12 \tilde \nabla_M A\tilde \nabla_N A
+ \tilde \nabla_M \tilde \nabla_N A
- \tilde g_{MN}
\left( \tilde \nabla_K\tilde\nabla^K A 
-
\frac{d-3}{4} \tilde \nabla_K A \tilde\nabla^K A \right)
\right].
\end{align}
When $A=A(z)$ this is straightforward to calculate by hand for $\tilde g_{MN} = \eta_{MN}$. Alternately, one may use a computer algebra system to geometric quantities for general metrics, e.g.\ \cite{HartleGRwebsite}.
We assume a bulk cosmological constant $\Lambda$ so that the 5D bulk Einstein action is
\begin{align}
S = -\int d^5x\; \sqrt{g}\; \left( M_*^3R + \Lambda\right).
\end{align}
The Einstein equation is $G_{MN} = (M_*)^{-3}T_{MN}$. The $MN=55$ component gives
\begin{align}
\frac 32 A'^2 = \frac{1}{2M_*^3} \Lambda e^{-A}.
\end{align}
This only has a solution for $\Lambda <0$ so that we're forced to consider \AdS spaces. %Using the substitution $f(z) = e^{-A(z)/2}$, one can solve this equation to obtain the general solution
	This equation is separable with the general solution,
\begin{align}
e^{-A(z)} &= \frac{1}{\left(kz + \text{constant}\right)^2}
&
k&=\frac{-\Lambda }{12 M_*^3}.
\label{eq:XD:RS:warp:factor:solution}
\end{align}
To recover (\ref{eq:RS:metric}) we identify $R= 1/k$ and impose $A=0$ at $z=R$, setting the constant to zero. The latter choice simply sets the warp factor at the \UV brane to be 1.
%% fix the constant to zero. The latter choice is an unphysical unit rescaling.  %, e.g.\ \cite{Csaki:2004ay} fixes it to one.
%One should be careful, however, since $\text{constant}=0$ makes $z=0$ singular. We know this isn't a physical singularity because we have assumed that the  the \RS extra dimension as an interval, $z \in [R,R']$. 

We must remember that the \RS space is finite---and has branes at its endpoints---when we solve the $MN=\mu\nu$ Einstein equations. These equations depend on the second derivative of $A(z)$ and one should be concerned that this may be sensitive to the energy densities on the branes. This is analogous to the Poisson equation in electrostatics where a second derivative picks up the $\delta$-function of a point charge. In general the branes carry tensions which appear as 4D cosmological constants, $\Lambda_{\textsc{ir},\textsc{uv}}$. Recalling the form of the induced metric $\sqrt{\hat g} = \sqrt{g/g_{55}}$, these appear in the action as
\begin{align}
\int d^5x  \; \sqrt{\frac{{g}}{{g_{55}}}}\; \Lambda_{\textsc{ir},\textsc{uv}} \delta(z-z_{\textsc{ir},\textsc{uv}})
\quad\Rightarrow\quad
T_{\mu\nu} = \frac{1}{\sqrt{g}} \frac{\delta S}{\delta g^{\mu\nu}} = \frac{g_{\mu\nu}}{2\sqrt{g_{55}}}  \left(\Lambda_{\textsc{ir}}\delta(z-R')+  \Lambda_{\textsc{uv}}\delta(z-R)\right).
\end{align}

\subsection{RS as an orbifold}
\label{app:XD:RS:Grav:BG:orbifold}

To better understand the physics of the brane cosmological constants, it is useful to represent the interval with an \textbf{orbifold} $S^1/\mathbbm{Z}_2$. This is simply the circle $y\in [-\pi,\pi]$ with the identification $y = -y$. While this may sound somewhat exotic, such compactifications are common in string theory, and was the original formulation of the \RS scenario. Note that $y$ can take any value due to the periodic identification of the circle, while the fixed points at $y = 0, \pi$ demarcate the physical \RS space. 
%
%% z = e^{-kR' y}/k

The orbifold identification forces us to modify (\ref{eq:XD:RS:warp:factor:solution}) by replacing $z\to |z|$ to preserve the $z\leftrightarrow -z$ symmetry.%\footnote{It is conventional that the $y$ and $z$ coordinates are related by $e^{-A(z)}dz^2 = dy^2$. If we set the constant in (\ref{eq:XD:RS:warp:factor:solution}) to zero, one may argue that $e^{A(z)}$ is already an even function of $z$. This, however, introduces a coordinate singularity at $z=0$. In our interval picture this isn't a problem since the physical space is $z\in [R,R']$, but from the orbifold perspective we are allowed to explore any value of $z$ due to the circle and orbifold identifications.}.
This absolute value, in turn, leads to $\delta$ functions in $A''(z)$ at the fixed points,
\begin{align}
A''(z) = -\frac{2k^2}{(k|z|+\text{const})^2}
+ \frac{4k}{k|z|+\text{const}} \left(\delta(z-R) - \delta(z-R')\right).
\end{align}
The $\mu\nu$ Einstein equation then implies
\begin{align}
-\frac 32 \eta_{\mu\nu}
\left[
-\frac{4k\left(\delta(z-R) - \delta(z-R')\right)}{k|z|+\text{const}}
\right]
= 
\frac{\eta_{\mu\nu}}{2M_*^3}
\left[
\frac{\Lambda_\textsc{uv}\delta(z-R) + \Lambda_\textsc{ir}\delta(z-R')}{k|z|+\text{const}}
\right].
\end{align}
From this we see that the brane cosmological constants must have opposite values,
\begin{align}
\Lambda_\textsc{uv} = -\Lambda_\textsc{ir} = 12 k M_*^3. 
\label{eq:XD:RS:tuning:for:cosmological:constant}
\end{align}
Recall, further, that $k$ is related to the bulk cosmological constant by (\ref{eq:XD:RS:warp:factor:solution}), so that this represents a tuning of the bulk and brane cosmological constants. This is a necessary condition for a static, gravitational solution. Physically, we see that the brane and the bulk cosmological constants are balanced against one another to cause the brane to be flat.

\subsection{Bulk Fermions in RS}
\label{sec:XD:RS:bulk:fermion}

The properties of fermions in a curved space can be  subtle. In particular, it's not clear how to generalize the usual Dirac operator,  $i\gamma^\mu\partial_\mu$. In this appendix we review properties of fermions in an extra dimension and then derive the form of the fermion action in \RS. 

\subsubsection{The fifth $\gamma$ matrix}
\label{sec:comp:fermions:in:5D}

Firstly, unlike in 4D where the fundamental fermion representation is a Weyl spinor, 5D Lorentz invariance requires that fermions appear as Dirac spinors. A simple heuristic way of seeing this is to note that in 4D one can construct a $\gamma^5 \sim \gamma^0\cdots \gamma^3$ as a linearly independent chirality operator. In 5D, however, $\gamma^5$, is part of the 5D Clifford algebra and is just a normal $\gamma$ matrix in the $z$-direction. Note that the normalization of $\gamma^5$ is fixed by $\{\gamma^5,\gamma^5\} = 2\eta^{55}$ and has a factor of $i$ compared to the usual definition in 4D. One should immediately be concerned: if the 5D fermions are Dirac, then how does one generate the \emph{chiral} spectrum of the Standard Model matter? As we show below, this follows from a choice of boundary conditions. An excellent reference for the properties of fermions in arbitrary dimension is \cite{Polchinski:1998rr}. 

\subsubsection{Vielbeins}

In order to write down the fermionic action, we first need to establish some differential geometry so that we may write the appropriate covariant derivative for the spinor representation. We will be necessarily brief here, but refer to \cite{Bertlmann:1996xk, Frankel:1997ec, Gockeler:1987an} for the interested reader\footnote{For a beginner-friendly introduction, see \cite{Collinucci:2006hx} or your favorite general relativity textbook.}. 

The familiar $\gamma$ matrices which obey the Clifford algebra are only defined for flat spaces. That is to say that they live on the tangent space (locally inertial frame) of our spacetime manifold. In order to define curved-space generalizations of objects like the Dirac operator, % $i\gamma^\mu\partial_\mu$, 
we need a way to convert spacetime indices $M$ to tangent space indices $a$. \textbf{Vielbeins}, $e^a_\mu(x)$, are the geometric objects which do this. The completeness relations associated with vielbeins allow them to be interpreted as a sort of ``square root'' of the metric in the sense that
\begin{align}
	g_{MN}(x) &= e^a_M(x)e^b_M(x)\eta_{ab},
\end{align}
where $\eta_{ab} = \text{diag}(+,-,\cdots,-)$ is the Minkowski metric on the tangent space. For our particular purposes we need the inverse vielbein, $e^M_a(x)$, defined such that
\begin{align}
	e^M_a(x)e^a_N(x) &= \delta^M_{\phantom{M}N}
	&
	e^M_a(x)e^b_N(x) &= \delta_a^{\phantom{a}b}.
\end{align}
Spacetime indices are raised and lowered using the spacetime metric $g_{MN}(x)$ while tangent space indices are raised and lowered using the flat (tangent space) metric $\eta_{ab}(x)$.

Physically we may think of the vielbein in terms of reference frames. The equivalence principle states that at any point one can always set up a coordinate system such that the metric is flat (Minkowski) at that point. Thus for each point $x$ in space there exists a family of coordinate systems that are flat at $x$. For each point we may choose one such coordinate system, which we call a frame. By general covariance one may define a map that transforms to this flat coordinate system at each point. This is the vielbein. One can see that it is a kind of local gauge transformation, and indeed this is the basis for treating gravity as a gauge theory built upon diffeomorphism invariance. %Mathematically, the vielbein represents the frame bundle on the spacetime. 

\subsubsection{Spin covariant derivative}

The covariant derivative is composed of a partial derivative term plus connection terms which depend on the particular object being differentiated. For example, the covariant derivative on a spacetime vector $V^\mu$ is
\begin{align}
	D_M V^N &= \partial_M V^N +\Gamma^N_{ML}V^L.
\end{align}
The vielbein allows us to work with objects with a tangent space index, $a$, instead of just spacetime indices, $\mu$.
The $\gamma$ matrices allow us to further convert tangent space indices to spinor indices.
We would then define a covariant derivative acting on the tangent space vector $V^a$,
\begin{align}
	D_M V^a &= \partial_M V^a + \omega_{M b}^{a}V^b,
\end{align}
where the quantity $\omega_{M b}^{a}$ is called the spin covariant derivative. Consistency of the two equations implies
\begin{align}
	D_M V^a &= e^a_N D_M V^N.
\end{align}
This is sufficient to determine the spin connection.
% We won't prove the result here, but one can look up the appropriate references (e.g. your favorite general relativity of differential geometry books). 
It is a fact from differential geometry that the spin connection is expressed in terms of the veilbeins via  \cite{Sundrum:1998sj}
\begin{align}
	\omega_M^{ab} &= \frac{1}{2} g^{RP}e^{[a}_R\partial_{[M}e^{b]}_{P]} + \frac{1}{4} g^{RP}g^{TS}e^{[a}_{R}e^{b]}_T\partial_{[S}e^c_{P]}e^d_M\eta_{cd}\\
	&= \frac{1}{2}e^{Na}\left(\partial_M e^b_N-\partial_N e^b_M\right)-\frac{1}{2}e^{Nb}\left(\partial_M e^a_N - \partial_N e^a_M\right) - \frac{1}{2}e^{Pa}e^{Rb}\left(\partial_P e_{Rc}-\partial_R e_{Rc}\right)e^c_M.
\end{align}

When acting on spinors one needs the appropriate structure to convert the $a,b$ tangent space indices into spinor indices. This is provided by
\begin{align}
	\sigma_{ab} &= \frac{1}{4}\left[\gamma_a,\gamma_b\right]
\end{align}
so that the appropriate spin covariant derivative is
\begin{align}
	D_M &= \partial_M + \frac 12 \omega^{ab}_M \sigma_{ab}.
\end{align}

\subsubsection{Antisymmetrization and Hermiticity}%\label{RS:sec:antisymmetrization}

The fermionic action on a $d$-dimensional curved background is\footnote{%
We write $\sqrt{|g|}$ to allow for a general sign of $g=\det g_{\mu\nu}$.
}
\begin{align}
	S &= \int d^d x\, \sqrt{|g|} \; \overline\Psi \left(ie^M_a\gamma^a\overleftrightarrow{D_M} - m \right)\Psi,\label{eq:fermionaction:basic}
\end{align}
where the antisymmetrized covariant derivative is defined by a difference of right- and left-acting derivatives
\begin{align}
	\overleftrightarrow{D_M} = \frac 12 D_M - \frac 12 \overleftarrow{D_M}.
\end{align}
This is somewhat subtle. The canonical form of the fermionic action must be antisymmetric in this derivative in order for the operator to be Hermitian and thus for the action to be real. In flat space we are free to integrate by parts in order to write the action exclusively in terms of a right-acting Dirac operator. 
%This is a very nice thing to say and `makes sense,' but the actual meaning is a little bit subtle. 
%
Hermiticity is defined with respect to an inner product. The inner product in this case is given by
\begin{align}
	\langle \Psi_1 | \mathcal O \Psi_2 \rangle &= \int d^dx\,\sqrt{|g|}\; \overline{\Psi_1} \mathcal O \Psi_2.
\end{align}
A manifestly Hermitian operator is $\mathcal O_H = \frac 12 \left(\mathcal O + \mathcal O^\dag\right)$, where we recall that
\begin{align}
	\langle \Psi_1 | \mathcal O^\dag \Psi_2\rangle &= \langle \mathcal O \Psi_1 | \Psi_2\rangle
	%\\
	%&
	=\int d^dx\,\sqrt{|g|}\; \overline{\mathcal O\Psi_1} \Psi_2.
\end{align}
The definition of an inner product on the Fock space of a quantum field theory can be nontrivial on curved spacetimes. However, since our spacetime is not warped in the time direction there is no ambiguity in picking a canonical Cauchy surface to quantize our fields and we may follow the usual procedure of Minkowski space quantization with the usual Minkowski spinor inner product. 

As a sanity-check, consider the case of the partial derivative operator $\partial_\mu$ on flat space time. The Hermitian conjugate of the operator is the left-acting derivative, $\overleftarrow{\partial_\mu}$, by which we really mean
\begin{align}
	\int d^dx \, \overline{\Psi_1}\partial^\dag \Psi_2 = \langle\Psi_1| \partial_\mu ^\dag \Psi_2\rangle = \langle \partial_\mu\Psi_1|\Psi_2\rangle = \int d^dx \, \overline{\partial_\mu\Psi_1} \Psi_2 =  \int d^dx \, \overline{\Psi_1}\overleftarrow{\partial_\mu}\Psi_2=  \int d^dx \, \overline{\Psi_1}\left(-\partial_\mu\right)\Psi_2.\nonumber
\end{align} 
In the last step we've integrated by parts and dropped the boundary term. We see that the Hermitian conjugate of the partial derivative is negative itself. Thus the partial derivative is not a Hermitian operator. This is why the momentum operator is given by $\hat P_\mu = i\partial_\mu$, since the above analysis then yields $\hat P_\mu^\dag = \hat P_\mu$, where we again drop the boundary term and recall that the $i$ flips sign under the bar. 

Now we can be explicit in what we mean by the left-acting derivative in (\ref{eq:fermionaction:basic}). The operator $ie^M_a\gamma^aD_M$ is not Hermitian and needs to be made Hermitian by writing it in the form $\mathcal O_H = \frac 12 \left(\mathcal O + \mathcal O^\dag\right)$. Thus we may write a manifestly Hermitian Dirac operator as,
\begin{align}
	\overline{\Psi}\left(\text{Dirac}\right)\Psi &= \overline{\Psi}\left[\frac{1}{2}\left(ie^M_a\gamma^a D_M \right)+\frac{1}{2}\left(ie^M_a\gamma^a D_M \right)^\dag\right]\Psi\\
%	&= \overline{\Psi}\frac i2 e^M_a\gamma^a D_M  \Psi + \overline{\frac i2 e^M_a\gamma^a D_M\Psi}\Psi\\
	&=  \overline{\Psi}\frac i2 e^M_a\gamma^a D_M  \Psi - \frac{i}{2}e^M_a \overline{\gamma^a D_M\Psi}\Psi,
\end{align}
where we've used the fact that $e^M_a$ is a real function with no spinor indices. The second term on the right-hand side can be massaged further,
\begin{align}
	\overline{\gamma^a D_M\Psi}\Psi &= \Psi^\dag \overleftarrow{D_M}^\dag \gamma^{a\dag}\gamma^0 \Psi
%	\\
%	&
%	= \Psi^\dag \overleftarrow{D_M}^\dag \left(\gamma^0\gamma^{a}\gamma^0\right)\gamma^0 \Psi
%	\\
%	&
	= \Psi^\dag (\overleftarrow{\partial_M}+\omega^{bc}_M\sigma^{bc\dag}) \gamma^0 \gamma^{a} \Psi
%	\\
%	&
	= \overline\Psi \overleftarrow{D_M} \gamma^{a} \Psi
%	\\
%	&
	= \overline\Psi \gamma^a \overleftarrow{D_M} \Psi.
\end{align}
Note that we have used that $\gamma^{M\dag} = \gamma^0\gamma^M\gamma^0$ and, in the last line, that $[\sigma^{bc},\gamma^a] = 0$. Putting this all together, we can write down our manifestly real fermion action %(i.e. manifestly Hermitian Dirac operator) 
as in (\ref{eq:fermionaction:basic}),
\begin{align}
	S &= \int d^d x\, \sqrt{|g|} \; \overline\Psi \left(ie^M_a\gamma^a\overleftrightarrow{D_M} - m \right)\Psi\\
	&= \int d^d x\, \sqrt{|g|} \left( \frac i2 \overline\Psi e^M_a\gamma^aD_M\Psi - \frac i2 \overline{D_M\Psi} e^M_a\gamma^a\Psi - m\overline\Psi\Psi\right).
\end{align}

All of this may seem overly pedantic since integration by parts allows one to go back and forth between the `canonical' form and the usual `right-acting only' form of the fermion kinetic operator. Our interest, however, is to apply this to the Randall-Sundrum background where integration by parts introduces boundary terms and so it is crucial to take the canonical form of the Dirac operator as the starting point.

\subsubsection{Application to the RS background}

We now apply this machinery to the \RS background. The vielbein and inverse vielbein are
\begin{align}
	e^a_M(z) = \frac Rz \delta^a_M \quad \quad \quad \quad e^M_a(z) = \frac zR \delta^M_a.
\end{align}
We may write out the spin connection term of the covariant derivative as
\begin{align}
	\omega_M^{ab} &= \underbrace{\frac{1}{2}g^{RP} e^{[a}_R\partial_{[M}e^{b]}_{P]}}_{\omega_M^{ab}(1)} + \underbrace{\frac 14 g^{RP}g^{TS} e^{[a}_Re^{b]}_T \partial_{[S}e^a_{P]} e^d_M \eta_{cd}}_{\omega_M^{ab}(2)}.
\end{align}
This can be simplified using the fact that the vielbein only depends on $z$. The first part is 
\begin{align}
	\omega_M^{ab}(1) 
%	&= \phantom{+} \frac{1}{2}g^{RP}e^a_R\partial_{[M}e^b_{P]}
%	- \frac{1}{2} g^{RP}e^b_R\partial_{[M}e^a_{P]}\\
%	&= \phantom{+}\frac 12 g^{RP}e^a_R\partial_M e^b_P - \frac 12 g^{RP}e^a_R \partial_P e^b_M
%	- \frac 12 g^{RP}e^b_R\partial_M e^a_P + \frac 12 g^{RP}e^b_R \partial_P e^a_M
%	\\
%	&= -\frac 1{2z} g^{RP}e^a_R e^b_P \delta^5_M  + \frac 1{2z} g^{RP}e^a_R  e^b_M \delta^5_P
%	+\frac 1{2z} g^{RP}e^b_R e^a_P \delta^5_M  - \frac 1{2z} g^{RP}e^b_R  e^a_M \delta^5_P
%	\\
%	&= -\frac 1{2z} \eta^{ab} \delta^5_M  + \frac 1{2z} g^{R5}e^a_R  e^b_M 
%	+\frac 1{2z} \eta^{ba} \delta^5_M  - \frac 1{2z} g^{R5}e^b_R e^a_M
%	\\
%	&= -\frac{1}{2z}\eta^{ab}(\delta^5_M-\delta^5_M) + \frac{1}{2z}g^{R5}\left(e^a_Re^b_M - e^b_Re^a_M\right)\\
%	&= -\frac{1}{2z}\delta^R_5\left(\delta^a_R\delta^b_M-\delta^b_R\delta^a_M\right)\\
	&= \phantom{+}\frac{1}{2z}\delta^{[a}_M\delta^{b]}_5,
\end{align}
where we've used $\partial_M e^b_P = -\frac 1z e^b_P \delta^5_M$ and the completeness relation $g^{MN}e^a_Me^b_M = \eta^{ab}$. Similarly, with some effort the second part is given by
\begin{align}
	\omega_M^{ab}(2)
	&=\phantom{+}\frac{1}{2z}\delta^{[a}_M\delta^{b]}_5.
\end{align}
These vanish identically for $M=5$. We can now write out the spin-connection part of the covariant derivative,
\begin{align}
	\frac 12 \omega^{ab}_M\sigma_{ab} &=  \frac 12 \left(\frac 1z \delta^{[a}_M\delta^{b]}_5\right)_{M\neq 5}\frac 14 \left[\gamma_a,\gamma_b\right]
	%\\
%	&= 
= \frac{1}{4z} \left(\gamma_M\gamma_5 + \delta^5_M\right),
\end{align}
where we've inserted a factor of $\delta^5_M$ to cancel the $(\gamma_5)^2$ when $M=5$. %(Note that the natural convention is that $(\gamma^5)^2=-\mathbbm{1}$ since this is what satisfies the 5D Clifford algebra.)
Finally, the spin connection part of the covariant derivative is
\begin{align}
	\frac 12 \omega^{ab}_M\sigma_{ab} &= \frac 1{4z}\left(\gamma_M\gamma_5+\delta^5_M\right) 
\end{align}
so that the spin covariant derivative is
\begin{align}
	D_M= 
	\begin{cases}
		\partial_\mu + \frac{1}{4z}\gamma_\mu\gamma_5 & \text{if } M=\mu\\
		\partial_5 & \text{if } M=5.
	\end{cases}
\end{align}

For all of the geometric heavy lifting we've done, we are led to an anticlimactic result: the spin connection drops out of the action,
\begin{align}
	S 
%	&= \int d^5x \,\left(\frac Rz\right)^4 \; \frac i2 \overline\Psi \delta^M_a \gamma^a D_M \Psi - \frac i2\delta^M_a \overline{D_M \gamma^a\Psi}  \Psi\\
	&= \int d^5x \, \frac i2\left(\frac Rz\right)^4 \;\left(\overline\Psi \gamma^M \overleftrightarrow{\partial_M}\Psi + \frac{1}{4z}\overline\Psi  \gamma_\mu\gamma_5\gamma^\mu\Psi - \frac{1}{4z}\overline{\gamma_\mu\gamma_5\gamma^\mu\Psi}\Psi\right),
\end{align}
The two spin connection terms cancel since $\overline{\gamma_\mu\gamma^5\gamma^\mu\Psi}\Psi = \overline\Psi\gamma_\mu\gamma_5\gamma^\mu\Psi$, so that upon including a bulk mass term,
\begin{align}
	S &= \int d^5x \,\frac{i}{2}\left(\frac Rz\right)^4 \, \overline\Psi\gamma^M\overleftrightarrow{\partial_M}\Psi - \int d^5x \,\frac{i}{2}\left(\frac Rz\right)^5 \, m\overline\Psi\Psi
%	\\
%	&
	= \int d^5x \,\frac{i}{2}\left(\frac Rz\right)^4 \, \overline\Psi\left(\gamma^M\overleftrightarrow{\partial_M}-\frac cz\right)\Psi,\label{eq:fermionaction:RS:leftright}
\end{align}
where $c=mR = m/k$ is a dimensionless parameter that is the ratio of the bulk mass to the curvature.
Before we can dimensionally reduce the action straightforwardly, 
we must write the Dirac operator to be right-acting, i.e.\ acting on $\Psi$, so that we can vary with respect to $\overline\Psi$ to get an operator equation for $\Psi$. Obtaining this is from (\ref{eq:fermionaction:RS:leftright}) is now a straightforward matter of integration by parts of the left-acting derivative term. Note that it is crucially important that we pick up a derivative acting on the metric/vielbein factor $(R/z)^4$. We would have missed this term if we had mistakenly written our original `canonical action,' (\ref{eq:fermionaction:basic}), as being right-acting only.

The integration by parts for the $M=\mu=0,\cdots,4$ terms proceeds trivially since these directions have no boundary and the metric/vielbein factor is independent of $x^\mu$. Performing the $M=5$ integration by parts we find
\begin{align}
	S
	% &=\int d^4x\int_{R'}^R dz\,\left(\frac Rz\right)^4 \overline\Psi\left(i\slashed{\partial} + \frac i2\gamma^5\overleftrightarrow{\partial_5} - \frac cz \right)\Psi
	%\\
	&
	= \int d^4x\int_{R}^{R'} dz\,\left(\frac Rz\right)^4 \overline\Psi\left(i\slashed{\partial} +  i\gamma^5\partial_z - i\frac 2z\gamma^5  - \frac cz \right)\Psi + \left.(\text{boundary term})\right|^{R'}_R.
\end{align}
The term in the parenthesis can be identified with the Dirac operator for the  Randall-Sundrum model with bulk fermions.
%This `definition' is up to conventions regarding the inclusion of the mass term and factors of $i$. 
The boundary term is
\begin{align}
	(\text{boundary}) &= \left.\left( R/z\right)^4\left(\psi\chi-\overline\chi\overline\psi\right)\right|_R^{R'},\label{eq:boundaryterm}
\end{align} 
where we've written out the Dirac spinor $\Psi$ in terms of two-component Weyl spinors $\chi$ and $\psi$. This term vanishes when we impose chiral boundary conditions, which we review in the next section. 
%The final form of the \RS fermion action is
%\begin{align}
%	S&=\int d^4x\int_{R'}^R dz\,\left(\frac Rz\right)^4 \overline\Psi\left(i\slashed{\partial} +  i\gamma^5\partial_5 - i\frac 2z\gamma^5  - \frac cz \right)\Psi.\label{eq:RS:fermion:action}
%\end{align}
In terms of Weyl spinors this gives
\begin{align}
	S&=\int d^4x\int^{R'}_R dz\,\left(\frac Rz\right)^4 
	\begin{pmatrix}
		\psi & \overline\chi
	\end{pmatrix}
	\begin{pmatrix}
		-\partial_z + \frac{2-c}{z} & i\slashed{\partial}\\
		i\overline{\slashed{\partial}} & \partial_z -\frac{2+c}{z}
	\end{pmatrix}
	\begin{pmatrix}
		\chi\\
		\overline\psi
	\end{pmatrix}
	,\label{eq:RS:fermion:action:Weyl}
\end{align}
where we use the two-component slash convention $\slashed{\overline{v}}=v_\mu \overline\sigma^\mu$, $\slashed{v}=v_\mu\sigma^\mu$. 

\subsubsection{Chiral boundary conditions}
\label{sec:xd:rs:chiral:bc}

The vector-like (Dirac) nature of 5D spinors is an immediate problem for model-building since the Standard Model is manifestly chiral and there appears to be no way to write down a chiral fermion without immediately introducing a partner fermion of opposite chirality and the same couplings. %\footnote{The same problem is found in $\mathcal N>1$ supersymmetric models.}. 
To get around this problem, we can require that only the zero modes of the 5D fermions---those which are identified with Standard Model states---to be chiral. We show that one chirality of zero modes can indeed be projected out, while the heavier Kaluza-Klein excitations  are vector-like but massive.

We can project out the zero modes of the `wrong-chirality' components of a bulk Dirac 5D fermion by imposing chiral boundary conditions that these states vanish on the branes. 
For left-chiral boundary conditions, $\psi=0$ on the branes, while for right-chiral boundary conditions, $\chi=0$ on the branes. These boundary conditions force the `wrong-chirality' zero mode to be identically zero everywhere. Thus we are guaranteed that both terms in (\ref{eq:boundaryterm}) vanish at $z=R,R'$ for either chirality.
Imposing these chiral boundary conditions is equivalent to the statement that the compactified extra dimension is an orbifold. 
This treatment of boundary conditions for interval compact spaces was first discussed from this viewpoint in \cite{Csaki:2003sh}.

\subsubsection{KK Decomposition of RS Fermions}
\label{app:XD:RS:KK:fermions}

Some care is necessary to dimensionally reduce the fermion equation of motion coming from (\ref{eq:RS:fermion:action:Weyl}): this is a matrix equation in Weyl spinor space with off-diagonal elements. Analogously to what we are used to in flat 4D space, one may `square' the warped 5D Dirac equation to obtain a scalar equation of motion that is diagonal. For convenience, define the conjugate Dirac operators
\begin{align}
	\mathcal D 
	&= 
%	\left(\frac Rz\right)^4 
%	\left(
	i \gamma^M \partial_M - i \frac 2z \gamma^5  - \frac cz
%	\right) 
	&
	\bar{\mathcal D}
	&= 
%	\left(\frac Rz\right)^4 
%	\left(
	i \gamma^M \partial_M - i \frac 2z \gamma^5  + \frac cz
%	\right) 
	\ .
\end{align}
One recognizes at $S = \int d^4x \, dz (R/z)^4 \bar\Psi \mathcal D \Psi$. Decompose the 5D Dirac fermion $\Psi(x,z)$ in the usual way,
\begin{align}
	\Psi(x,z) &= \frac{1}{\sqrt{R}} \sum_n^\infty \Psi^{(n)}(x) 
f^{(n)}(z) \ . 
\label{eq:app:xd:rs:psi:kk:decomp}
\end{align}
Our trick is to apply $\bar{\mathcal D}$ to the equation of motion $\mathcal D \Psi(x,z) = 0$. The combined `squared' operator $\bar{\mathcal D} \mathcal D$ is
\begin{align}
	\bar{\mathcal D} \mathcal D &=
	\begin{pmatrix}
	(\bar{\mathcal D} \mathcal D)_- & \\
	& (\bar{\mathcal D} \mathcal D)_+	
	\end{pmatrix}
	&
(\bar{\mathcal D} \mathcal D)_\pm &= 
\partial^2 - \partial_z^2 + \frac 4z \partial_z 
+ \frac{c^2 \pm c - 6}{z^2} \ .
\end{align}
Applying this to a \KK mode\footnote{%
We have implicitly applied to the entire \KK tower in (\ref{eq:app:xd:rs:psi:kk:decomp}) and then used the orthogonality of solutions.
} $f^{(n)}(z)$ gives an expression that is equivalent to the equation of motion. 
Multiplying by $(R/z)^4$---which is non-singular over the \RS interval $(R,R')$---for convenience and using $\partial^2 = -p^2 = -m_n^2$, we have
\begin{align}
	\left( \frac Rz \right)^4
	\left(
	-m_n^2 - \partial_z^2 + \frac{4}{z}\partial_z + \frac{c^2 \pm c - 6}{z^2}
	\right)f^{(n)}(z) &= 0 \ .
	\label{eq:xd:rs:scalar:eom}
\end{align}
This can now be solved to yield the \KK profiles $f^{(n)}(z)$.
One may confirm that this is a Sturm-Liouville equation of general form (\ref{eq:xd:rs:sturm:liouville:box}) with the following parameters:
\begin{align}
	p(z) &= \left( \frac Rz \right)^4
	&
	q(z) &= - \left( \frac Rz \right)^4 \frac{c^2 \pm c - 6}{z^2}
	&
	w(z) &= \left( \frac Rz \right)^4
	&
	\lambda &= m_n^2 \ .
\end{align}
From the Sturm-Liouville orthogonality relation (\ref{eq:xd:rs:sturm:liouville:orthogonality}), we find that the fermion profiles satisfy 
\begin{align}
	\int \frac{dz}{R} \left( \frac Rz \right)^4 f^{(n)}(z) f^{(m)}(z) &= \delta_{mn} \ .
	\label{eq:xd:rs:fermion:orthog}
\end{align}
Compare this to the orthonormality condition for the scalar field, (\ref{eq:XD:RS:scalar:orthonormal}).

The general solution of the equation of motion (\ref{eq:xd:rs:scalar:eom}) is then fixed to a unique solution using the normalization condition (\ref{eq:xd:rs:fermion:orthog}) and boundary conditions on the \UV and \IR branes.
For example, the profile of a bulk fermion in \RS,
\begin{align}
\Psi_c^{(0)}(x,z) = \frac{1}{\sqrt{R'}} 
\left(\frac zR \right)^2
\left(\frac{z}{R'}\right)^{-c}
\sqrt{
    \frac{1-2c}{1-(R/R')^{1-2c}}
    } 
P_L \Psi_c^{(0)}(x),
\label{eq:XD:RS:fermion:profile}
\end{align}
where $\Psi_c^{(0)}(x)$ is a canonically normalized 4D field and $P_L$ is the usual left-chiral projector. The term in the square root is a flavor factor that is often written as $f_c$.

\subsection{Gauge fields in RS}
\label{sec:XD:RS:bulk:gauge}

We now move on to the case of bulk gauge fields. See Section~2 of \cite{Sundrum:2005jf} for a brief, pedagogical discussion of gauge fields in a compact, flat extra dimension. We follow the approach of \cite{Randall:2001gb}, though we adapt it to follow the same type of derivation espoused above for the fermion propagator. 
The bulk action is
\begin{align}
S_5 &= \int d^4x dz\, \sqrt{g} \; \left[
-\frac 14 F_{MN}F^{MN} + (\text{brane}) + (\text{gauge fixing})
\right]
\end{align}

%\subsubsection{Inverting the quadratic term}

To derive the propagator, we would like to write the kinetic term in the form $A_M\mathcal{O}^{MN} A_N$ so that we may invert the quadratic differential operator $\mathcal{O}^{MN}$. This require judicious integration by parts including the $(R/z)$ factors from the metric and the measure, $\sqrt{g}$. The relevant integration is
\begin{align}
%\frac{R}{4z}F^{MN}F_{MN} &= 
%-\frac{R}{2} A^N\partial^M\left(\frac 1z \partial_M\right) A_N
%+ \frac{R}{2} A^N\partial^M\left(\frac 1z \partial_N\right) A_N
%+  \frac{R}{2} \partial^M\left(\frac 1z A^N \partial_{[M}A_{N]}\right),
%\label{RS:eq:bulk:gauge:kinetic}
%\\
\frac{-R}{4z}F^{MN}F_{MN} &= 
\frac 12 \left[
A^N\partial^M\left(\frac Rz \partial_M\right) A_N
- A^N\partial^M\left(\frac Rz \partial_N\right) A_N
-  \partial^M\left(\frac Rz A^N \partial_{[M}A_{N]}\right)
\right]  ,
\label{RS:eq:bulk:gauge:kinetic}
\intertext{%
where the last term integrates to a boundary term. Observe that this boundary term vanishes for both Dirichlet and Neumann boundary conditions so that it vanishes for $\mu\to\nu$ and 5th component scalar propagators. It does not vanish, however, for the case of vector--scalar mixing. For simplicity, we will drop the term here in anticipation that it will be removed by gauge fixing. With this caveat, the above integration becomes}
%\frac{R}{4z}F^{MN}F_{MN} &= 
%A_\mu 
%\left[
%\frac{R}{2z} \partial^2 \eta^{\mu\nu} 
%- \frac{R}{2}\partial_z\left(\frac 1z \partial_z\right) \eta^{\mu\nu}
%-\frac{R}{2z} \partial^\mu\partial^\nu
%\right]
%A_\nu 
%+
%A_5 \frac Rz \partial_z \partial^\mu A_\mu
%- A_5 \frac{R}{2z} \partial^2 A_5.
%\label{RS:eq:bulk:gauge:kinetic:nice} 
%\\
\frac{-R}{4z}F^{MN}F_{MN} &= 
A_\mu \frac{R}{2z}
\left[
 \partial^2 \eta^{\mu\nu} 
- z \partial_z\left(\frac 1z \partial_z\right) \eta^{\mu\nu}
- \partial^\mu\partial^\nu
\right]
A_\nu 
+
A_5 \frac Rz \partial_z \partial^\mu A_\mu
- A_5 \frac{R}{2z} \partial^2 A_5\ .
\label{RS:eq:bulk:gauge:kinetic:nice}
\end{align}
This is now in the desired form: we can read off the quadratic differential operators which encode the propagation of the 5D gauge bosons. Observe that we have a term that connects the 4D vector $A_\mu$ to the 4D scalar $A_5$. We prefer to work with these as separate fields. This term is removed by a judicious choice of gauge fixing.

\subsubsection{Gauge fixing}

We must now gauge fix to remove the gauge redundancy which otherwise appears as unphysical states in the propagator. Ideally we would like to pick a gauge where the scalar vanishes $A_5=0$ and the vector has a convenient gauge, say, Lorenz gauge $\partial_\mu A^\mu=0$. Unfortunately, these gauges are incompatible. Intuitively this is because we only have a single gauge fixing functional to work with in the path integral so that we are allowed to set at most one expression to vanish. Instead, motivated by the desire to cancel the vector--scalar mixing in (\ref{RS:eq:bulk:gauge:kinetic:nice}) and to recover the usual $R_\xi$ gauge in 4D, we choose a gauge fixing functional
\begin{align}
\mathcal L_\text{gauge fix} &=
-\left( \frac Rz \right) \frac{1}{2\xi}
\left[
\partial_\mu A^\mu - \xi z \partial_z \left(\frac 1z A_5 \right)
\right]^2
\end{align}
We have introduced a gauge fixing parameter $\xi$ which will play the role of the ordinary $R_\xi$ gauge fixing parameter in 4D. We can integrate by parts to convert this to the form $A_M\mathcal{O}^{MN}_\text{gauge fix} A_N$,
\begin{align}
\mathcal L_\text{gauge fix} &=
 A_\mu \frac{1}{2\xi} \frac{R}{z} \partial^\mu \partial_\nu A_\nu - A_5 \frac{R}{z} \partial_z\partial^\mu A_\mu + A_5 \frac{\xi}{2}\frac{R}{z} \partial_z \left[
 z \partial_z \left(\frac 1z A_5\right)
 \right].
\end{align}
Observe that the second term here cancels the unwanted mixing term in (\ref{RS:eq:bulk:gauge:kinetic:nice}). Summing this together with the gauge kinetic term gives a clean separation for the kinetic terms for the gauge vector and scalar:
\begin{align}
%\mathcal L_\text{gauge} + \mathcal L_\text{gauge fix} =&
%\phantom{+}A_\mu \left[
%\frac R{2z} \partial^2 \eta^{\mu\nu} 
%- \frac{R}{2} \partial_z \left(\frac 1z \partial_z \right) \eta^{\mu\nu}
%-\left(1-\frac 1\xi\right) \frac{R}{2z} \partial^\mu\partial^\nu \right] A_\nu\nonumber \\
%&+ A_5 \frac{R}{2z} \left[
%-\partial^2 + \xi
%\left(
%\frac{1}{z^2} - \frac{1}{z}\partial_z + \partial_z^2
%\right)
%\right] A_5\\
%\equiv & \phantom{+} A_\mu \mathcal{O}^{\mu\nu} A_\nu + A_5 \mathcal{O}_5 A_5.
%\label{RS:eq:quadratic:gauge:operators}
%
\mathcal L_\text{gauge} + \mathcal L_\text{gauge fix} =&
\phantom{+}A_\mu \frac{R}{2z}\left[
\eta^{\mu\nu} \partial^2 
-\left(1-\frac 1\xi\right) \partial^\mu\partial^\nu 
- \eta^{\mu\nu} z \partial_z \left(\frac 1z \partial_z \right) 
\right] A_\nu \nonumber 
\\
& 
- A_5 \frac{R}{2z} \partial^2 A_5
+ A_5 \frac{R \xi}{2z} \partial_z \left[
 z \partial_z \left(\frac 1z A_5\right)
 \right] 
%
%+ A_5 \frac{R}{2z} \left[
%-\partial^2 + \xi
%\left(
%\frac{1}{z^2} - \frac{1}{z}\partial_z + \partial_z^2
%\right)
%\right] A_5
\\
\equiv & \phantom{+} A_\mu \frac 12 \mathcal{O}^{\mu\nu} A_\nu + A_5 \frac 12 \mathcal{O}_5 A_5.
\label{RS:eq:quadratic:gauge:operators}
\end{align}
As above, now that we have the action written in terms of right-acting operators on the gauge fields. The 5D equations of motion are simply
\begin{align}
	\mathcal O^{\mu\nu} A_\nu &= 0
	&
	\mathcal O_5 A_5 &= 0 \ .
\end{align}
We now proceed to do a \KK reduction to determine the \KK mode properties. %, $A^{(n)}_\mu(x,z) = A^{(n)}_\mu(x)h^{(n)}(z)$. 

%\subsubsection{KK Reduction}
\subsubsection{KK Decomposition of RS Gauge Bosons}

Define the \KK reductions of the 5D $A_\mu$ and $A_5$ fields,
\begin{align}
	A_\mu(x,z) &= \frac{1}{\sqrt{R}} \sum_n A_\mu^{(n)}(x) h^{(n)}(z)
	&
	A_5(x,z) &= \frac{1}{\sqrt{R}} \sum_n A_5^{(n)}(x) h_5^{(n)}(z) \ .
\end{align}
For the $n^\text{th}$ \KK mode, we know $\partial^2 A_\mu^{(n)}(x) = - p^2 A_\mu^{(n)}(x) = - m_n^2 A_\mu^{(n)}(x)$, defining the mass of the KK mode, $m_n$. Taking the $\xi = 1$ gauge for simplicity, the equation of motion for the $n^\text{th}$ \KK mode is a differential equation for the \KK masses and profiles,
\begin{align}
\left( \frac Rz \right)
\left[
	- m_n^2 - z \partial_z \left(\frac 1z \partial_z \right)
	\right]h^{(n)}(z) &= 0 \ .
	\label{eq:xd:app:A4:eom}
\end{align}
This is Sturm-Liouville equation of generic form (\ref{eq:xd:rs:sturm:liouville:box}).
%\begin{align}
%	\partial_z \left[ p(z) \partial_z h(z) \right] + q(z) h(z) = - \lambda w(z) h(z) \ ,
%	\label{eq:xd:app:sturm}
%\end{align}
where in the case of $h^{(n)}$ we identify
\begin{align}
	p(z) &= \frac Rz 
	&
	q(z) &= 0 
	&
	w(z) &= \frac Rz
	&
	\lambda &= m_n^2 \ .
\end{align}
The Sturm-Liouville weight, $w(z)$, defines the orthonormality relation (\ref{eq:xd:rs:sturm:liouville:orthogonality}),
\begin{align}
%	\int h^{(n)}(z) h^{(m)}(z) w(z) dz = 
	\int \frac{dz}{R} \frac{R}{z} h^{(n)}(z) h^{(m)}(z) = \delta_{mn} \ .
\end{align}
Observe that the weight differs from that of a scalar field, (\ref{eq:XD:RS:scalar:orthonormal}), or a fermion (\ref{eq:xd:rs:fermion:orthog}).
The general solution for the $n^\text{th}$ \textsc{kk} mode profile of a bulk gauge field is
\begin{align}
h^{(n)}(z) = a z J_1(m_n z) + b z Y_1(m_n z),
\label{eq:xd:app:gauge:profile}
\end{align}
where $J_\alpha$ and $Y_\alpha$ are Bessel functions. The $n=0$  modes are constant.

\subsubsection{The $A_5$ scalar gauge boson}

We now turn to the $A_5$ piece of the 5D gauge field. The equation of motion for the $n^\text{th}$ \KK mode of the $A_5$ is
\begin{align}
	\partial_z\left[z \partial_z \left(\frac 1z h_5^{(n)}\right)\right]
	+ \frac{m_{n,5}^2}{\xi} h_5^{(n)} \ .
\end{align}
One could follow the same analysis as above, but it turns out that there is a convenient shortcut if we
compare this to (\ref{eq:xd:app:A4:eom}). Let us make the ansatz that
\begin{align}
	h_5^{(n)}(z) &= \frac{1}{m_n} \partial_z h^{(n)}(z) \ 
	&
	m_{n,5}^2 = m_n^2 \ .
\end{align}
One can confirm that this is a solution to the equation of motion by applying (\ref{eq:xd:app:A4:eom}) and choosing $\xi =1$ for consistency with our gauge choice for $A_\mu$. 

The results are general for any $\xi$, and indeed one recognizes that the equations of motion for the $A_\mu$ and $A_5$ \KK modes mirror that of massive gauge bosons and the Goldstone modes that they eat in a spontaneously broken gauge theory. This is manifest, for example, in the $\xi$-dependence in the $A_5$ \KK mass. This behavior is not a coincidence and is, indeed, predicted from the perspective of the extra dimension as a deconstruction with many copies of the gauge group spontaneously breaking into the symmetric combination, see Section~\ref{sec:deconstruction}. 
To be explicit: the $n^\text{th}$ \KK vector, $A_\mu^{(n)}$ with $n>0$, can be understood as the gauge boson of its very own copy of the gauge symmetry. This gauge boson, however, is massive because it picks up the \KK mass $m_n$. We know that massive gauge bosons have three polarizations and so expect that the longitudinal polarization came from eating a Goldstone boson. These Goldstone bosons are precisely the $A_5^{(n)}$ states. The symmetry that was broken spontaneously is the product of gauge symmetries in the deconstruction. In other words: a 5D gauge symmetry can be thought of as a product of 4D gauge symmetries: $G_\text{5D} \approx G_\text{4D}\times G_\text{4D} \times \cdots  G_\text{4D}$. The compactification of the extra dimension breaks $G_\text{4D}^n \to G_\text{4D}$ in such a way that the $A_5^{(n)}$s are the Goldstone bosons. In unitary gauge, $\xi \to \infty$, the $A^{(n)}_5$ fields decouple, but in general one must include them as internal states in calculations. 

For the Standard Model gauge fields one selects boundary conditions where the scalar zero mode $A_5^{(0)}$ vanishes. However, one can also realize the Higgs as the fifth component of a 5D gauge field whose 4D vector piece has no zero mode~\cite{Manton:1979kb,Hatanaka:1998yp,Hosotani:2004wv}.
The realization of this gauge-Higgs unification scenario in Randall-Sundrum is dual to composite Higgs models~\cite{Contino:2003ve, Agashe:2004rs, Hosotani:2005nz} in the sense described in Section~\ref{sec:xd:RS:meaning:of:5d}. With this application in mind, the zero mode profile of the $A_5$ is 
\begin{align}
h_5^{(0)}(z)
&= 	az + b z\log z \ .
\end{align}
The proportionality to $z$ means that such a field is peaked sharply on the \IR brane and is indeed phenomenologically suitable to be a Higgs.

\subsubsection{Gauge boson masses and profiles}

A \SM gauge field must have a zero mode (which is identified with the \SM state) so that it must have Neumann boundary conditions (\textsc{bc}). Using the formulae for derivatives of Bessel functions and (\ref{eq:xd:app:gauge:profile}), we find
\begin{align}
Y_0(m_n R) J_0(m_nR') = J_0(m_n R) Y_0(m_n R'),
\label{eq:XD:RS:gauge:KK:bc}
\end{align}
where $m_n$ is the mass of the $n^\text{th}$ \KK mode. 
We know that $m_n \sim n/R'$ and that $R \ll R'$. Thus $m_n R \approx 0$ for reasonable $n$. Now invoke two important properties of the $J_0$ and $Y_0$ Bessel functions:
\begin{enumerate}
\item $J_0(0) = 1$ and  $|J_0(x)|<1$.
\item $Y_0(x\to 0) \to -\infty$ and $|Y_0(x>y_1)| <1$  where $y_1$ is the first zero of $Y_0(x)$.
\end{enumerate}
From this we see that the left-hand side of (\ref{eq:XD:RS:gauge:KK:bc}) is very large and negative due to the $Y_0(m_n R)$ term while the right-hand side is a product of terms that are $\mathcal O(1)$ or less. This implies that $J_0(m_n R') \approx 0$. In other words, the \KK masses are given by the zeros of $J_0$.  The first zero is $x_1 = 2.405$ so that the first \KK gauge boson excitation has mass $m_n \approx 2.4/R'$. The solution for the  the $n^\text{th}$ \textsc{kk} mode profile of a \SM gauge field is thus 
\begin{align}
h^{(n)}(z) &= \mathcal N z \left[
Y_0(m_n R) J_1(m_n z)
-
J_0(m_n R) Y_1(m_n z)
\right].
\end{align}
The normalization is fixed by performing the $dz$ integral and requiring canonical normalization of the zero mode 4D kinetic term,
\begin{align}
\int d^4x \, dz \sqrt{g} F_{MN}F_{PQ}g^{MP}g^{NQ}
=
\int d^4x \, dz \; \frac Rz
F^{(0)}(x)_{\mu\nu}F^{(0)}(x)^{\mu\nu} \left[\frac{h^{(0)}(z)}{\sqrt{R}}\right]^2 +  \cdots.
\end{align}
This gives $\mathcal N^{-2} = \log R'/R$.

Finally, for the $W$ and $Z$ bosons, the Higgs \vev on the \IR brane changes the boundary conditions so that the zero mode profile is not flat. Heuristically it introduces a kink on the profile near the \IR brane. Since $M_Z \ll m_1$, we may treat this as a perturbation to $m_0 =0$ so that the $Z$ boson profile is~\cite{Csaki:2002gy} % see hep-ph/0203034
\begin{align}
h_Z^{(0)}(z) = \frac{1}{\sqrt{R\log R'/R}}\left[
1 -
\frac{M_Z^2}{4} \left(z^2 - 2z^2 \log \frac zR\right)
\right],
%\label{eq:XD:RS:Z:profile}
\end{align}
and similarly for the $W$.

%\begin{framed}
%\noindent\footnotesize \textbf{Caution with finite loops}. 
\subsection{Caution with finite loops}
One should be careful when calculating loop diagrams in theories with extra dimensions. When one calculates a finite loop, say a dipole operator, na\"ive application of effective field theory suggests taking only the lowest \KK mode and letting the 4D loop momentum go to $k\to\infty$. This, however, can lead to erroneous results since the loop integral runs over all momenta, including those in the fifth dimension. Only integrating over the 4D directions removes terms that scale like $k^2/M_\text{\KK}^2$ which would otherwise make an $\mathcal O(1)$ finite contribution.
This can appear as a dependence on the order in which one does the 4D loop integral versus \KK sum; this discrepancy has appeared in the \RS $gg\to h$ production calculations \cite{Carena:2012fk}.
One way to avoid this problem is to work in mixed position-momentum space \cite{Randall:2001gb}. This was used to calculate \RS constraints from $f\to f' \gamma$ \cite{Csaki:2010aj, Blanke:2012tv} and the muon magnetic moment in \cite{Beneke:2012ie}. These references include Feynman rules for performing mixed space calculations.
For a recent explanation of the subtleties of 5D dipoles and the resolution to puzzles in the previous literature, see~\cite{Agashe:2014jca}. In particular, Section 3 of that paper shows how to quickly estimate the size of 4D couplings from overlap integrals.
%\end{framed}

\section{The CCWZ Construction}
\label{sec:comp:CCWZ}

The general theory of Goldstone bosons is described in the papers by Callan, Coleman, Wess, and Zumino (\CCWZ) \cite{Coleman:1969sm, Callan:1969sn}. 
In this appendix we summarize the \CCWZ procedure and identify key aspects that are often referred to in the composite Higgs literature. 
See \S 19.5 -- 19.7 of \cite{Weinberg:1996kr}, \S 2.3 of \cite{Panico:2015jxa}, or the introductions of \cite{coserthesis, budinekthesis} for more pedagogical and explicit discussions, the relevant sections of \cite{Bando:1987br}, or \cite{Donoghue:1994fk, Burgess:1998ku} for more depth on how this procedure is applied to the chiral Lagrangian.
Since this discussion can be somewhat abstract, we shall include boxes relating each section to the example of chiral perturbation theory ($\chi$\textsc{pt}) from Section~\ref{sec:comp:chipt}.

The \CCWZ construction is a systematic way to write down the interactions of a theory in which a global symmetry $G$ spontaneously broken to $H$. The $G$ symmetry is thus nonlinearly realized and one can write a theory of Goldstone bosons, as we reviewed for $\chi$\textsc{pt} in Section~\ref{sec:comp:chipt}. \CCWZ goes beyond this by also providing the ingredients for how to couple non-Goldstone fields to such a theory.

%%%%%%
\subsection{Preliminaries}

Suppose a Lagrangian is invariant under a global symmetry $G$, but that $G$ is spontaneously broken to a subgroup $H\subset G$ by some order parameter $\psi_0$. We assume that $\psi_0$ is the \vev of some field, $\psi(x)$, that transforms as a linear representation of $G$,
\begin{align}
	\psi(x) \to g \psi(x) \ .
	\label{eq:comp:app:linear:rep:def}
\end{align}
The statement that $\psi_0$ spontaneously breaks $G \to G/H$ means that for any $h\in H$, $h\psi_0 = \psi_0$. %We shall assume that $\psi_0$ is the \vev of some field, $\langle\psi(x)\rangle = \psi_0$.

The spontaneous symmetry breaking pattern $G\to H$ implies the existence of $\text{dim}\, G -\text{dim}\, H$ Goldstone bosons that take values on a vacuum manifold. This manifold is the \textbf{coset space} $G/H$ (`$G$ mod $H$'). In particular, the \emph{left} coset space $G/H$ is an equivalence class of elements $g\in G$ modulo elements $h\in H$, $g\sim gh$. In other words, any element in $g$ is equivalent to another element $g'$ if there exists an $h$ such that $g'=gh$. 
%One may evocatively write an element of $G/H$ as $gH$. 
Note that in general $G/H$ is not a group.

\begin{framed}
\noindent \footnotesize
$\chi$\textsc{pt} \textbf{coset space}. For case of two flavors,  $G=\text{SU}(2)_\text{L}\times\text{SU}(2)_\text{R}$ and $H = \text{SU}(2)_\text{V}$, the diagonal subgroup.
Let us choose $\psi(x)$ to be a field transforming in the bifundamental of SU(2)$_\text{L}\times$SU(2)$_\text{R}$. This means that it can be  represented as a $2\times 2$ matrix where the linear transformation (\ref{eq:comp:app:linear:rep:def}) is represented by
\begin{align}
	\psi(x) \to g_L \psi(x) g_R^{-1} \ ,
	\label{eq:comp:app:SU2:bifundamental:transform}
\end{align}
for $g_{L,R}$ are matrices in the defining representation of $\text{SU(2)}_{L,R}$. One could also have written this as a matrix acting on a column vector, $\psi^i(x)\to g(g_L, g_R)_{ij} \psi^j(x)$, which more closely matches (\ref{eq:comp:app:linear:rep:def}). However, the bifundamental representation is convenient since the transformation is more intuitive.
\end{framed}

\subsection{Decomposition of the Algebra}

The generators of $G$ can be divided between two classes: $T^i$ which generate the unbroken group $H$, and $X^a$ which do not. 
This is called the \textbf{Cartan decomposition}. The generators satisfy the following commutation relations for some structure constants $f$:
\begin{align}
\left[T^i,T^j\right] &= i f^{ijk} T^k \\
\left[T^i,X^\alpha\right] &= i f^{i\alpha\beta} X^\beta \label{eq:comp:app:cartan:alg:2} \\
\left[X^\alpha, X^\beta\right] &= i f^{\alpha\beta k} T^k + i f^{\alpha\beta\gamma} X^\gamma \ .
\label{eq:comp:app:cartan:alg:3}
\end{align}
The non-trivial commutator (\ref{eq:comp:app:cartan:alg:2}) is derived by making use of the Cartan inner product for two generators,  $\langle A|B \rangle \equiv \text{Tr}(AB)$, and the fact that the $H$ and $G/H$ algebras are orthogonal, $\langle T | X \rangle = 0$. Using the cyclicity of the trace, one finds
\begin{align}
\langle T^i | [T^j,X^\alpha]\rangle
=
\langle T^i | T^j X^\alpha \rangle
-
\langle T^i | X^\alpha T^j \rangle
=
\langle [T^i,T^j] |  X^\alpha \rangle
=
if^{ijk}\langle T^k |  X^\alpha \rangle
=0 \ .
\end{align}
This proves that $[T^i,X^\alpha]$ is a series of only the $X$ generators. The commutation relation (\ref{eq:comp:app:cartan:alg:2}) can be interpreted as defining the action of the subgroup $H$ on a set of matrices $X$; it thus implies that the $X$s furnish a linear representation of $H$. 

If, additionally, there exists a parity transformation $P$ such that $P^2=1$ and $P([g_1,g_2])=[P(g_1),P(g_2)]$ and further such that $P(X) = -X$ and $P(T)=+T$, then one can further restrict
\begin{align}
[X^\alpha, X^\beta] = i f^{\alpha\beta k}T^k.
\end{align}
In this case, the coset $G/H$ is a \textbf{symmetric space}.

\begin{framed}
\noindent \footnotesize
$\chi$\textsc{pt} \textbf{algebra}. The generators of the SU(2) algebra are $\tau^i = \frac 12 \sigma^i$, for $i=1,2,3$. These satisfy
\begin{align}
	\left[ \tau_L^i , \tau_L^j \right] & = i\varepsilon^{ijk} \tau^k_L
	&
	\left[ \tau_R^i , \tau_R^j \right] & = i\varepsilon^{ijk} \tau^k_R
	&
	\left[ \tau_L^i , \tau_R^j \right] & = 0 \ .
\end{align}
We may re-organize these into generators of the broken symmetry, $X=\tau_A^i = \tau_L^i - \tau_R^i$, and the unbroken symmetry, $T=\tau_V^i = \tau_L^i + \tau_R^i$. For convenience we only use lowercase Roman indices. One may check that
\begin{align}
	\left[ \tau_V^i , \tau_V^j \right] & = i\varepsilon^{ijk} \tau^k_V
	&
	\left[ \tau_A^i , \tau_A^j \right] & = i\varepsilon^{ijk} \tau^k_V
	&
	\left[ \tau_A^i , \tau_V^j \right] & = i\varepsilon^{ijk} \tau^k_A \ .
\end{align}
We see that a special result of chiral symmetry breaking is that $[X,X]\sim T$ without any component in the broken algebra. This is because SU(2)$_\text{L}\times$SU(2)$_\text{R}$ is a symmetric space with parity transformation $P = \tau_L \leftrightarrow \tau_R$. The observation that the algebra of the broken generators does not close is the reason why SU(2)$_\text{A}$ is not a properly subgroup of of SU(2)$_\text{L}\times$SU(2)$_\text{R}$.
A general transformation is parameterized by pairs of 3-vectors, $(\mathbf{v},\mathbf{a})$. The action on the bifundamental field $\psi(x)$ is
\begin{align}
\psi(x) \to 
e^{i(\mathbf{v}+\mathbf{a})\cdot \mathbf\tau} 
\psi(x)
e^{-i(\mathbf{v}-\mathbf{a})\cdot \mathbf\tau} \ .
\label{eq:comp:app:bifundamental:AV:trans}
\end{align}
\end{framed}

\subsection{Decomposition of the Group}

The distinct $G$ elements $g h_1, g h_2, g h_3, \ldots \in G$ are all identified with the same element of $G/H$. For each element of $G/H$, it is useful to pick a representative element of $G$, which we denote $\hat g$. 
Then any group element $g \in G$ may be written in the form $g = \hat g h$, for some $h \in H$. Further, for a compact and connected group, one may further write each of $\hat g$ and $h$ as an exponentiation of elements of the algebra, so that for any $g\in G$,
\begin{align}
	g &= e^{i\xi^\alpha X^\alpha} e^{i u^i T^i} \equiv \hat g(\xi) h(u) \ .
	\label{eq:comp:app:CCWZ:gdecomp}
\end{align}

\subsection{Decomposition of the Linear Representation}

Suppose we have a field $\psi(x)$ which transforms as a non-trivial linear representation of the group $G$. This field contains the Goldstone degrees of freedom associated with the broken generator directions in field space; that is, the field directions with a flat potential that transform $\psi_0$. Let us define an object $\gamma(x)\in G$ that factorizes $\psi(x)$,
\begin{align}
	\psi(x) \equiv \gamma(x) \tilde{\psi}(x) \ ,
	\label{eq:comp:CCWZ:non:linear:from:linear:psi}
\end{align}
in such a way that $\tilde\psi(x)$ contains no Goldstone degrees of freedom. Indeed, if one were to ignore all `radial' (massive) excitations of $\psi(x)$, one may pick this to be the \vev, $\tilde\psi(x) = \psi_0$, so that $\gamma(x)$ is simply the transformation from $\psi_0 \to \psi(x)$.
%,
By the invariance of $\psi_0$ under $H$ transformations, $\gamma(x)$ is only defined up to right multiplication by any $h\in H$. In other words, we may identify $\gamma(x)$ with the representative element $\hat \gamma(x)$ which is chosen to be the exponentiation of only broken generators, analogously to $\hat g$ above. We may now drop the hat on $\hat \gamma$ for notational clarity. Let us suggestively call the transformation parameter $\pi(x)$,
\begin{align}
	\gamma(x) \equiv \gamma\left[\pi \right] = e^{i\pi^\alpha(x)  X^\alpha}.
	\label{eq:comp:CCWZ:non:linear:from:linear:gam}
\end{align}
The $\pi^a(x)$ are to be identified with the Goldstone bosons. We leave it dimensionless, remembering that the pion field with canonical mass dimension can be restored by taking $\pi^a(x) \to \pi^a(x)_\text{can}/f$.

Suppose the Lagrangian of the theory with respect to the linearly represented field $\psi(x)$ is written in terms of $\psi(x)$ and $\partial \psi(x)$. $G$-invariants formed out of only $\psi(x)$ don't contain the Goldstone fields, while those made of $\partial \psi(x)$ do. $\partial \psi(x)$ can be written in terms of $\psi_0$ and the Goldstone fields using (\ref{eq:comp:CCWZ:non:linear:from:linear:psi}) and (\ref{eq:comp:CCWZ:non:linear:from:linear:gam}),
\begin{align}
	\partial_\mu \psi(x) &= \gamma 
	\left[
		\partial_\mu
		+ \gamma^{-1}\left(\partial_\mu \gamma \right) 
	\right] \tilde\psi,
	\label{eq:comp:CCWZ:derivative:psi}
\end{align}
where we've suppressed the $x$ dependence of $\gamma$. Without loss of generality, we can write $\gamma^{-1}\partial_\mu \gamma = \gamma^\dagger \partial_\mu \gamma$ in terms of the broken and unbroken generators,
\begin{align}
\gamma^{-1}\partial_\mu \gamma &= i D^\alpha_\mu X^\alpha + i E^i_\mu T^i
\label{eq:gaminvdgam:gen}
\\
D^\alpha_\mu &= D^{\alpha\beta}(\pi) \partial_\mu \pi^\beta
\\
E^i_\mu &= E^{i\beta}(\pi)\partial_\mu \pi^\beta.
\end{align}
In the case of a symmetric space, the parity operator, $P$, gives a short-cut to express the $D^\alpha_\mu(x)$ and $E^i_\mu(x)$. $P$ takes $X\to - X$ and $T\to T$. From this we see that it takes, $\gamma[\pi] \to \gamma[-\pi] = \gamma^{-1}[\pi]$ and thus $\gamma^{-1}\partial_\mu \gamma \to \gamma\partial_\mu \gamma^{-1}$. We may thus take the sum and difference of (\ref{eq:gaminvdgam:gen}) with its parity conjugate to derive
\begin{align}
	iD_\mu^\alpha X^{\alpha} 
	&= \frac{1}{2}\left(
	\gamma^{-1}\partial_\mu \gamma - \gamma\partial_\mu \gamma^{-1}
	\right)
	&
	iE_\mu^i T^{i} 
	&= \frac{1}{2}\left(
	\gamma^{-1}\partial_\mu \gamma + \gamma\partial_\mu \gamma^{-1}
	\right) \ .
	\label{eq:app:comp:D:and:E:sym}
\end{align}

\begin{framed}
\noindent \footnotesize
\textbf{General decomposition of $\gamma^{-1}\partial\gamma$ into $D$ and $E$}.
The general decomposition (\ref{eq:gaminvdgam:gen}) uses the identity
\begin{align}
\partial_\mu e^{i\pi\cdot X} = i\partial_\mu \pi^\alpha \int_0^1 ds\, e^{i(1-s)\pi\cdot X}
 X^\alpha e^{is\pi\cdot X}
 \label{eq:app:comp:matrix:derivative:trick}
\end{align}
and the Baker--Campbell--Hausdorff (\textsc{bch}) relation,
\begin{align}
	e^A B e^{-A} = B + [A,B] + \frac{1}{2!}\left[A,[A,B]\right] + \cdots \ ,
	\label{eq:app:comp:BCH}
\end{align}
 to show that you end up with an expansion in the $X$s and $T$s:
\begin{align}
	\gamma^{-1}\partial_\mu \gamma &= 
	i\partial_\mu\pi^{\alpha}(x)\, 
	e^{-i\pi(x)\cdot X}
	\int_0^1 ds\, e^{i(1-s) \pi(x)\cdot X}
	X^\alpha
	e^{i s \pi(x)\cdot X}
	\\
	&= i\partial_\mu\pi^{\alpha}(x)\, 
	\int_0^1 ds\, e^{-i s \pi(x)\cdot X}
	X^\alpha
	e^{i s \pi(x)\cdot X}
	\\
	&= i\partial_\mu\pi^{\alpha}(x)\, 
	\int_0^1 ds\,
	X^\alpha 
%	- is\pi^\beta(x)[X^\beta,X^\alpha] 
%	- \frac{s^2}{2!}\pi^\beta(x) \pi^\delta(x)\left[X^\delta,[X^\beta,X^\alpha]\right]+ \cdots \ .
	- is[\pi(x)\cdot X,X^\alpha] 
	- \frac{s^2}{2!}\Big[\pi(x)\cdot X,[\pi(x)\cdot X,X^\alpha]\Big]+ \cdots \ .
\end{align}
From this last line we invoke the algebra with respect to the Cartan decomposition, (\ref{eq:comp:app:cartan:alg:3}), which tells us that $\gamma^{-1}\partial_\mu \gamma$ is indeed composed of a series of terms in both the broken and unbroken algebras. See \S 2.8 of \cite{budinekthesis} for a compact expression. These identities are derived in textbooks on the representations of Lie groups.
\end{framed}

\begin{framed}
\noindent \footnotesize
$\chi$\textsc{pt} \textbf{linear representation}. We use the field $\psi(x)$ transforming in the bifundamental of SU(2)$_\text{L}\times$SU(2)$_\text{R}$, (\ref{eq:comp:app:SU2:bifundamental:transform}). The factorization of the linearly realized field $\psi(x)$ into a Goldstone piece and a `radial' piece, (\ref{eq:comp:CCWZ:non:linear:from:linear:psi}), is
\begin{align}
	\gamma(x) \tilde\psi(x) = 
	e^{i\mathbf{\pi}(x)\cdot \mathbf{\tau}}
	\begin{pmatrix}
		r(x) & 0 \\
		0 & r(x)
	\end{pmatrix}
	e^{i\mathbf{\pi}(x)\cdot \mathbf{\tau}} \ ,
	\label{eq:comp:app:SU2:gamma:tilde:psi}
\end{align}
where we represent $\tilde\psi(x)$ as the unit $2\times 2$ matrix times a scalar function, $r(x) \, \mathbbm{1}_{2\times 2}$, which represents the radial (massive) degrees of freedom. We may replace $\tilde \psi(x)$ with the \vev, $\psi_0 = \langle r(x)\rangle \, \mathbbm{1}_{2\times 2} = f  \mathbbm{1}_{2\times 2}$. Observe that we have explicitly written the left- and right-acting parts of the transformation in the broken (axial) direction so that this is simply (\ref{eq:comp:app:bifundamental:AV:trans}) with $\mathbf a = \mathbf \pi(x)$. 
Taking 
$\tilde \psi(x) \to \psi_0$, 
and peeling off the $\psi_0$ and corresponding $f$,
one obtains $\gamma(x) = \text{exp}\left(2i\mathbf\pi(x)\cdot \tau\right)$. Since this is a symmetric space, (\ref{eq:app:comp:D:and:E:sym}) straightforwardly gives an expression for the $D$ and $E$.
\end{framed}

\subsection{Transformation of the Goldstones}

\subsubsection{Transformation under a general group element}

We would like to see how the $\pi^a(x)$ transform under the global group $G$. We can derive this implicitly from the transformation of the linear field $\psi(x) \to g \psi(x)$ using
 (\ref{eq:comp:CCWZ:non:linear:from:linear:psi}) and (\ref{eq:comp:CCWZ:non:linear:from:linear:gam}). The decomposition of a general group element $g$ in (\ref{eq:comp:app:CCWZ:gdecomp}) tells us that there exist $\pi'$ and $u'$ such that
\begin{align}
g\psi(x) = g e^{i\pi\cdot X} \tilde\psi(x) \equiv e^{i\pi'\cdot X} e^{iu'\cdot T}\tilde\psi(x),\label{app:geipiX}
\end{align}
where the primed fields are, in general, \emph{nonlinear} functions of $g$ and $\pi(x)$, that is $\pi' = \pi'(g,\pi)$ and $u' = u'(g,\pi)$.
%%%
We may write this as 
%$g\gamma[\pi] \tilde \psi(x) = \gamma[\pi']h(u')\tilde\psi(x)$ so that
%It is cleaner to write this after peeling off the $\tilde\psi(x)$, 
\begin{align}
%	g \gamma(\pi) = \hat g(\pi') h(u').
%	g \gamma[\pi] = \gamma[\pi'] h(u').
	g\gamma[\pi] \tilde \psi(x) = \gamma[\pi']h[u']\tilde\psi(x) \ , 
	\label{eq:comp:CCWZ:gen:transform}
\end{align}
from which we may interpret an induced $H$-transformation on the Goldstone-less field $\tilde\psi(x)$
\begin{align}
	\tilde\psi(x) \to \tilde\psi'(x) =  h[u'] \tilde \psi(x) \ ,
	\label{eq:comp:CCWZ:linear:field:in:H}
\end{align}
and an accompanying implicitly defined transformation of the Goldstones, $\pi(x)\to \pi'(x)$. Both the $\tilde\psi(x)$ and $\pi(x)$ transformations have a messy dependence on $\pi$ and $g$. We can now appreciate what the non-linear\footnote{%
The meaning of `non-linear' is more clear when contrasted with the case of a linear transformation, for which
\begin{align}
	\gamma[\pi] \to g_\text{lin} \gamma[\pi] g^{-1}_\text{lin} = \gamma[R_{ab}\pi^a] \ ,
\end{align}
for some linear representation $R$ of the algebra of a linearly realized group $G_\text{lin}$.} realization has bought us: we now have a transformation rule for the Goldstone-less field $\tilde\psi(x)$ which realizes the full symmetry group $G$ as a non-linear representation of the unbroken symmetry group $H$. If $\tilde\psi(x) = \psi_0$ is the order parameter for this symmetry breaking, then it is $H$-invariant. However, the decomposition $\psi(x) = \gamma[\pi] \tilde\psi(x)$ holds for general fields and the transformation law (\ref{eq:comp:CCWZ:linear:field:in:H}) is thus way to write the transformation law of a field $\tilde \psi$ without reference to its linearization $\psi$. In low energy \QCD, for example, this can be used to describe the interactions of nucleons with pions.

\subsubsection{Transformation under $H$}

The story is different for a transformation under the unbroken group, $H$. In this case, the Goldstones transform linearly with respect to the algebra of $H$. This serves as a useful counterpoint to the non-linear representation of general $G$ transformations above. To see this, let us transform the field $\psi(x)$, a linear representation of $G$, by an element $h\in H$. We again make use of the decomposition (\ref{eq:comp:CCWZ:non:linear:from:linear:psi}),
\begin{align}
	h\psi(x) = h \, \gamma[\pi]\tilde\psi(x) = h\, \gamma[\pi]\,  h^{-1} h \, \tilde\psi(x) \ , 
\end{align}
where in the last step we have inserted $\mathbbm 1 = h^{-1} h$. Observe that this is of the form (\ref{eq:comp:CCWZ:gen:transform}) with 
\begin{align}
	\tilde\psi(x) & \to \tilde\psi'(x) = h \tilde\psi(x) 
	&
	\gamma[\pi] &\to \gamma[\pi'] = h \gamma[\pi] h^{-1} \ .
	\label{eq:comp:CCWZ:transform:unbroken:H}
\end{align}
The $\pi$ transformation, in particular, is a \emph{linear} representation of $H$. This can be seen by invoking the algebra of the Cartan decomposition, (\ref{eq:comp:app:cartan:alg:3}), which states that the commutator of a broken and unbroken generator is proportional to the broken generator, $[T,X]\sim T$. Using this relation in the \textsc{bch} formula (\ref{eq:app:comp:BCH}) shows that
\begin{align}
	h X^a h^{-1} = R_{ab} X^b \in G/H \ .
	\label{eq:comp:app:CCWZ:hXh:transform}
\end{align}
From this it is clear that (\ref{eq:comp:CCWZ:transform:unbroken:H}) is a linear representation of the unbroken group $H$ acting on the Goldstone fields that live on the space of broken generators\footnote{To make this more clear, one may expand $\gamma[\pi]$ as a power series and insert $\mathbbm{1} = h^{-1} h$ so that $X^n = Xh^{-1} h X \cdots h^{-1} hX$, from which point one may use (\ref{eq:comp:app:CCWZ:hXh:transform}) to prove the linearity of (\ref{eq:comp:CCWZ:transform:unbroken:H}).}.

\begin{framed}
\noindent \footnotesize
$\chi$\textsc{pt} \textbf{pion transformations}. Let us demonstrate this for SU(2)$_\text{L}\times$SU(2)$_\text{R} \to $SU(2)$_\text{V}$. We showed above that the (un-)broken generators are parameterized by 3-vectors $\mathbf{a}$ ($\mathbf{v}$) such that the transformation of a bifundamental field $\psi(x)$ by $(\mathbf{v}, \mathbf{a})$ is (\ref{eq:comp:app:bifundamental:AV:trans}). Further, 
the decomposition $\psi(x) = \gamma[\pi] \tilde\psi(x)$ corresponds to (\ref{eq:comp:app:SU2:gamma:tilde:psi}). The transformation $\psi(x)\to g\psi(x)$ is thus identified with
\begin{align}
g_{\mathbf{v},\mathbf{a}}\gamma[\pi]\tilde\psi(x) &= 
	e^{i(\mathbf{v}+\mathbf{a})\cdot\mathbf\tau}
	e^{i\mathbf\pi\cdot\mathbf\tau}
	\begin{pmatrix}
		r(x) & 0 \\
		0 & r(x)
	\end{pmatrix}
	e^{i\mathbf\pi\cdot\mathbf\tau}
	e^{-i(\mathbf{v}-\mathbf{a})\cdot\mathbf\tau} \ .
\end{align}
When we specialize to the case of a purely vectorlike (unbroken) transformation, $\mathbf a = 0$, we may insert factors of $\mathbbm 1 = e^{-i\mathbf{v}\cdot \tau} e^{i\mathbf{v}\cdot \tau}$ to see that the $\pi$ transforms linearly,
\begin{align}
h_\mathbf{v}\gamma[\pi]\tilde\psi(x) &= 
\left[
	e^{i\mathbf{v}\cdot\mathbf\tau}
	e^{i\mathbf\pi\cdot\mathbf\tau}
	e^{-i\mathbf{v}\cdot\mathbf\tau}
	\right]
	\;
	e^{i\mathbf{v}\cdot\mathbf\tau}
	\begin{pmatrix}
		r(x) & 0 \\
		0 & r(x)
	\end{pmatrix}
	e^{-i\mathbf{v}\cdot\mathbf\tau}
	\;
	\left[
	e^{i\mathbf{v}\cdot\mathbf\tau}
	e^{i\mathbf\pi\cdot\mathbf\tau}
	e^{-i\mathbf{v}\cdot\mathbf\tau} \right] \ ,
\end{align}
where each bracketed term can be written as $[\cdots] = e^{i\pi'\cdot \tau}$, and the matrix in the center is unchanged. By comparison, for a general transformation $g_{\mathbf{v},\mathbf{a}}$ the best we can do is to insert $\mathbbm 1 = h_{\tilde{\mathbf{v}}}^{-1} h_{\tilde{\mathbf{v}}}$ for some $\tilde{\mathbf{v}}(\pi, g)$ such that it implicitly defines $\pi'$ by
\begin{align}
g_{\mathbf{v},\mathbf{a}}\gamma[\pi]\tilde\psi(x) &= 
	e^{i(\mathbf{v}+\mathbf{a})\cdot\mathbf\tau}
	e^{i\mathbf\pi\cdot\mathbf\tau}
	e^{-i\tilde{\mathbf{v}}\cdot\mathbf\tau}
	e^{i\tilde{\mathbf{v}}\cdot\mathbf\tau}
	\begin{pmatrix}
		r(x) & 0 \\
		0 & r(x)
	\end{pmatrix}
	e^{-i\tilde{\mathbf{v}}\cdot\mathbf\tau}
	e^{i\tilde{\mathbf{v}}\cdot\mathbf\tau}
	e^{i\mathbf\pi\cdot\mathbf\tau}
	e^{-i(\mathbf{v}-\mathbf{a})\cdot\mathbf\tau} \ 
	\\
	&\equiv
	e^{i\mathbf\pi'\cdot\mathbf\tau}
	\begin{pmatrix}
		r(x) & 0 \\
		0 & r(x)
	\end{pmatrix}
	e^{i\mathbf\pi'\cdot\mathbf\tau} \ .
\end{align}
Here $\tilde{\mathbf{v}}(\pi,g)$ is \emph{defined} such that this relation is true. Thus our general transformation is realized non-linearly, where $\pi \to \pi'$ is now complicated and $h_{\tilde{\mathbf{v}}}$ is the $H$ transformation that non-linearly realizes $G$.
\end{framed}

\subsection{From Linear to Non-Linear}\label{sec:app:comp:CCWZ:UV:to:IR}

From a linear \UV theory, we have identified the Goldstone fields and can integrate out the massive `radial' modes to obtain a low-energy Lagrangian by taking $\tilde\psi(x)\to \psi_0$. Effective field theory tells us that this \UV theory wasn't necessary to construct the low energy theory of Goldstone bosons. So once we have a theory of Goldstone interactions, we may remain agnostic about the specific \UV completion of the theory\footnote{%
You can also use this in the opposite direction: if you need an aide to write down non-linear interactions of Goldstone bosons, you can always construct a linear \UV theory and decouple the radial excitations. 
}. Prior to the discovery of the Higgs boson---a linear \UV completion of the theory of the Goldstone bosons eaten by $W^\pm$ and $Z$---the reason why experiments like \LEP could make precision measurements of the \SM without knowing the details of the Higgs is simply that the precision measurements asked precise questions about the non-linear sigma model (\NLSM) of Goldstones that were insensitive to the particular \UV completion, linear or otherwise.

There are also reasons why one might be interested in keeping around a field like $\tilde\psi(x)$ in the effective theory. These radial fields can be used to introduce, say, nucleon excitations in the chiral Lagrangian. 
The radial modes are identified with excitations along the \vev direction $\psi_0$. %So let us define the radial field $\tilde\psi(x) = r(x) \psi_0$ with a \vev $\langle r(x)\rangle = 1$. 
From (\ref{app:geipiX}), we see that the radial field transforms as 
%$\psi_r \to h(u') \psi_r$. 
\begin{align}
	\tilde\psi \to h\left[ u'(g,\pi) \right] \tilde\psi \ .\label{eq:comp:CCWZ:induced:H:rep}
\end{align}
Thus in order to build $G$-invariants out of the radial fields $\tilde\psi$, it is sufficient to construct $H$ invariants. Said differently, the decomposition $\psi(x) = \gamma[\pi] \tilde\psi(x)$ converts  $G$-linear representations, $\psi$, into non-linear realizations of $G$, $\pi$ and $\tilde\psi$. %$H$-linear representations $\tilde\psi$ that non-linearly realize the full $G$ symmetry.

\subsection{A Low-Energy Lagrangian without the UV}

Let us now return to (\ref{eq:comp:CCWZ:derivative:psi}) since we know from the Goldstone shift symmetry that the Goldstones only appear in derivative interactions. The object $g^{-1}\partial g$, where $g=\gamma$ in (\ref{eq:comp:CCWZ:derivative:psi}), is called the \textbf{Maurer-Cartan form}, it takes an element of the group $g\in G$, differentiates it---pulling out the Lie algebra element based at $g$---and pulls that generator back to the group identity so that one can compare elements of the algebra on the same tangent space.  

The expansion of the Maurer-Cartan form into broken and unbroken generators is given in (\ref{eq:gaminvdgam:gen}). Differentiating the transformation rule (\ref{eq:comp:CCWZ:gen:transform}), 
\begin{align}
%	g \partial_\mu\gamma(\pi) = \left[\partial_\mu\gamma(\pi')\right] h + \gamma(\pi') \partial_\mu h.
	g \partial_\mu\gamma[\pi]\tilde\psi + \cancel{g \gamma[\pi] \partial_\mu\tilde\psi} = \left(\partial_\mu\gamma[\pi']\right) h \tilde \psi + \gamma[\pi'] \left( \partial_\mu h\right) \tilde\psi + \cancel{\gamma[\pi'] h \partial_\mu\tilde\psi} \ .
\end{align}
We have crossed out terms on each side which are identical 
%since $g\gamma[\pi] = \gamma[\pi'] h[u']$ 
by (\ref{eq:comp:CCWZ:gen:transform}).
Now peel off the common factor of $\tilde\gamma$ on each term and multiply each side of this equation from the left by 
$\gamma^{-1}[\pi] g^{-1} = h^{-1}[u'] \gamma^{-1}[\pi']$ to find,
%the respective side in the inverse of (\ref{eq:comp:CCWZ:gen:transform}),
\begin{align}
	\gamma^{-1}[\pi] \partial_\mu\gamma[\pi] = h^{-1} \gamma^{-1}[\pi']\Big( \big(\partial_\mu\gamma[\pi']\big) h + \gamma[\pi'] \partial_\mu h\Big).
\end{align}
Comparing this to (\ref{eq:gaminvdgam:gen}), we find 
\begin{align}
	i D_\mu^\alpha[\pi] X^\alpha + iE^i_\mu[\pi] T^i
	= ih^{-1} D^\alpha_\mu[\pi'] X^\alpha h^{-1}
	+ i h\left(E_\mu^i[\pi']T^i + i\partial_\mu\right)h^{-1}.
\end{align}
In other words, the objects $D$ and $E$ defined in (\ref{eq:gaminvdgam:gen}) transform under $g\in G$ as
\begin{align}
	D_\mu^\alpha &\to h D_\mu^\alpha h^{-1}
	\\
	E^i_\mu &\to h E^i_\mu h^{-1} - i h \partial_\mu h^{-1},
\end{align}
where $h = h\left[u'(\pi,g)\right]$ as in (\ref{eq:comp:CCWZ:gen:transform}). This should look very familiar: $D$ transforms linearly and $E$ transforms like a gauge field. Both transform under $G$ through representations of $H$ rather than the whole group $G$. This realizes the observation in Section~\ref{sec:app:comp:CCWZ:UV:to:IR}: to write Lagrangians for nonlinear realizations of $G/H$, we need to construct invariants with respect to only $H$. The linear object $D$ can indeed be used to construct a simple lowest-order Lagrangian,
\begin{align}
	\mathcal L = \frac{f^2}{4}\text{Tr}(D_\mu D^\mu),
\end{align}
where we've introduced the symmetry breaking scale $f$ to preserve dimensionality. This should be compared to (\ref{eq:comp:chiral:pert}), our derivative expansion for in chiral perturbation theory. 
%% ADDED IN BOX BELOW

\begin{framed}
\noindent \footnotesize
$\chi$\textsc{pt} \textbf{vs}.\ \textsc{ccwz}. See \S 2.8 of \cite{budinekthesis} for an explicit calculation showing that the \textsc{ccwz} and chiral perturbation theory leading-order effective Lagrangians are indeed the same.
\end{framed}

%% SEE below 2.3.33 of Panico to see why this is better than \ref{eq:comp:chiral:pert}, the usual derivative expansion

What about the curious object $E_\mu$? This appears to transform as the gauge field of a local symmetry. The locality of this symmetry is inherited from the $x$-dependence of the Goldstone fields $\pi(x)$ and is unsurprising since the coset identification $g\sim gh$ is local. $E_\mu$ is thus a `gauge potential' with respect to the unbroken symmetry $H$. It encodes the fact that while all symmetries are global, the $H$ symmetry secretly is a subgroup of the larger, spontaneously broken $G$ symmetry. Indeed, differentiating (\ref{eq:comp:CCWZ:induced:H:rep})---recalling that $h(u')$ depends on $x$ through its implicit dependence on $\pi(x)$---shows that derivatives of the non-Goldstone fields transform inhomogeneously under $G$. Promoting the partial derivative to a covariant derivative, $\mathcal D_\mu = \partial_\mu \to \partial_\mu + i  E_\mu$, ensures that $\mathcal D_\mu \tilde\psi(x)$ transforms homogeneously under $H$.

\begin{framed}
\noindent \footnotesize
\textbf{When did $H$ become gauged?} The appearance of a covariant derivative and a gauge symmetry may seem surprising in a system where \emph{global} symmetry $G$ is spontaneously broken to a subgroup $H$. The appearance of a local symmetry, however, is not surprising since the resulting coset space $G/H$ precisely describes a gauge redundancy. Mathematically, the description of a `gauged' symmetry is identical to that of a spontaneously broken global symmetry.
See, for example, \cite{Burgess:1998ku} or chapter 7 of~\cite{Banks:2008fk}.
For the mathematically inclined, details of the geometric structure of these theories are presented in \cite{Coquereaux:1986ua} and \cite{Yastremiz:1991jn}.
\end{framed}

The punchline is that one can construct a Goldstone boson Lagrangian which is invariant under the full, nonlinearly realized group $G$, by constructing an $H$-invariant Lagrangian out of $D_\mu$. One can further introduce non-Goldstone fields $\tilde\psi$ (not necessarily related to the linear field that gets a \vev) so long as one uses the appropriate $H$ covariant derivative, $\mathcal D_\mu$, with the corresponding `gauge field' $E_\mu$. In this way one may include, for example, `nucleon' excitations to the effective theory.

The description above is based on a `standard realization' of the nonlinearly realized symmetry, (\ref{app:geipiX}). One of the main results of the \CCWZ papers was the observation that every non-linear realization can be brought to this standard realization \cite{Coleman:1969sm, Callan:1969sn}. Physically, this means that no matter how one imposes the $G/H$ restriction, the $S$-matrix elements for the low-energy dynamics will be identical. Explicit examples of this are presented in chapter \textsc{iv} of \cite{Donoghue:1994fk}

\begin{framed}
\noindent \footnotesize
$\chi$\textsc{pt} \textbf{including nucleons}. Explicit examples of how one may invoke the \textsc{ccwz} formalism to extend the effective theory to include heavy particles can be found in chapter \textsc{iv}-7 of \cite{Donoghue:1994fk}, \S 2.3 of \cite{coserthesis}, or \S 2.3 of \cite{Bando:1987br}.
\end{framed}

\small
\bibliographystyle{utphys}
%\bibliography{ESHEP_Csaki, references, FlipSUSY, Seibergology,  RPV, GFDM, contino}
\bibliography{ESHEP_Csaki}

\end{document}